\documentclass[%
 reprint,
superscriptaddress,
 amsmath,amssymb,
 aps,
]{revtex4-2}

\usepackage{graphicx}% Include figure files
\usepackage{dcolumn}% Align table columns on decimal point
\usepackage{bm}% bold math
\usepackage{color}
\usepackage[caption=false]{subfig}
\begin{document}

\preprint{APS/123-QED}

\title{Transport Barriers in Magnetized Plasmas- General Theory with Dynamical Constraints}

\author{M. Kotschenreuther}
\email[Email: ]{mtk@austin.utexas.edu}
\affiliation{University of Texas, Austin, Texas, USA}

\author{X. Liu}%
\affiliation{University of Texas, Austin, Texas, USA}%

\author{S. M. Mahajan}%
\affiliation{University of Texas, Austin, Texas, USA}%

%\author{M. Zarnstorff}%
%\affiliation{Princeton Plasma Physics Laboratory, Princeton, New Jersey, USA}%

\author{D. R. Hatch}%
\affiliation{University of Texas, Austin, Texas, USA}%

\author{G. Merlo}%
\affiliation{University of Texas, Austin, Texas, USA}%

\date{\today}% It is always \today, today,

\begin{abstract}
A fundamental dynamical constraint - that fluctuation induced charge-weighted particle flux must vanish- can prevent instabilities from accessing the free energy in the strong gradients characteristic of Transport Barriers (TBs). Density gradients, when larger than a certain threshold, lead to a violation of the constraint and emerge as a stabilizing force.  This mechanism, then, broadens the class of  configurations (in magnetized plasmas) where these high confinement  states can be formed and sustained. The need for velocity shear, the conventional agent for TB formation, is obviated. The most important ramifications of the constraint is to permit a charting out of the domains conducive to TB formation and hence to optimally confined fusion worthy states; the detailed investigation is conducted through new analytic methods and extensive gyrokinetic simulations. 

\begin{description}
\item[DOI]
???
\end{description}
\end{abstract}

\maketitle

\section{Introduction}

This paper is an investigation into the fundamental physics of Transport Barriers (TBs) in magnetically confined plasmas ~\cite{synakowski_98}, \cite{connor2004}, \cite{garbet_04},~\cite{wagner_07},~\cite{ida2018}. Over a distance of centimeters, experimental TBs routinely sustain temperature differences exceeding those from the core to edge of the sun. Crucially,  the rates of heat loss in these enormous gradients are surprisingly small. This is the spectacular property that is relied upon in many proposals to produce fusion gain in terrestrial plasmas.  \emph {By conducting  an enquiry into the physical mechanisms responsible for the existence of such a desirable but  ``unwarranted'' thermodynamic state, we plan to chart out parameter regimes that are most hospitable to TBs with highest confinement.} Such regimes, and paths to accessing them, will be described in the last section of the paper. 

In the conventional gyrokinetic stability analysis, the TB formation is attributed to the suppression of turbulent transport associated with the dominant instability-the coupled Ion Temperature Gradient and Trapped Electron Mode $ITG/TEM$~\cite{coppi_74,dannert_05,waltz_07,ernst_09,merz_10}; the suppressing agent is the ExB (velocity) shear. As experiments progress to fusion conditions, it was/is feared that the paucity of velocity shear could eliminate/weaken transport suppression. Fortunately, however, there are many experiments that do observe TBs precisely in the low shear regime \cite{ding} \cite{pan} \cite{frig2007}-What mechanism, then, causes and maintains this fusion relevant low transport state?   

\subsection {The charge-weighted Flux constraint}

It is quite amazing that the answer lies in a very general property of strongly magnetized plasmas; there is a fundaments physical constraint that gyrokinetic fluctuations must obey to exist: the fluctuation driven net charged-weighted flux must be zero (s is the species index)
\begin{equation}
\sum_s q_s \Gamma_{rs} = 0, 
\label{eq:one}
\end{equation}
where $q_s$ is the charge, and $\Gamma_{rs}$ is the usual radial transport flux of particles (i.e., the flux averaged over space scales larger than the fluctuations). For linear eigenmodes, this is the conventional quasi-linear flux. This "flux constraint" (FC) is satisfied by low frequency plasma fluctuations that satisfy Maxwell’s equations  to the lowest order in the gyrokinetic expansion.  

A small digression on the origin of the FC, epitomized in Eq(\ref{eq:one}), is in order. It could be derived, for instance, from: a) collective momentum conservation after summing over fluctuation interactions in space, as in Coulomb collisional transport,  b) local frame invariance of local fluctuations, c) primarily electrodynamic considerations together with their invariances. (See Refs.~\cite{constraint_arxiv} \cite{kot1}, \cite{kot2}, \cite{Waltz}, \cite{Mahajan1}).

Although the dynamical constraint has been known for some time, it is for the first time that its enormous power and effectiveness will be explored in the context of TB formation.  We will focus on the suppression of two important instabilities in the gyrokinetic system---the coupled Ion Temperature Gradient and Trapped Electron Mode $ITG/TEM$---since these instabilities are typically the limiting factor for confinement.   If these instabilities can be tamed, drastic improvements to confinement can be achieved.  

%For historical reasons, we will focus on the suppression of two important instabilities in the gyrokinetic system: the coupled Ion Temperature Gradient and Trapped Electron Mode $ITG/TEM$

A simple translation of the content of Eq(\ref{eq:one}) reads---if it is not possible for a class of fluctuations to satisfy Eq(\ref{eq:one})---no instability is possible, no matter how much free energy may be available in the equilibrium gradients. In essence, the violation of FC will not let plasma relax the gradients via instability- configurations with large gradients (TBs) will be stable! 

Tracing the amazing TB phenomenon to these fundamental roots is not only beautiful, it also greatly clarifies the conditions necessary for these crucially important, thermodynamically counter-intuitive states to be sustainable. The fluctuation dynamics are much more clearly understood as a consequence of these two very different dynamics operating simultaneously: the aforementioned constraint on the total charge- weighted flux, and the well-known relationship of fluctuation growth with the plasma free energy.  A major accomplishments of this work lies in \emph{distinguishing} these two different dynamics, their separate roles, and their very different characteristic.

Oftentimes, we are able to analytically derive approximate expressions for conditions when the constraint is soluble/insoluble. These are simple and useful: specifically, boundaries, beyond which the constraint cannot be satisfied for any growing fluctuation, and hence regions with no instabilities are identified. Extensive gyrokinetic simulations show that these analytic boundaries explain a great deal of the observed behavior, for both linear instabilities, and also, nonlinear fluctuations. They are enormously helpful in revealing the central role of the charge flux constraint in the simulation behavior. By comparing these relations to the simulation behavior, the action of the constraint becomes "visible" for the first time. 

The invocation of the FC, thus,  brings the gross stability of the very complex gyrokinetic system to the realm of (partial) analytical tractability. Combined with detailed simulations, we, then, have a powerful tool to deeply investigate fluctuations. 

\subsection{Free Energy and Instability}

It is now time to review the other dynamics, the conventional free energy controlled mechanism for instability- \emph {whenever there is free energy (in equilibrium gradients in magnetically confined plasmas, for instance), growing instabilities can arise if they reduce the free energy in the equilibrium}.  For the gyrokinetic system, one can derive an evolution equation for the fluctuation induced free energy~\cite{constraint_arxiv} \cite{abel} \cite{hel1} \cite{hel2}, \cite{Lee}, \cite{Krommes}, \cite{Sugama}; the resulting entropy equation, or perhaps more precisely, the free energy equation reads: 

\begin{eqnarray} 
&&\frac{\partial}{\partial t} \sum_s \big[ \frac{T_s}{2} \int d\bm{x} d\bm{v}
\frac{\delta f_s^2}{F_M} + \frac{\bm{m}_s \bm{n}_s \delta V^2_{E\times B}}{2}
+ \frac{\delta E^2}{8 \pi} + \frac{\delta B^2}{8 \pi}\big] =  \nonumber \\
&&\sum_s \big[\bm{n}_s (\bm{Q}_s \frac{1}{T_s} \frac{ d T_s}{d x} +
\Gamma_s \frac{1}{n_s} \frac{d n_s}{d x}) \big] - \sum_s \int d\bm{x} d\bm{v}
\frac{\delta f_s}{F_M} \bm{C} (\delta f_s)   \nonumber \\
\label{eq:two}
&& 
\end{eqnarray}

Where $\delta f$ is the perturbed distribution function, and 
$\Gamma_s = \int d\bm{x} d\bm{v}\overrightarrow{x} \cdot \delta v_{E\times 
B} \delta f_s$ and $Q_s=\int d\bm{x} d\bm{v}\overrightarrow{x} \cdot \delta 
v_{E\times B} \delta f_s (\frac{1}{2 T_s}m_s v^2 - \frac{3}{2})$ are, 
respectively, the particle and heat fluxes, and $\bm{C}$ is the collision operator.  For simplicity of presentation, we have assumed $k_{\perp} \rho $ is small in Eq(\ref{eq:two}), but this is not necessary. 

In Eq(\ref{eq:two}), fluctuation free energy grows because the free energy in the macro (equilibrium) scale is reduced by the \emph{same amount} (due to quasi-linear fluxes), minus collisional entropy production (as in Boltzmann's H theorem). The Left Hand Side (LHS) is the rate of change of the free energy in the fluctuations. (This expression can be shown to be closely related to the familiar definition U-TS for small fluctuations, where U is the kinetic plus field energy, and S is the entropy $\sim f ln f$ for small fluctuations about a Maxwellian, as in the gyrokinetic ordering). The RHS is the rate of change of the equilibrium free energy, and has a classic structure: a product of thermodynamic forces $dT_s/dx$ and $dn_s/dx$ times the corresponding thermodynamic fluxes $Q_s$ and $\Gamma_s$ , respectively. Only the collisional term changes entropy (increasing it).

The growth rate of an instability is determined by how efficiently the instability causes this transfer. As expected, for the gyrokinetic system, the free energy transfer can be greatly increased by familiar factors such as "bad" magnetic curvature drifts of ions and electrons, resonances in phase space, etc. Such specific physics is contained in the very general Eq(\ref{eq:two}), although some manipulations are needed to render it more apparent. A specific and important example is magnetic field line curvature, which is well known to play a very important role in gyrokinetic instabilities. Especially for simplified models, it can be shown explicitly that "bad" curvature affects Eq(\ref{eq:two}) to give higher growth rates; its manifest effects will be seen in our calculations. For now, we simply note that considerations of curvature have their primary direct effect upon the free energy equation, in the expected qualitative way. 

It is obvious, then, that if the free energy dynamics were operating unchecked, steeper gradients "would" engender faster instability growth rates  for linear modes. And nonlinearly, larger fluctuation amplitudes should result from larger gradients, giving larger fluxes. This behavior is indeed seen in a truly vast variety of physical systems, including in realms far outside of magnetized plasma dynamics. 

\emph{Fluctuations in TBs, however, behave otherwise! This behavior is both theoretically striking and of great practical importance:  it is crucial to many approaches to attain thermonuclear energy gain}

The main mandate of this paper is to demonstrate that this "deviant''  behavior is a consequence of the constraining effects of the flux constraint epitomized in Eq(\ref{eq:one}). The story behind the 
``why and how of a Transport Barrier'' turns out to both deep and elegant. 

This story will be built by investigating, analytically as well as by detailed simulations, the most pertinent instability for TB formation, the ITG/TEM. We will proceed in several steps starting with its simplest manifestation---ITG with adiabatic electrons. Not only does the adiabatic electron approximation enable in-depth analytic probing, it is also highly relevant for realistic TB scenarios.  An accomplishment of this work is to show that the fundamental considerations that lead to TBs are qualitatively the same for the adiabatic electron case and with full electron dynamics (and to verify this with extensive simulations and analysis). In fact the parameter domains where TBs arise when full electron dynamics are included are typically those where the non-adiabatic electron response becomes small for reasons we describe here for the first time. And because of this, there is sometimes even \emph{quantitative} relevance of \emph{some} of the conditions derived for the adiabatic electrons case, and the conditions found in simulations for TB formation with full electron dynamics. The simplicity of the adiabatic electron case will make it an appropriate starting point when we examine simulation behavior. We show in Sec.~\ref{sec:kinetic_electrons} that one key to enabling this TB regime is arranging the magnetic geometry to minimize the impact of kinetic electrons.  When this is achieved, as it frequently is in experimental tokamak scenarios, the system approaches the adiabatic electron regime.  Hence, the early focus on adiabatic electrons is fully justified and ultimately we demonstrate the same TB behavior with fully electromagnetic, kinetic electron simulations in Sec.~\ref{sec:kinetic_electrons}. 

% As we add additional physics, deeper understanding follows along with new ways to affect TB formation.

%This story will be built by investigating, analytically as well as by detailed simulations, the historically most pertinent instability for TB formation, the ITG/TEM. We will proceed in several steps starting with its 
%simplest manifestation-ITG with adiabatic electrons. As we add additional physics, deeper understanding follows along with new ways to affect TB formation.

\section { General Background - Context for ITG/TEM }

Finding how $ITG/TEM$ behaves under the simultaneous influence of the two contending dynamics: the charge flux constraint Eq(\ref{eq:one}) and the free energy equation Eq(\ref{eq:two}), is crucial to understand much of the fluctuation behavior seen in simulations.  Investigating the character of instability, via the tension between these two \emph{qualitatively} different kinds of dynamics, leads to powerful new results and insights. It will be possible to distinguish, \emph{remarkably cleanly}, between what simulation behavior is owed to FC, and what originates in the free energy drive.

For linear modes, the dual dynamics analysis can be shown to be equivalent to a dispersion relation (see Ref.~\cite{constraint_arxiv} for details). This equivalence with the traditional mode of analysis shows that the new (FC) methodology does, indeed, capture crucial aspects of fluctuation dynamics. 

Of course, dispersion relations are the traditional way to theoretically investigate and understand instabilities. But by splitting out the FC and the free energy equation, one can obtain results (relatively easily) that have never been obtained by dispersion relations. Some examples are:

1) Analytic derivations of bounds on the stability for cases in realistic geometry, where the dispersion relation is analytically quite intractable. The bounds are verified by simulations with surprising accuracy.

2) Analytic descriptions of complex mode structure changes that have direct implications for non-linear heat fluxes. An accurate description of this is otherwise incomprehensible, except by using the solubility of the FC. Then, it is easily derived and understood. Again, this applies to realistic geometries where a kinetic dispersion relation is intractable.

3) Qualitative insights into the stability behavior when gradients are strong, regarding the relevant control variables and system behavior. Attempts to do this without using the dual dynamics approach fall subject to multiple contradictions, as is demonstrated especially clearly when trapped electrons are included. 

For the TB physics, the most important role of the FC is to suppress $ITG/TEM$ growth when free energy dynamics favors it, even strongly. 

There is a particularly striking property of the gyrokinetic system, revealed by extensive simulations (given in later chapters), that serves as an essential conceptual bridge between the dual dynamic analysis and the behavior of extensive and diverse simulations - the ``adaptability'' of fluctuations. Like many similar complex systems, the gyrokinetic system adapts to produce the most unstable mode (to decrease the free energy). For example, simulations show that the ITG/TEM in realistic geometry usually "finds a way" to tap free energy \emph{until very close to the boundary for which the constraint becomes insoluble}. In other words, the system does in fact behave just like the vast variety of physical systems with many degrees of freedom, it is just that there is a "hard" dynamical constraint which was not previously apparent and which, indeed, allows a seemingly "non-thermodynamic'' state to be sustained. The apparent violation of ubiquitous thermodynamic behavior of ITG/TEM in TBs is actually just a failure to appreciate the presence of the constraint. Nonlinear transport for ITG/TEM is also subject to the action of the constraint, as the simulations will show. 

The extra understanding  accruing from this fundamental statistical property of mechanical systems with large number of degrees of freedom is one of the distinguishing features of this work and will be an essential element in building what we call the statistical mechanical ansatz that will guide our attempts to understand the essential behavior of the system.

We end this section with several general explanatory remarks before we begin to present detailed calculations.

The crucial insight that animates this work is that there are indeed equilibrium conditions for which it is impossible for fluctuation induced flux to satisfy the FC. This is true in a strong sense: even when we broaden our consideration to include a much larger class of field fluctuations than just eigenmodes. 

We also emphasize that not all linear fluctuations are analyzable as eigenmodes, the conventional entity that we investigate in linear stability studies. Since the constraint has no particular knowledge of eigenmodes, results for the insolubility of the FC obtained in this paper (investigating constrained dynamics) apply to all fluctuations, eigenmodes are not. 

 A particularly dramatic example is the ITG mode with adiabatic electrons (subject of next section), where we can consider a fluctuating potential $\phi$ with any spatial structure, with any frequency broadly in the range of the instability, but which is growing. Even for this very large class of fluctuations, there are equilibrium conditions for which it is impossible to satisfy Eq(\ref{eq:one}). Such a statement does not just apply to eigenmodes. 

%So the following simple logical syllogism holds: 
%\vspace{12pt}
% 
% A) all eigenmodes must satisfy the FC, and 
% \vspace{12pt}
% 
% B) no growing gyrokinetic fluctuations of a broad class that includes eigenmodes can satisfy the FC
% \vspace{12pt}
% 
% so we ineluctably conclude that 
% \vspace{12pt}
%
% C) there can be no growing eigenmodes. 
%\vspace{12pt}
%
Crucially, deriving general conditions when no gyrokinetic fluctuations can grow turns out to be \emph{enormously easier  than a corresponding eigenmode calculation}. This is because one does not have to impose the complexities of the eigenmode condition. Hence we are able to obtain analytic bounds for the $ITG_{ae}$ in realistic geometries, for instance. \emph{Extensive} gyrokinetic simulations follow these analytically derived bounds, \emph{remarkably} closely. 

An important conceptual ramification of our approach is that an inability to find an unstable solution could originate in circumstances that are much broader than the eigenmode problem could predict.

These stability predictions often defy expectations from "commonsense" heuristic viewpoints. They also predict completely new types of mode behavior, found in the simulations, that has not been previously appreciated. \emph {And finally, simulations show that these considerations apply to nonlinear fluctuations and transport just as for linear instabilities}. 

%There is an important conceptual ramification to the logic of argument above. The origin of the situation where  Eq(\ref{eq:one}) has no solution \emph{cannot} be due to considerations that pertain \emph{specifically} to eigenmodes, since it is true of \emph{a much larger class} of fluctuations. Nearly all previous analysis of stability have, naturally enough, approached the problem by solving the eigenmode equations. They have therefore missed the fact that an inability to find an unstable solution derives from circumstances that are much broader than the eigenmode problem per se. 
%

Exploring conditions for which the fluctuation driven transport  cannot satisfy FC will be our route to  figuring out  what  makes TBs possible when, for example, velocity shear is low; the latter is, of course, a matter of great practical importance. From basic statistical mechanical principles, we know that thermodynamic forces drive their corresponding thermodynamic fluxes. One may then expect that there will exist parameter regimes where increasing equilibrium density gradients will eventually compel the driven fluxes to be non-zero violating the FC.  Since no growing fluctuations could exist, we could get a stable equilibrium state with high gradients, the transport barrier.

The reader will likely find that this rather different mode of probing stability is unfamiliar, primarily because we transcend here  the  methods of conventional plasma instability theory.  There are good reasons, however, to stick to and rely on the FC based stability analysis: 1) simulations in complex geometries follow criterion derived by simple methods 2) results can be qualitatively traced to fundamental statistical mechanical concepts that are far more general than gyrokinetic stability theory. Rather, those general considerations are applied to gyrokinetics as a specific context of interest.

A practical consequence is that equilibrium density gradients above a threshold emerge as a critical ingredient to TB formation without velocity shear (except in unusual cases), at least insofar as the ITG/TEM is the controlling fluctuation. 

Since equilibrium gradients are also a measure of the free energy drive (for instabilities), it might appear somewhat paradoxical that a system with  gradients greater than a threshold can become fluctuation free. This reflects, in fact, the very power of the deeper physics which tells us that no matter how large the free energy, the insolubility of the FC guarantees an essentially instability free state. The transport barrier, thus, is a perfectly warranted high gradient equilibrium!

When trapped electrons are included, the class of fluctuations above are ones where energetic considerations are not overwhelmingly stabilizing. This is true when electrons are included in tokamaks with negative magnetic shear or strong Shafranov shift. Stellarator cases with somewhat geometric features also follow this pattern, but we do not describe this in detail here for space.

%{\color {blue }The conditions for which FC is not satisfied} have essentially nothing to do with the absolute strength of equilibrium gradients, which also measure the strength of the free energy drive. So it is possible to have strong gradients and no instabilities. 

We will see that there is an important case of interest where FC insolubility may be limited: it may, for example, be insoluble only for fluctuations with low $k_y \rho_i$ (or $k_y \rho_e$ for ETG modes). In that case, the higher $k$ instabilities can cause some transport. However, for these conditions, the simulations show  that transport is reduced by two orders of magnitude consistent with the expectation that only the low $k$ modes cause strong transport. And in these cases of partial suppression, too, a TB becomes possible, even without velocity shear.

\section{A Model Calculation: the ITG instability with adiabatic electrons}

\begin{figure*}
%\iffalse
\subfloat[]{%
  \includegraphics[width=.4\linewidth]{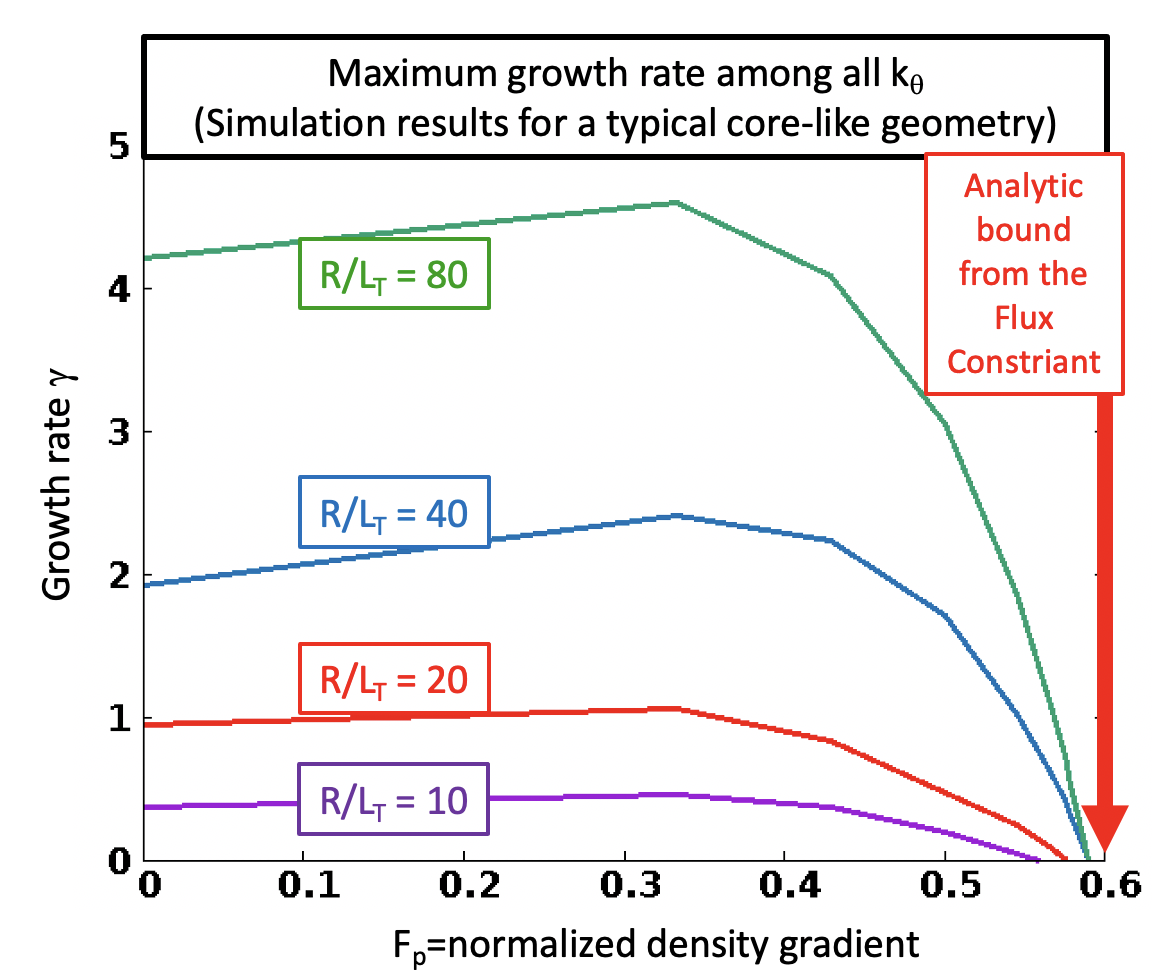}%
}\hfill
\subfloat[]{%
  \includegraphics[width=.4\linewidth]{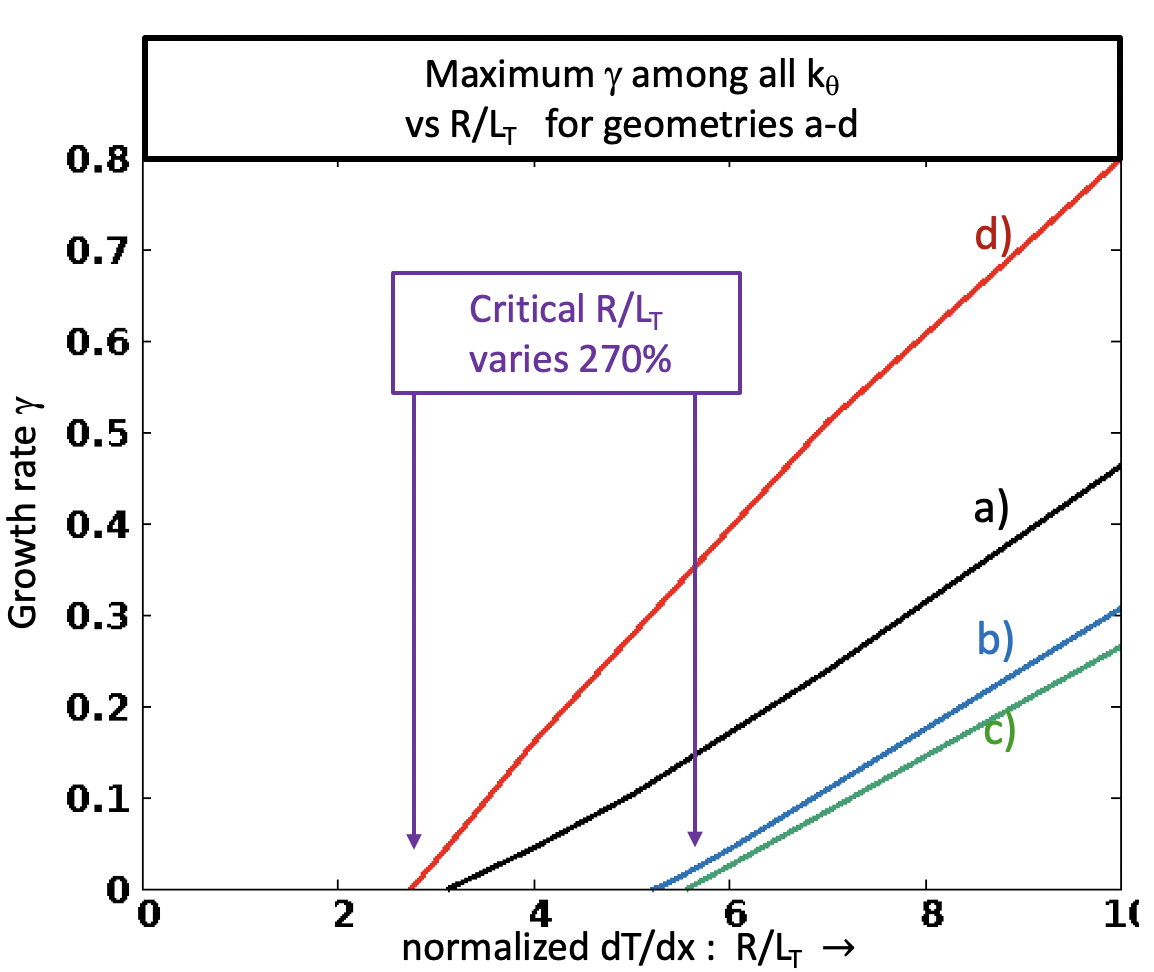}%
}
\vfill
\subfloat[]{%
  \includegraphics[width=.4\linewidth]{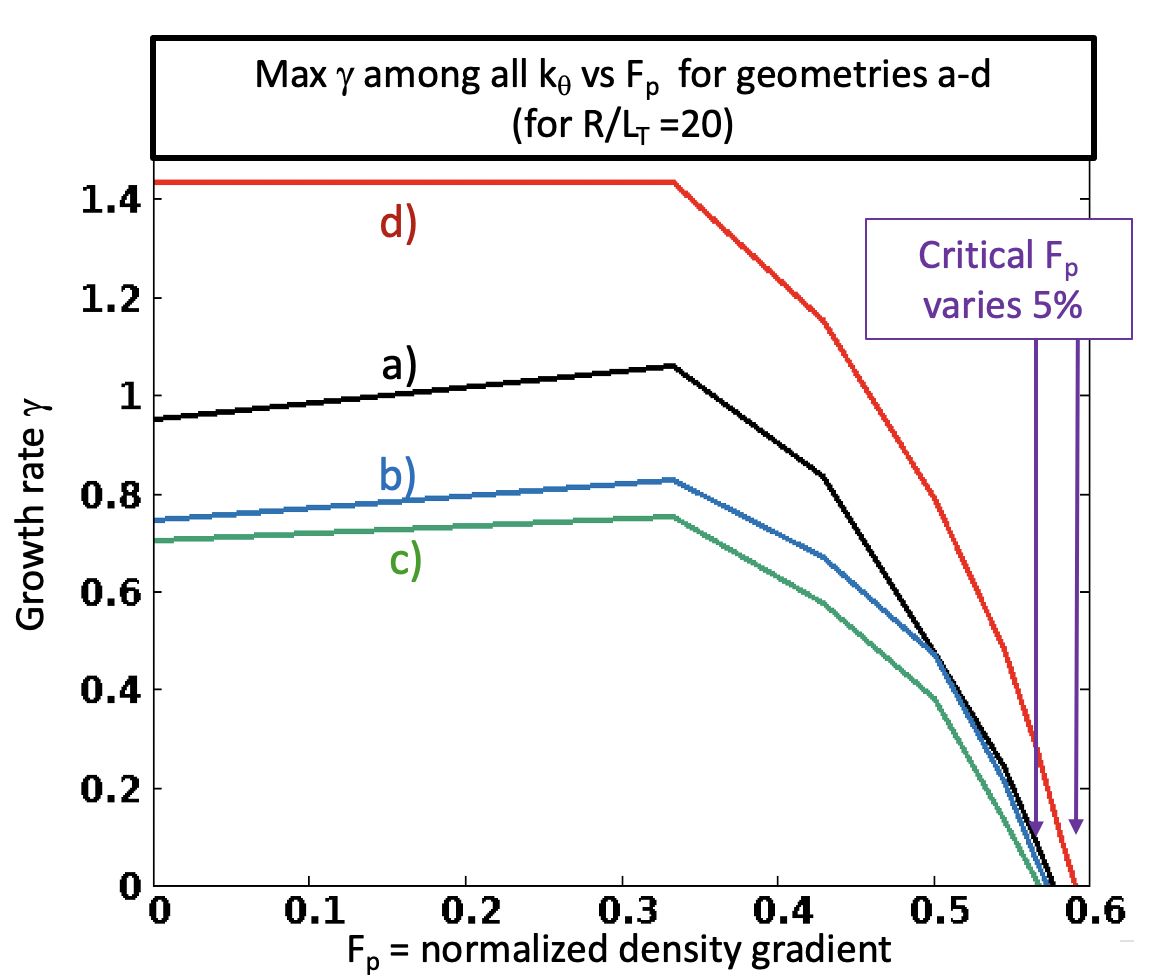}%
}\hfill
\subfloat[]{%
  \includegraphics[width=.4\linewidth]{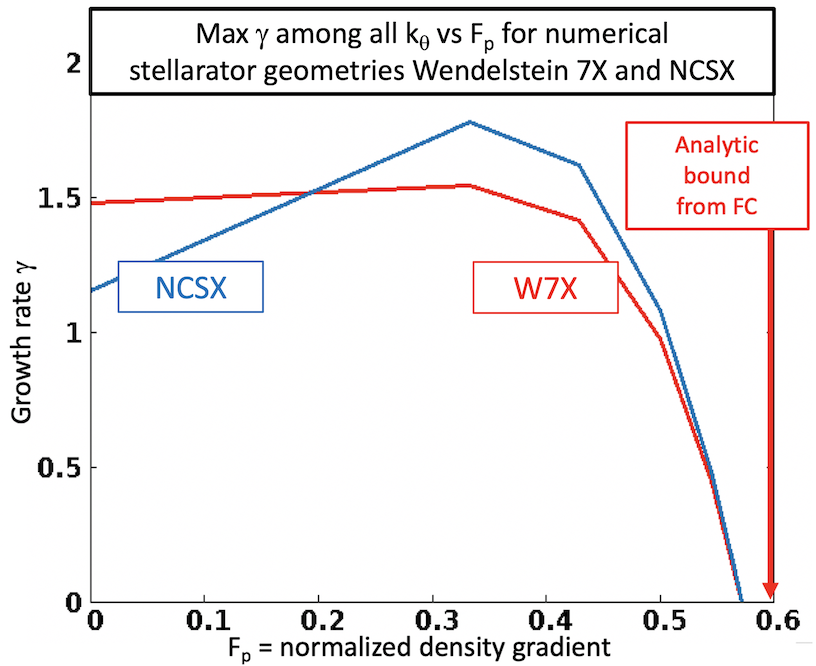}%
}
\vfill
\subfloat[]{%
  \includegraphics[width=.4\linewidth]{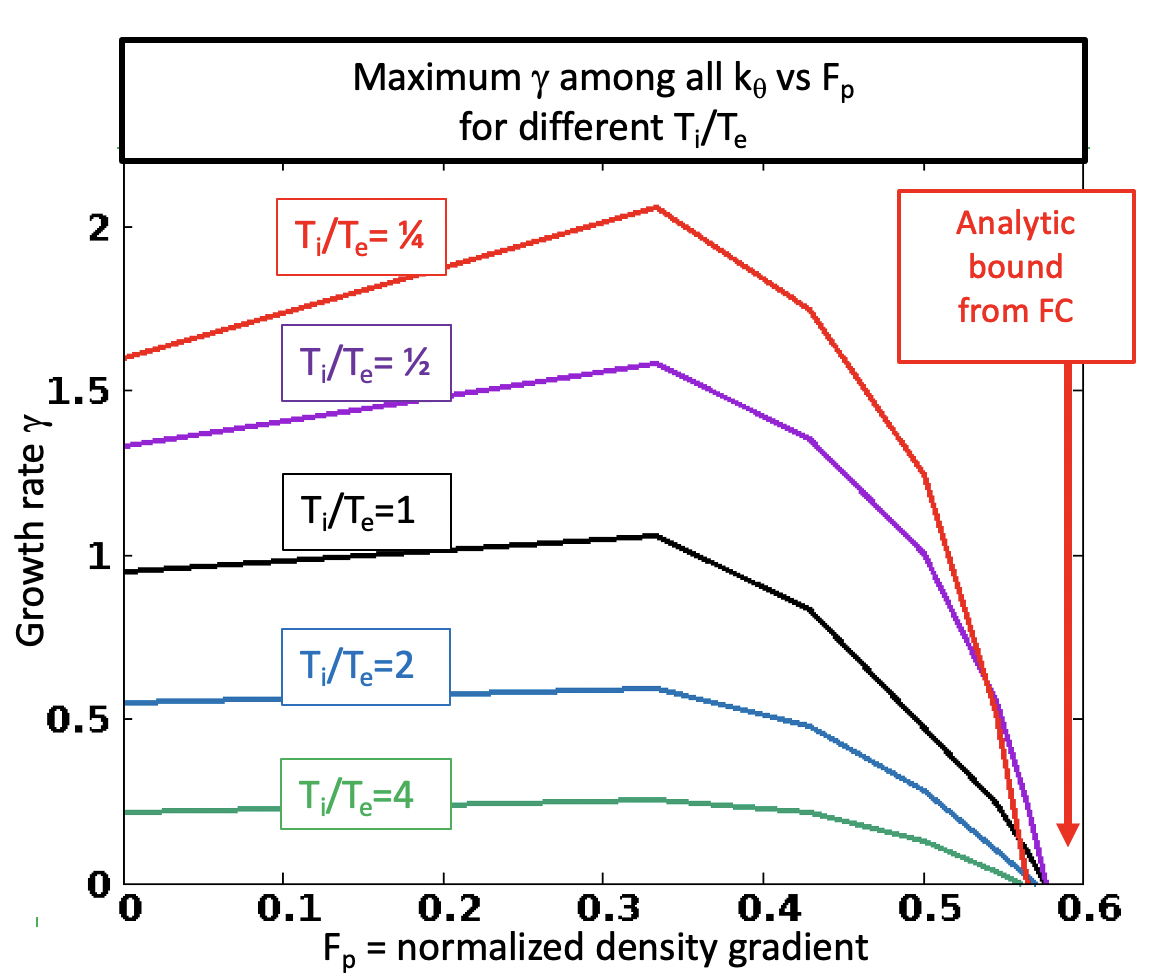}%
}\hfill
\subfloat[]{%
  \includegraphics[width=.4\linewidth]{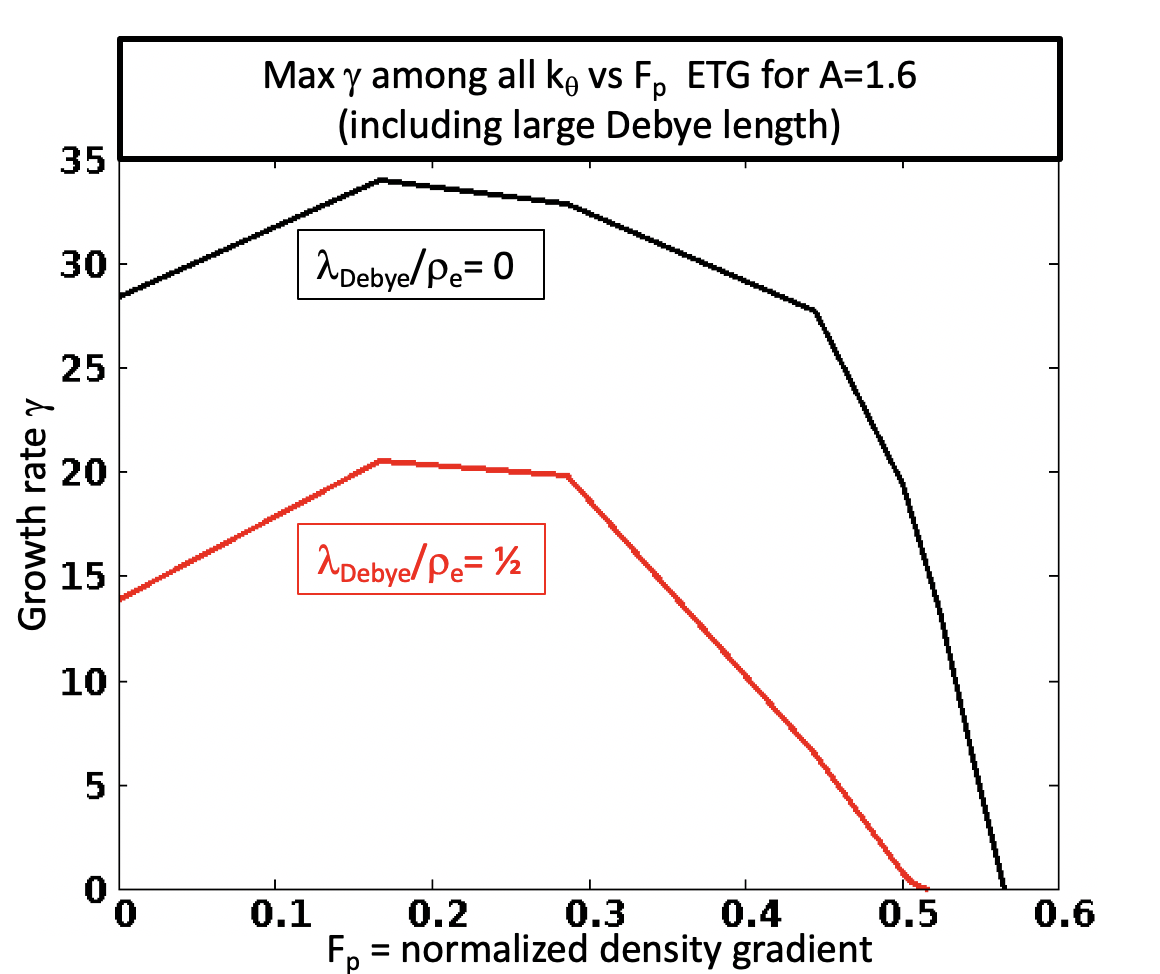}%
}
%\fi
\caption{\label{fig:one}  The growth rate of the most unstable mode (maximum for all $k_{\theta} \rho_i$) is plotted a) vs $F_P$ for a standard geometry like the cyclone base case) b) for four diverse geometries vs $R/L_T$ showing that the critical temperature gradient varies by $\sim 270\%$ c)  for those same geometries, the critical $F_P$ varies by $\sim 5\%$ d) for two stellarator geometries e) for an ITB geometry, for many values of $T_i/T_e$ f) for a low aspect ratio A geometry, an ETG for two different $\lambda_{Debye}$ } 

\end{figure*}

The ITG mode is a major instability that is often the dominant  source of energy loss in tokamaks. With the approximation of adiabatic electrons, its physics is simple enough that solubility of the flux constraint can be tested analytically. We call this mode the $ITG_{ae}$. The adiabatic electron approximation is clearly limited in its \emph{quantitative} applicability.  However, it \emph{qualitatively displays all the  relevant  consideration for the interplay of the FC, free energy balance and adaptivity that pertain to the full electron case}: how the FC leads to stabilization, how it compels serious modifications of eigenfunction structure, how the gyrokinetic system responds adaptively to the FC and to the spatial structure of curvature, and how such adaptivity implies that only the FC results in stability for all but the most extreme geometries.  We demonstrate in Sec.~\ref{kinetic_electrons} that \emph{all these same dynamics are operative for the full electron case and are crucial to TB formation.} These considerations have never been described before now. And they are easier to derive, explain and demonstrate in simulations for the $ITG_{ae}$. And, as we will see, even some \emph{quantitative} aspects of our considerations for the $ITG_{ae}$ are to be found in the full electron simulations. Thus, the present focus on adiabatic electrons is perfectly justified.

%The ITG mode is a major instability that is often the dominant  source of energy loss in tokamaks. With the approximation of adiabatic electrons, its physics is simple enough that solubility of the flux constraint can be derived analytically. We call this mode the $ITG_{ae}$. In this section, we compare the simplest such analytic result to gyrokinetic simulations of the mode in realistic geometry. 

%In this section, we compare the simplest such analytic result to gyrokinetic simulations of the mode in realistic geometry.  The adiabatic electron approximation is clearly limited in its applicability.  However, we subsequently demonstrate in Sec.~\ref{kinetic_electrons} that the same dynamics persist even with kinetic electrons for certain magnetic geometries.  Thus, the present focus on adiabatic electrons is perfectly justified.

For over half a century, it has been known that for the slab-like $ITG_{ae}$, mode stability is based upon the parameter $\eta = \frac{1/L_T}{1/L_n}$,where $1/L_n $ and $1/L_T$ are the usual inverse gradient scale lengths of density and temperature. %This $\eta$ is arguably the single best known parameter in gyrokinetic instability theory for magnetically confined plasmas. 
The mode is unstable only when $\eta$ exceeds some critical value $\eta_{crit}$.  So for a given temperature gradient, the slab $ITG_{ae}$ can be stabilized by a sufficient density gradient. 

We will show that stabilization of the slab $ITG_{ae}$ by low $\eta$ is a consequence of the insolubility of the FC. Once this connection is recognized, a large variety of fluctuation behaviors can be simply understood- even aspects that were previously not. 

In toroidal geometries, the curvature driven $ITG_{ae}$ is both more unstable and more complex. In the next section, we derive a very simple and general criterion for the amount of density gradient beyond which the FC is insoluble. Since the density gradient is the primary cause of stability, it is convenient to define a more suitable normalized density gradient parameter (i.e., other than $\eta$):  

\begin{equation}
F_P = (Tdn/dx)/[d(nT)/dx]=\frac{1/L_n}{(1/L_n+1/L_T)} \label{eq:three},
\end{equation}
the Fraction of density gradient in the Pressure gradient. $F_P = 1/(1+\eta)$ will be the parameter of choice in this study. We will soon derive that, for $ITG_{ae}$, the FC is insoluble when the density gradient is large enough to make

\begin{equation}
F_P > 0.6 \label{eq:four}
\end{equation}

Note that the density gradient magnitudes corresponding to Eq(\ref{eq:four}) do actually arise in ITBs and pedestals, and, therefore, are experimentally relevant. Such density gradients are crucial to allowing the existence of the TB if velocity shear is low, as we will see.

Simulations in realistic geometry adhere to this criterion \emph{remarkably} accurately.

One must emphasize here that Eq(\ref{eq:four}) is not just another way of writing the conventional $\eta$ criterion for linear stability; it represents altogether different physics.
 
In fact, when examined  in the context of our past experience with linear theory, this result is \emph{extraordinarily} surprising. Let us contrast the simplicity of Eq\eqref{eq:four} with well known critical \emph{temperature} gradient criterion in terms of $1/L_{T crit}$; the latter is found by decreasing the temperature gradient, and scanning over all $k_{\theta} \rho_i$, until the maximum $\gamma$ vanishes. In toroidal geometries $1/L_{T crit}$ varies by a factor 2-4 or more depending upon the safety factor, magnetic shear, the temperature ratio $T_i/T_e$, etc., etc. The fact that the constrained dynamics leads to a stability criterion that involves only the fractional density gradient and has no cognizance of the multidimensional parameter space that determines the conventional stability attests to its simplicity, and, as we will verify shortly, "universality ''.

%One would surely have expected that the stability point for density gradients would be substantially affected by similar variations of parameters. This would be the heuristic expectation based upon widespread experience with the critical temperature gradient.

\emph{ To test the fractional density stability criterion by simulations, one repeats the procedure for the linear stability.} If one scans over all $k_{\theta} \rho_i$ and plots the maximum growth rate, stability is always reached for $F_P$ close to the analytic upper bound. \emph{Variations in parameters that strongly change $1/L_{T crit}$ , and that otherwise strongly change the instability strength and $\gamma$, have very little effect upon the $F_P$ for stability. }. This is the first example where simulation behavior cleanly distinguishes between considerations of the FC and of energetics. Energetics crucially determine $\gamma$ and  $1/L_{T crit}$, but not the critical $F_P$.

Let us put this in an even larger physical perspective. Consider the following thought experiment: we maintain a strong temperature gradient at a constant value, and increase the density gradient. Eventually, stability results. Up until now, stabilization of the $ITG_{ae}$ by density gradients has been regarded as a "quirk" of linear theory- a very specific property of a particular mode type: the dispersion relation of simplified analytic models show this behavior. Simulations, qualitatively, also show this behavior. This qualitative trend has been known for such a long time that it is regarded as "natural". But one should keep in mind that from the general perspective of statistical mechanics, it is anything but this straightforward: despite there being an extremely large number of degrees of freedom, and hence, a very large number of potential mode structures, with widely varying space and time scales, none of these succeed in tapping the free energy in the equilibrium. Even when gradient magnitudes are extremely high, and when factors like the curvature drive are highly destabilizing. 

It will be a misconception to think that adiabatic electron assumption is, somehow, responsible for this unexpected result. Since adiabatic electrons carry no density fluxes, the $ITG_{ae}$ cannot relax density gradients; density gradients, thus, are not a drive for the instability. This merely implies \emph{that increasing the density gradient should not increase the growth rate. In no way, can it explain why density gradients drive the growth rate to zero. In no way, can it explain why density gradients prevent a complex system (with a large number of degrees of freedom) from accessing the free energy in temperature gradients to feed instabilities and turbulence, that, in turn, will relax  temperature gradient strength in any equilibrium geometry.} 

What explains this amazing phenomenon is the insolubility of the FC (Eq(\ref{eq:four})- No fluctuations can grow (in any geometry for any plasma parameters) when FC is violated-no relaxation routes are accessible.

We will now reinforce the power of the FC route to stability by presenting gyrokinetic simulations for different and realistic geometries and parameters. In particular, we show
\begin{itemize}
\item the factors that \emph{strongly} affect the free energy equation may have \emph{little impact} upon the FC solubility. 
\item the response of fluctuations to each dynamic is ``cleanly" distinguished. 
\end{itemize}
  
To the best of our knowledge the results below are new. 

Simulations were done choosing a standard type of magnetic geometry (so-called "Miller geometry" with representative shaping) for different  temperature gradients, geometric and other equilibrium parameters. In all cases we fix the temperature gradient at some value and scan the density gradient until stability is found. The growth rate of the most unstable mode (scanning $k_{\theta} \rho_i$) to resolve the peak $\gamma$, is plotted.  

We start with a standard case in Fig\ref{fig:one}a (the cyclone base case but with elongation and triangularity). We choose a wide range of temperature gradients ($R/L_T$) up to very high values ($R/L_{T_i}$= 10, 20,40 and 80 ,where R is the torus major radius.) The magnitude of the equilibrium gradient obviously strongly affects the free energy equation, but not the solubility bound in terms of $F_P$. The simulation results reflect this fact:  growth rates vary by an order of magnitude, but the critical $F_P$ varies by $\sim 5\%$ (a hundred times less), and is always quite close to the analytic upper bound for the solubility of the FC. 

Without an understanding based upon the solubility of the flux constraint, this result would be quite perplexing. The same pattern is repeated by many different parameter variations. 

We now display stability curves  for a variety of magnetic geometry parameters -magnetic shear, safety factor q, and Shafranov shift. As we traverse this space, the critical temperature gradient $R/L_{Tcrit}$ varies by a factor of $\sim 270\%$ (fig(\ref{fig:one}b)) while the critical $F_P$ varies only by $~5\%$(fig(\ref{fig:one}c)) - about two orders of magnitude less. And to boot, the stabilizing $F_P$ is always close to the analytic upper imposed by FC.

 In fig(\ref{fig:one}d), we show that the stability curves for representative stellarator geometries from- Wendelstein 7X~\cite{stellarator2} and the proposed NCSX~\cite{stellarator1}, are similar to that of tokamaks. 
 
% ******
%{ \color{red} pls put this a footnote reference}
% (These stellarator results were for VMEC geometries for NCSX and W7X courtesy of M. Zarnstorff, P. Xanthopoulos, I. McKinney and M.J. Pueschel.)
%*******
Vast experience with ITG has told us that the temperature ratio $T_i/T_e$ strongly affects the growth rate and critical temperature gradient. In the run of geometries shown in fig(\ref{fig:one}e), the $R/L_{Tcrit}$ varies by $~400\%$ as $T_i/T_e$ is varied between $1/4$ and $4$.  The growth rate (normalized to the the ion transit frequency $v_{th i}/R$ varies by $\sim 500\%$. This is because  $T_i/T_e$ affects the relative magnitude of the adiabatic electron contribution to the free energy in Eq(\ref{eq:two}).  But since adiabatic electrons contribute no particle flux, we expect no $T_i/T_e$ effect upon the upper bound of $F_P$. Simulations find exactly that- the critical $F_p$ varies by $\sim 4\%$ (again about two orders of magnitude less that the variation in $\gamma$ or $R/L_{Tcrit}$) and is close to the computed solubility limit. 

The insensitivity to $T_i/T_e$ variation further continues the pattern that what effects  the free energy equation has little effect upon the stability limit imposed by the FC. 

Finally, we consider the ETG, which is well known to be nearly isomorphic to the $ITG_{ae}$. We compare a case with finite Debye length, $\lambda_D \rho_e = 1/2$ , to one with $\lambda_D \rho_e = 0$ in fig(\ref{fig:one}f). Somewhat similarly to $T_i/T_e$, the finite Debye length gives an additional contribution to the free energy equation ($ \sim \delta E^2$) that lowers the growth rate by a factor $\sim 2$, but this does not contribute to the particle flux, and the FC bound is unaffected. The critical $F_P$ is affected by only $\sim 10\%$,  far less than the growth rate. 

Let us summarize the simulation results reported above. They illuminate, for the first time, a crucial point about the deep underlying physics of the gyrokinetic system: effects that strongly affect the free energetic dynamics may not /often do not impact the one parameter solubility limit due to the FC, Eq\eqref{eq:four}. The latter is what controls the stabilization of the $ITG_{ae}$ by density gradients; the predicted threshold is totally insensitive to  numerous effects that strongly affect free energy dynamics. 

Since the solubility of the FC emerges as the key dynamic that commands the simultaneous existence of steep gradients and weak instabilities, and hence of TB formation, it is this dynamic that we will continue \emph{exploring and testing} in this study. 

The simulations also highlight ``adaptability'' of the gyrokinetic system. When temperature gradients are high, the modes become stable only quite close to the solubility boundary of the FC, despite huge variation in parameters. One does intuitively expect that a system with many degrees of freedom would  adapt- "find a way" to tap strong equilibrium free energy to relax it if that were at all possible. The gyrokinetic system, just like other similarly complex physical systems, does exactly that but for a "hard" constraint in the dynamics that precludes accessibility in some regions of the equilibrium parameter space. Even then, the system remains unstable until very close to the FC boundary.

This latter picture is further reinforced by simulation results that follow, which are in many ways \emph{much more} striking than the ones in fig 1. But first, we indicate how the solubility criterion for the FC can be obtained analytically.

\section{Mean field theory for computing FC bounds: Simplified Kinetic Model}

Given the complexity of the gyrokinetic system, we make a simplification that is conceptually similar to one that is used, very effectively, in other branches of physics: mean field theory. In this approximation, many degrees of freedom are treated as a single, averaged one. We call the gyrokinetic version given here the Simplified Kinetic Model (SKiM).  In Appendix B we give a somewhat more accurate version of this model, but it leads to exactly the same solubility bounds as this simpler version. The simplicity of this version makes it easier to see various important conceptual points.

%This has often proven to be a very illuminating approximation in statistical physics. 

We begin by quantifying bounds on the solubility of the FC in this simple system, to arrive at analytic expressions.  We can then use gyrokinetic simulations for realistic parameters to examine how well these bounds describe the behavior of the system. We find that bounds derived using SKiM have a remarkable agreement with realistic simulations. The ultimate justification for this model is the high degree of agreement of its predictions have with simulation. However, it adds value beyond simulations, particularly in elucidating  basic conceptual issues.

Consider electrostatic modes with potential $\phi$. The conventional gyrokinetic equation in ballooning coordinates has a ``large number of degrees of freedom'' associated with variations with the distance $l$ along a field line. We wish to reduce these. It is straightforward to systematically expand the gyrokinetic equation in large $\omega$ and small $k_{\perp} \rho_i$ to derive a dispersion relation that will contain certain eigenfunction averages:

\begin{equation}
<k_{\parallel}>^2 =\frac {\int  \mathrm{d}l |\frac{\partial \phi}{\partial l}|^2} {\int  \mathrm{d}l |\phi|^2|}
\label{eq:kpar}
\end{equation}

\begin{equation}
<k_{\perp}>^2 =\frac {\int  \mathrm{d}l |\phi|^2 k_{\perp}^2 } {\int  \mathrm{d}l |\phi|^2},
\label{eq:kperp}
\end{equation}

\begin{equation}
<\omega_{di}>=\frac {\int  \mathrm{d}l |\phi|^2 \omega_{di} } {\int  \mathrm{d}l |\phi|^2}
\label{eq:omdi}
\end{equation}

These have a transparent interpretation as an eigenfunction average ion drift frequency $<\omega_{di}>$, parallel wave number $<k_{\parallel}>$, and perpendicular wavenumber  $<k_{\perp}>$.  We can consider these eigenfunction averages to encapsulate many degrees of freedom in a single average quantity, conceptually like in mean field theory. 

%We can consider these eigenfunction averages to encapsulate many degrees of freedom in a single average quantity, conceptually like in mean field theory. 

The gyrokinetic distribution function in a geometry with constant $k_{\parallel}$, $k_{\perp}$ and $<\omega_{di}>$ is easy to solve. (It is equivalent to a straight field line geometry with constant curvature which can be Fourier transformed in space.) In terms  eigenfunction averages  defined above, the non-adiabatic part of the distribution function (s is the ion species index) is

\begin{equation}
h_s=\frac{\frac{-i q_s \phi}{T_s}J_0( <k_{\perp}> \rho_s)(\omega-\omega_s^{\star})}{-i(\omega-<\omega_{ds}>)+v_{\parallel} <k_{\parallel}> + C}
\label{eq:eight}
\end{equation}

where $\omega_s^{\star}$ is diamagnetic frequency, and is the sum of a contribution from density gradients $\omega_{ns}^{\star}$ and from temperature gradients $\omega_{Ts}^{\star}(mv^2/2T -3/2)$. 

The $ITG_{ae}$ dispersion relation, obtained by using the usual quasineutrality condition 
\begin{eqnarray}
&&\sum_{ion \ species} \left( \int \mathrm{d} \vec{v} \frac{ F_{Ms} J_0( <k_{\perp}>\rho_i)^2(\omega-\omega_s^{\star})}{\omega-<k_{\parallel}>v_{\parallel}-<\omega_{ds}>}-1 \right) \frac{n_s q_s^2}{T_s} - 
\label{eq:SKiMdr} \nonumber \\
&&   \frac{n_e e^2}{T_e} = 0 \nonumber \\,
\end{eqnarray}
must be solved numerically, even for a single species. We can compare the analytic SKIM dispersion relation to simulation results only after using the eigenfunction provided by simulations. There is good qualitative, often, semi-quantitative agreement between SKIM and the simulation growth rates. 
%This dispersion relation is only partially explicit because we still need the wave function to evaluate $<k_{\parallel}>$,   $<k_{\perp}>$ and $<\omega_{di}>$. 

We can derive more elaborate eigenfunction averages than Eq(\ref{eq:kpar})-Eq(\ref{eq:omdi}), to lead to better agreement with simulations for given simulation eigenfunctions. But the intuitively clear expressions in Eq(\ref{eq:kpar})-Eq(\ref{eq:omdi}) will suffice for our present purposes; they encapsulate in a tractable way the average effect of the dynamics in many degrees of freedom along a field line.

%And it is likely that for any particular case, there exist \emph{some} value of  $<k_{\parallel}>$,   $<k_{\perp}>$ and $<\omega_{di}>$ whereby the dispersion relation will reproduce the simulation result for $omega$ (since there are three real quantities to vary to match the two real quantities $\gamma$ and $\omega$); such a value could be said to constitute an "accurate" eigenfunction average. 
%

 The particle flux from the non-adiabatic response Eq(\ref{eq:eight}) may be computed to be (for a single ion species),
\begin{eqnarray}
&&\int d\bm{v} f_M J_0(<k_{\perp}>\rho_i)^2  \big[ \frac {\gamma} {\gamma^2+(\omega_r-<k_{\parallel}>v_{\parallel}-<\omega_d>)^2}\big] \nonumber\\
&&\big[ 1/L_n+(v^2-3/2)/L_T\big] = 0 \nonumber \\ 
\label{eq:FC0}
\end{eqnarray}
for instability growth rate $\gamma > 0$ and real frequency $\omega_r$. We have made one small simplifying assumption above: that gradients are steep, as in a TB, so that  $\omega_s^{\star}>>\omega$. This is quite well satisfied for $ITG_{ae}$ (and also the ITG/TEM in following sections).

Crucially, the resonance factor above presumes $\gamma > 0$ . If $\gamma \le 0$ there are additional well-known terms from the Landau contour that must be included in Eq(\ref{eq:FC0}). Here we consider only exponentially growing fluctuations.

 Since the eigenfunctions vary considerably among the cases in fig 1, the averages  $<k_{\parallel}>$,   $<k_{\perp}>$ and $<\omega_{di}>$ do as well. And the real mode frequency $\omega_r$ varies too. 
Now we proceed to derive an FC bound that is independent of these quantities.

It is trivial to note that if $1/L_n > 3/ (2 L_T)$, the RHS of Eq\eqref{eq:FC0} is positive definite; consequently the flux condition cannot be satisfied. This condition, translating as $F_P>0.6$ guarantees stability of all fluctuations ( for any $<k_{\perp}>$ , $<k_{\parallel}>$, $<\omega_d>$), and any real frequency $\omega_r$ and $\gamma > 0$). It is equally evident that stability could be possible for $F_P< 0.6$ because of the term proportional to $v^2$; the latter contribution, however, will depend on the mode details. 

\emph{The stability criterion $F_P>0.6$ is to be contrasted with its counterpart from a  dispersion relation. The latter is complicated enough that numerical solution is required, even for this simplified model, and there are many well-known dependencies of the eigenvalue upon a host of parameters. On the other hand, the criterion $F_P>0.6$ is easily obtained from very brief analytical arguments, is far simpler in character, takes no cognizance of geometrical complexity or fluctuation attributes and is true for all fluctuations, eigenmodes or not; it is universal, in particular, it holds for fluctuations with any temporal or spatial structures, any spectrum in $\omega$  and $k_y$.} 

And the criterion also predicts the behavior of \emph{far} more complex simulations in realistic geometry as shown above and also below. 

At a deeper level, the insolubility of the FC cannot be due to considerations that pertain specifically to eigenmodes whose study constitutes the bulk of conventional stability theory. \emph{The reasons why the  particle flux can be always positive for sufficiently large density gradients, must be found in physical considerations more fundamental.}

\emph{Note that the particle flux in the FC is the charge weighted transport flux of particles}. Since the\emph{thermodynamic forces drive the corresponding thermodynamic fluxes}, a sufficiently large density gradient will, eventually, cause it to be nonzero (in the direction down the gradient). Hence the FC will be insoluble. Hence, the $ITG_{ae}$ is stable. Particle flux for the $ITG_{ae}$ refers to only ion flux; there is no flux for adiabatic electrons. More detailed exposition of this thermodynamic force -flux relationship will be given in section XI.

The FC insolubility criterion is a \emph {very robust result}-takes no cognizance  of plasma, magnetic or fluctuation properties. This is, perhaps, why the insolubility bound of SKiM agrees so well with simulations of much more complicated geometries: the insolubility criterion is independent of many of the details where SKiM is different from the more complex system. The results displayed in Fig 1 illustrate this "universal'' nature of the stability limit.  SKiM model is equivalent to a straight field line system which is, after all, a valid gyrokinetic system, it is just in a simple geometry.

Before closing this section, let us consider several details that are pertinent to particular results in fig 1. Note that the parameter $T_i/T_e$ does affect the dispersion relation (Eq(\ref{eq:SKiMdr})) via the adiabatic electron response while it is totally absent in the FC Eq(\ref{eq:FC0}) as adiabatic electrons do not carry any flux. Since the dispersion relation pertains to eigenmodes, the eigenmodes will also depend upon $T_i/T_e$. The fact that this particular attribute of the eigenmode is absent in the insolubility criterion is a particular manifestation of the general property we emphasized earlier- the FC bound samples all fluctuations, eigenmodes or not. As the results in fig 1e show, this analytical FC based approach predicts the behavior of the simulations quite well. 

 Even though the FC does not depend upon $T_i/T_e$, the free energy \emph{does} through the adiabatic electron contribution on the LHS of Eq(\ref{eq:two}). It is well known that large $T_i/T_e$ is stabilizing for the $ITG_{ae}$ and larger $T_i/T_e$  reduces the growth rate ($\gamma \sim \partial/\partial t$) and, consequently, has a strong affect upon the critical temperature gradient. 

All these effects are perfectly captured by simulations (fig 1e). Varying $T_i/T_e$ strongly alters the growth rate and the critical temperature gradient, but simulation stability boundary to $F_P$ is nearly independent of $T_i/T_e$ -another example of effects that impact the free energy equation but not the FC bound. 

For the SKIM dispersion relation for ETG, Debye length $\lambda_D$ enters as an additional term $\sim \lambda_D^2 k^2 \phi$. This term, just like large $T_i/T_e$, decrease free energy (the growth rate).  However, it causes no particle flux, so the $F_P$ bound is hardly affected( Fig 1f).  

In following sections, we will consider a more nuanced bound on the FC (derived from SKiM along similar lines ), and its comparison to simulations. This comparison is in many ways more striking evidence of the validity of the approach above. And it has more practical relevance to nonlinear simulation results. But first, we consider in more detail one of the basic conceptual issues raised by the results described above.

\section{Interpretations of FC bounds in conjunction with the simulation results: a statistical mechanical ansatz}

The criterion $F_P>0.6$ defines the absolute limit for stability.  There is nothing in the logic of the argument that implies that simulations must find stability only for $F_P$ \emph{close to} $0.6$. 

In fact, it is easily shown by solving either the 0D dispersion numerically, or the FC solubility numerically, that many values of $<k_{\perp}>$ , $<k_{\parallel}>$ and $<\omega_d>$ lead to stability or FC insolubility at $F_P$ values much less than $0.6$. And this can be true by a margin of $~50\%$ or more. Similarly, the simulation results for \emph{low} values of $k_y \rho_i$ modes attain stability for $F_P$  values that are significantly less than $0.6$ (described in the next section).

\emph{Nonetheless, the remarkable fact is that the simulations, when scanned over a wide range of $k_y \rho_i$, consistently find stability when $F_P \cong 0.6$.}

We believe that the likely explanation of this surprising simulation behavior must be found, at least partially, in the dynamic complexity of the gyrokinetic system. A system with large number of degrees of freedom, and a large number of eigenmodes, contrives  to find some mode that has values of $<k_{\perp}>$ , $<k_{\parallel}>$ and $<\omega_d>$ close to the ones which approach the absolute maximum $F_P = 0.6$. And possibly, the number of possible unstable eigenmodes is particularly large when the temperature gradients are high. 

"The system finds a way" to be maximally unstable.

The preceding statement ought to be viewed as a metaphor describing  a class of behavior displayed in many complex systems. Such systems often manifest "self-organization" that is widely acknowledged as behaving in an "adaptive" manner, that is, organized structures in the system autonomously respond to changing or challenging conditions- which in this case, would mean the fluctuations adapt to avoid extinction as the solubility bound is approached. "Adaptivity" is an emergent property that is widely observed in complex systems. 

This adaptivity may originate from the system's capacity to support large number of eigenmodes, with a wide range of  $<k_{\perp}>$ , $<k_{\parallel}>$ and $<\omega_d>$. Even if only a small fraction of the total number of modes have the requisite structure to approach the absolute bound on $F_P$, there will be some that do. 

As we proceed with our enquiry, we will attempt to organize our findings in a somewhat systematic manner.
This organization will be done under the banner of a what we call a "statistical mechanical ansatz": its first element is that \emph{gyrokinetic systems have apparent adaptivity- for any specified plasma conditions, there will be some class of fluctuations that will remain unstable until close to the upper bound on FC solubility.}

Upper bounds on stability, therefore, have not only the virtue of simplicity, but are very meaningful indicators of the overall stability of very complex systems. Like all descriptions in statistical mechanics, the current approach forgoes an attempt to describe details of the system (the specific eigenmodes), in favor of getting at the most important pertinent, general properties (the stability boundary). 

Our approach to gyrokinetic stability is quite different from the usual one where one attempts to solve complex eigenmode relations. It is novel, powerful, and far reaching.

In sections below, we will see that simulations show that the adaptability of gyrokinetic instabilites applies to more than just the FC. It also applies to energetic effects such as stabilizing curvature. So the ansatz above can be generalized to be more inclusive.

Gyrokinetic simulations, as in fig 1, are a test of this ansatz. Like many theories in statistical mechanics, it is very difficult to know \textit{a priori} whether a given particular system has enough degrees of freedom for the statistical description to be acceptably accurate. We have tested if the observed behavior follows the predictions pertinent to a complex system. We displayed some results in Fig.1, and we will present many more in future sections.

The preceding ansatz might also explain why it is that an approximate theory such as the SKiM model is able to predict the behavior of realistic simulations so accurately. The bounds derived from SKiM agree with the stability in gyrokinetic simulations to roughly $< 10\%$ accuracy. This is considerably more accurate than the typical level of agreement of SKiM predicted growth rates with simulation growth rates using Eq(\ref{eq:kpar})-Eq(\ref{eq:omdi}) and the dispersion relation Eq.~(\ref{eq:SKiMdr}).  Thus the utility of SKIM to predict the characteristics of a specific eigenmode is limited; it, however, excels in predicting the  general (independent of the nature of fluctuations, magnetic geometry, and detailed plasma parameters) phenomenon like the upper bound on FC solubility. The SKIM upper bounds are in excellent agreement with the simulations results in Fig1, calculated, for example, for complex magnetic geometries. Some additional analysis on this subject are given in Appendix B. 

The FC will shortly be derived and examined under more specific conditions; it will no longer be universal. Agreements with simulations, however,  becomes even more striking, and reinforce the validity of these concepts.

\section{The Flux Constraint for low k modes, and implications for nonlinear simulations}

%\begin{figure*}
%\subfloat[\label{sfig:2a}]{%
%  \includegraphics[width=.56\linewidth]{Picture2a.png}%
%}\hfill
%\subfloat[\label{sfig:2b}]{%
%  \includegraphics[width=.43\linewidth]{Picture2c.png}%
%}

\begin{figure*}
\subfloat[]{%
  \includegraphics[width=.56\linewidth]{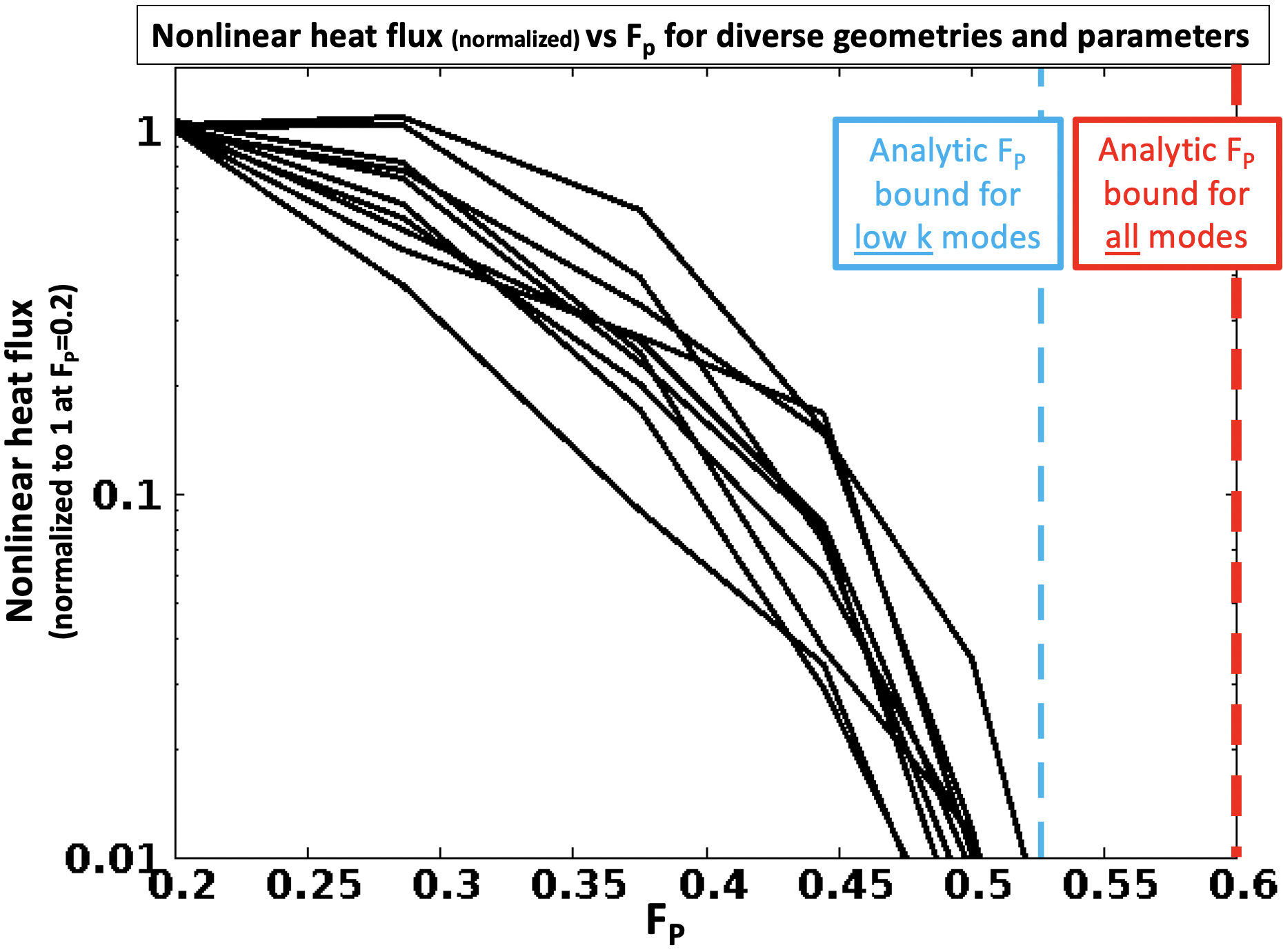}%
}\hfill
\subfloat[]{%
  \includegraphics[width=.43\linewidth]{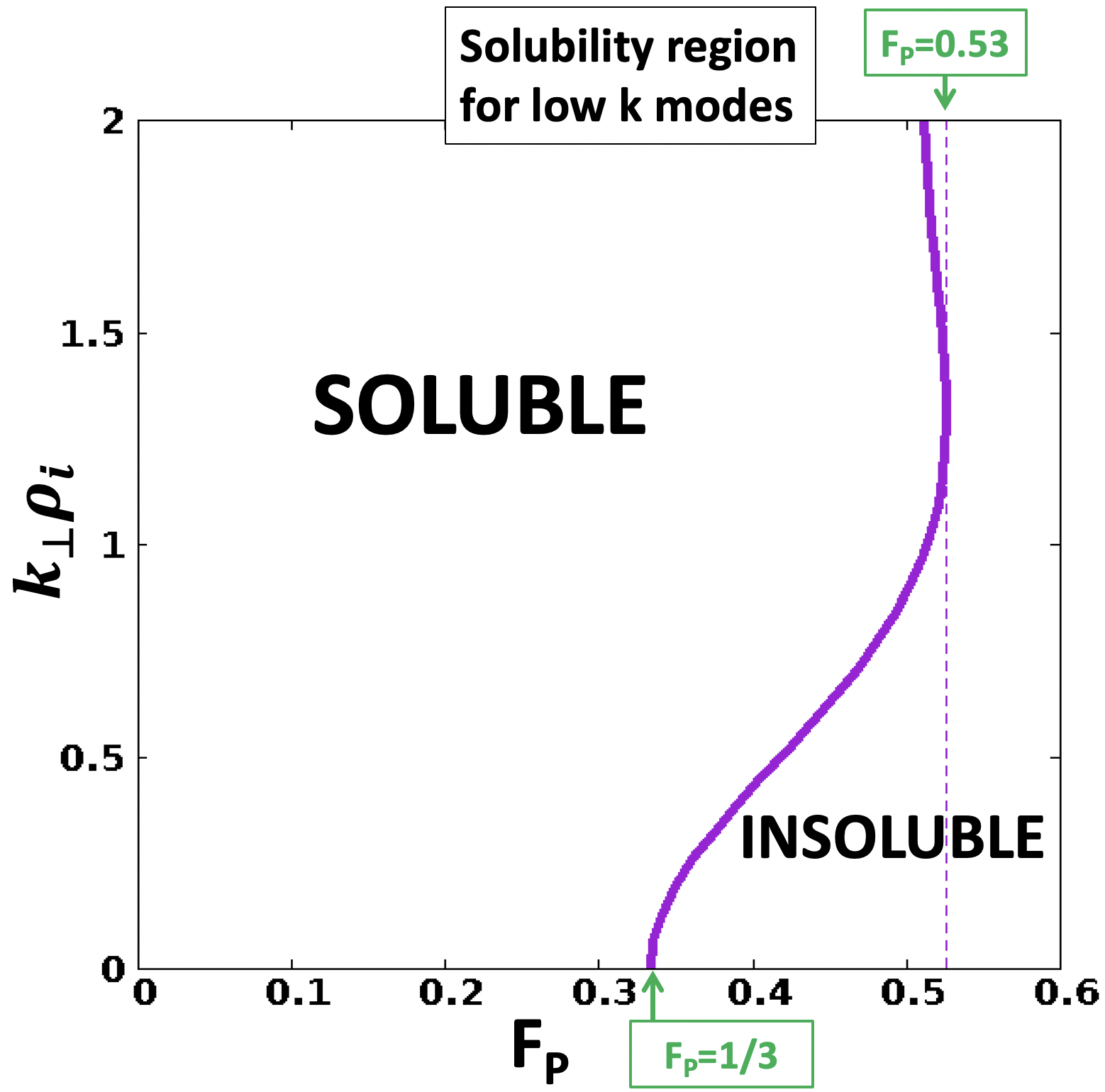}%
}
\caption{\label{fig:two} a) Heat flux vs $F_P$ for ten diverse nonlinear simulations  b) Plot of the soluble and insoluble regions of the FC for low k modes: the$F_P$ solubility limit depends upon $k_\perp \rho_i$}

\end{figure*}

We begin this section with several nonlinear simulation results for the $ITG_{ae}$, for diverse geometries and parameters. This will motivate the need to create a framework specifically for low $k$ modes. This will uncover arguable the most extraordinary consequence of the flux constraint. 

\begin{figure*}
\subfloat[]{%
  \includegraphics[width=.61\linewidth]{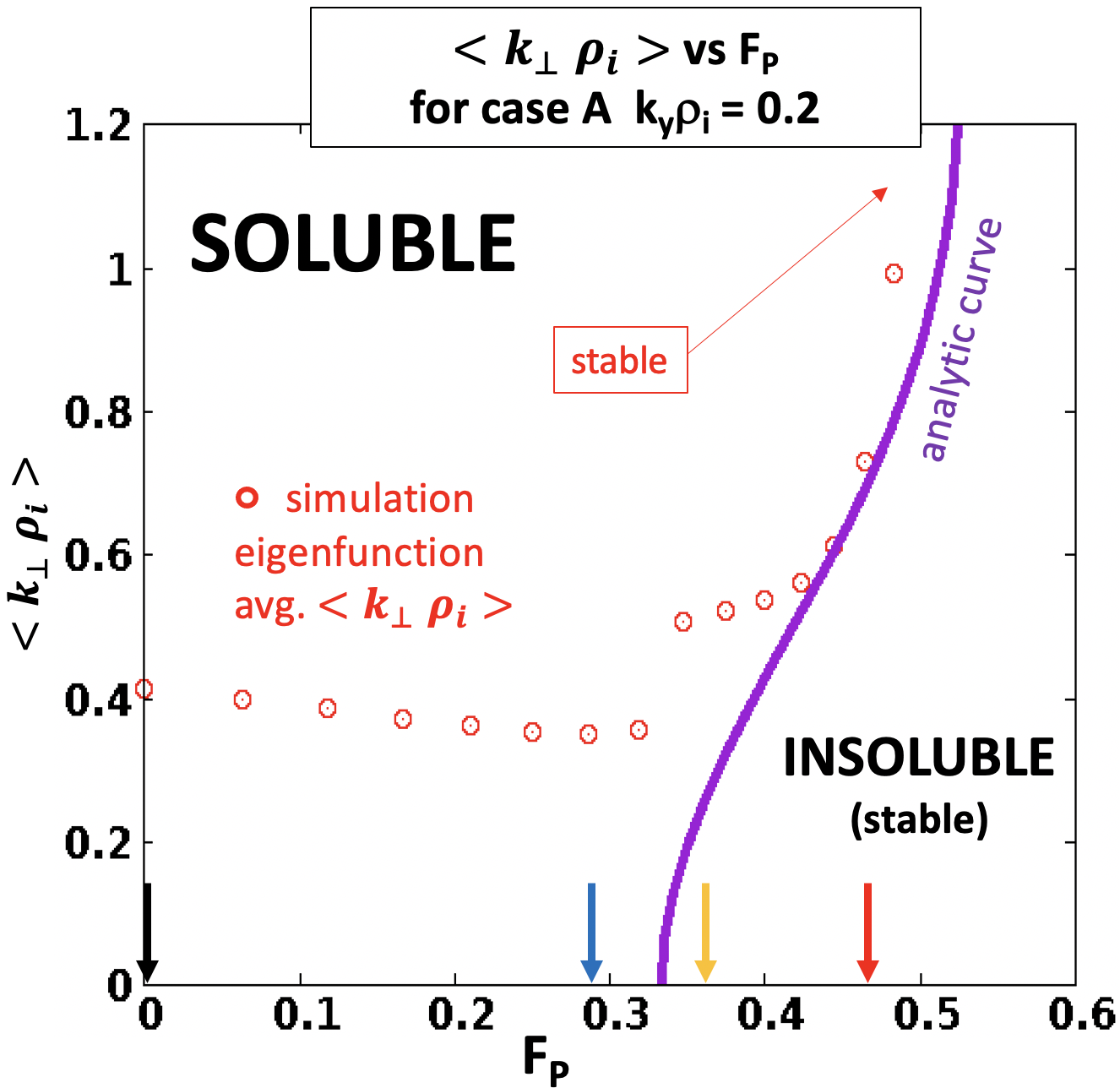}%
}\hfill
\subfloat[]{%
  \includegraphics[width=.39\linewidth]{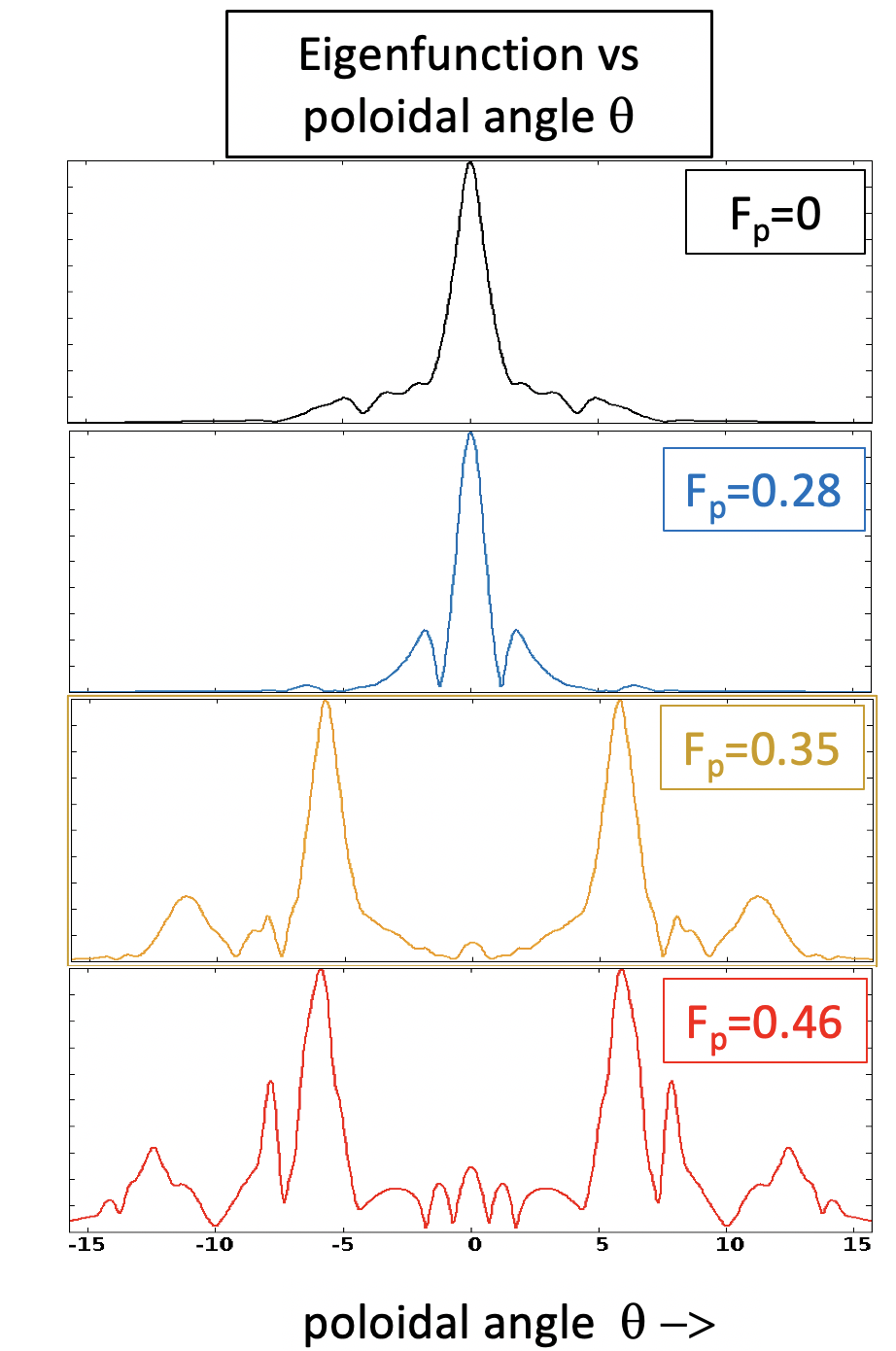}%
}

\caption{\label{fig:three} A) The solubility regions together with the simulation results for the $<k_{\perp} \rho_i>$ from the eigenfunction. The unstable eigenfunctions track the analytically computed boundary to stay in the soluble region, by increasing $<k_{\perp} \rho_i>$ as the boundary is approached B) The eigenfunction for various $F_P$ values. To increase the $<k_{\perp} \rho_i>$, the eigenfunctions progressively broaden to larger $\theta$ }
\end{figure*}

\begin{figure*}
\subfloat[]{%
  \includegraphics[width=.5\linewidth]{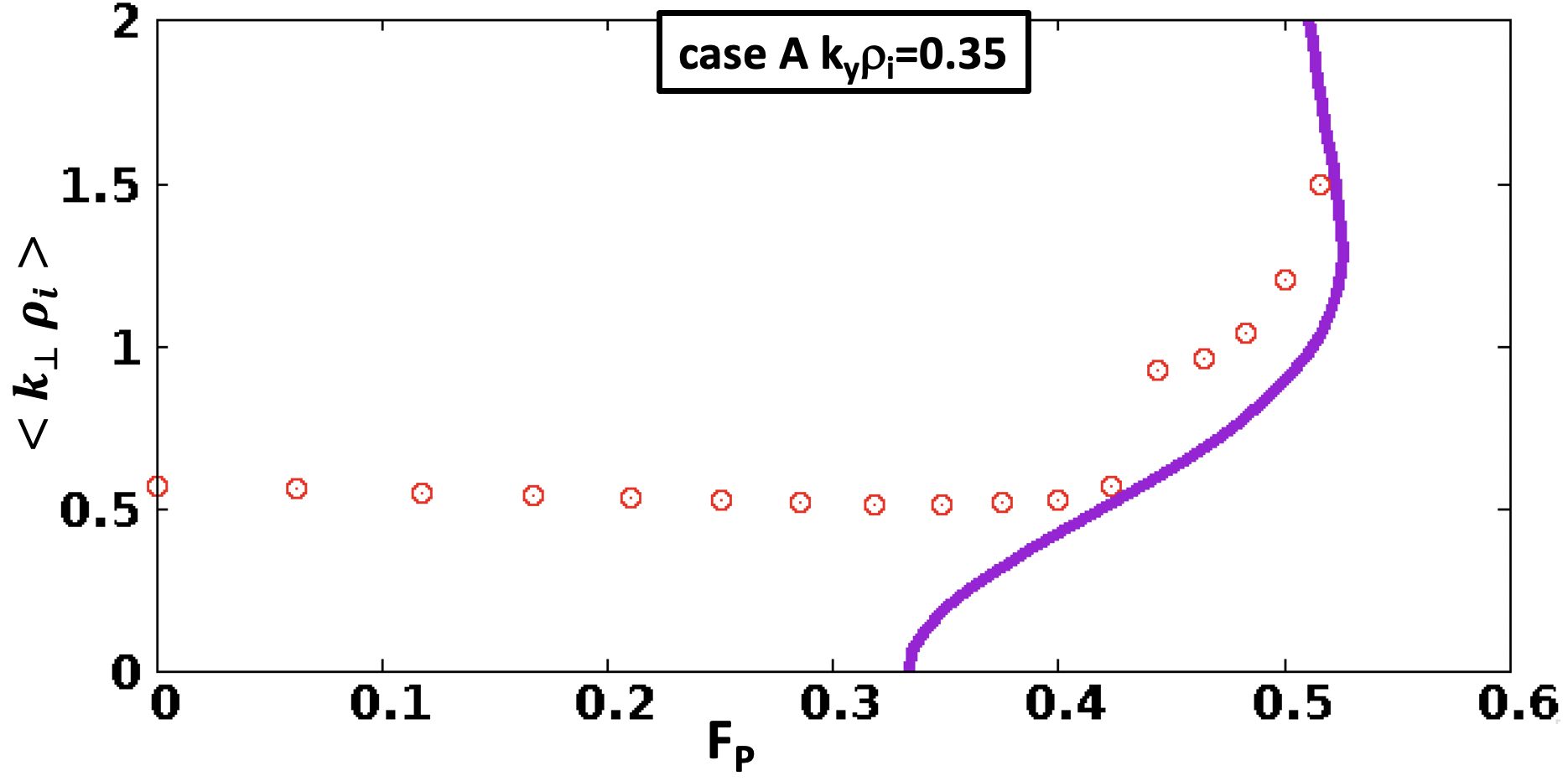}%
}\hfill
\subfloat[]{%
  \includegraphics[width=.5\linewidth]{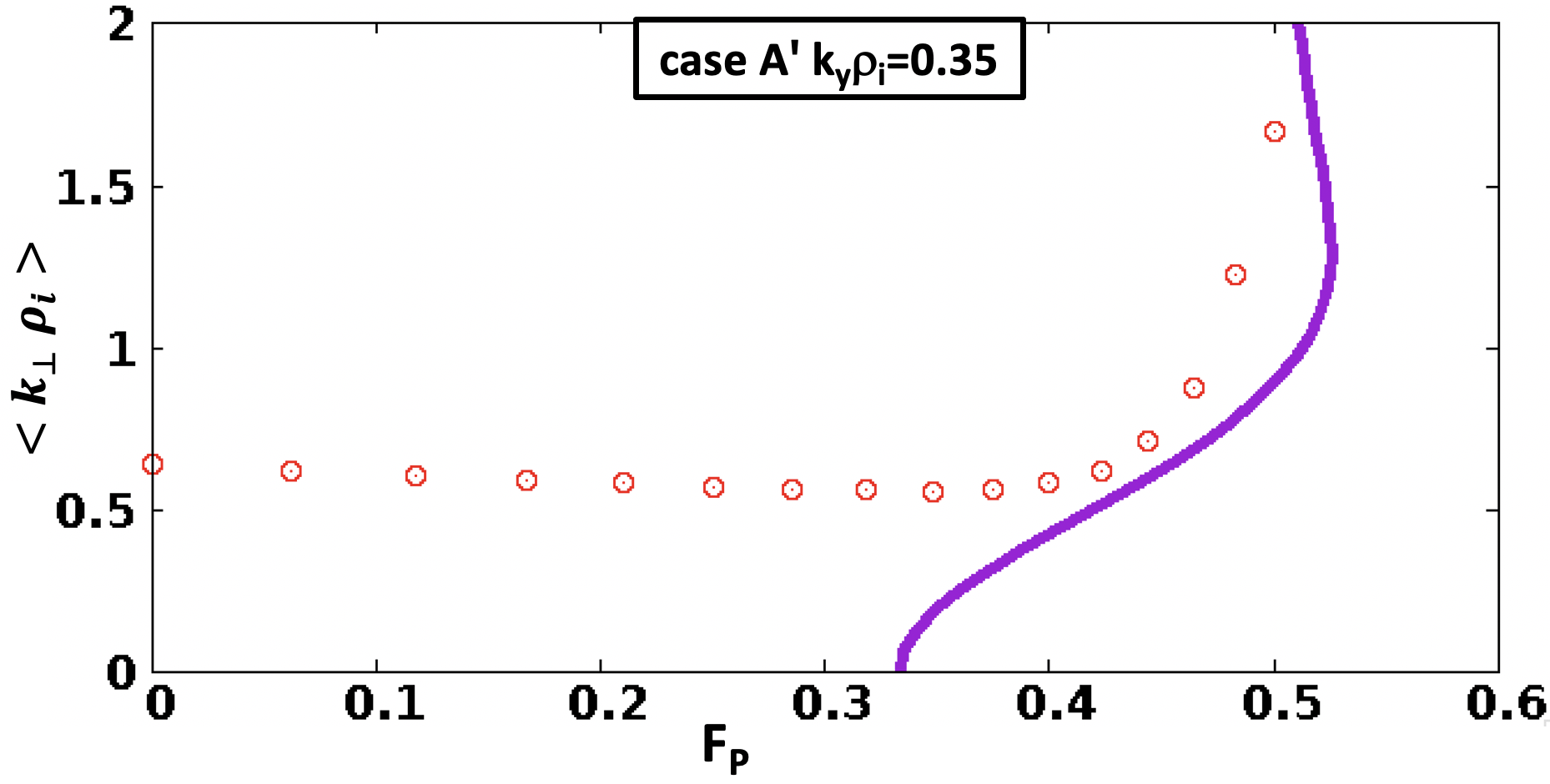}%
}
\vfill
\subfloat[]{%
  \includegraphics[width=.5\linewidth]{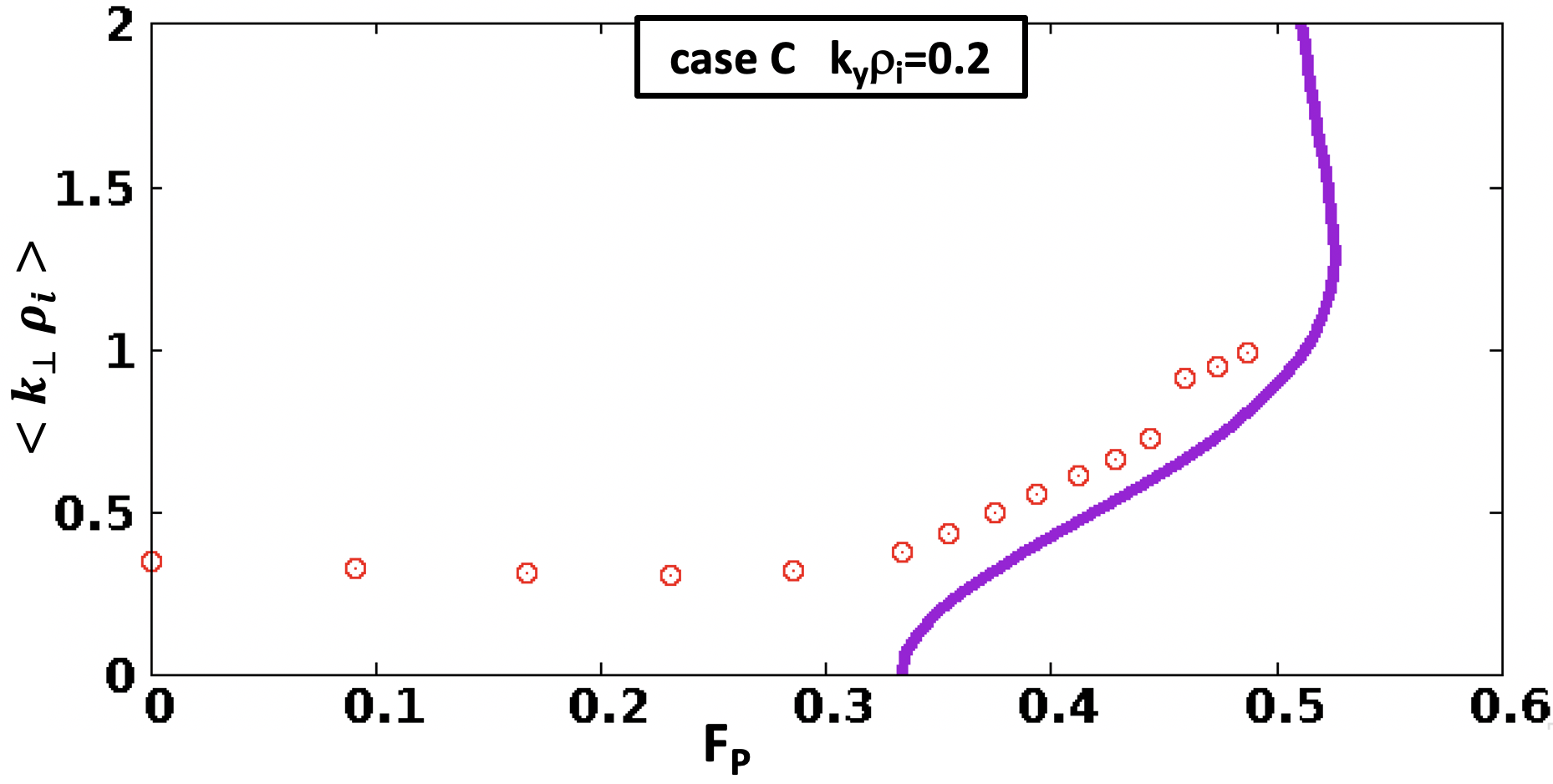}%
}\hfill
\subfloat[]{%
  \includegraphics[width=.5\linewidth]{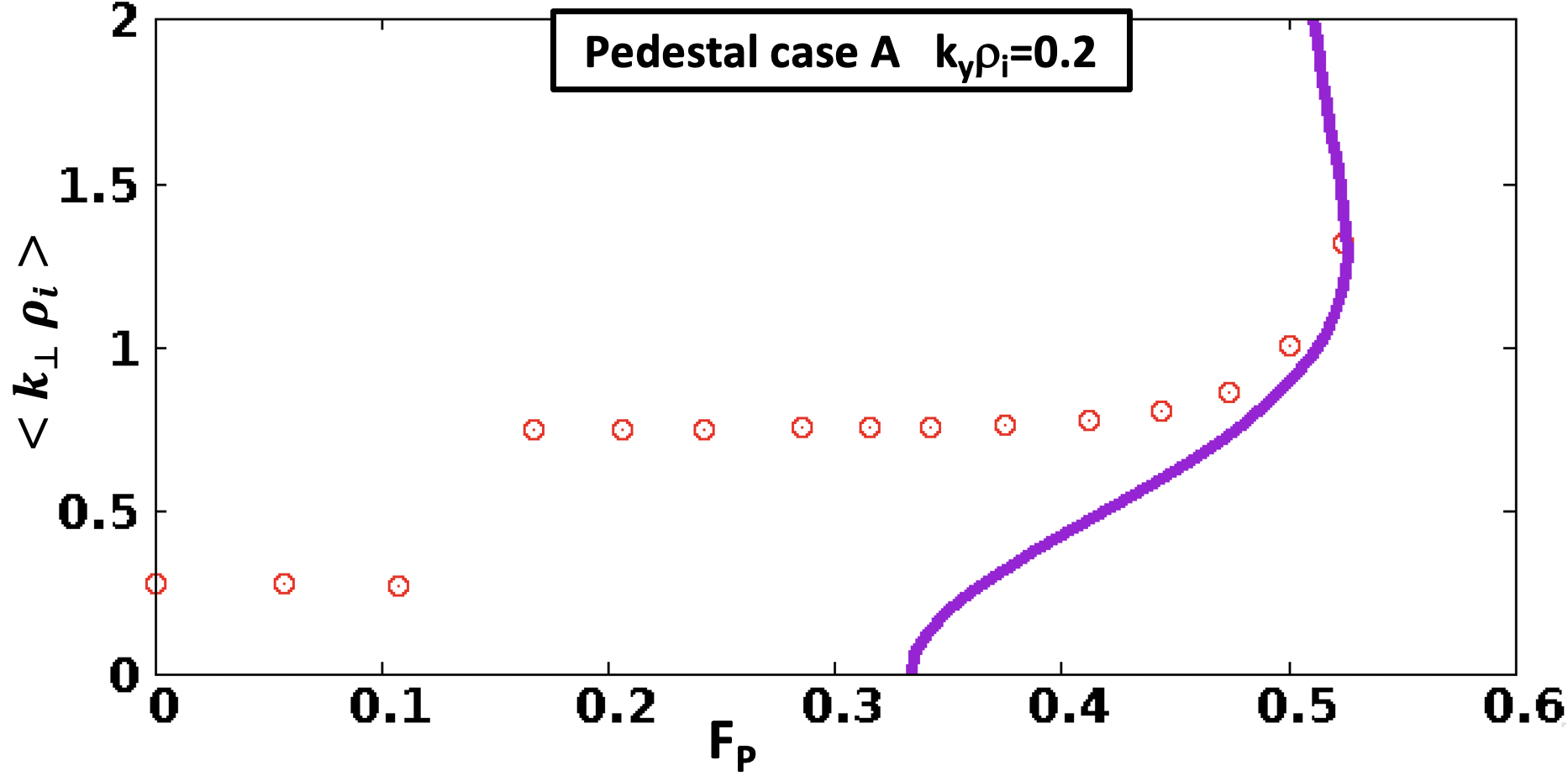}%
}
\caption{\label{fig:kpFp1} $<k_{\perp} \rho_i>$ from simulation eigenfunction vs $F_P$ for representative cases. The unstable eigenfunctions track the analytically computed boundary to stay in the soluble region, by increasing $<k_{\perp} \rho_i>$ as the boundary is approached }
\end{figure*}

\begin{figure*}
\subfloat[]{%
  \includegraphics[width=.5\linewidth]{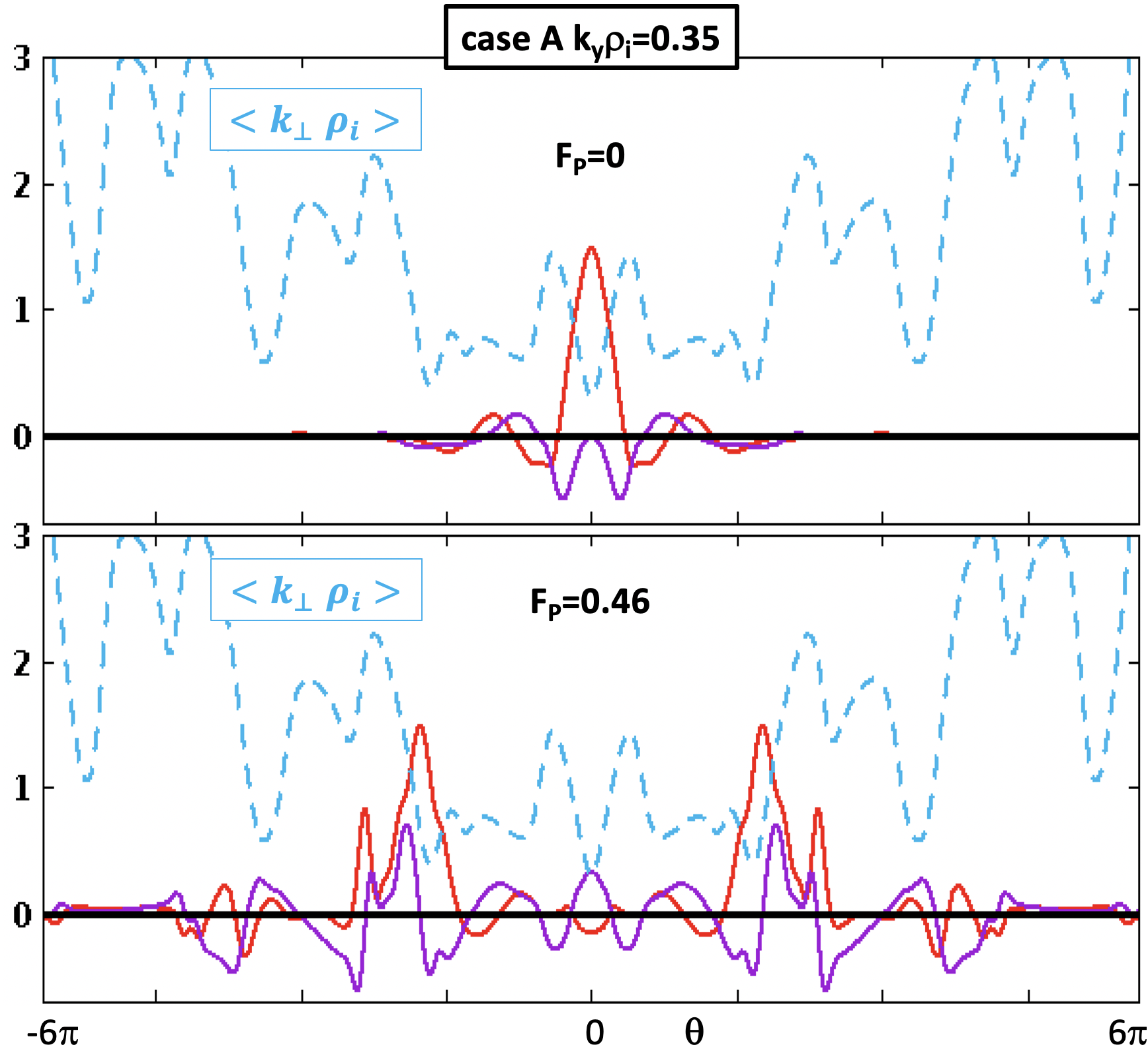}%
}\hfill
\subfloat[]{%
  \includegraphics[width=.5\linewidth]{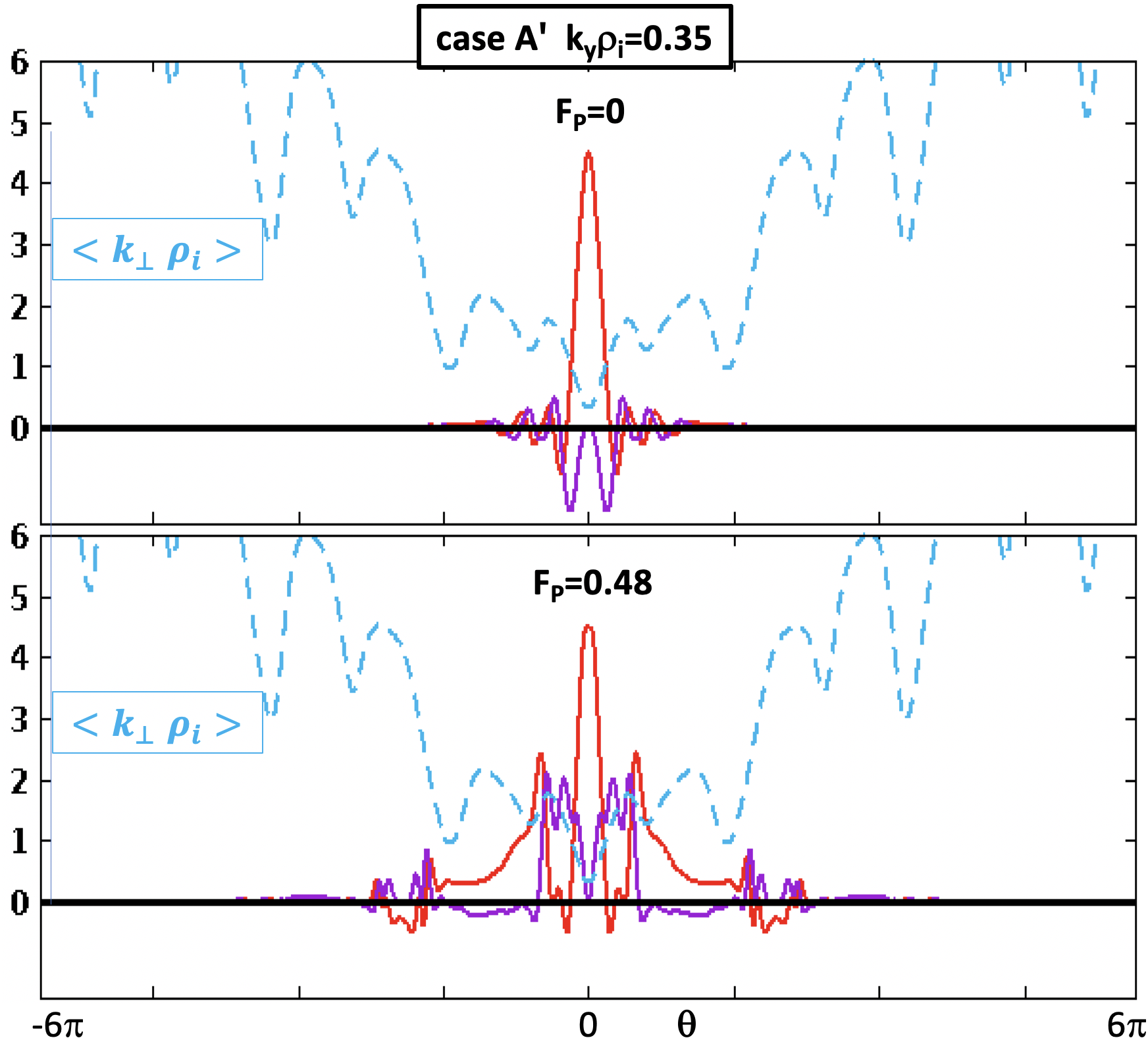}%
}
\vfill
\subfloat[]{%
  \includegraphics[width=.5\linewidth]{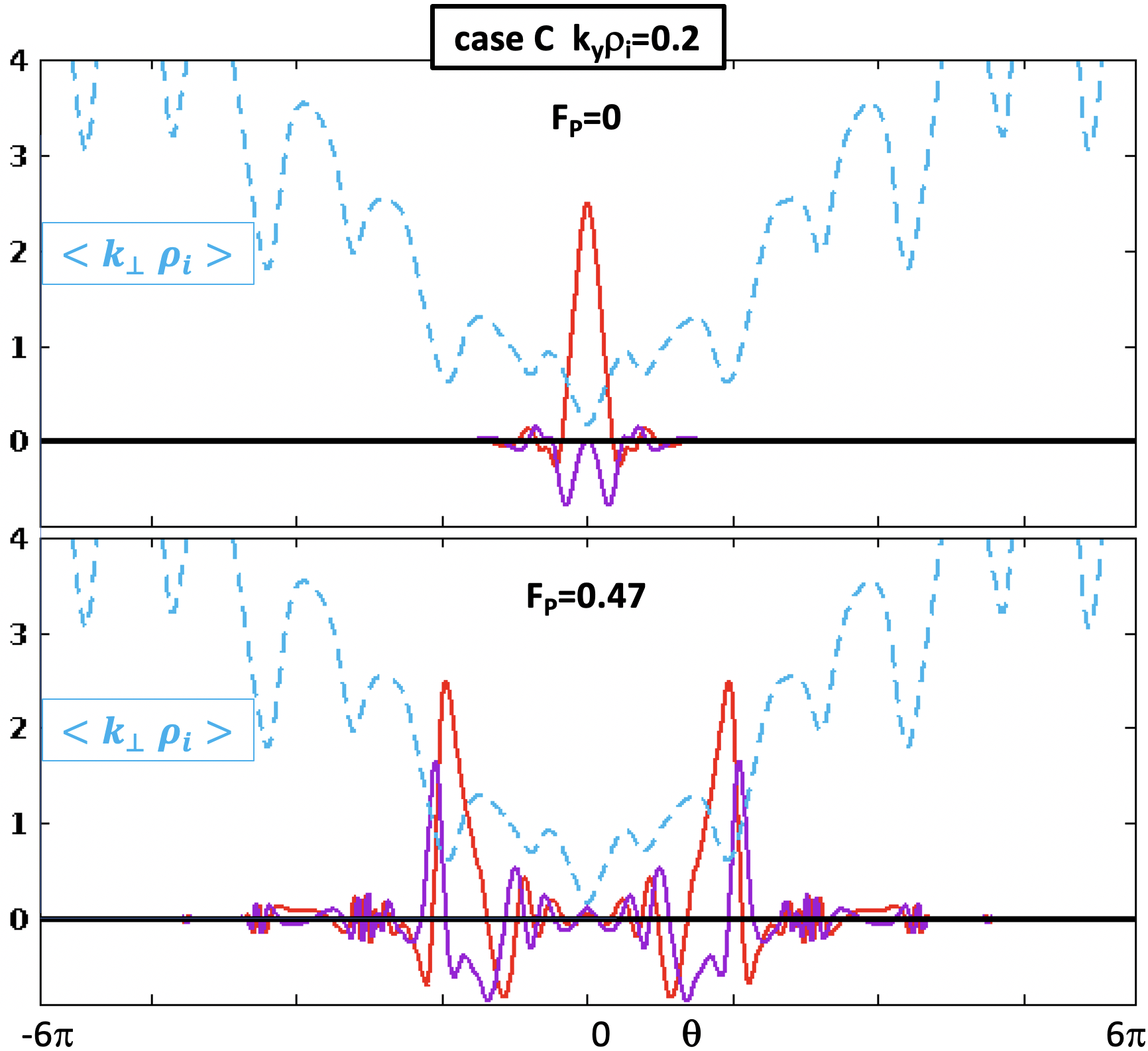}%
}\hfill
\subfloat[]{%
  \includegraphics[width=.5\linewidth]{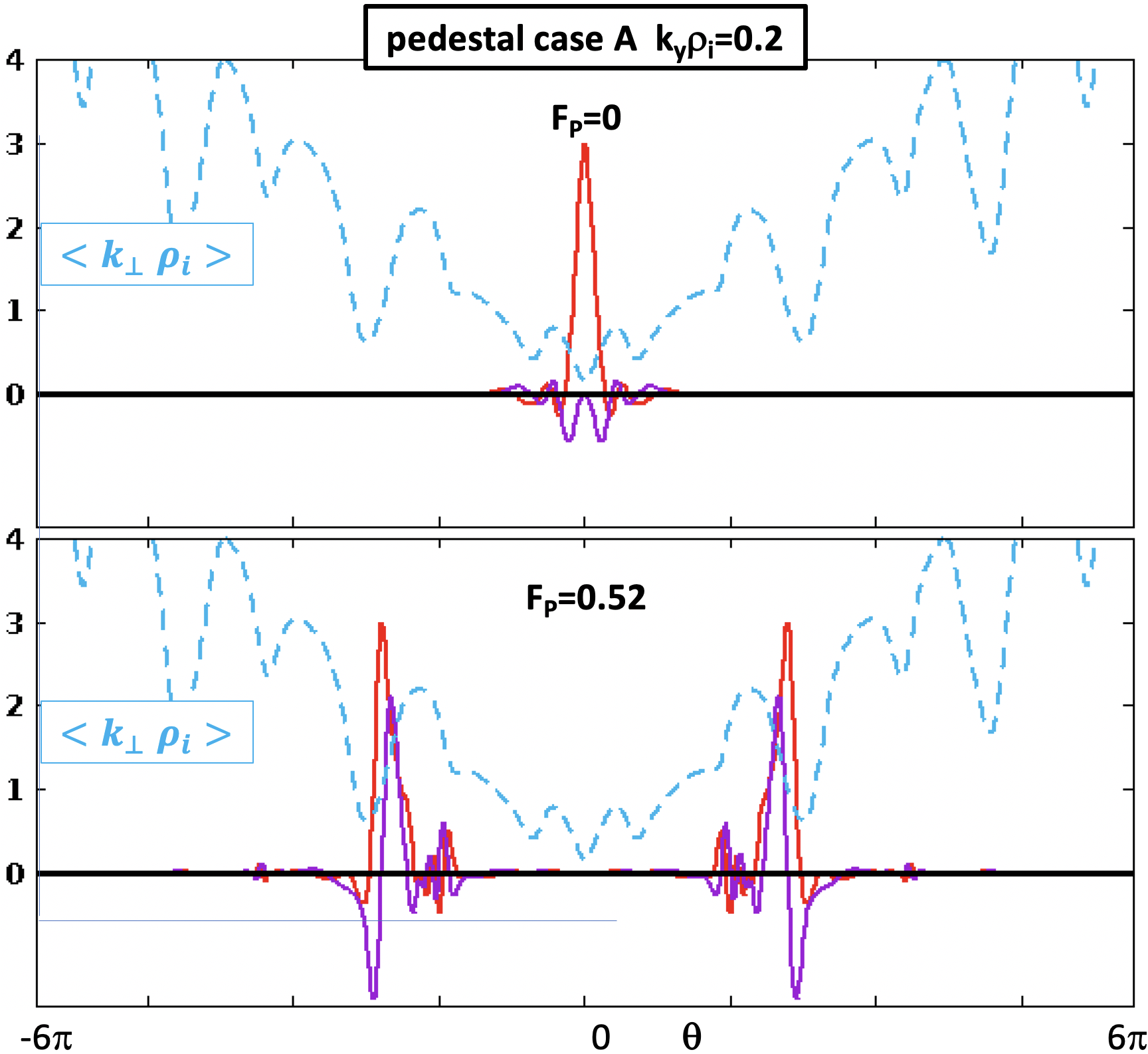}%
}
\caption{\label{fig:kpphi1} The simulation eigenfunctions for the four cases in fig(\ref{fig:kpFp1}) for $F_P=0$ and $F_P$ near the stability limit. The value of $<k_{\perp} \rho_i>(\theta)$ for the geometry is also shown. Near the solubility limit, the eigenfunction structure shifts to reside in regions of larger  $<k_{\perp} \rho_i>$. The eigenfunction structure is often quite complex, and differs strongly from case to case. Nonetheless, the eigenfunction averaged $<k_{\perp} \rho_i>(\theta)$ tracks the analytic expression.}
\end{figure*}

\begin{figure*}
\subfloat[]{%
  \includegraphics[width=.33\linewidth]{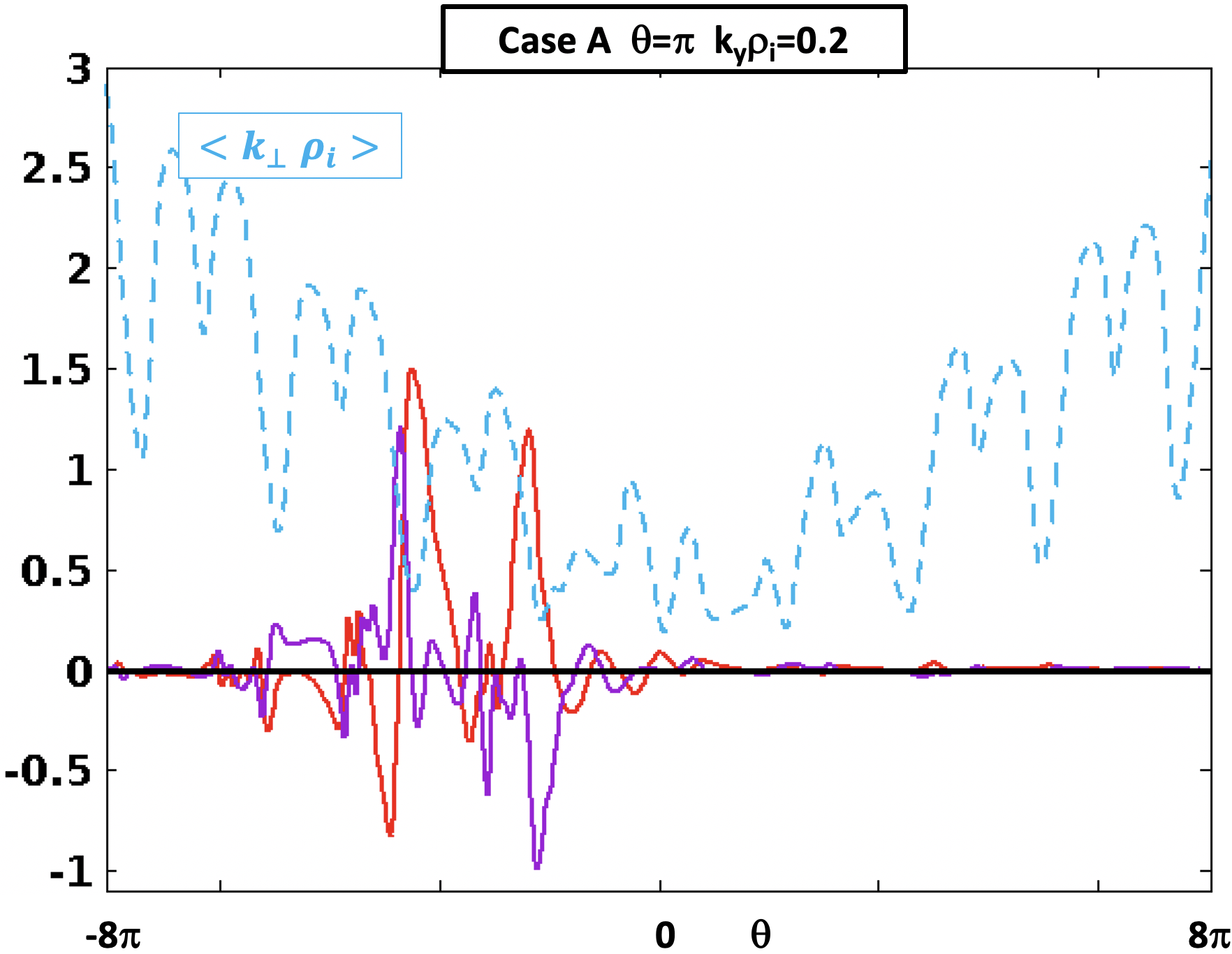}%
}\hfill
\subfloat[]{%
  \includegraphics[width=.33\linewidth]{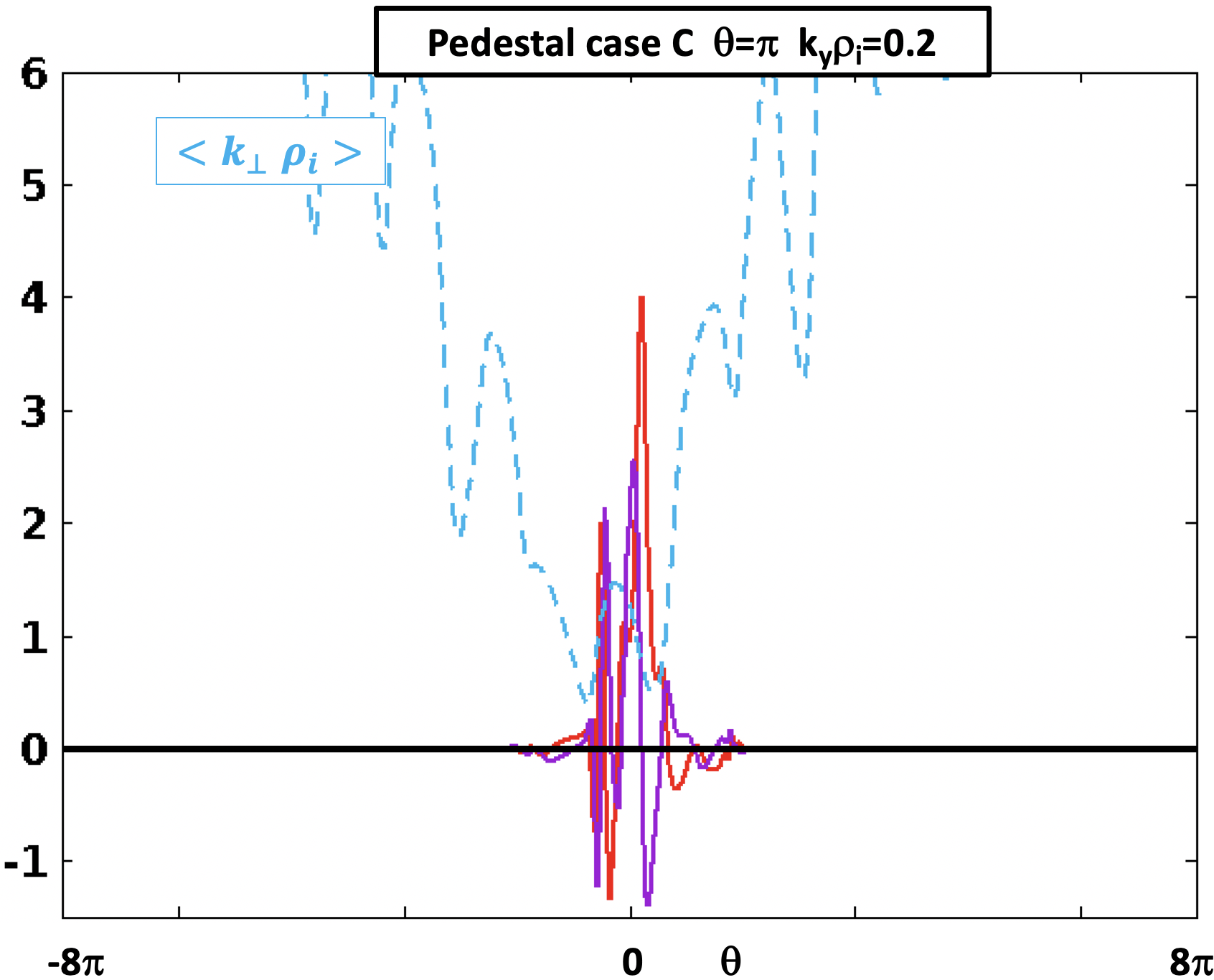}%
}
\hfill
\subfloat[]{%
  \includegraphics[width=.33\linewidth]{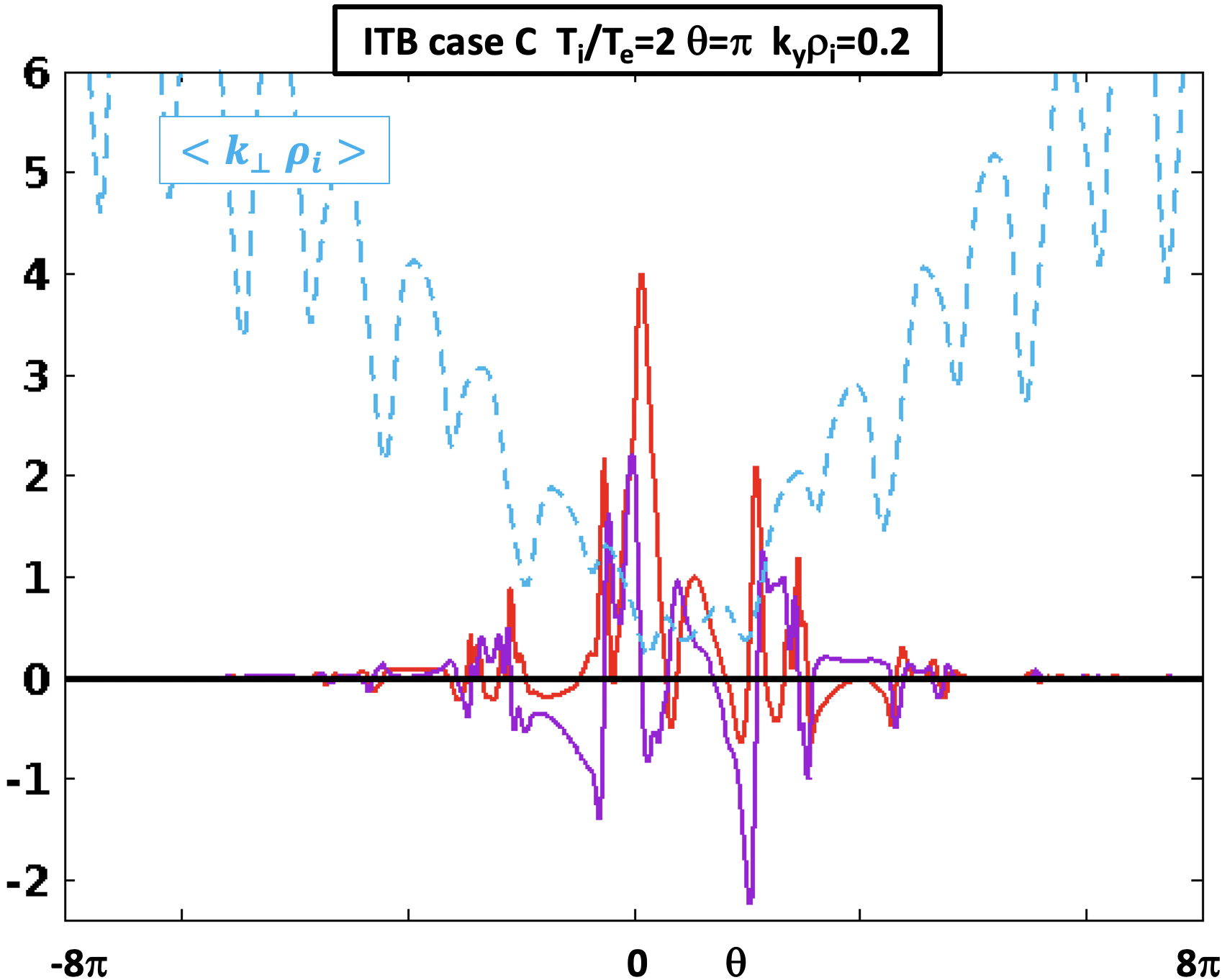}%
}\vfill
\subfloat[]{%
  \includegraphics[width=.33\linewidth]{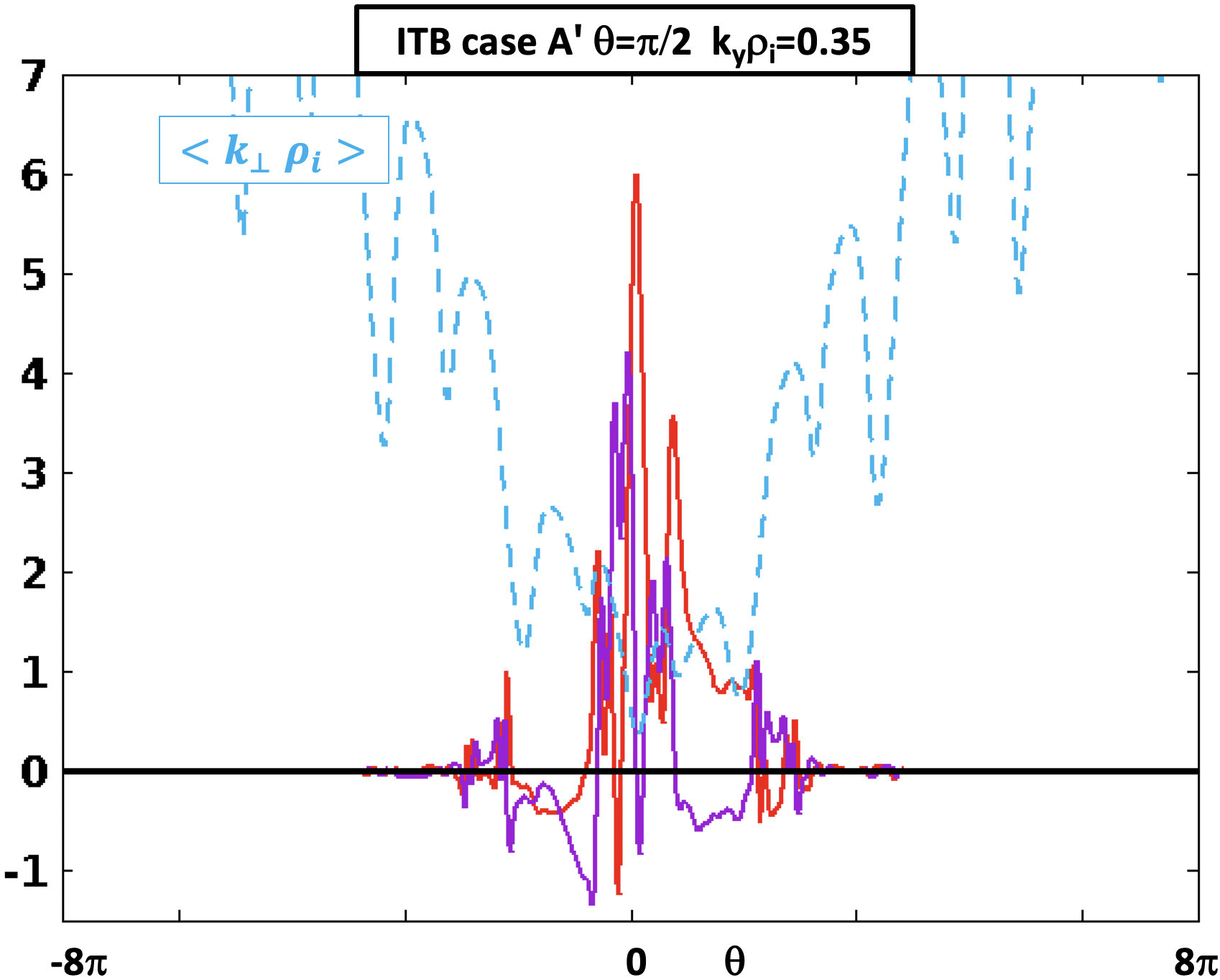}%
}
\hfill
\subfloat[]{%
  \includegraphics[width=.33\linewidth]{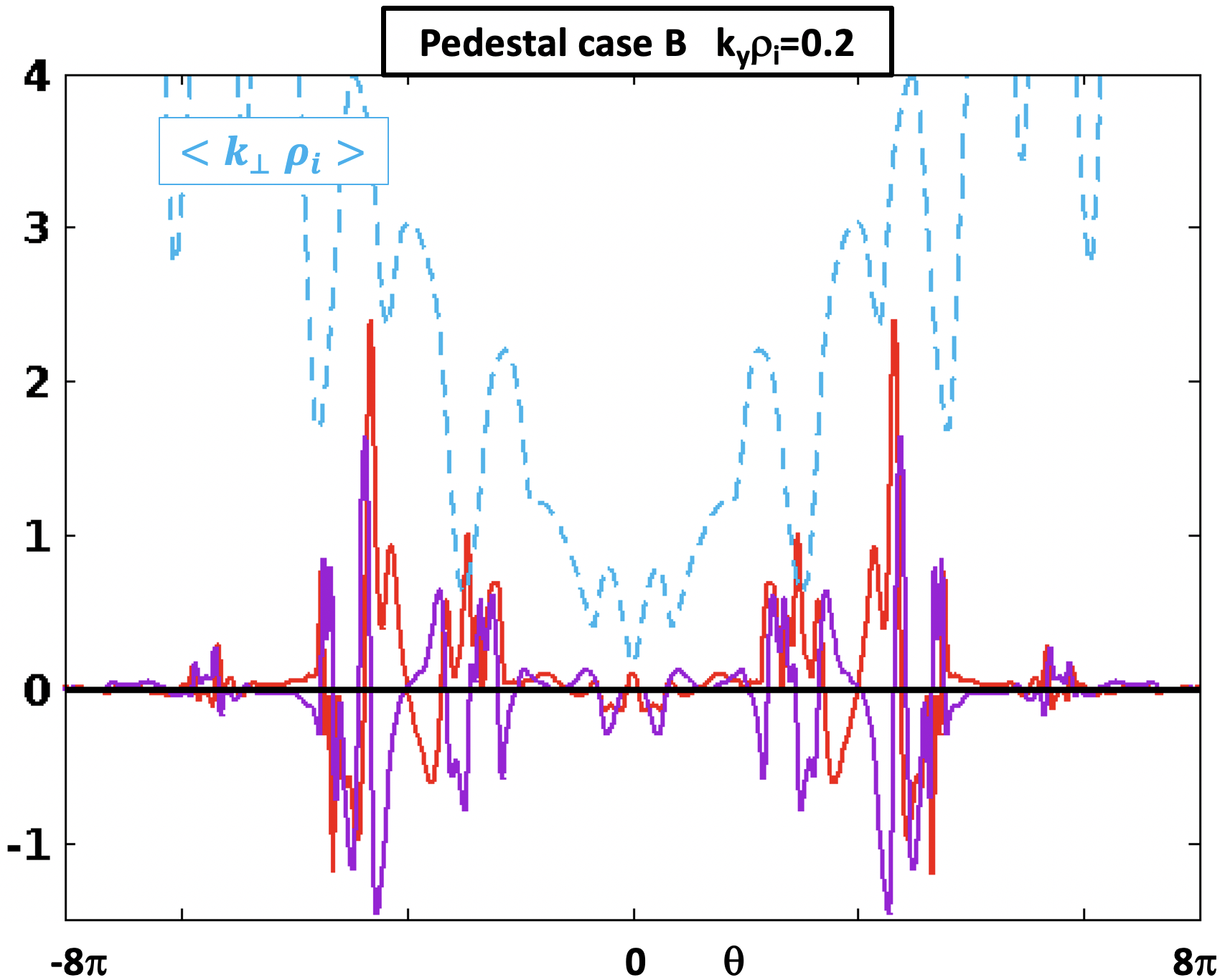}%
}\hfill
\subfloat[]{%
  \includegraphics[width=.33\linewidth]{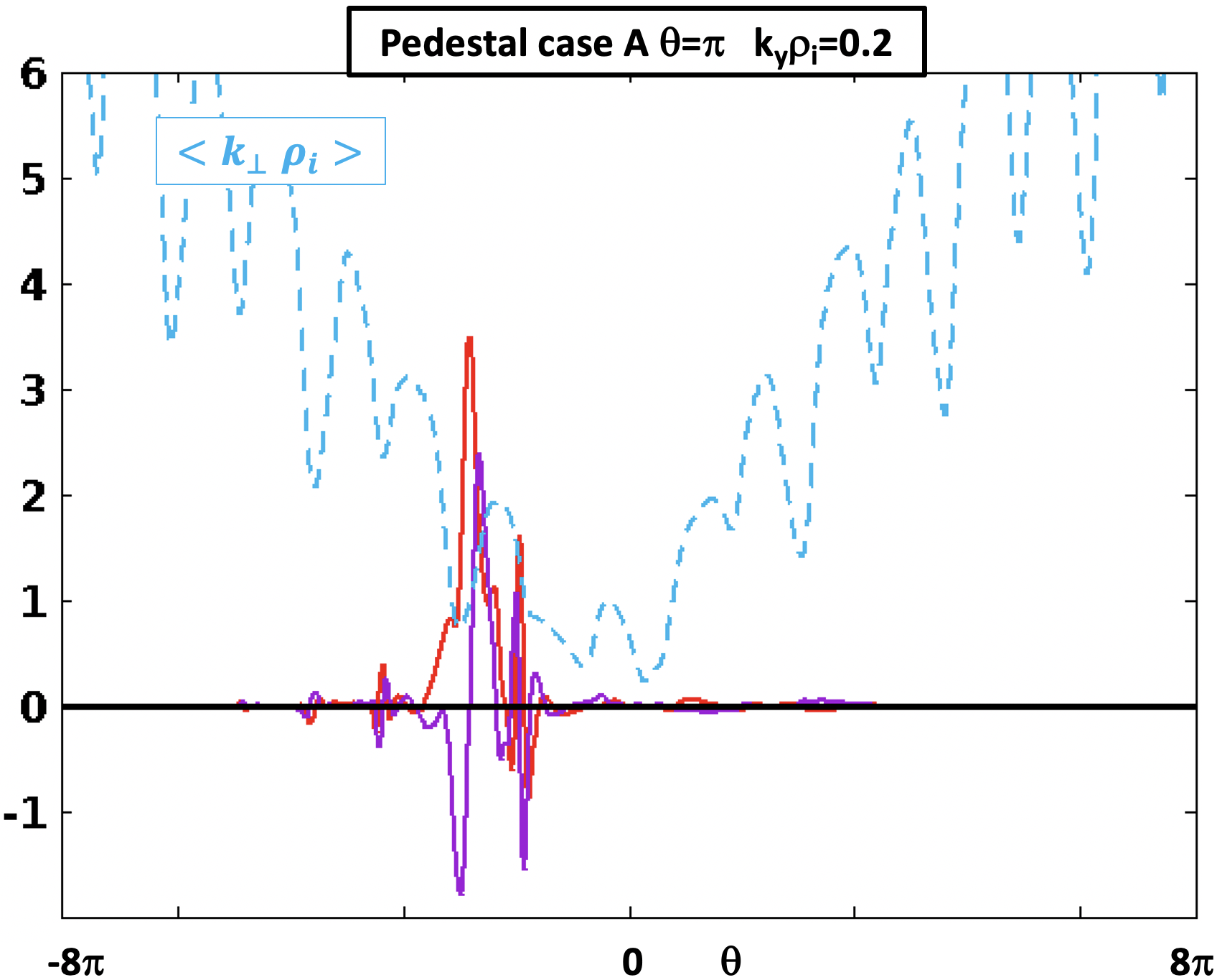}%
}
\caption{\label{fig:kpphi2} Simulation eigenfunction for representative cases, for $F_P$ near the stability limit. We include cases where $k_x$ is not zero on the outer midplane, i.e., $\eta_0$ is not zero. The value of $<k_{\perp} \rho_i>(\theta)$ for the geometry is also shown. Near the solubility limit, the eigenfunction structure shifts to reside in regions of larger  $<k_{\perp} \rho_i>$. The eigenfunction structure is often quite complex, and differs strongly from case to case. Nonetheless, the eigenfunction averaged $<k_{\perp} \rho_i>(\theta)$ tracks the analytic expression.}
\end{figure*}

In fig \ref{fig:three}a, we plot the nonlinear heat flux (normalized to its value at $F_P=0.2$) verses $F_P$; nonlinear simulations were carried out for very diverse geometries and parameters,

\begin{itemize}
\item Using Miller geometry, we vary safety factor $q$ ($1.4 \le q \le 4$), $\hat{s}$ ($-0.5 \le \hat{s} \le 2$),  Shafranov parameter $\alpha$ ($0.18 \le \alpha \le 6$), and local inverse aspect ratio $\epsilon$ ($0.1 \le \epsilon \le 0.3$).

\item Gradients vary substantially as well ($15\le R/L_T \le 50$ and $1 \le T_i/T_e \le 3$).) 

\item There is an almost "universal" behavior for these cases: The nonlinear heat flux for all cases falls by about two orders of magnitude for  $F_P \sim 0.5$. This is significantly before $F_P = 0.6$. 

\item The drop in heat flux with $F_P$ is much faster than the growth rates; in the examples in fig 1, the maximum growth rates typically fell by only a factor $\sim 2$, at $F_P \sim 0.5$, not the two orders of magnitude found in nonlinear simulations.

\item The nonlinear drop is very close to an analytic bound for solubility of the FC that we will derive for \emph{low $k_y$} modes. 

\end{itemize}

We now turn our attention to understanding this behavior.

It is well known that nonlinear heat fluxes for $ITG_{ae}$ are dominated by lower $k_y \rho_i$ fluctuations. In the next section, we compute an analytical bound for solubility of the FC for this class of fluctuations. \emph{The new analytic bound gives a good approximation for the $F_P$ value at which flux drops 2 orders of magnitude or more.}

A reduction in the heat flux by about two orders of magnitude makes a TB possible, even if the flux is not zero. This new FC bound implies that a lower density gradient may be enough for a TB. When impurity effects are included, this bound extends to considerably lower $F_P$ yet, i.e, even less density gradient is needed. With experimentally relevant impurity levels, in fact, a substantial proportion of observed TBs satisfy this new bound. Hence TBs are possible, at least insofar as $ITG_{ae}$ is concerned, even without velocity shear; insolubility of the FC assures it.

This linear bound for solubility of the FC for low $k$ modes is very relevant.      

We will derive this bound from SKiM in detail in next section: we start with Eq(\ref{eq:FC0}), make a particular expansion, and perform simple and straightforward manipulations. In this section, we simply describe it and give extensive comparisons to simulations. Notice that though the maximal bound $F_P =0.6$ was "universal'', the new bound depends upon $<k_{\perp} \rho_i>$. Thus, it depends upon the mode structure of the fluctuation. 

This new FC bound can be read from fig(\ref{fig:two}b) where regions of solubility are demarcated in the $<k_{\perp} \rho_i>$- $F_P$ plot. The FC  solvability range depends upon the eigenfunction average of  $<k_{\perp} \rho_i>$. For low values of $<k_{\perp} \rho_i>$, stability results when $F_P > 1/3$. The maximum soluble $F_P\cong 0.53$ occurs at $<k_{\perp} \rho_i> \cong 1.3$. (In all these cases, $Z_{eff}=1$; the impurity modifications will be described in later sections).

\emph{The behavior of low $k_y \rho_i$ eigenmodes as $F_P$ exceeds $\sim 1/3$ is a dramatic example of adaptivity in the gyrokinetic system.} Consider an example for ITB geometry in figure 3a. The parameters are similar to a JET campaign with slightly reversed shear, and pellets to increase density gradients. It is also similar to some high $\beta_{poloidal}$ cases on DIII-D with beam fueling. The eigenmode shape dramatically changes as $F_P$ is increased, so that the eigenfunction average $<k_\perp \rho_i>$ is increased to stay in the soluble (= unstable) range, computed analytically. The simulation eigenfunction has $<k_\perp \rho_i>$ that clearly tracks the analytic bound. The change in the shape of the eigenfunction are shown in Fig 3b. Initially, the eigenfunction is quite conventional: a "bump" localized at the outboard midplane. It keeps this basic shape at $F_P = 0.28$, that is, less than $1/3$. But as $F_P$ exceeds this, the eigenfunction broadens. At $F_P = 0.38$, it has considerable amplitude at higher $\theta$, where $k_{\perp}$ is larger. In the conventional ballooning representation used here, higher $k_\perp$ occurs at higher $\theta$. In other words, the eigenfunction is evolving in the way it must in order to remain in the region of soluble FC. At higher $F_P$, the mode must have even higher $<k_\perp \rho_i>$ to remain unstable. We see that at $F_P=0.46$, the eigenfunction has even higher amplitude at high $\theta$, and the $<k_\perp \rho_i>$ is still in the soluble region. Increasing $F_P$ beyond $0.46$ results in stability for this case. However, before becoming stable, the eigenfunction has undergone enormous evolution to remain in the soluble region of the FC, by increasing the eigenfunction average $<k_\perp \rho_i>$.

\emph{We have found that this qualitative behavior is essentially universal for $ITG_{ae}$. This is the most definitive behavior that demonstrates that FC solubility is a controlling influence on the mode dynamics as $F_P$ is increased}. There are multiple crucial ramifications of this result, which we describe in this section. Extensive simulations support them.

Firstly, we describe how this phenomenon provides an explanation for the issue we began this chapter with: the roughly universal behavior of nonlinear simulations of the $ITG_{ae}$ as $F_P$ is increased, in diverse geometries and parameters. 

The eigenfunction behavior where $k_{\perp}$ increases is expected to strongly influence the nonlinear heat flux. An increase in $k_{\perp}$ implies a reduction in the size of turbulent eddies. The turbulent diffusion has a smaller "step size" which should reduce the transport. Also, increases in $k$ increase the nonlinear coupling to stable modes in the system, and this should decrease the turbulent amplitudes. This should also reduce the turbulent heat flux. A common dimensional estimate of the heat diffusivity from linear eigenmodes is the "mixing length" rule $D_{mix}=\chi \sim \gamma/ k_{\perp}^2$. The new feature of the FC bound in fig(\ref{fig:three}) is that $D_{mix}$ should be reduced by increases in the denominator. We'll compare $D_{mix}$ to the nonlinear simulation $\chi$ below, and their behavior is quite similar.

So as $F_P$ increases beyond $1/3$, either the numerator vanishes ($\gamma \rightarrow 0$) or the denominator increases substantially or a combination of these. In view of fig(\ref{fig:three}b), this behavior due to the FC should be universal.  The nonlinear results follow precisely this pattern, despite huge changes in parameters between the simulation cases. For all cases, the turbulent flux has sharp reduction as $F_P$ increases through the range from $1/3 -0.53$, as predicted by the analytic FC solubility boundary. And finally, at $F_P=0.53$, the low $k_y$ modes become stable, and the heat flux is very low. Hence, the character of the nonlinear results can be understood as a "universal" consequence of the FC for low k modes. 

Hence, simulations indicate that the FC bound for low $k$ modes has considerable nonlinear significance. Let us turn to other remarkable consequences. 

In this and following sections, we employ simulations in diverse geometries and parameters. These are summarized in the table below. 

\begin{center}
  \includegraphics[width=.95\linewidth]{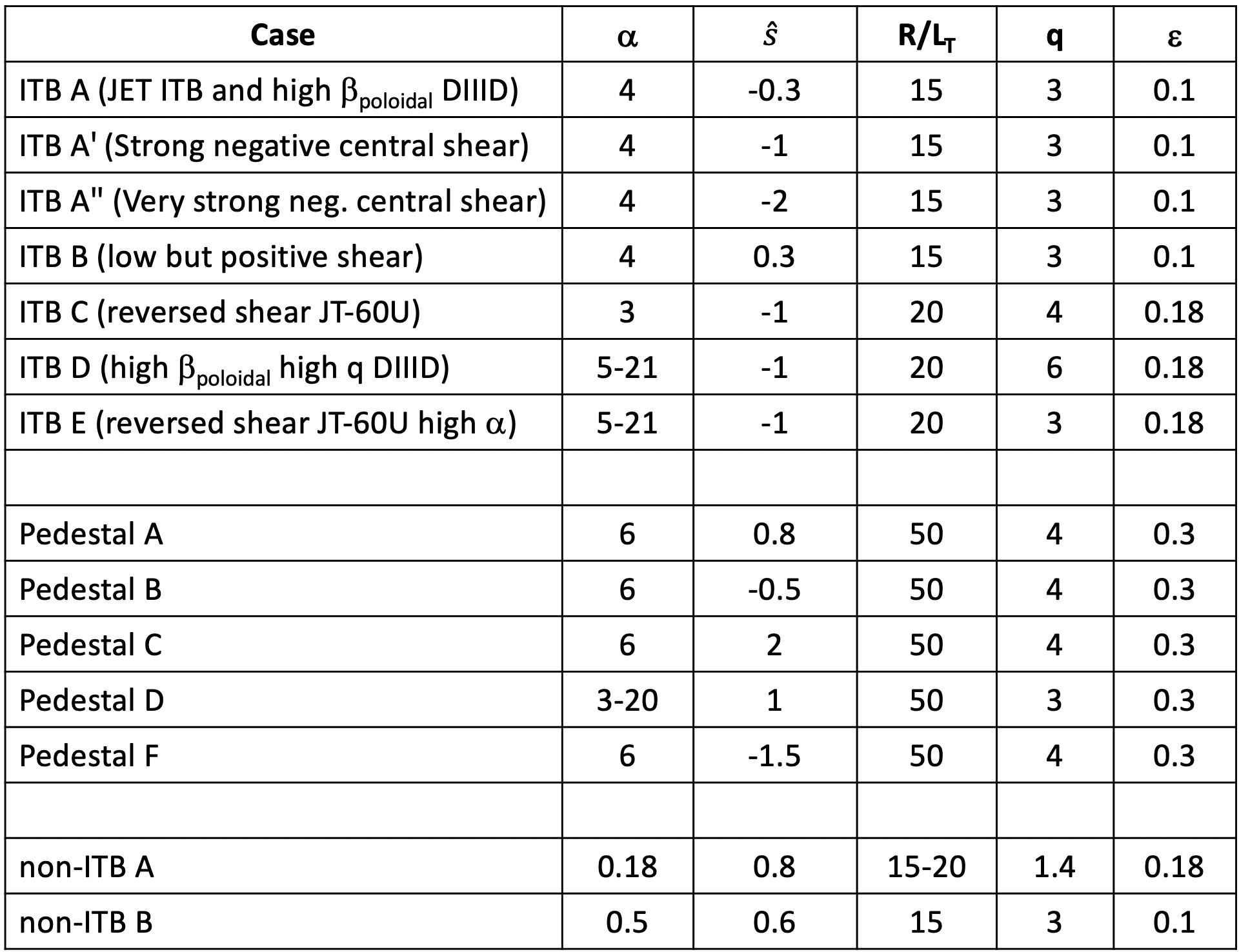}\\
Table 1: Parameters used for the simulations
\end{center}

In the four diverse cases in fig(\ref{fig:kpFp1}), the $<k_{\perp} \rho_i>$ from all the unstable simulation eigenfunctions stays in the soluble region. There are two important aspect of these results: 1) Most obviously, the modes become stable as the analytic boundary is approached. This is qualitatively like the results in fig(\ref{fig:kpFp1}),  2)  as the boundary is approached, the eigenfunction structure changes very strongly to increase $<k_{\perp} \rho_i>$. The eigenfunctions are given in fig(\ref{fig:kpphi1}), for $F_P=0$ and for $F_P$ very close to the stability limit found in the simulations. 

The cases with $F_P=0$ have a very familiar structure: a bump on the outboard midplane,  in the region of "bad" curvature. This also is in a region of low $k_{\perp} (\theta)$. For $F_P$ near the solubility boundary, the structure changes radically, and is concentrated in a region of much larger $k_{\perp} (\theta)$. 

\emph{Note that each case has eigenfunctions that differ substantially in detail near the stability limit. Nonetheless, the eigenfunction average $<k_{\perp} \rho_i>$ tracks the analytic curve quite well.}

 Let us take stock of what we have learnt:

\begin{itemize}

\item     \emph{The eigenfunction undergoes very severe structural modifications, that allows it to stay in the soluble region.} 
\item      There is absolutely no apparent reason for such extreme changes, \emph{other than to stay in the soluble region of the FC}
\item 	If it was not for the analytic solubility curve and the eigenfunction average $<k_{\perp} \rho_i>$, these modifications would \emph{seem inexplicable and even random.}
\item 	At high $F_P$, \emph{each geometry's eigenfunctions are both quite complex and uniquely different.} The only common factor among them \emph{is that their $<k_{\perp} \rho_i>$ is rather near the analytic solubility boundary for the $F_P$ value}.
\item.      This "adaptive" behavior is highly reminiscent of diverse examples of simulations of complex systems with many degrees of freedom, where \emph{the adaptivity is an emergent behavior that would not be predicated, and, the specifics of the adaptation are nearly impossible to predict ahead of time. }

\end{itemize}

And perhaps most important of all, the eigenfunctions are incomprehensibly complex to understand on their own. But their behavior becomes coherent in a reduced description in terms of eigenfunction averages. This is the hallmark of  a system when a statistical mechanical approach is justified. The analogy is with a gas. A description that accounts for individual particle orbits is too complex; descriptions that are much more useful are in terms of quantities that average over such details, (e.g. pressure and temperature). Although we are not aware of such a qualitative argument being made for eigenfunctions in gyrokinetics before now, the present results clearly have such a character.

The FC calculation has brought into focus the adaptivity of the mode structure to keep the mode unstable. We are \emph{only} able to detect this property because the analytically-derived FC solubility, now, depends upon the eigenmode structure, compelling changes to that structure for the mode to remain viable. The observed  changes in the complexity/ detailed structure of eigenmode would defy understanding if it were not as an adaptation to preserve the FC solubility.

We will see in following sections, that adaptivity has important ramifications of its own, independent of the FC.  

In view of the importance of this property, we probe it more thoroughly by including an additional element. All the figures for linear modes up until now have been for $k_x=0$, or in different terminology for ballooning coordinates with their particular periodicity ($k_x=k_y \hat{s} \eta_0$), for $\eta_0 =0$. With additional degrees of freedom added via finite $k_x$, we would expect some $\eta_0$ to be able to approach the solubility limit more closely. Actually, there are two contrasting regions: For cases where $F_P \sim 0$, finite $k_x$ modes are almost always more stable than $k_x=0$. However, as the analytic FC bound in fig(\ref{fig:three}b) is approached, we find that some modes with $k_x \neq 0$ can be more unstable, and solubility limit can be approached more closely.

In the following, for each configuration and $F_P$ value, we perform simulations for $\eta_0 =0$, $\pi/2$ and $\pi $. Among these, we plot the value of $<k_{\perp} \rho_i>$ that is closest to the bound, or, below it.  If it is in fact below the curve, that is, it shows a disagreement with the analytic bound. So plotting the lowest $<k_{\perp} \rho_i>$ from multiple $\eta_0$ is a more stringent test of the accuracy of that bound. 

Results are plotted in fig(\ref{fig:four}) for multiple cases. The minimum value of $<k_{\perp} \rho_i>$ is usually distinctly closer to the boundary between soluble and insoluble regions for the FC.  In a few cases the point penetrate very slightly into the insoluble region. One must keep in mind that the analytically computed expression is an expansion to lowest order in a parameter, so there are some corrections to it. The results strongly support the analytic bound. 

In  fig(\ref{fig:four}), we have labeled points as hollow circles when their $D_{mix} < 2\%$ of the maximum $D_{mix}$ for that scan. Such low $D_{mix}$ occur only when the   $k_{\perp}$ has increased by several fold. Most of those points that extend slightly into the analytic insoluble region have very low  $D_{mix}$, so they are quite feeble modes.

For the cases in fig(\ref{fig:four}), the overall character of the eigenfunctions just before stabilization fits the description above. Representative eigenfunctions are shown in  fig(\ref{fig:kpphi2}). All cases have $<k_{\perp} \rho_i>$ quite close to the analytic boundary (including for $k_x \neq 0$), but their mode structure is too complex to understand otherwise. 

\begin{figure*}
\subfloat[]{%
  \includegraphics[width=.33\linewidth]{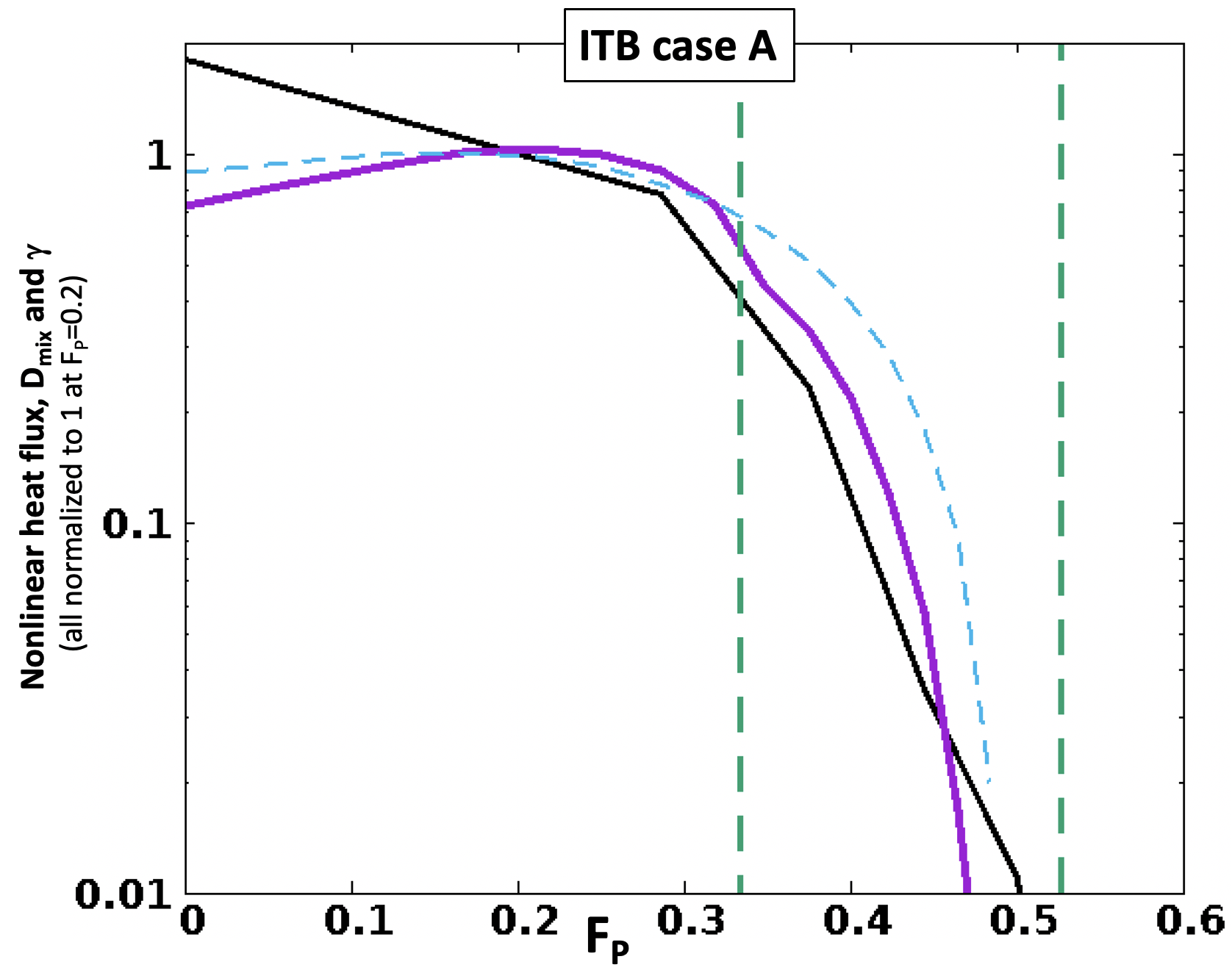}%
}\hfill
\subfloat[]{%
  \includegraphics[width=.33\linewidth]{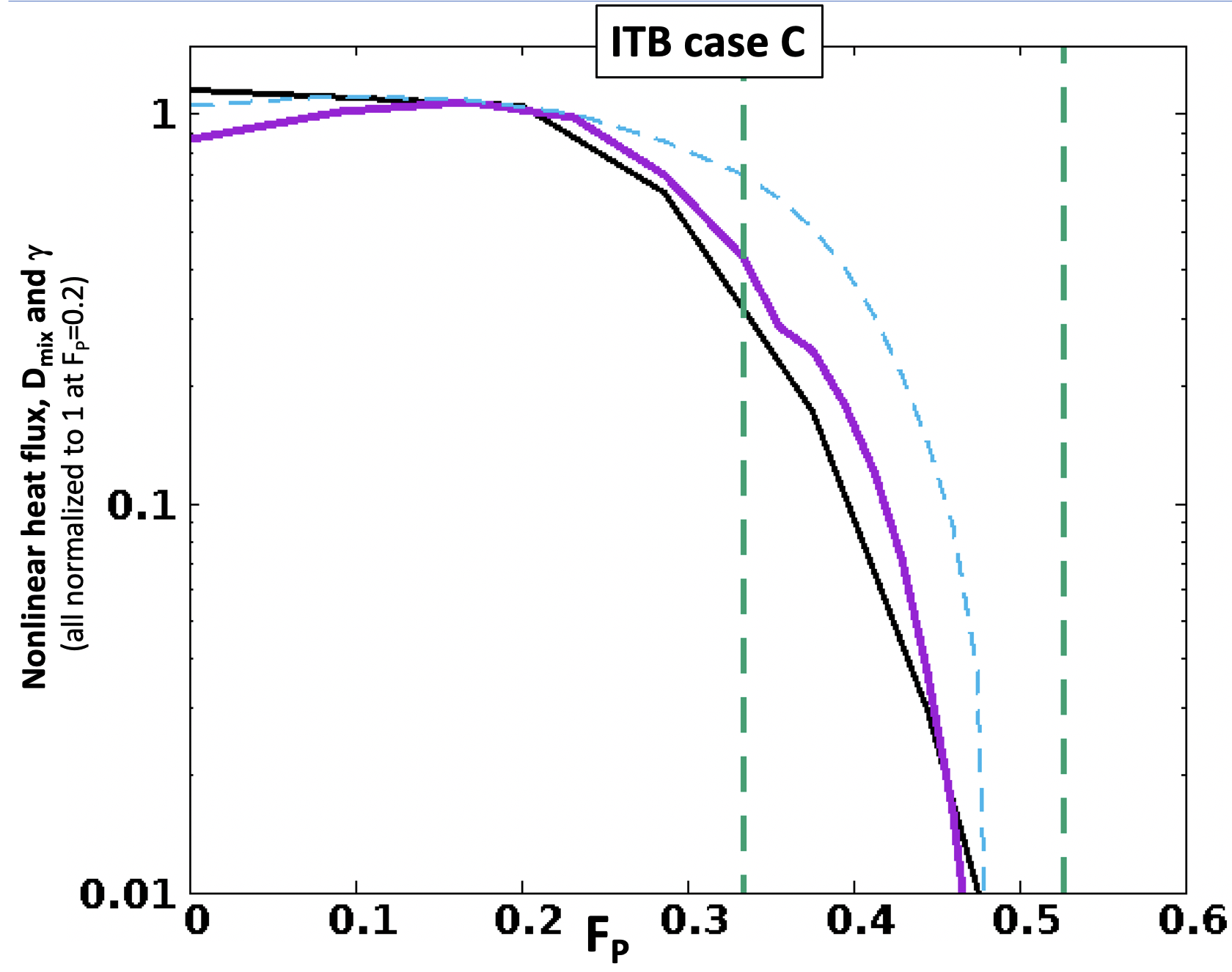}%
}
\hfill
\subfloat[]{%
  \includegraphics[width=.33\linewidth]{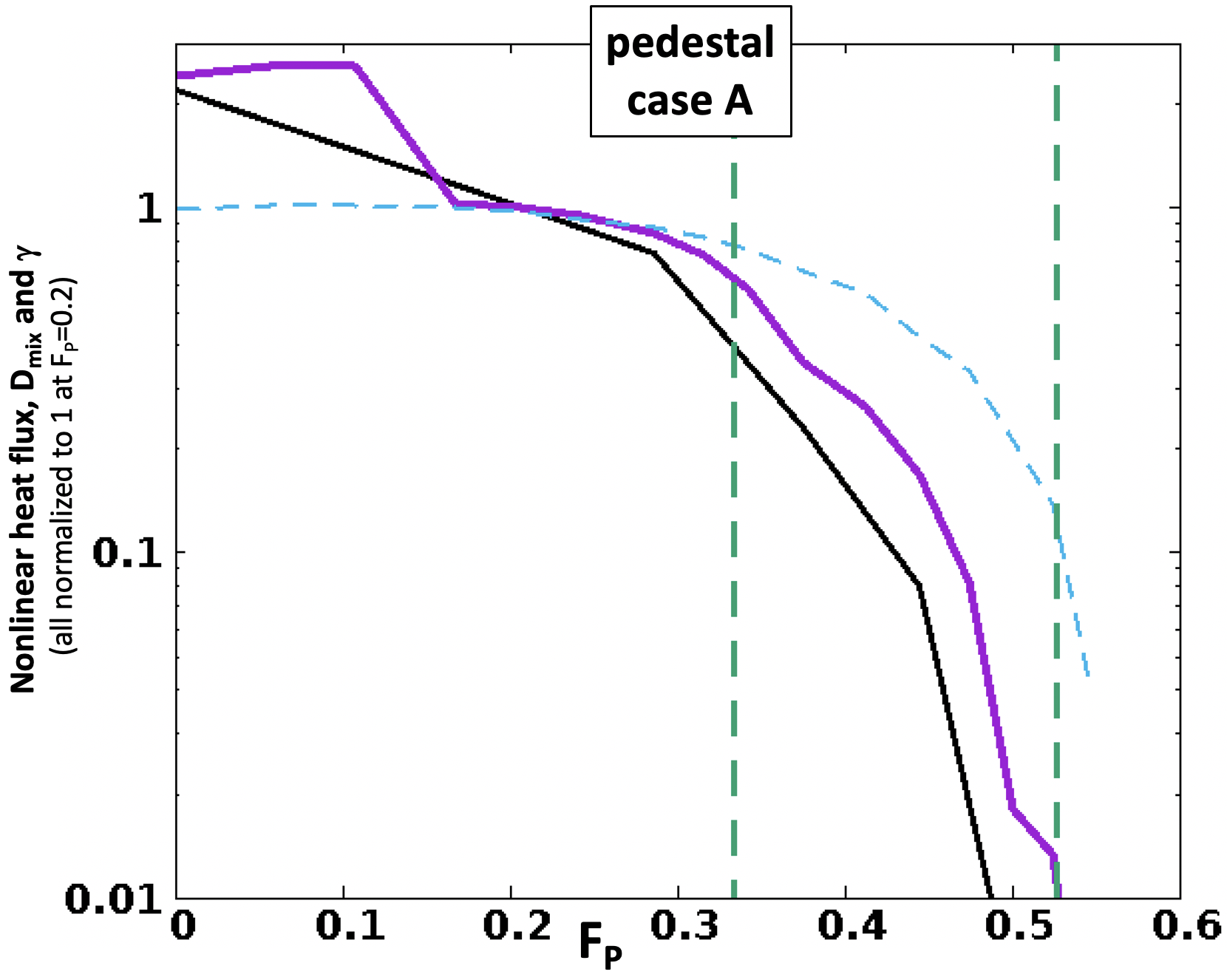}%
}\vfill
\subfloat[]{%
  \includegraphics[width=.33\linewidth]{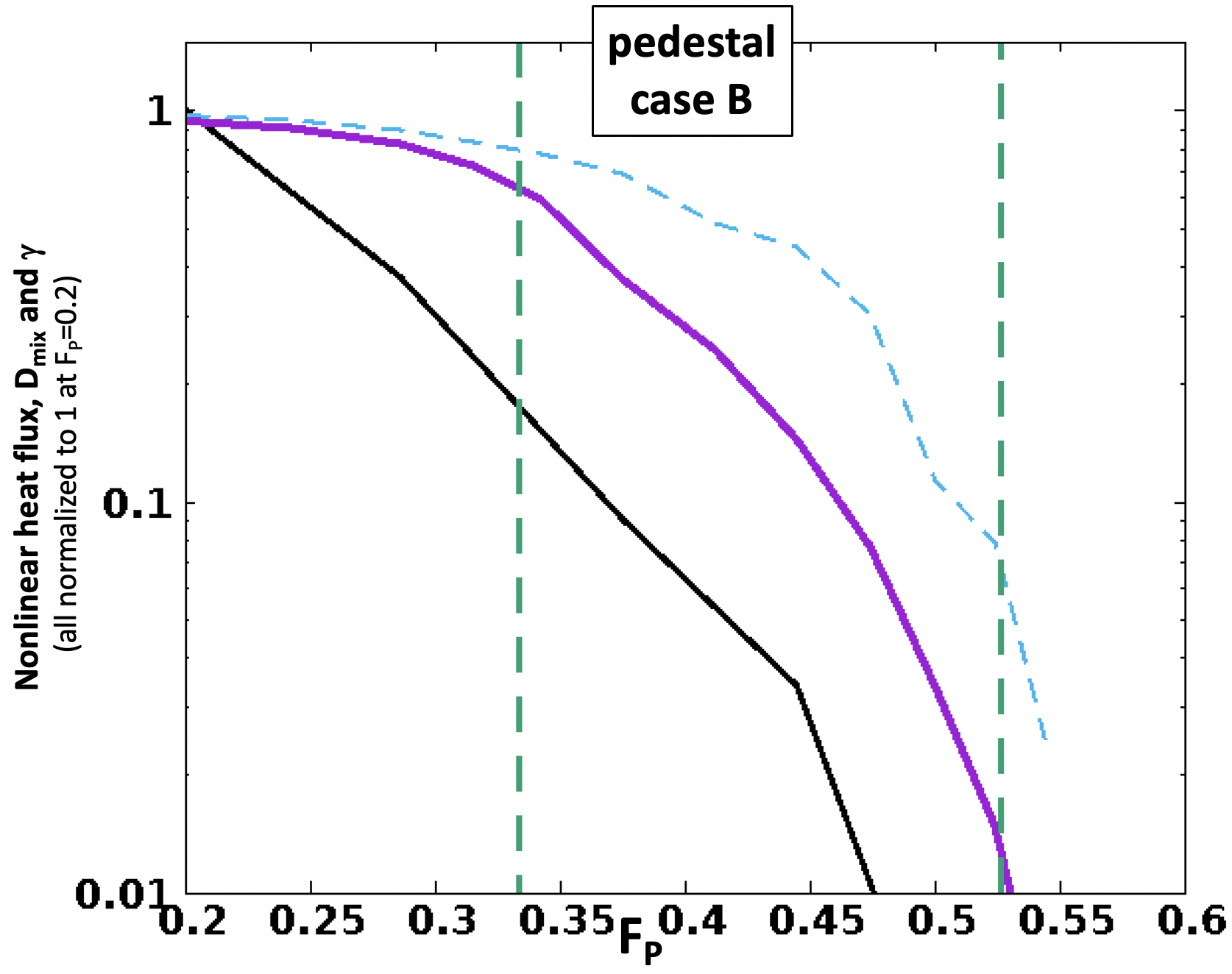}%
}
\hfill
\subfloat[]{%
  \includegraphics[width=.33\linewidth]{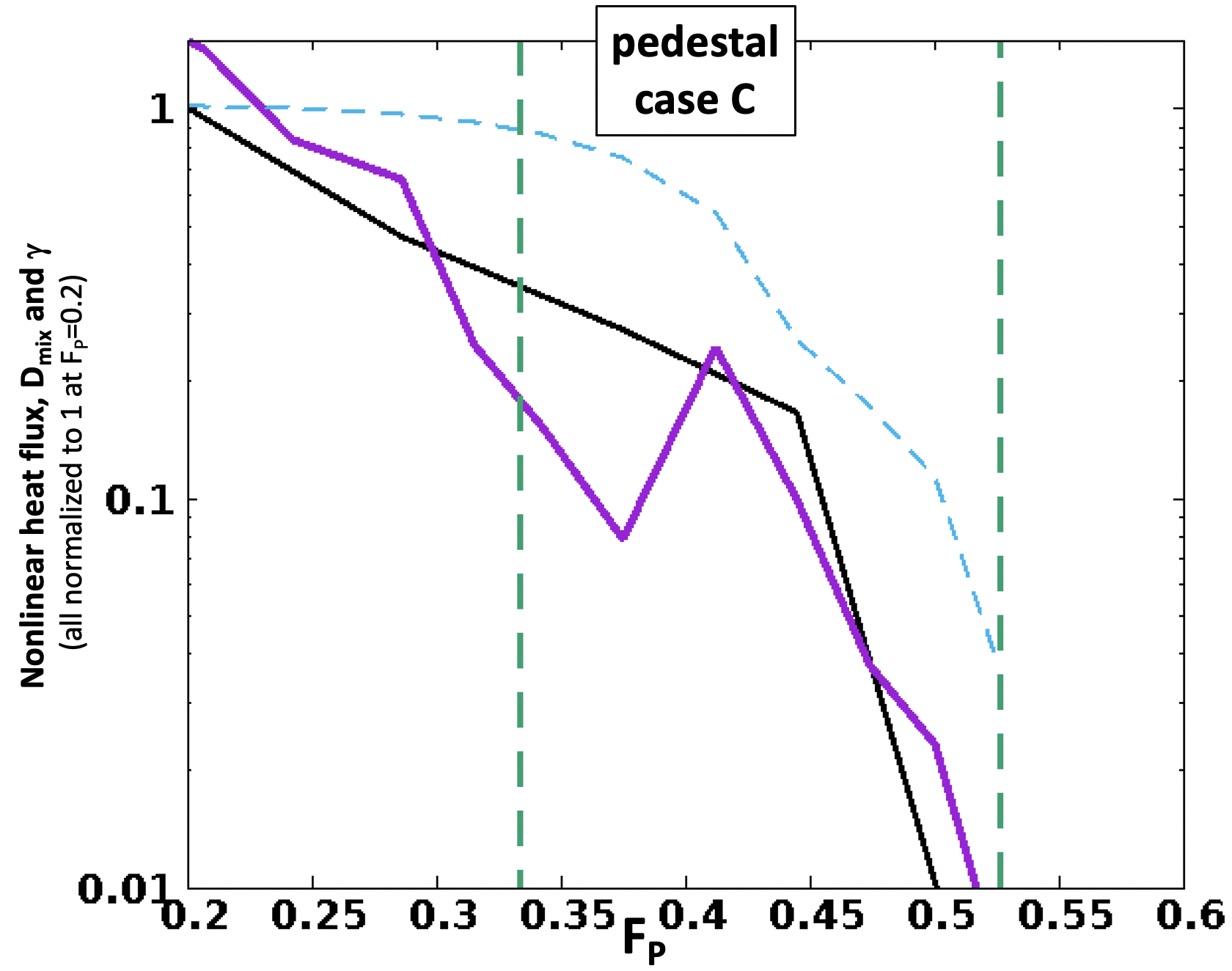}%
}\hfill
\subfloat[]{%
  \includegraphics[width=.33\linewidth]{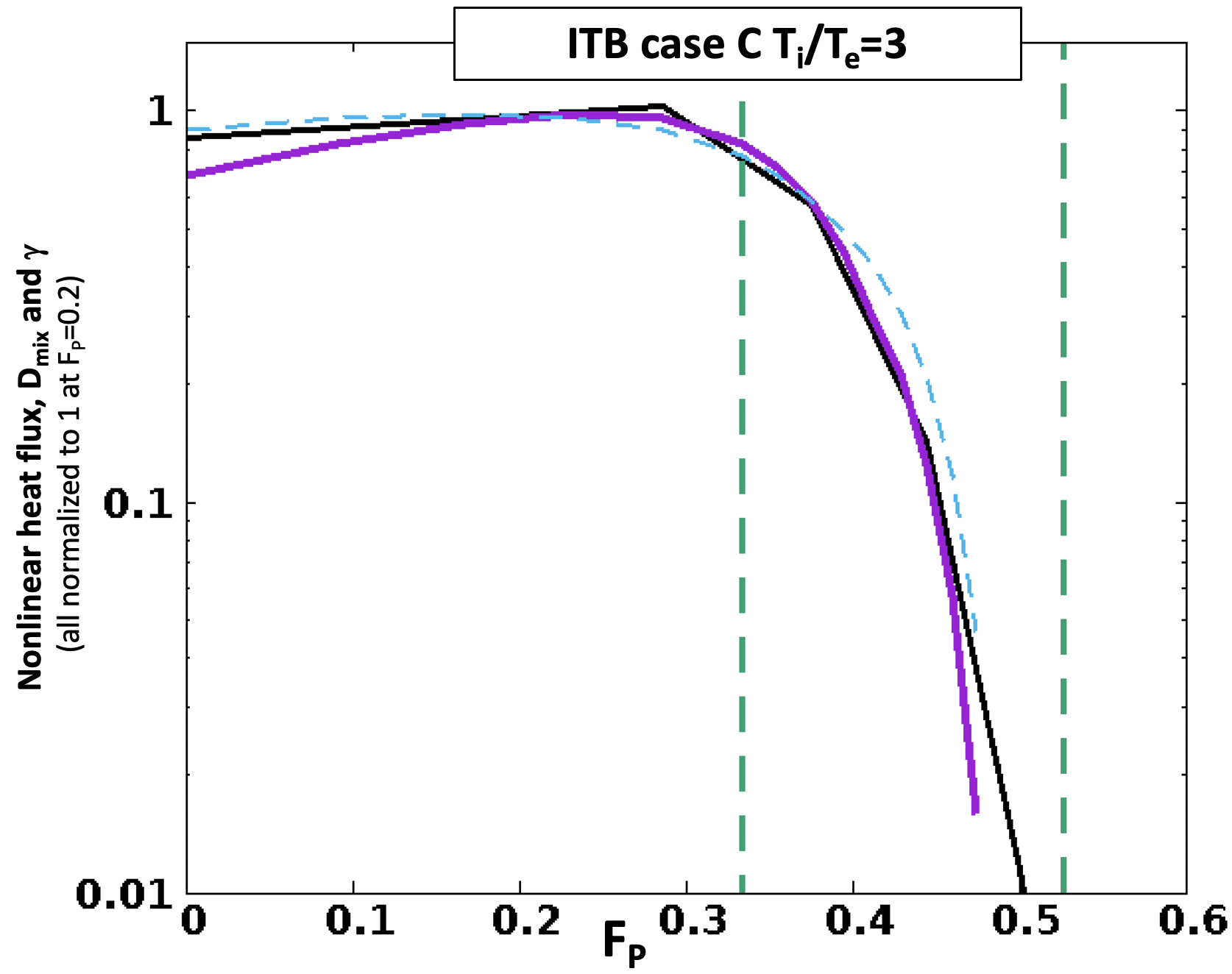}%
}\vfill
\subfloat[]{%
  \includegraphics[width=.33\linewidth]{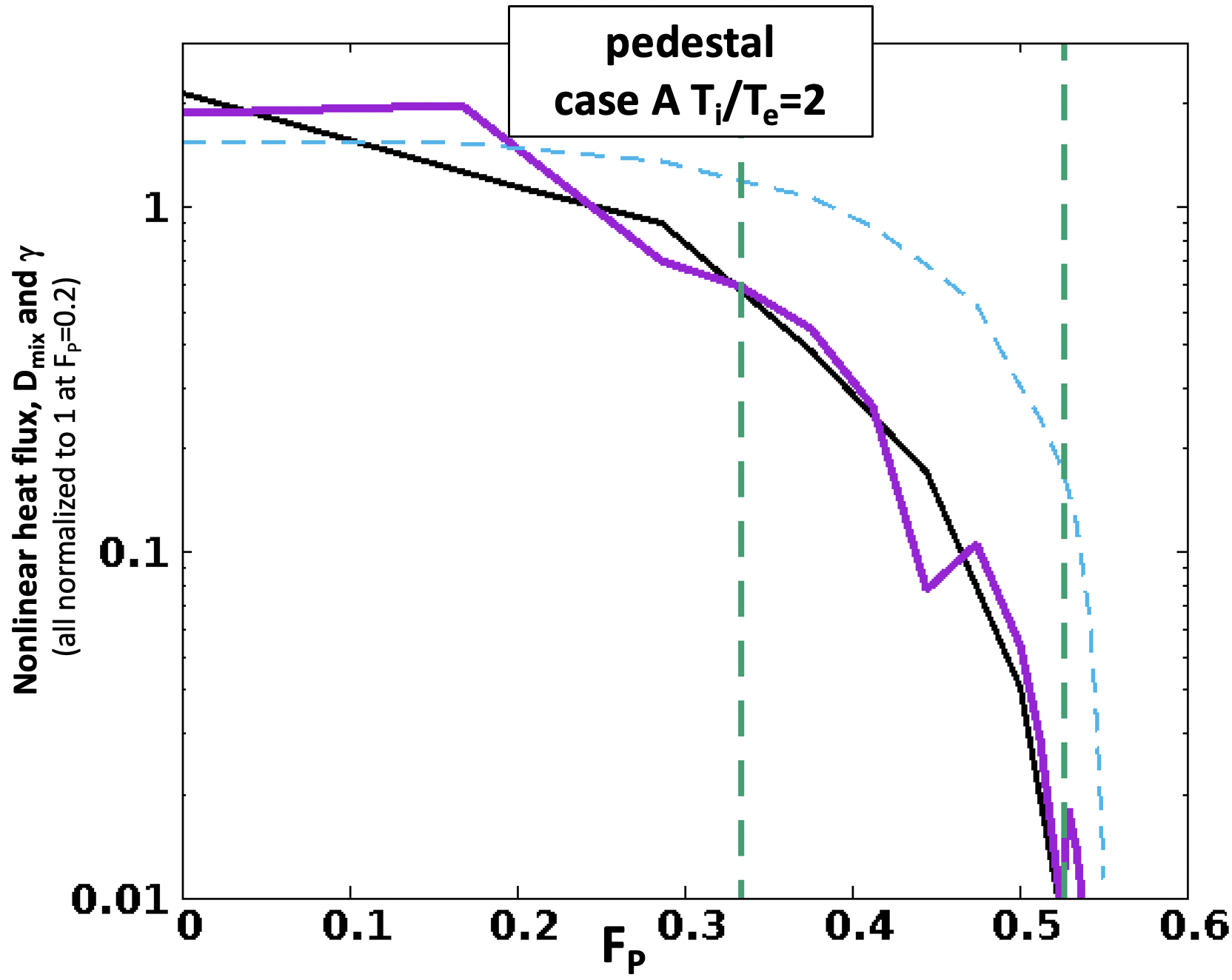}%
}\hfill
\subfloat[]{%
  \includegraphics[width=.33\linewidth]{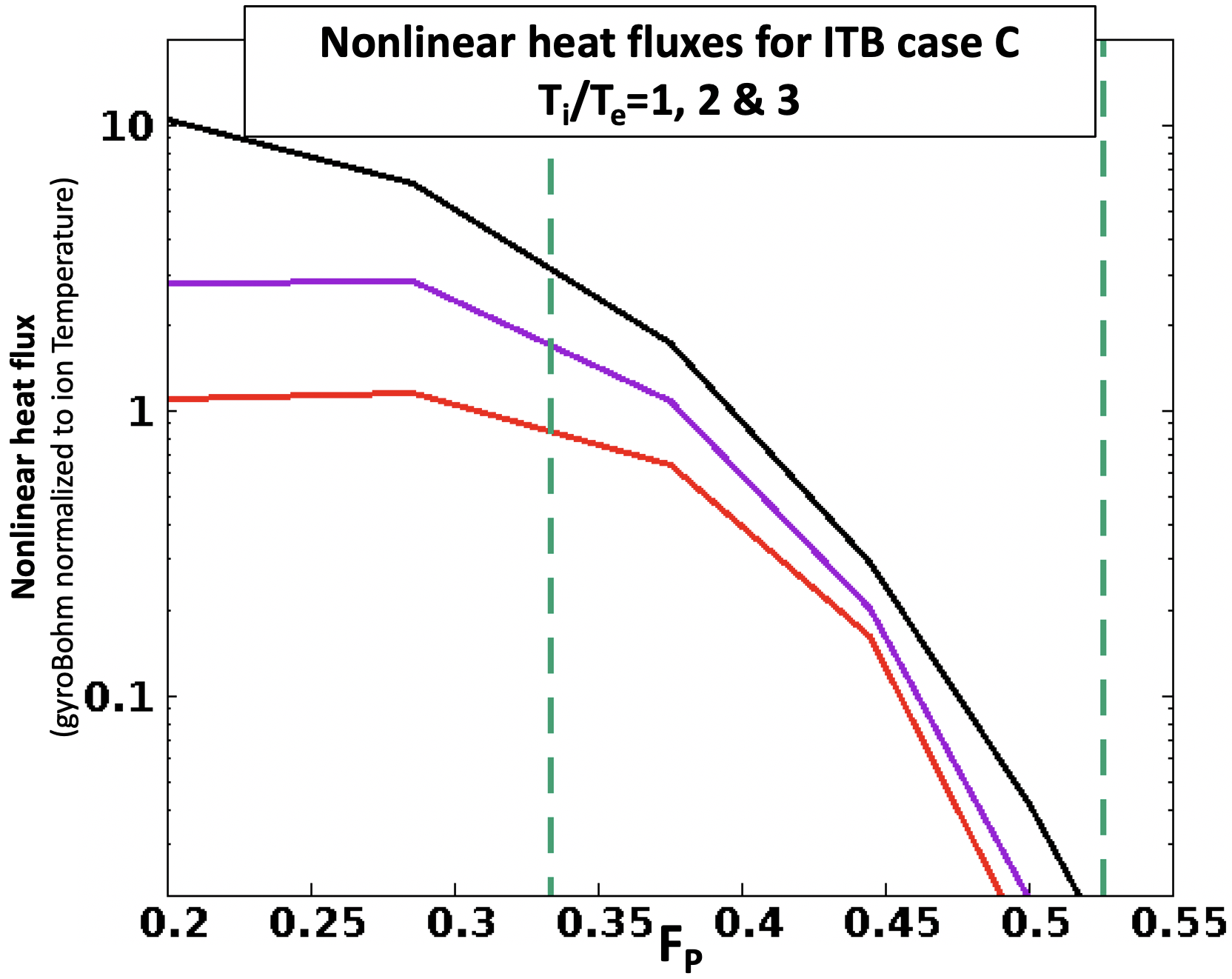}%
}\hfill
\subfloat[]{%
  \includegraphics[width=.33\linewidth]{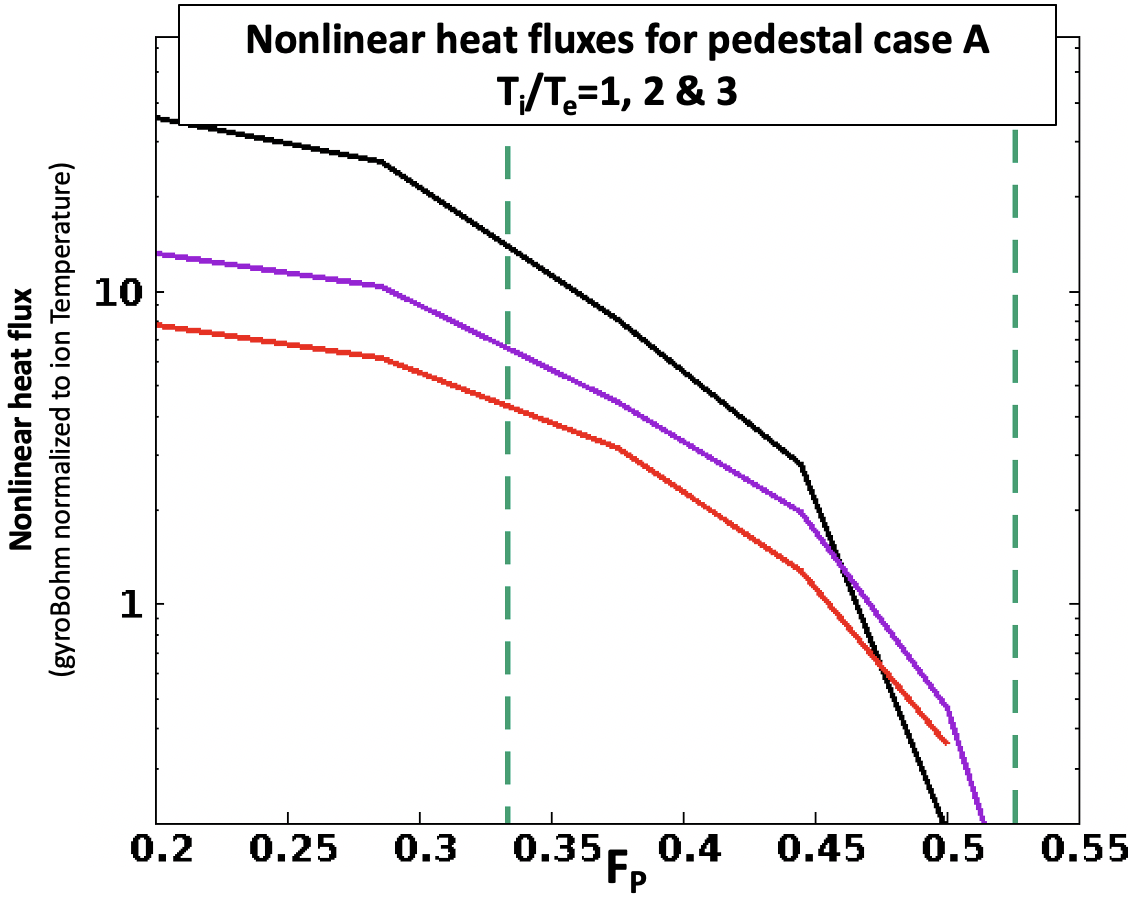}%
}
\caption{\label{fig:five} $D_{mix}$ (the maximum for all all three $\eta_0$) and the growth rate (also the maximum for all all three $\eta_0$) is compared to the nonlinear heat flux for many diverse cases. For clarity, results are normalized to unity at $F_P=0$. The $D_{mix}$ follows the nonlinear result considerably more closely than $gammma$, indicating that the increase in $<k_\perp \rho_i>$ plays an important role in the nonlinear flux reduction. }
\end{figure*}

\begin{figure*}
\subfloat[]{%
  \includegraphics[width=.33\linewidth]{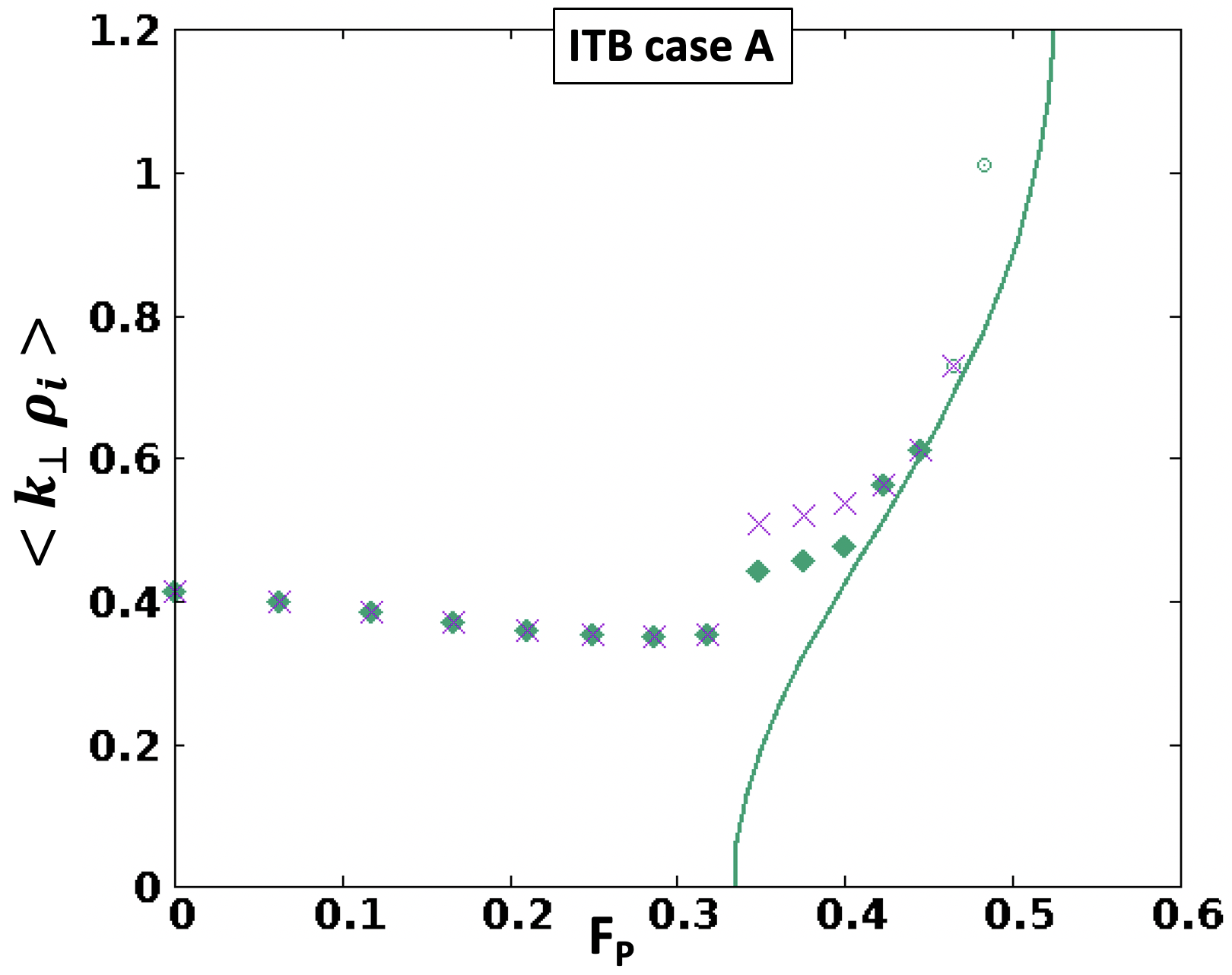}%
}\hfill
\subfloat[]{%
  \includegraphics[width=.33\linewidth]{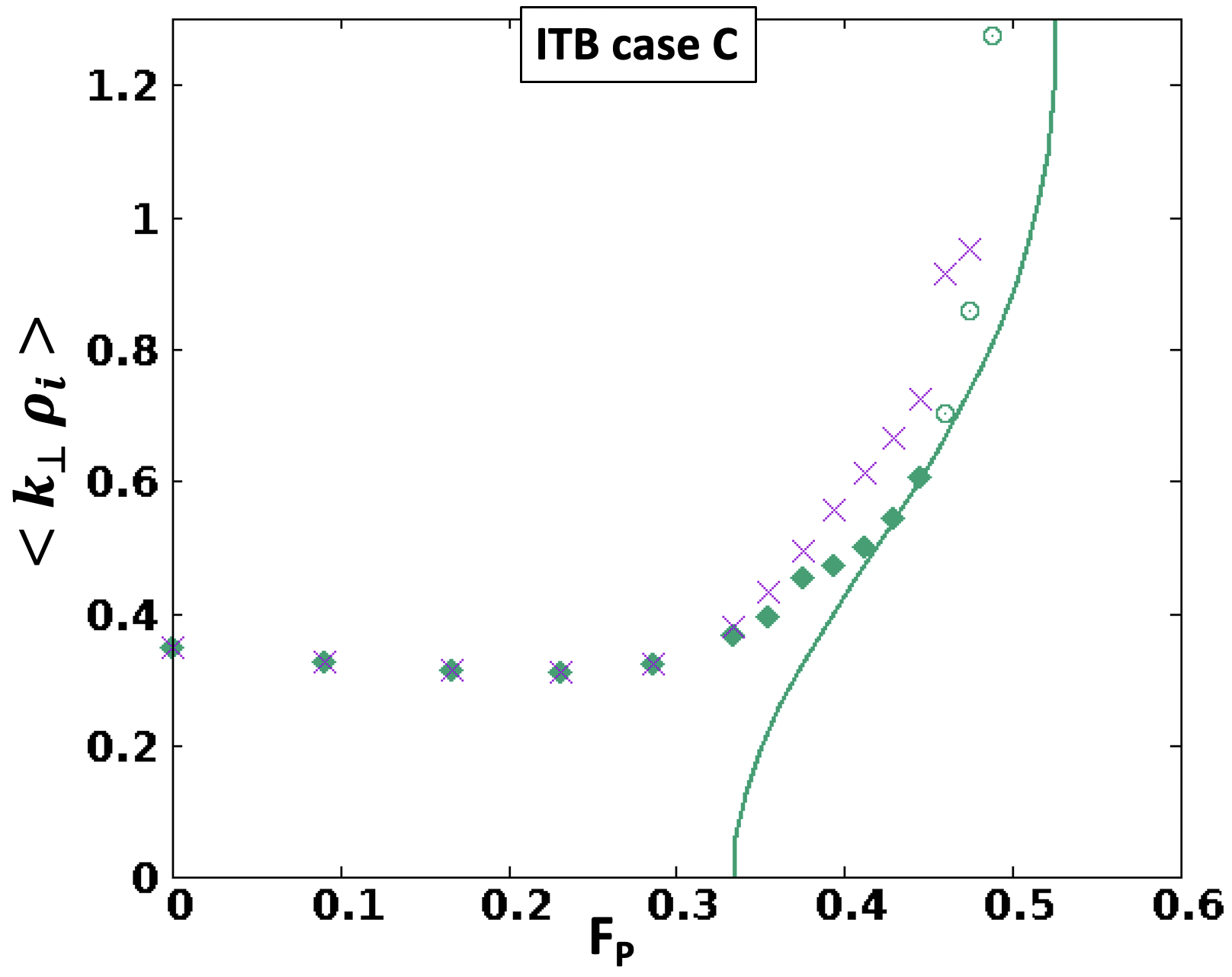}%
}
\hfill
\subfloat[]{%
  \includegraphics[width=.33\linewidth]{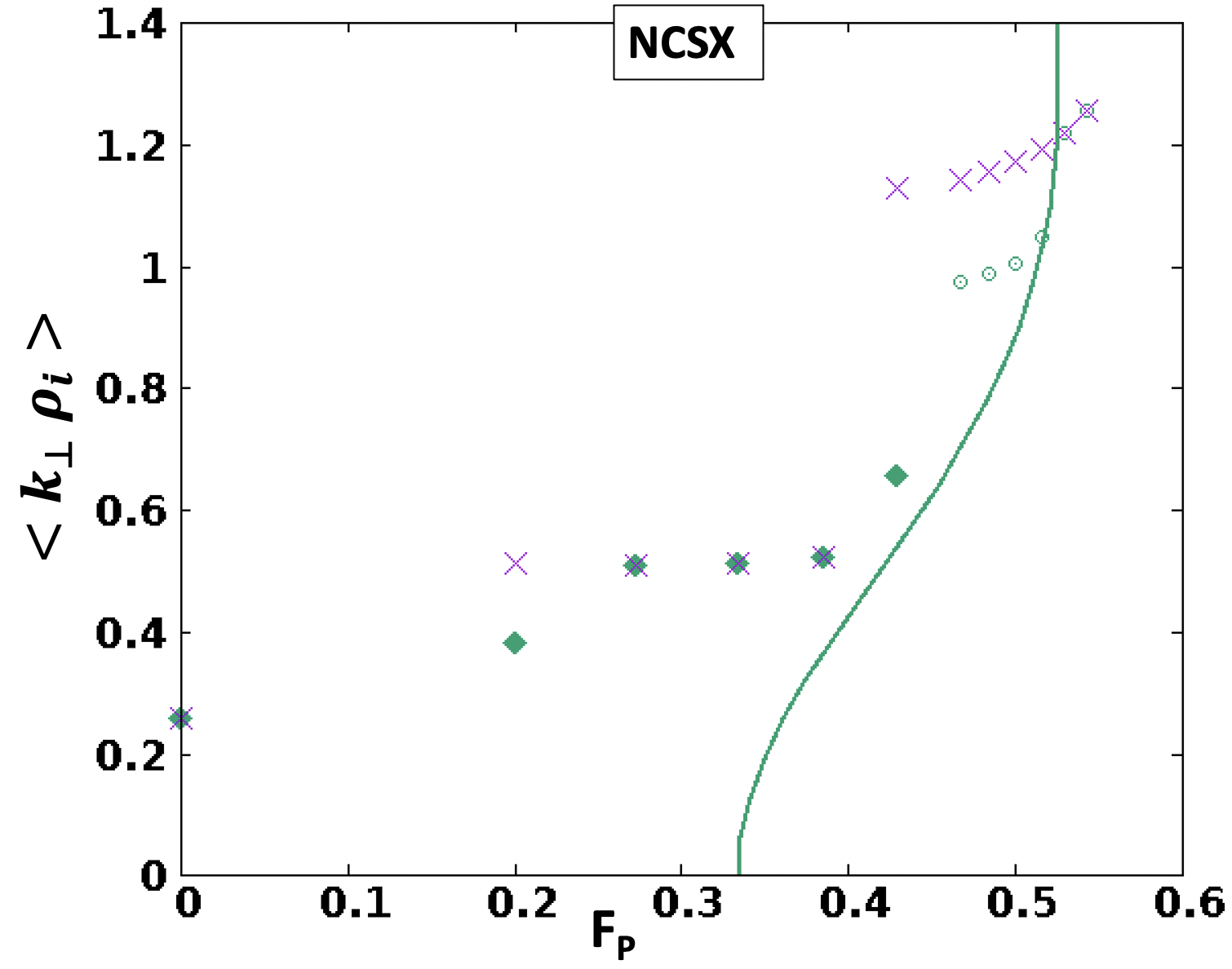}%
}\vfill
\subfloat[]{%
  \includegraphics[width=.33\linewidth]{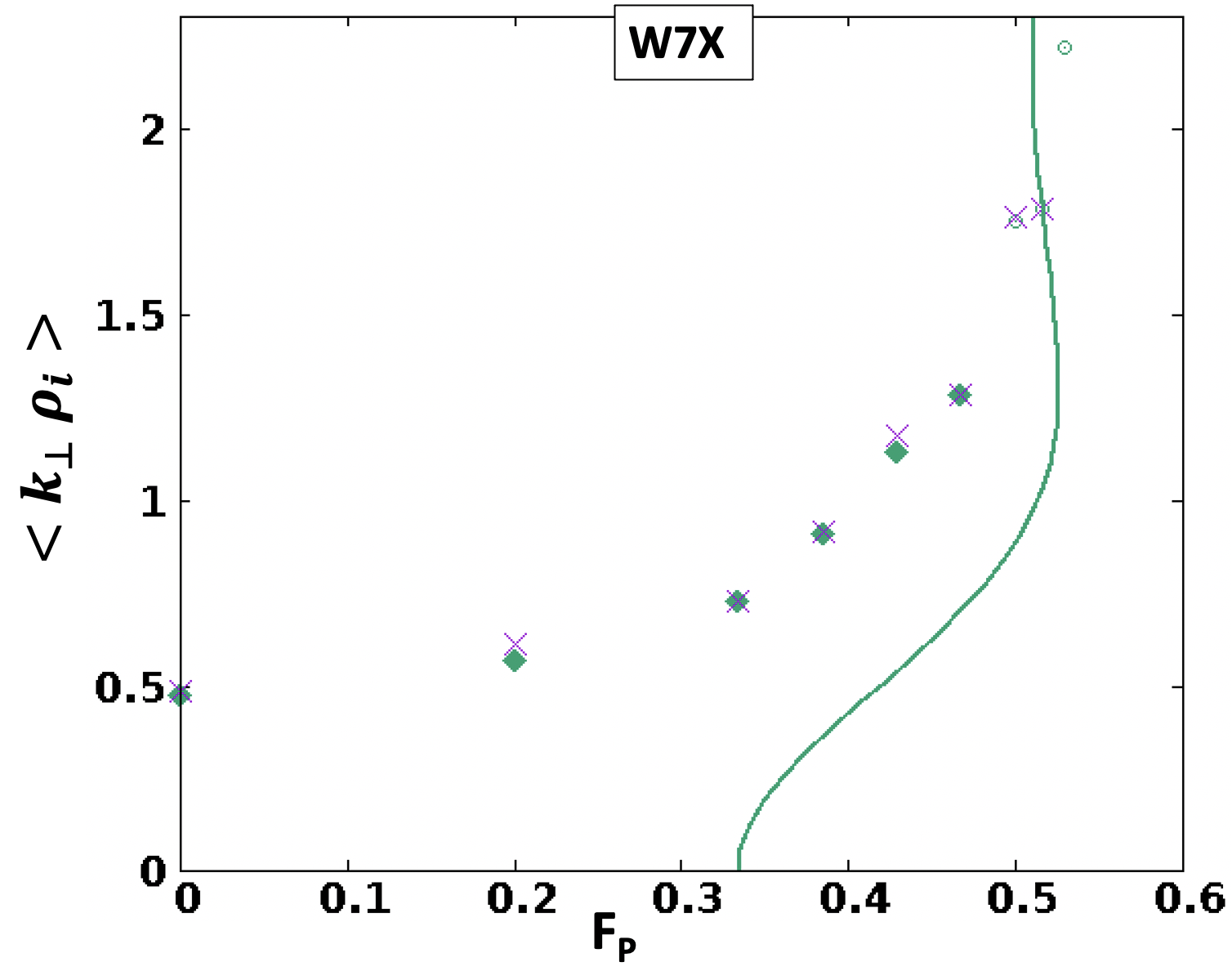}%
}
\hfill
\subfloat[]{%
  \includegraphics[width=.33\linewidth]{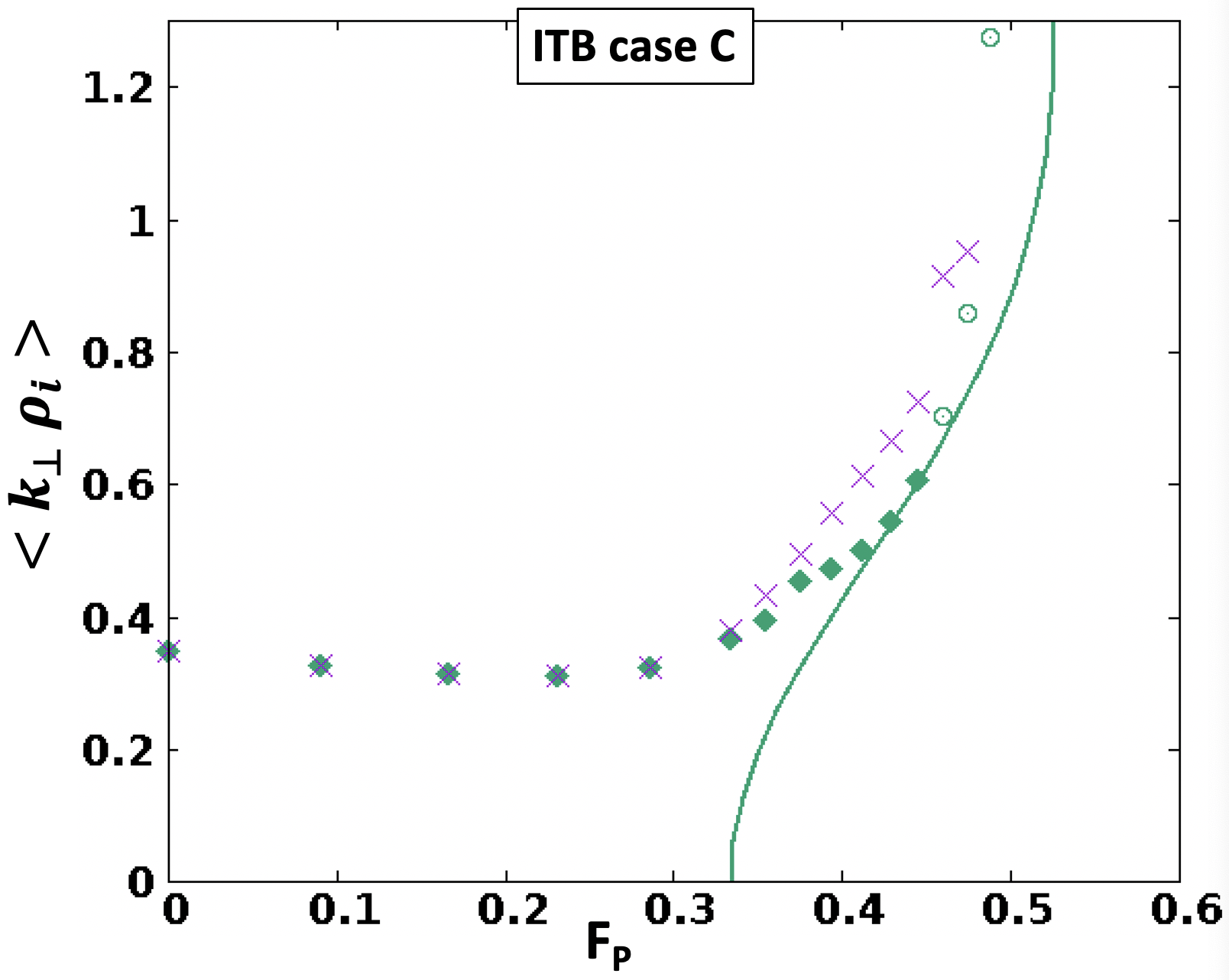}%
}\hfill
\subfloat[]{%
  \includegraphics[width=.33\linewidth]{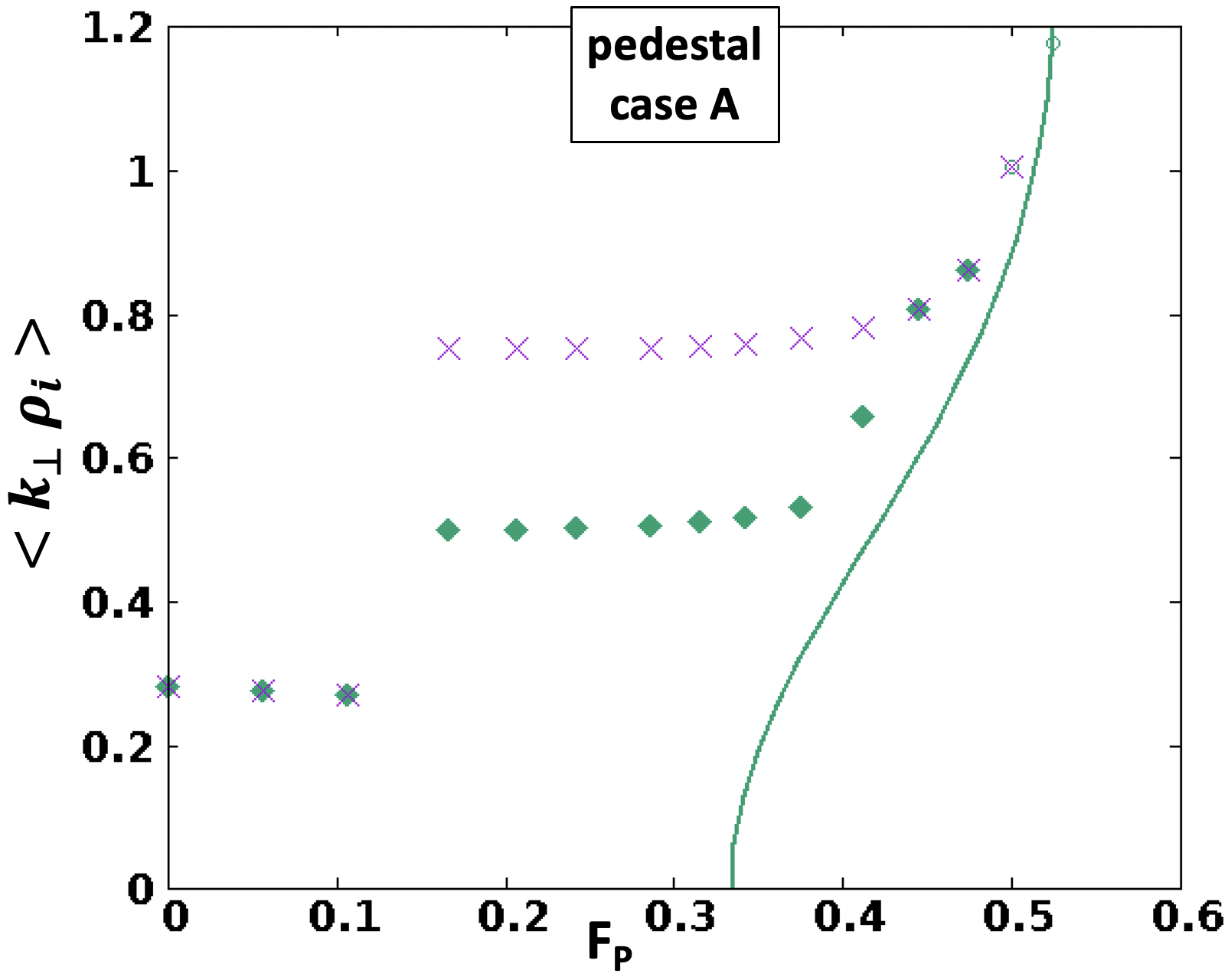}%
}\vfill
\subfloat[]{%
  \includegraphics[width=.33\linewidth]{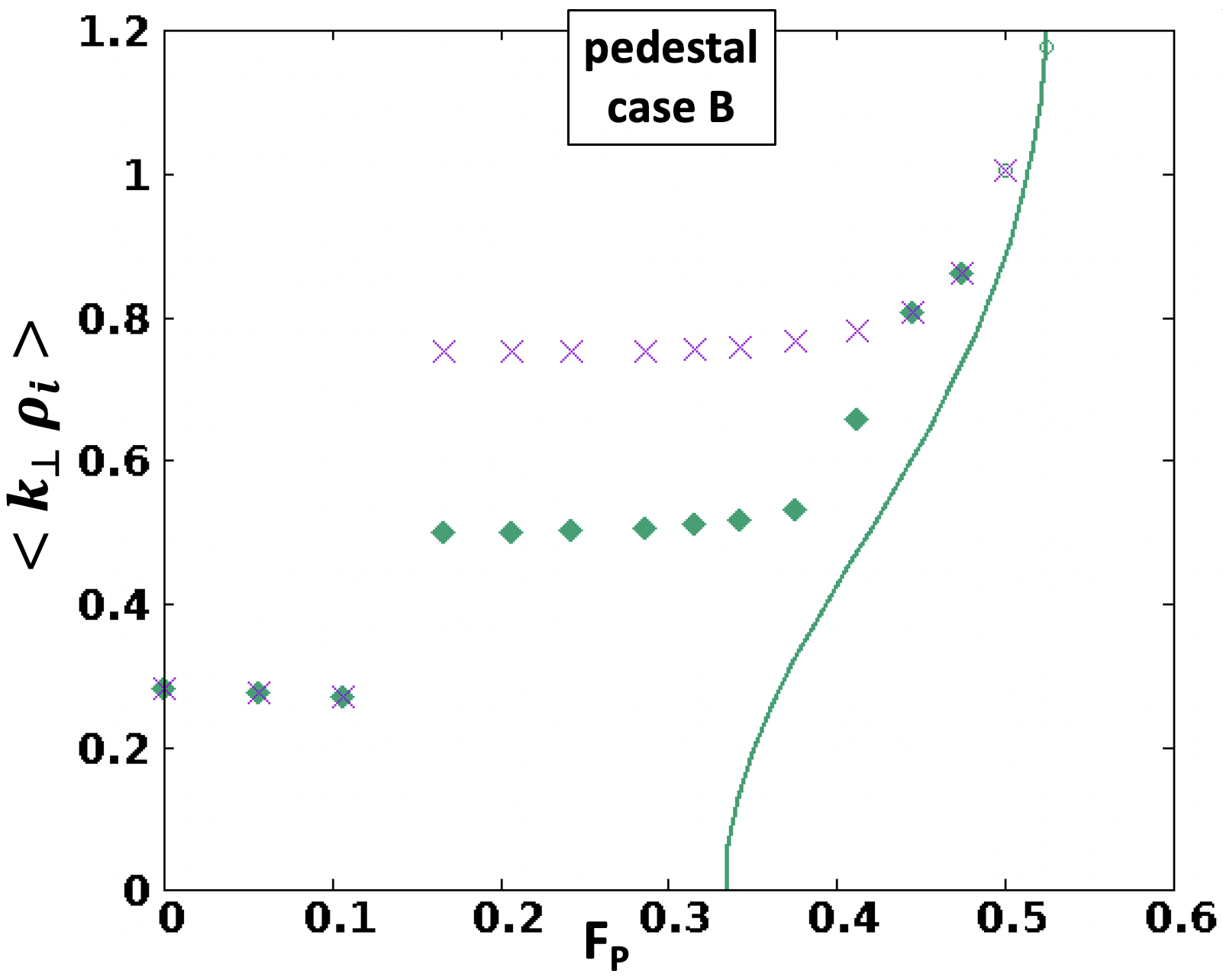}%
}\hfill
\subfloat[]{%
  \includegraphics[width=.33\linewidth]{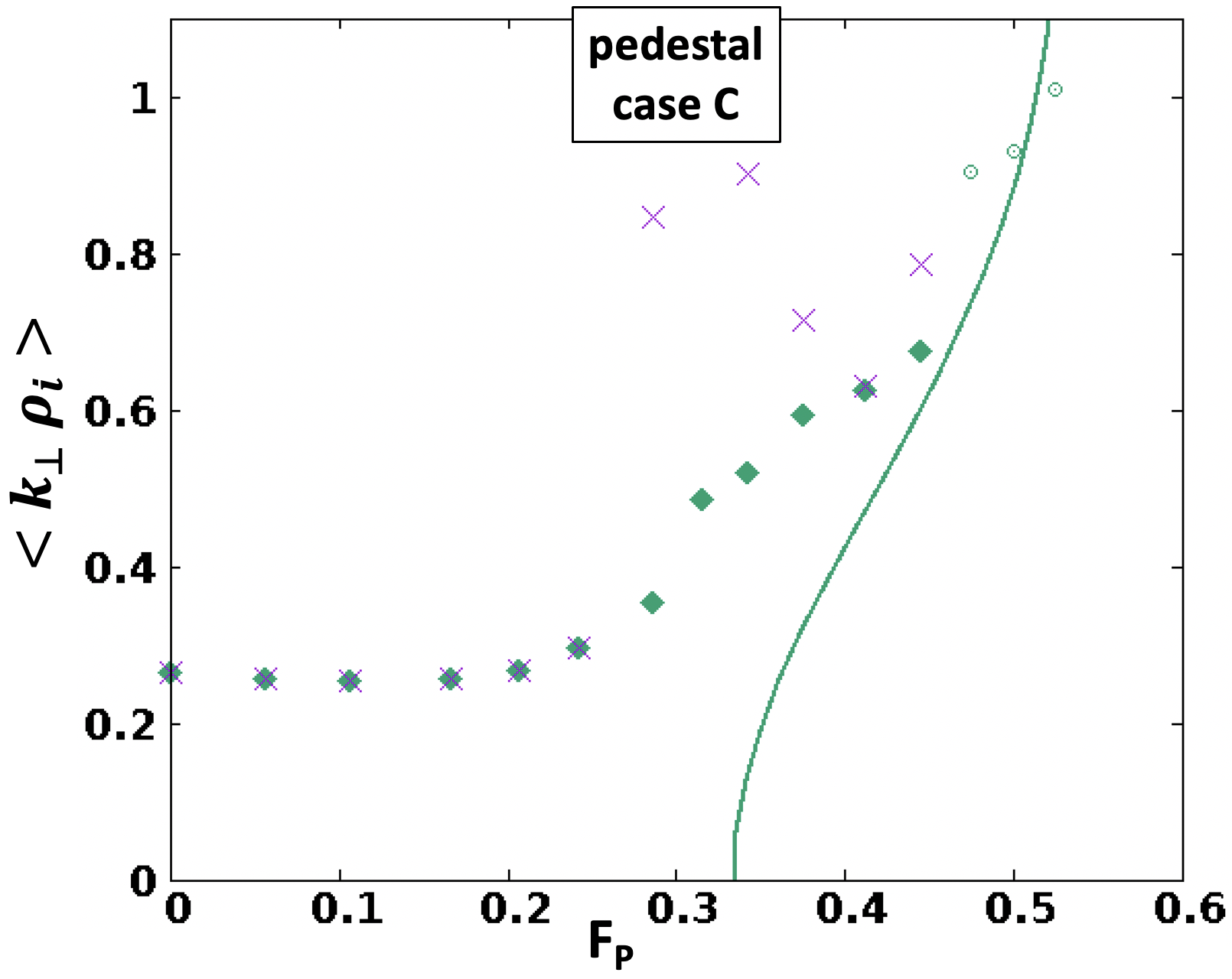}%
}\hfill
\subfloat[]{%
  \includegraphics[width=.33\linewidth]{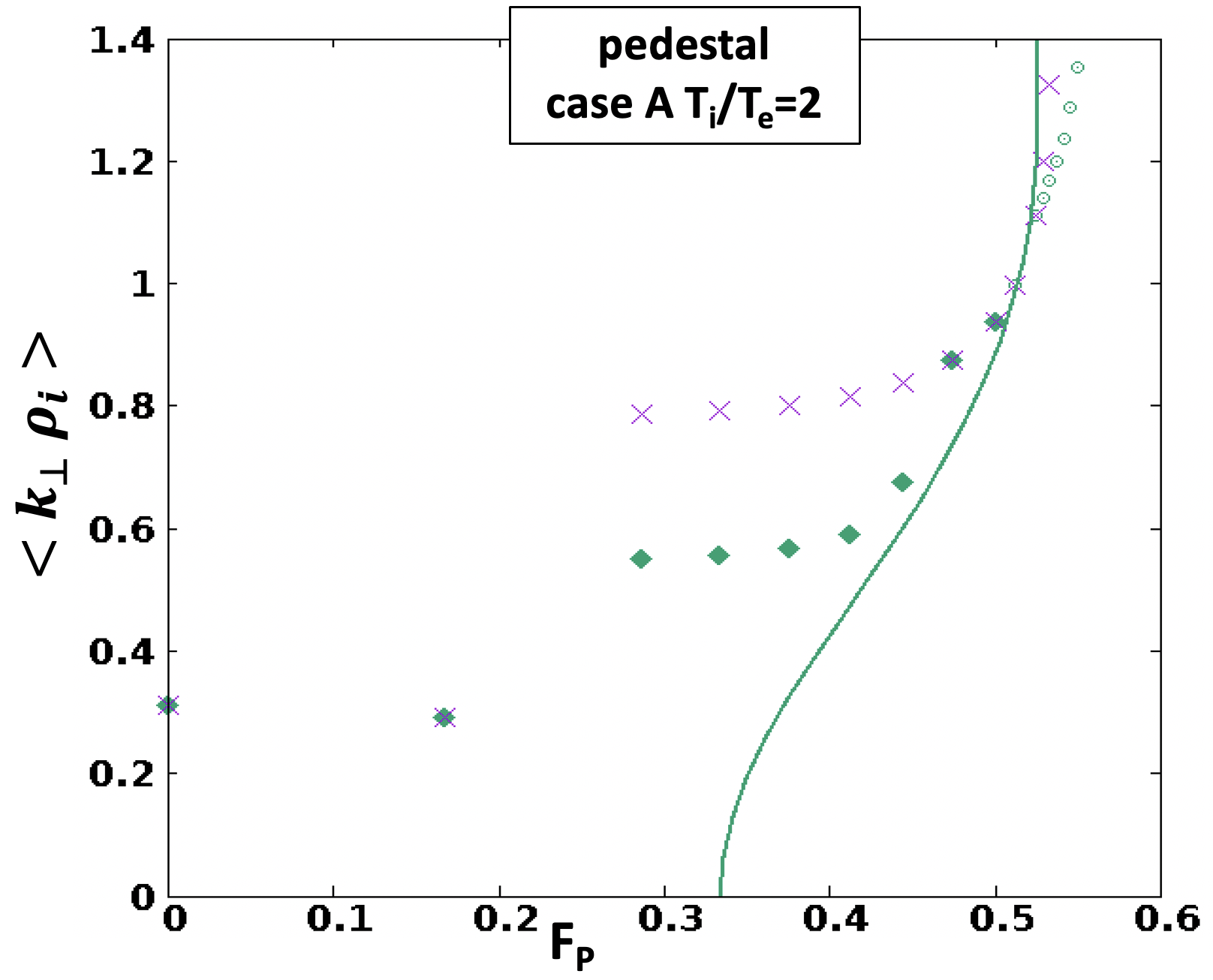}%
}
\caption{\label{fig:four} The minimum value of $<k_\perp \rho_i>$ is plotted for equilibrium from Table 1, along with the analyticaly computed solubility boundary. The crosses are the $<k_\perp \rho_i>$ for $\eta_0 =0$, and the points are the minimum $<k_\perp \rho_i>$ among the three $\eta_0$ simulated (for unstable modes). Hollow points have  $D_{mix} < 2\%$ of the maximum for the scan, that is, they are unstable but very feeble. As expected, the addition of more degrees of freedom (multiple $\eta_0$) allows some modes to approach the solubility boundary more closely, and sometime, slightly cross it. Despite feeble modes crossing the boundary, the overall agreement is very good, especially considering that approximations must be used to obtain the analytic results.}
\end{figure*}

Let us revisit the nonlinear relevance of the analytic bound. Compare the trends in $D_{mix}$ to the trends in the nonlinear heat flux (fig(\ref{fig:five})). Between $F_P$ = 0.2 and $F_P$=0.53, the reductions in heat flux are a reasonably good match to the trends in $D_{mix}$. The reduction in $\gamma$, however, is usually considerably weaker than in the nonlinear $\chi$. This is evidence that the increase in $<k_{\perp}>$ is playing an important role in the reduction of the nonlinear heat flux. And as we have seen above, the increase in $<k_{\perp}>$ is due to the FC solubility boundary plus the adaptivity of the gyrokinetic system.

Finally, in fig\ref{fig:five}(h-i) , the variation of nonlinear heat flux with $F_P$ for $T_i/T_e= (1,2$ and $3$) is displayed. We normalize to gyroBohm units in the ion temperature. We can consider the results as a thought experiment where we keep the ion temperature and gradients constant, and only change the $T_e$ for the adiabatic electrons. The FC solubility bound is unaffected by this, but, the energetic equation is strongly affected by the stabilizing effect of higher $T_i/T_e$. This is manifest as a strong reduction in heat flux by almost an order of magnitude. Despite such a major change, the $F_P$ value at which flux is reduced by roughly two orders of magnitude is almost unchanged. This correlated perfectly with the simulation results for $D_{mix}$. This is yet more evidence that the gyrokinetic system is ruled by two separate dynamics, the free energy equation and the FC, and the simulation behavior distinguishes between these quite cleanly, and this applies nonlinearly as well.

\emph{The simulation results in this section, even more than previous sections, impel a statistical mechanical approach.} Due to complexity and major variations in eigenfunctions ( particularly near solubility limit of the FC ), the conventional stability theory becomes both intractable and opaque. However, when analyzed as a mean field theory (dealing with statistically averaged quantities), the gyrokinetic system becomes much more transparent.  The large number of degrees of freedom that allows a statistical description, also imbues the system with a strong adaptability; stability behavior becomes predictable - the modes becomes stable only near the solubility limit of the FC.

\emph{In other words, if one chooses a description in terms of eigenfunction averages, instead of eigenmodes, and posits the system with adaptability, the behavior of the system is vastly simplified, and becomes quite tractable- and simulations support this approach}.

We, thus, add another element to the statistical mechanical ansatz: 

 \begin{itemize}

\item 	In situations with large free energy (large gradients), the insolubility of the FC is often the operative dynamic that leads to stability. 
\item 	As is observed in very many systems with a large number of degrees of freedom, the gyrokinetic system manifests apparent adaptive behavior, and "finds a way" to dissipate free energy until a "hard" dynamical constrain prevents it- here, the FC.

\end{itemize}

After analyzing simulation and theoretical results in the following sections, we will add new elements to the ansatz.

\section{Graphical visualization of the two interacting dynamics- the constraint and free energy}

Exposing physics through a simultaneous graphical representation of the FC and of free energy (like the curvature drive) is very elucidating. 

\begin{figure*}
\subfloat[\label{sfig:7a}]{%
  \includegraphics[width=.5\linewidth]{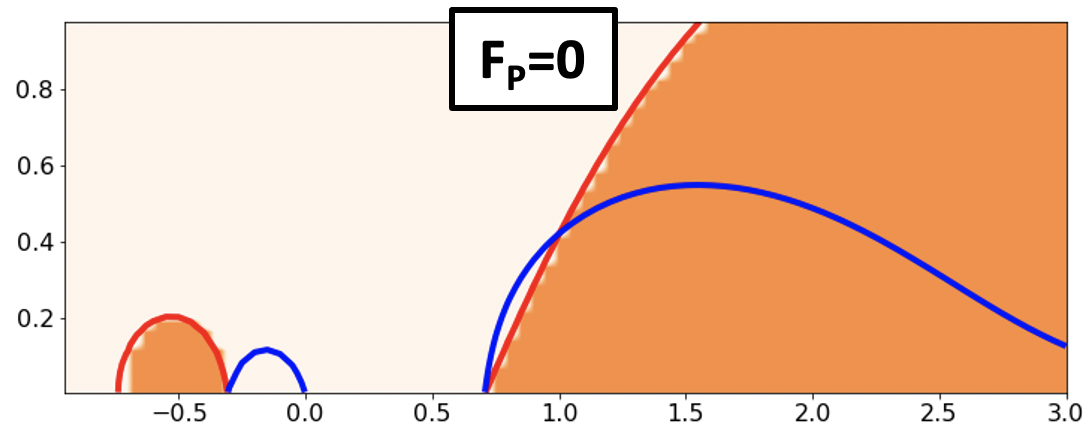}%
}\hfill
\subfloat[\label{sfig:7b}]{%
  \includegraphics[width=.5\linewidth]{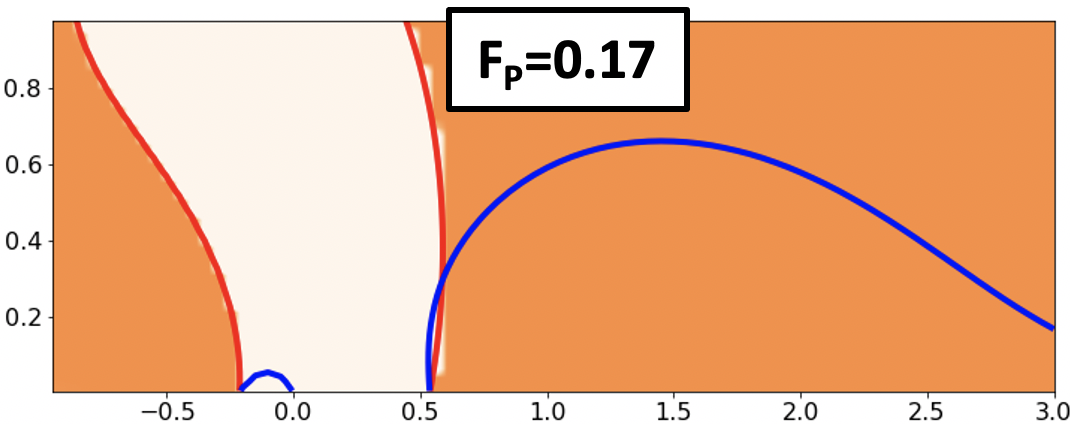}%
}
\vfill
\subfloat[\label{sfig:7c}]{%
  \includegraphics[width=.5\linewidth]{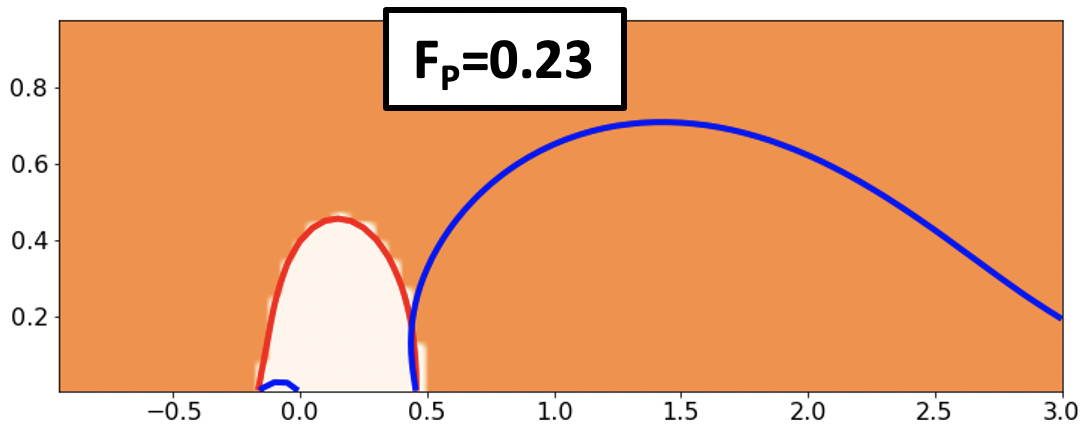}%
}\hfill
\subfloat[\label{sfig:7d}]{%
  \includegraphics[width=.5\linewidth]{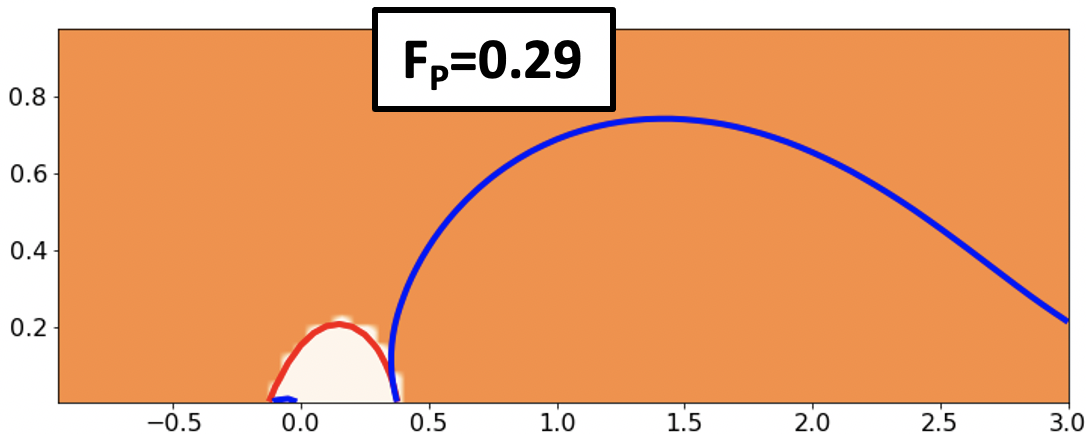}%
}
\vfill
\subfloat[\label{sfig:7e}]{%
  \includegraphics[width=.5\linewidth]{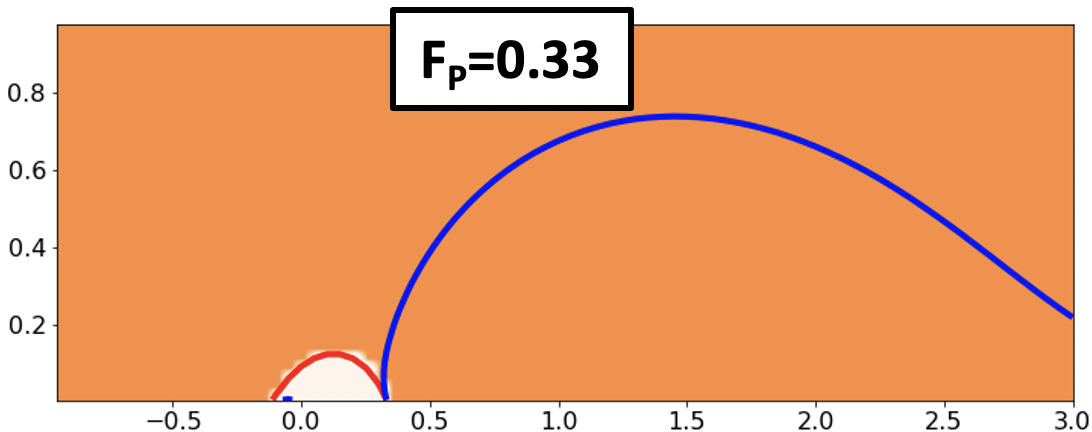}%
}\hfill
\subfloat[\label{sfig:7f}]{%
  \includegraphics[width=.5\linewidth]{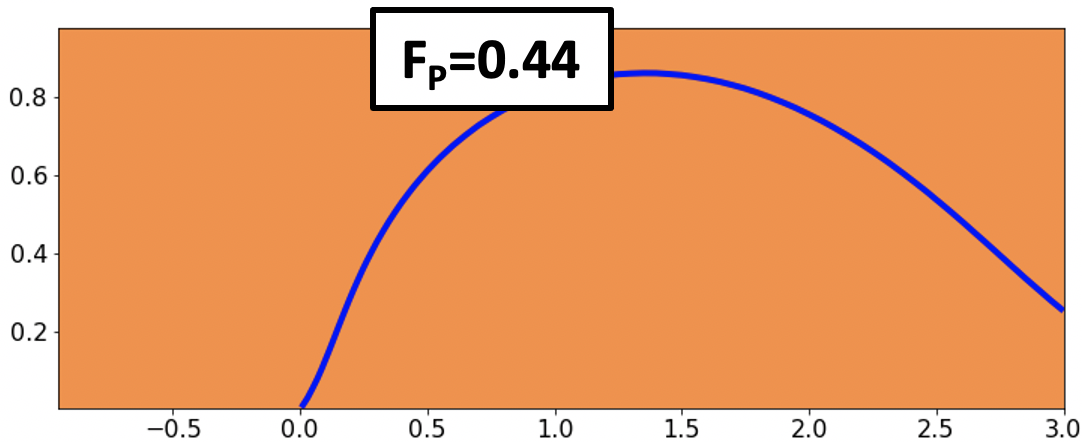}%
}
\caption{\label{fig:BP1} The complex plane with various plots scanning a range of $F_p$ from 0 to 0.44.  The blue contour denotes the line satisfying the energy equation and the red contour denotes the line satisfying zero charge flux.  Eigenvalues lie at the intersection of the two.  Dark orange regions denote areas with positive charge flux and light regions denote areas of negative charge flux.  Note the progressively lower line satisfying the zero charge flux constraint, forcing growth rates to small values and indicating access to the transport barrier regime.}
\end{figure*}

\begin{figure*}
\subfloat[\label{sfig:7a}]{%
  \includegraphics[width=.5\linewidth]{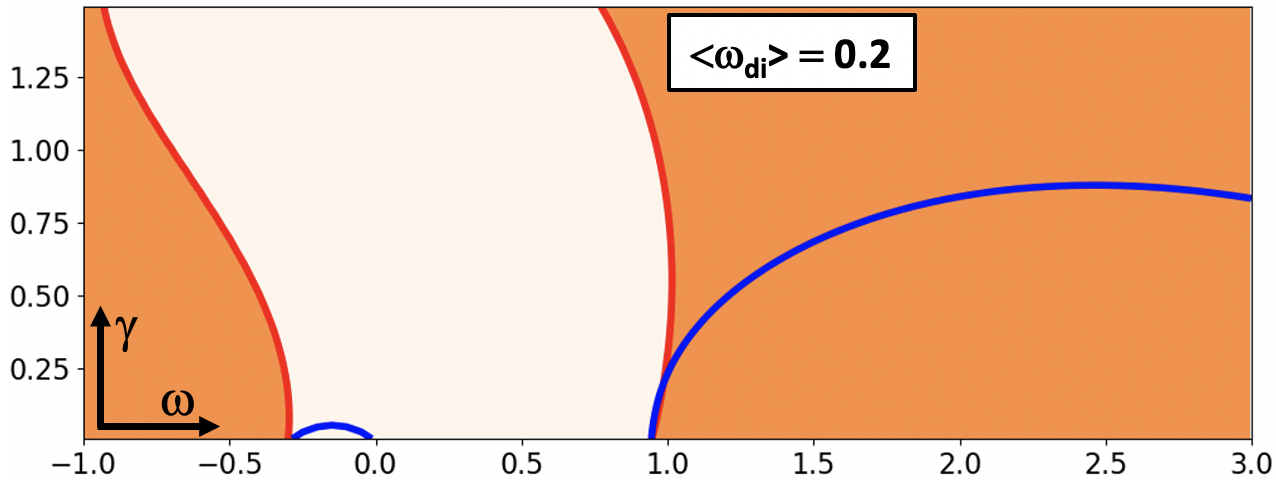}%
}\hfill
\subfloat[\label{sfig:7b}]{%
  \includegraphics[width=.5\linewidth]{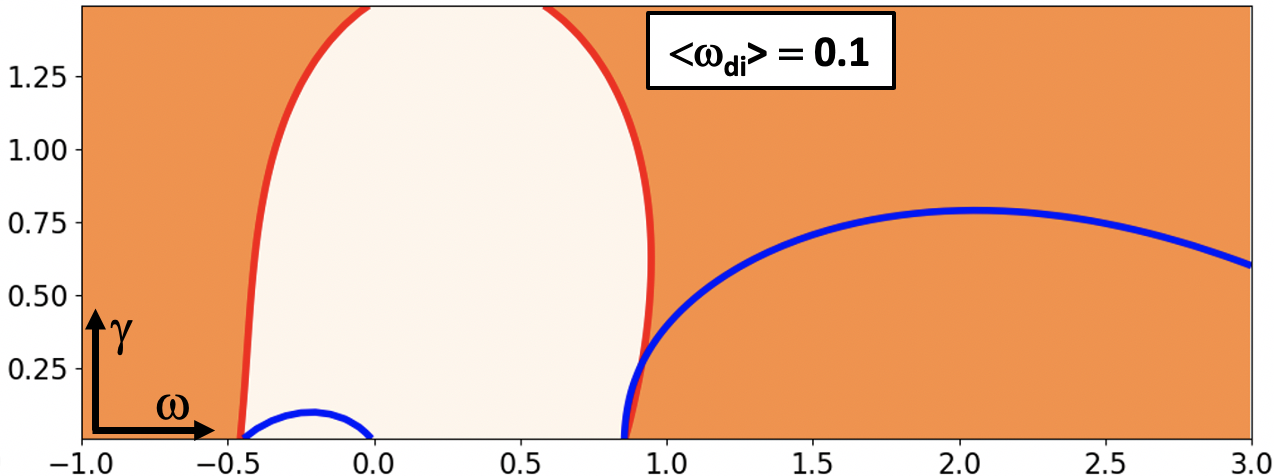}%
}
\vfill
\subfloat[\label{sfig:7c}]{%
  \includegraphics[width=.5\linewidth]{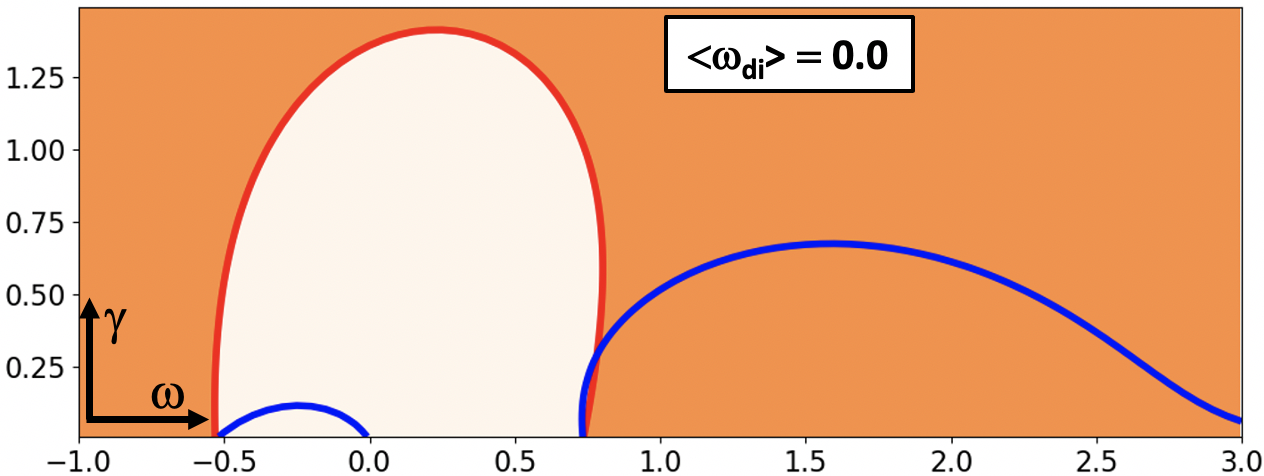}%
}\hfill
\subfloat[\label{sfig:7d}]{%
  \includegraphics[width=.5\linewidth]{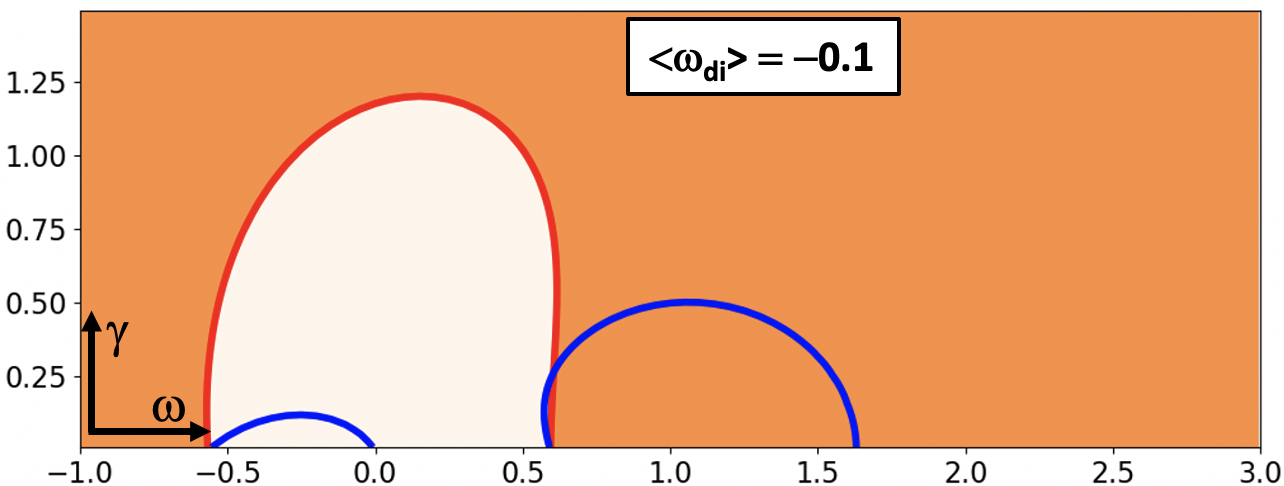}%
}
\vfill
\subfloat[\label{sfig:7e}]{%
  \includegraphics[width=.5\linewidth]{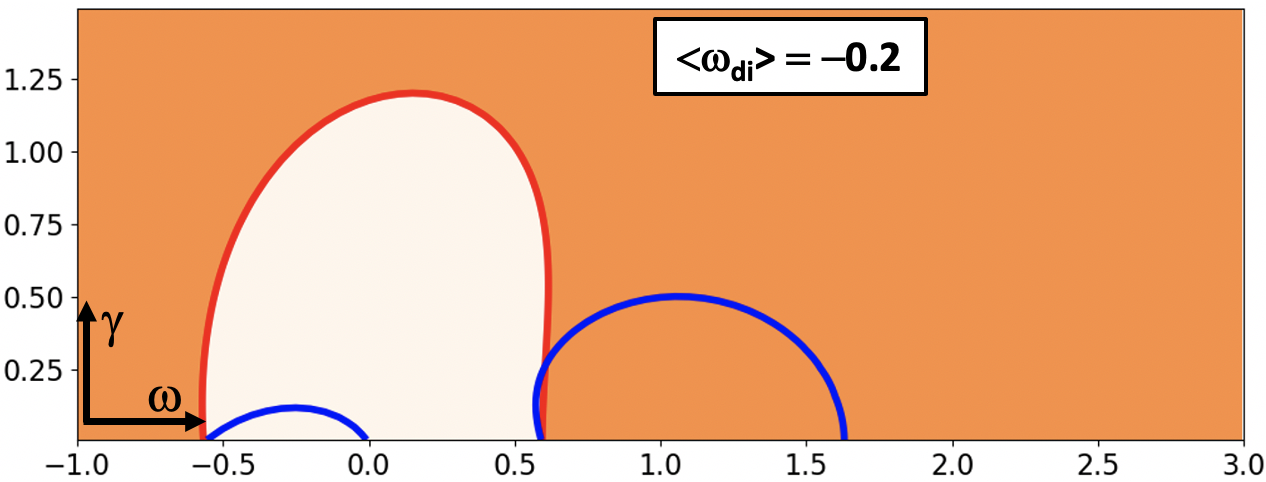}%
}\hfill
\subfloat[\label{sfig:7f}]{%
  \includegraphics[width=.5\linewidth]{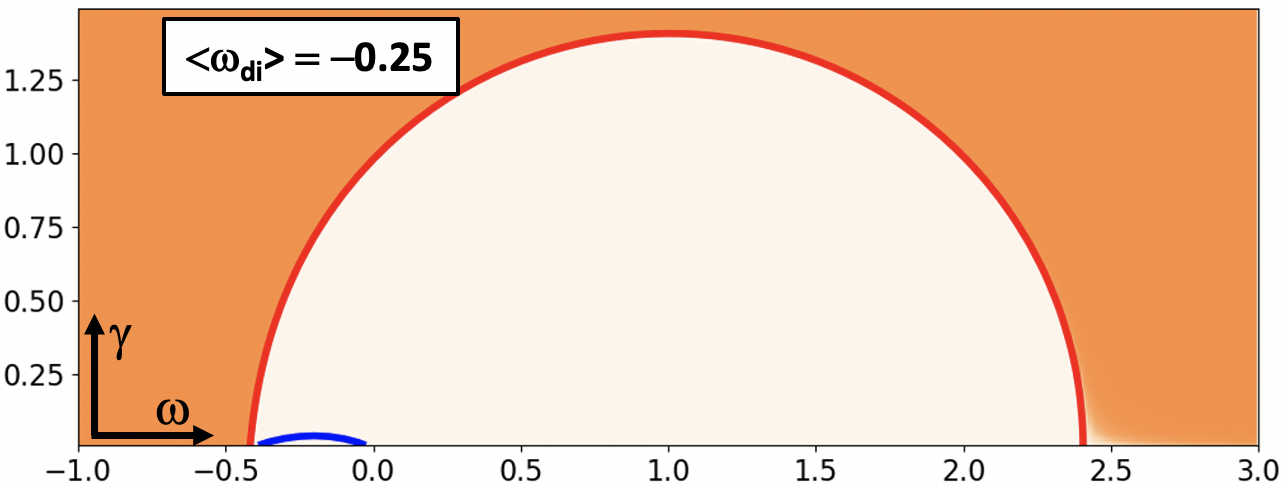}%
}
\caption{\label{fig:BP2} The complex plane with various plots scanning a range of eigenmode averaged drift frequencies.  The blue contour denotes the line satisfying the energy equation and the red contour denotes the line satisfying zero charge flux.  Eigenvalues lie at the intersection of the two.  Dark orange regions denote areas with positive charge flux and light regions denote areas of negative charge flux.  Note the progressively lower line satisfying the zero charge flux constraint, forcing growth rates to small values and indicating access to the transport barrier regime. }
\end{figure*}

Consider the SKiM dispersion relation Eq.~(\ref{eq:SKiMdr}), which we can write as $D(\omega)=0$. It can be shown that the FC and the free energy equation are closely related to this. The charge flux is proportional to $Im(D(\omega))$ and the free energy equation is proportional to $Re(i\omega D(\omega))$. Hence the FC is  $Im(D(\omega))=0$ and the free energy balance is $Re(i\omega D(\omega))=0$. Each of these equations corresponds to a curve in the upper half plane $(\omega_r,\gamma)$.  We gain a great deal of insight by plotting each of these two curves; the simultaneous solution to the two is equivalent to the dispersion relation. 

In fig(\ref{fig:BP1}), drawn for $ITG_{ae}$ in an ITB, we shade the region with positive particle flux (in the direction driven by the density gradient) orange. The region of negative particle flux is white. The {\emph red} boundary between these is where the FC is satisfied. The free energy balance is the blue curve.

As $F_P$ is increased, the FC curve undergoes a profound change. As expected, the region of positive particle flux grows, since a greater particle flux is driven by the density gradient (thermodynamic forces drive their respective thermodynamic fluxes). Also, the region of negative particle flux shrinks. Hence, the curve where the FC is satisfied becomes smaller and smaller. Eventually, for large enough $F_P$, the entire upper half plane has positive flux: the FC has no solution, and there is no instability. 

Notice that the free energy curve does not change qualitatively during this process. Stability for higher $F_P$  comes because the red curve shrinks(and disappears); the solution will have progressively lower $\gamma$, until no positive value $\gamma$ solves the FC. 

For a very different sequence, consider progressively decreasing the curvature drift, going from positive to negative values. This will also stabilize the mode. See fig(\ref{fig:BP2}). We keep $F_P=0.2$, well away from the solubility limit. As we see, the free energy curve shrinks until it has a solution only near $\gamma \sim 0$. The FC, on the other hand, always has a considerable extent at high $\gamma$. The reason for stability in this case is clearly that the free energy balance equation cannot be satisfied for strongly negative (stabilizing) curvature.

Hence, these plots make it very clear that stability can result from two very different physical origins.  This is the essence of the dual dynamics approach. These same type of plots will be very clarifying for cases with non-adiabatic electrons.

We now turn to the stabilizing effects of curvature, i.e., energetic stabilization.

\section{ Curvature and Adaptability for {\bf $ITG_{ae}$} }

We will derive the analytic FC bound in the next section. But first, having introduced the concept of adaptability, let us show that it extends to considerations beyond the FC alone. 

In Section VI, we observed a truly remarkably adaptive behavior of $ITG_{ae}$: as the FC constraint tends towards insolubility, the mode structure changes, often radically, to remain unstable by staying in the soluble region of the FC. Response to curvature stabilization is similarly adaptive-simulations show that $ITG_{ae}$ mode ``fights'' \emph{the influence of stabilizing curvature by residing, preferentially, in the ``bad'' curvature region.} The eigenfunction averages are crucial in rendering such changes comprehensible. 

%As we will see next, the mode is similarly adaptive in its structure to avoid being stabilized due to curvature. We have found  as the adaptive behavior to the FC in the section above. And the eigenfunction averaged curvature is crucial to elucidating this. 

The results of this section will allow us to extend the scope of the "statistical mechanical ansatz'' by adding another item to the list:

\begin{itemize}

\item 	The apparent adaptivity also applies to the curvature stabilization -the eigenmode has a strong ability to stay out of the regions of strongly stabilizing curvature.   

\end{itemize}

Note that this adaptation circumvents curvature stabilization, and the imposition of the FC is crucial to stabilization.  %Note that without the crucial imposition of the FC, this adaptation could have have circumvented curvature stabilization. 
Since adaptation to prevent stabilization by curvature persists even when non-adiabatic electron effects are included in the ITG/TEM, the role of  FC dynamics for stabilization will remain crucial. 

Configurations favorable for TB formation, often, have strongly stabilizing curvature at most positions of the poloidal angle. This is well known~\cite{bour} for tokamaks, especially, where TBs are often found under conditions of negative magnetic shear or strong Shafranov shift. It is appropriate to investigate whether the stabilization seen in the simulations is also significantly affected by such energetic stabilization factors, in addition to the clearly established stabilization due to $F_P$. 

We first consider an equilibrium sequence similar to the ITB case A (with parameters given in Table 2), but make the magnetic shear much more negative. Recall that case A is roughly representative of a JET campaign with controlled density profiles in the ITB region using pellet injection. Also, quite similar parameters are often attained in high $\beta_{poloidal}$ cases on DIII-D. We consider these parameters, and also make the shear more negative yet. The cases $\#3$ and $\#4$ correspond, roughly, to the actual experimental ITBs on JET and many cases on DIII-D. Cases 5 and 6 have much more highly negative shear. However, they are roughly representative of some other experimental discharges with strongly reversed shear (e.g. NCS shots on DIIID and strongly reversed shear shots on JT60U). For completeness, cases 1 and 2 are included, and are roughly representative of situations \emph{before} an ITB, where $\alpha$ is not high and  $\hat{s}$ is not very low. 

\begin{center}
  \includegraphics[width=.95\linewidth]{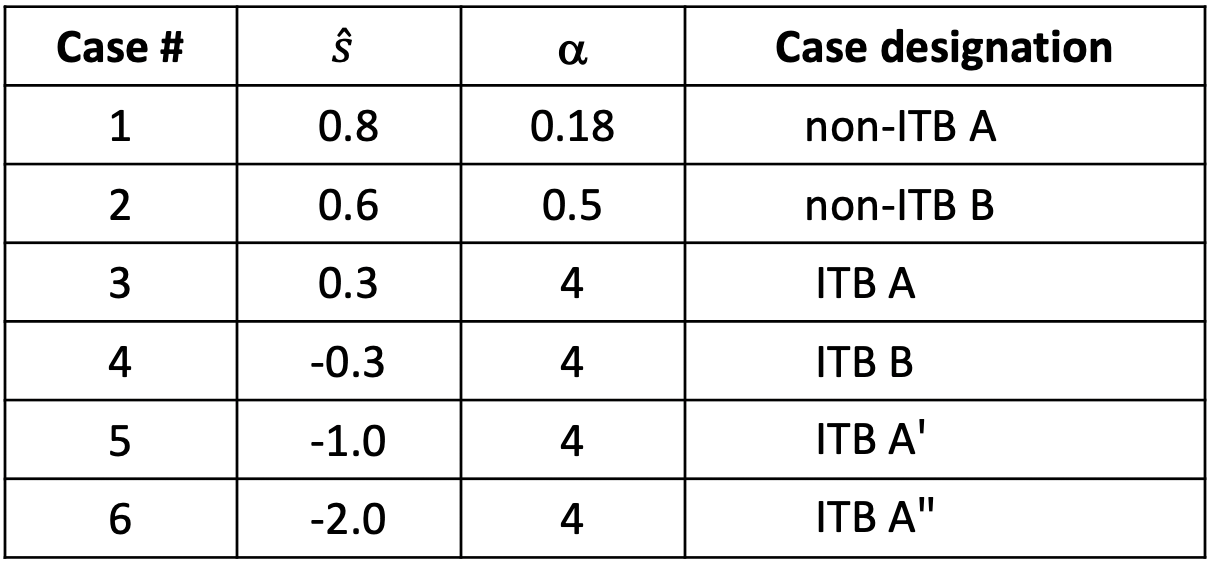}\\
Table 2: Parameters for simulations
\end{center}

\begin{figure*}
\subfloat[]{%
  \includegraphics[width=.51\linewidth]{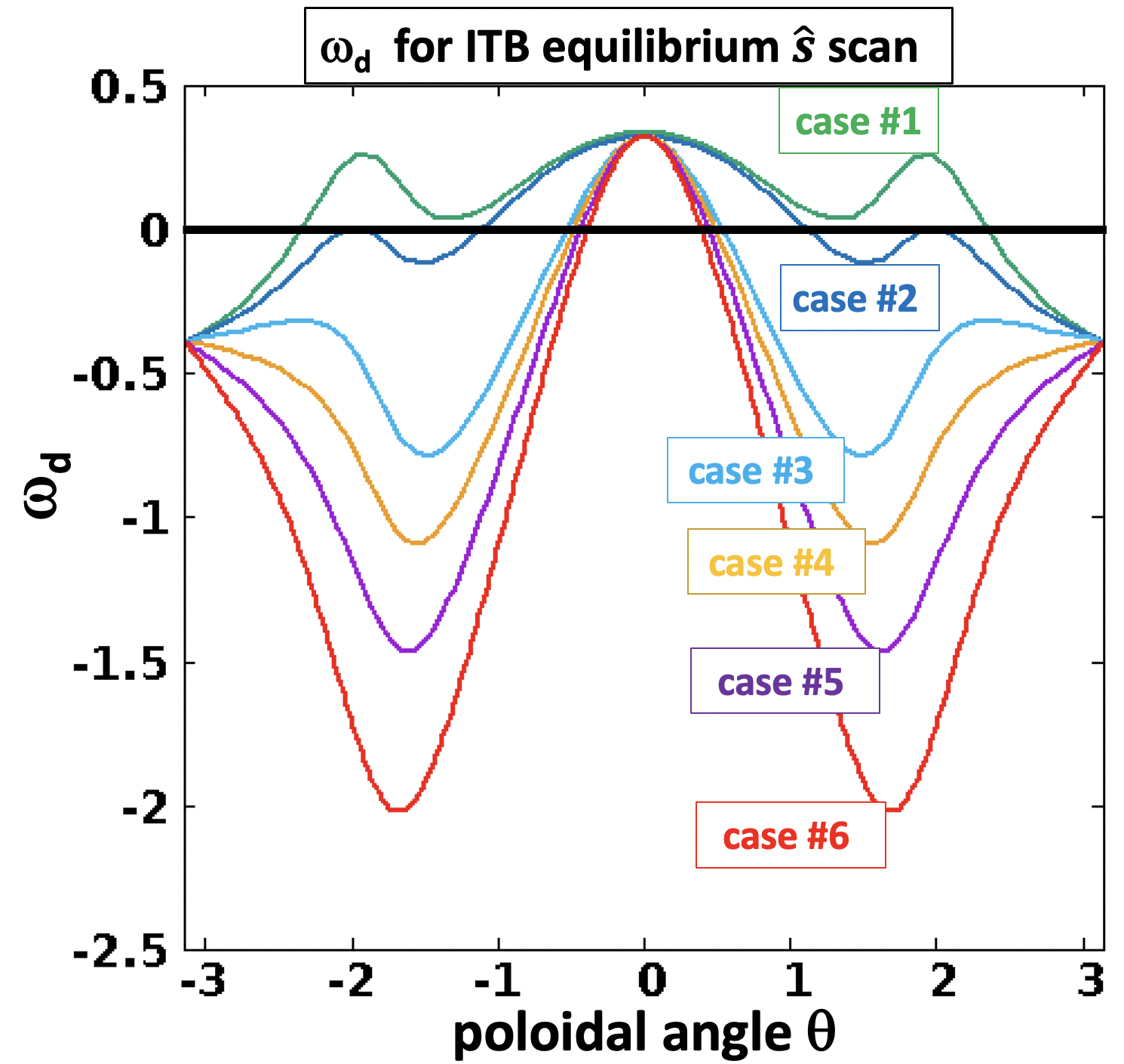}%
}\hfill
\subfloat[]{%
  \includegraphics[width=.49\linewidth]{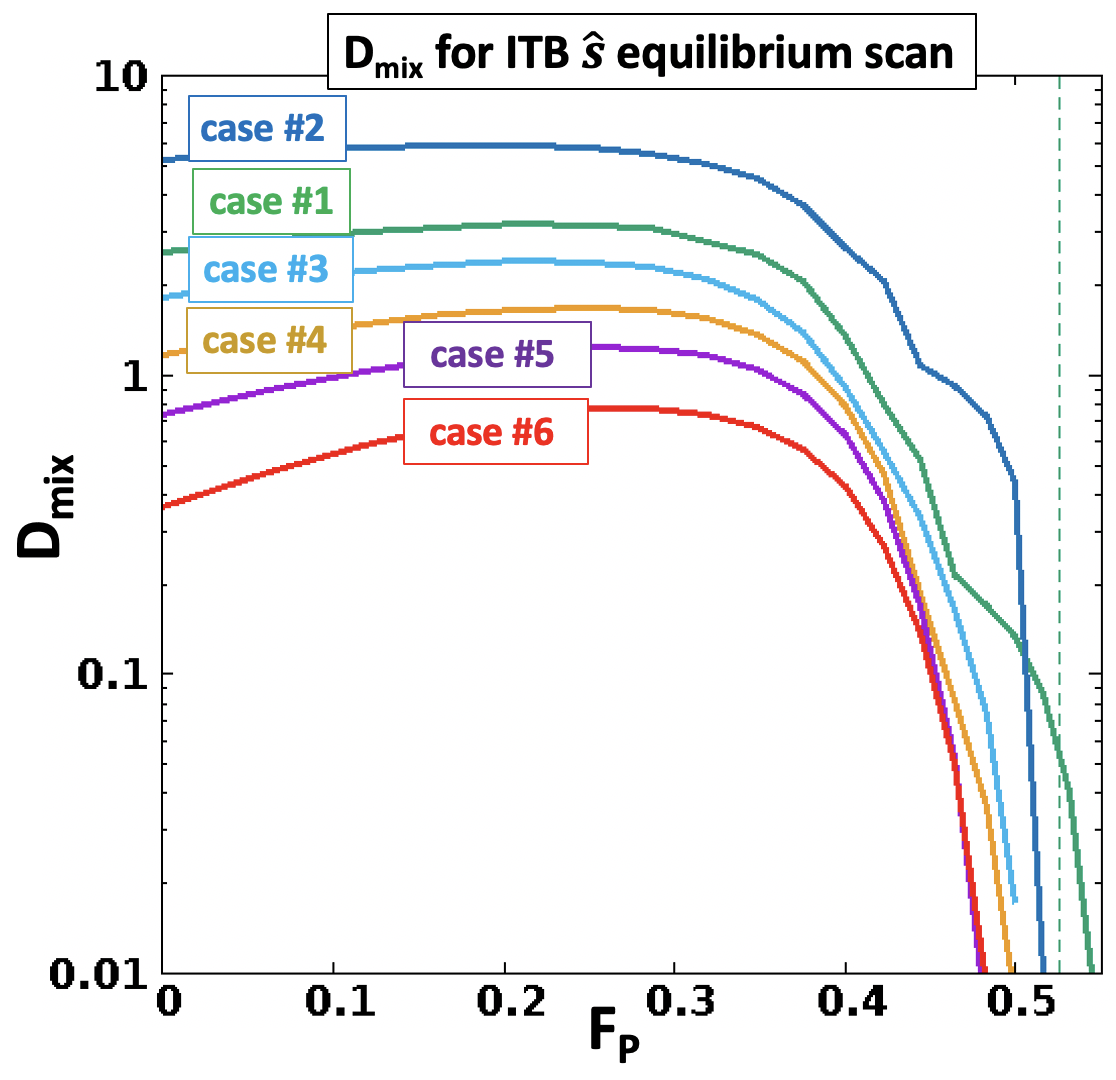}%
}
\vfill
\subfloat[]{%
  \includegraphics[width=.5\linewidth]{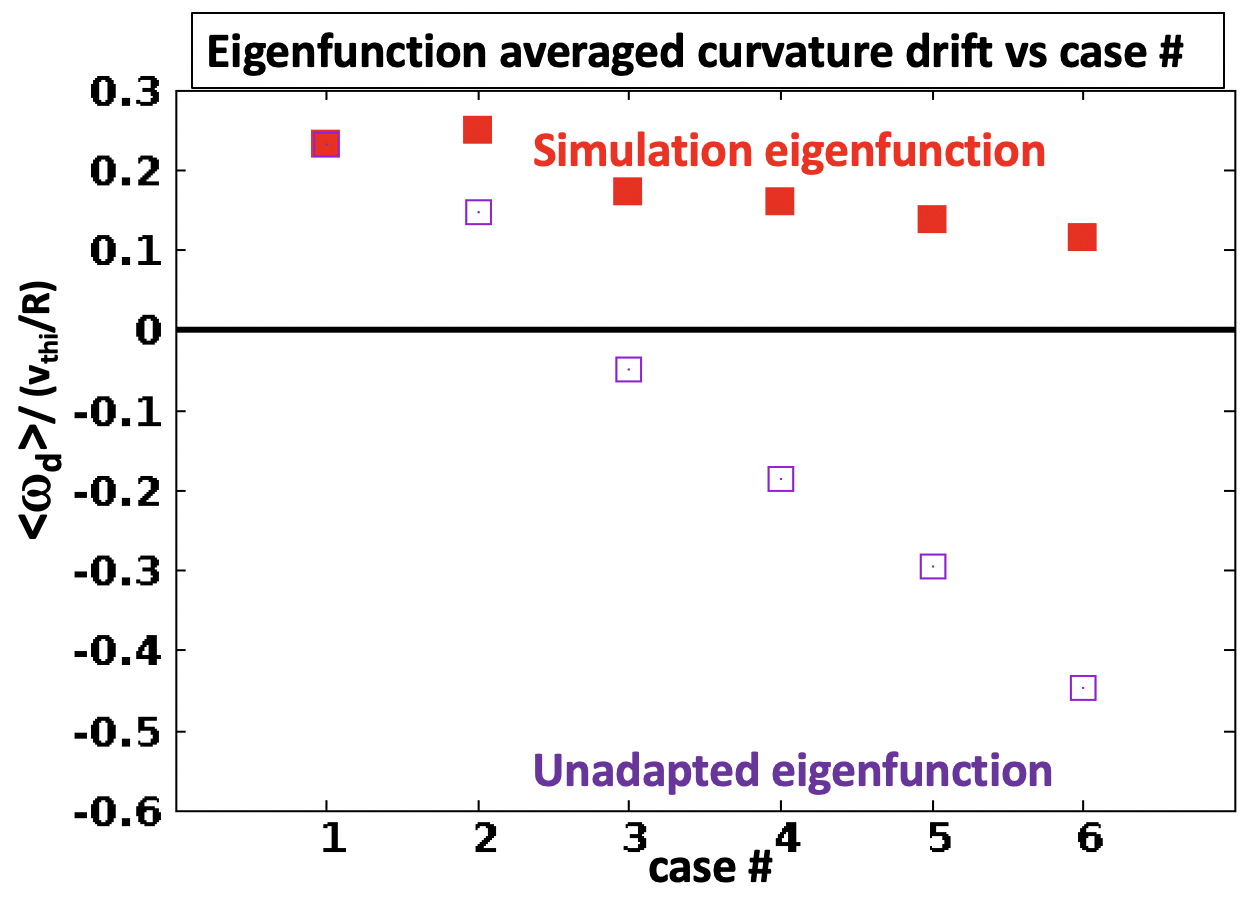}%
}\hfill
\subfloat[]{%
  \includegraphics[width=.5\linewidth]{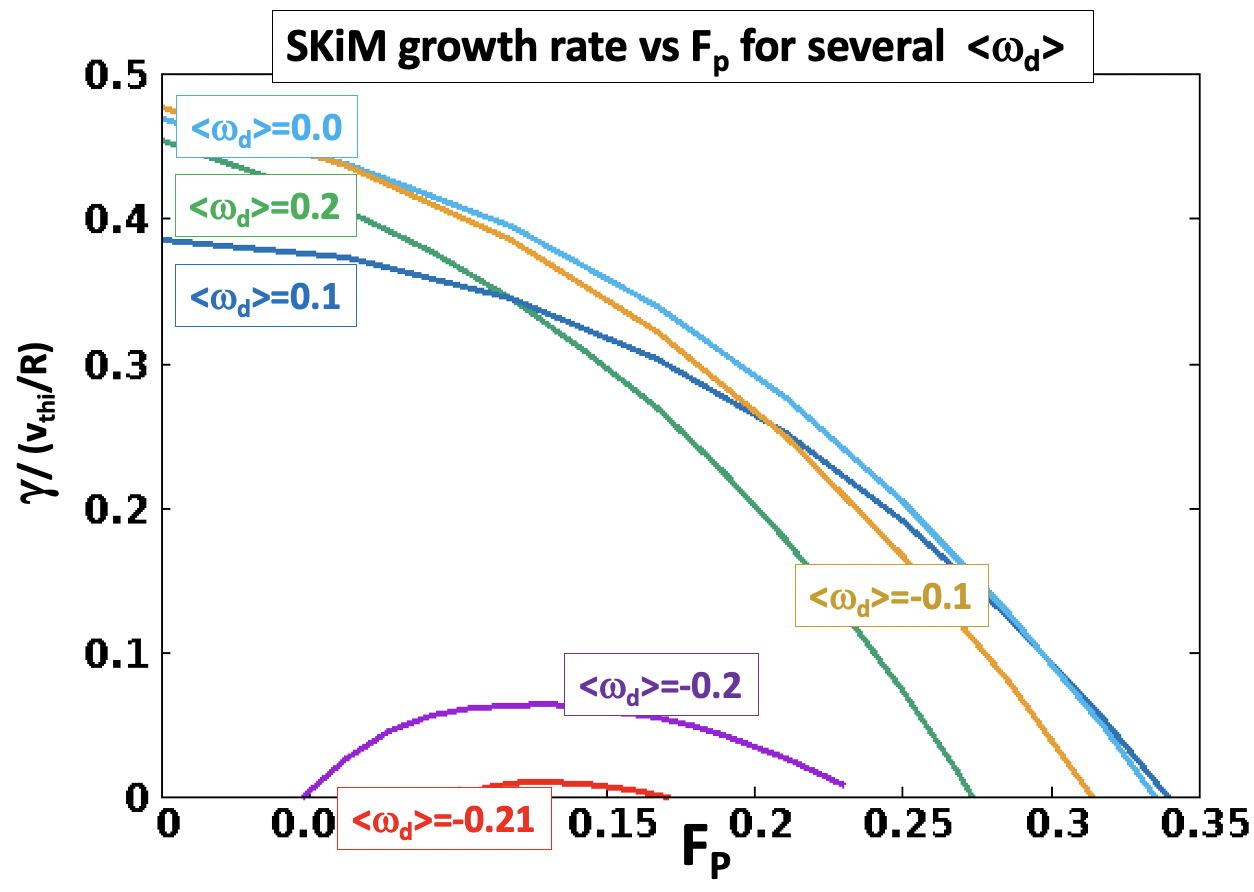}%
}
\caption{\label{fig:curv_ae1} a) The curvature $\omega_di(\theta)$ for the equilibrium sequences shown, for decreasing $\hat{s}$. It becomes overwhelmingly stabilizing at strongly negative  $\hat{s}$ b) $D_{mix}$ from simulations for $k_{\theta} \rho_i=0.35$; the maximum $D_{mix}$ for three values of $\eta_0$ are shown c) the eigenfunction averaged curvature for $F_P=0$ for both the true eigenfunction and the "unadapted" eignfunction for case 1; the simulation eigenfunction adapts to maintain destabilizing curvature, and if it had not, average curvature would become very stabilizing d) Growth rates from SKiM for parameters representative of these eigenfunctions, but for various $<\omega_d>$. Without adaptation, the curvature would stabilize the mode independent of the $F_P$ bound. }
\end{figure*}

The resulting curvature drive in the gyrokinetic equation, $\omega_d$, is plotted as a function of poloidal angle $\theta$ in fig \ref{fig:curv_ae1}(a) (for the standard ballooning coordinate description). Positive values are destabilizing and negative values are stabilizing. For experimental ITBs on JET, the curvature is very preponderantly negative. And as shear is even more negative, it becomes \emph{extraordinarily} stabilizing, except for a small region of "bad" curvature around the outboard mid plane. 

We run simulations on these equilibria as in the section above (with the same three different values of $k_x$), and plot the maximum $D_{mix}$ versus $F_P$ in fig 11b. The $D_{mix}$ is reduced an order of magnitude by the energetic effects, but the \emph{the stability point in $F_P$ is hardly affected}. 

The results of this parametric scan are yet another example of the separation between energetic effects and effects of the FC. 

But it is \emph{also an example of the strong adaptability of these modes to avoid outright stabilization by overwhelming "good" curvature}. We turn to this aspect now and use the eigenfunction average curvature,  Eq.~(\ref{eq:omdi}). 

\begin{figure*}
\subfloat[]{%
  \includegraphics[width=.33\linewidth]{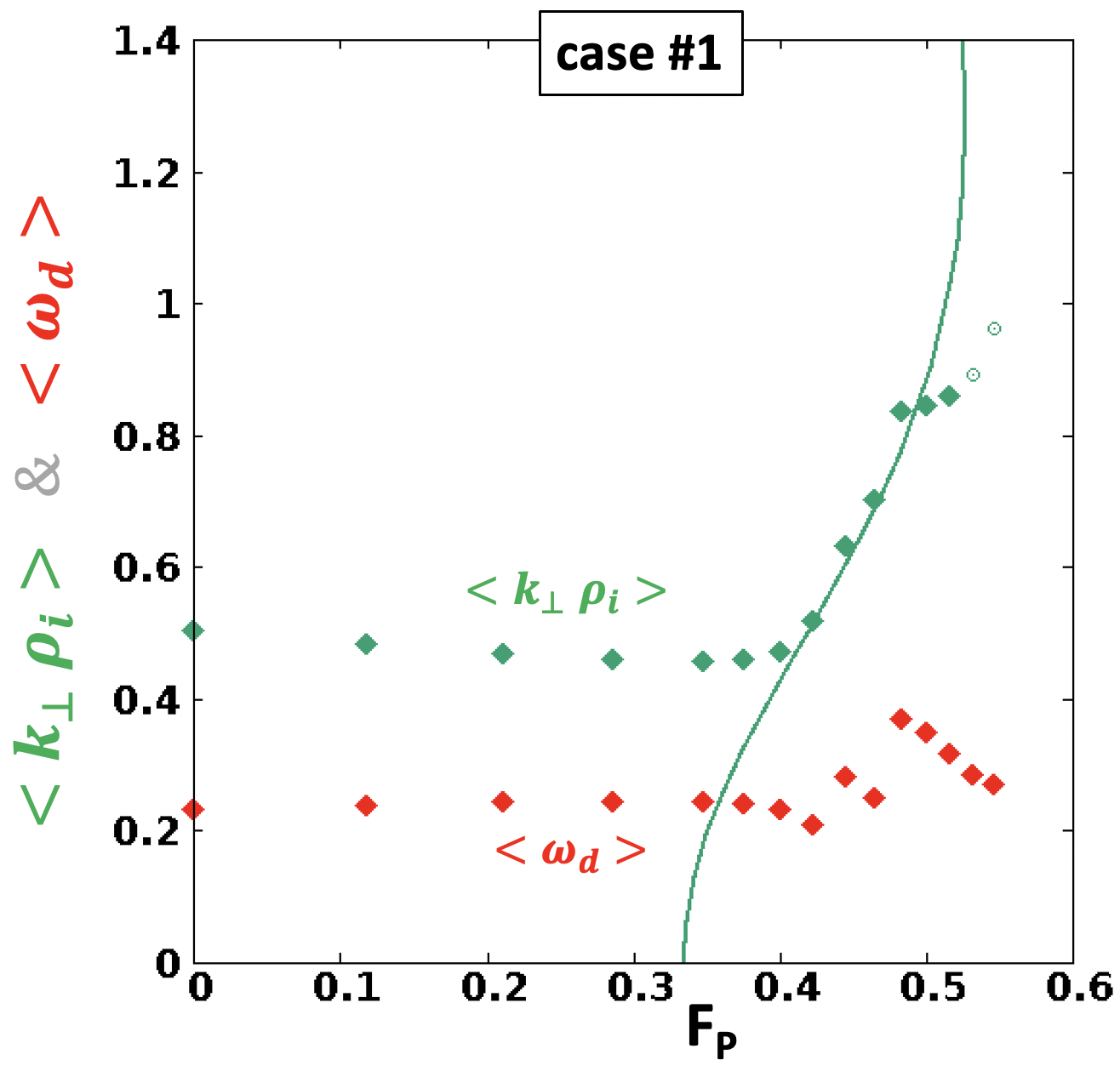}%
}\hfill
\subfloat[]{%
  \includegraphics[width=.33\linewidth]{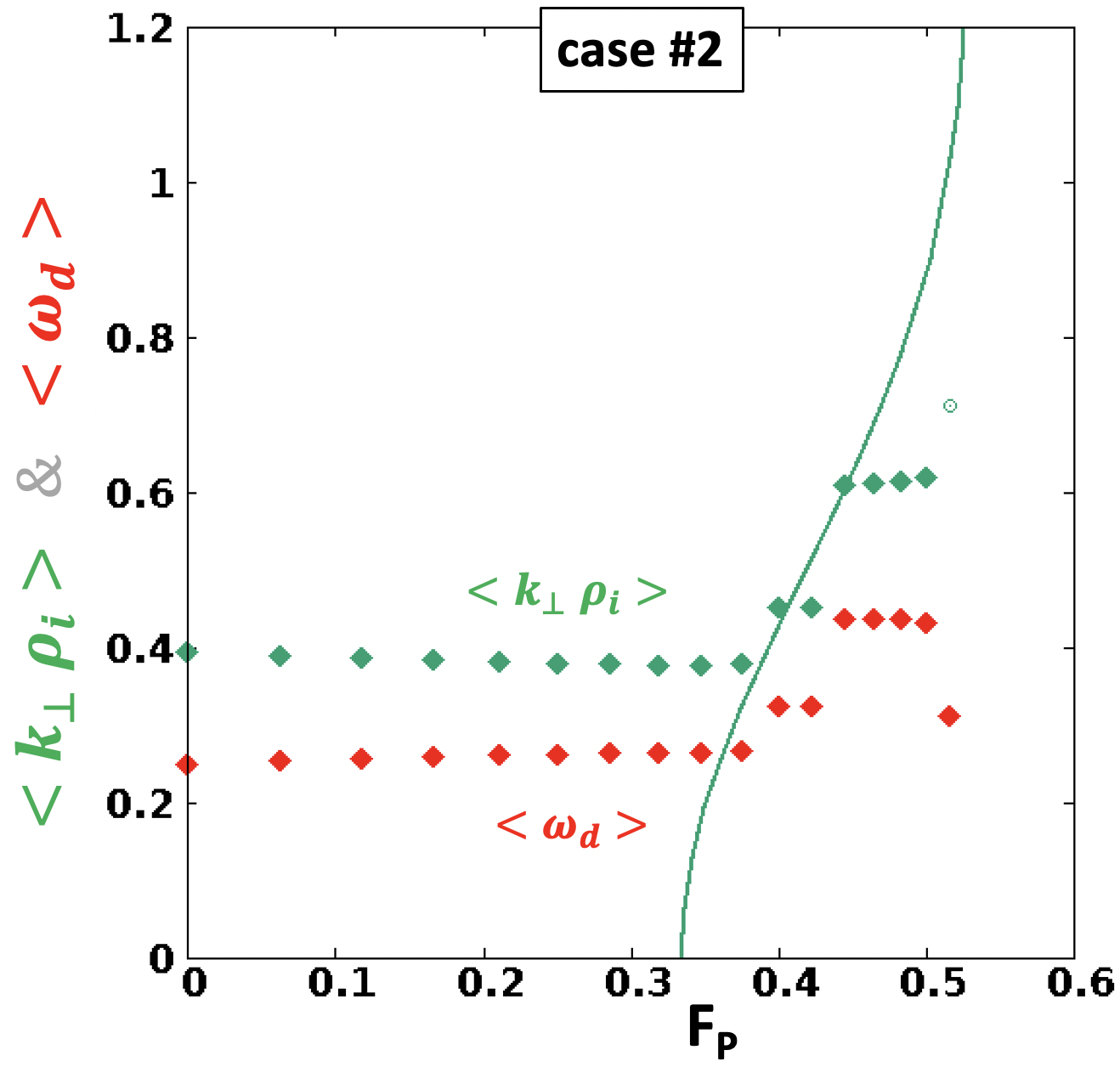}%
}\hfill
\subfloat[]{%
  \includegraphics[width=.33\linewidth]{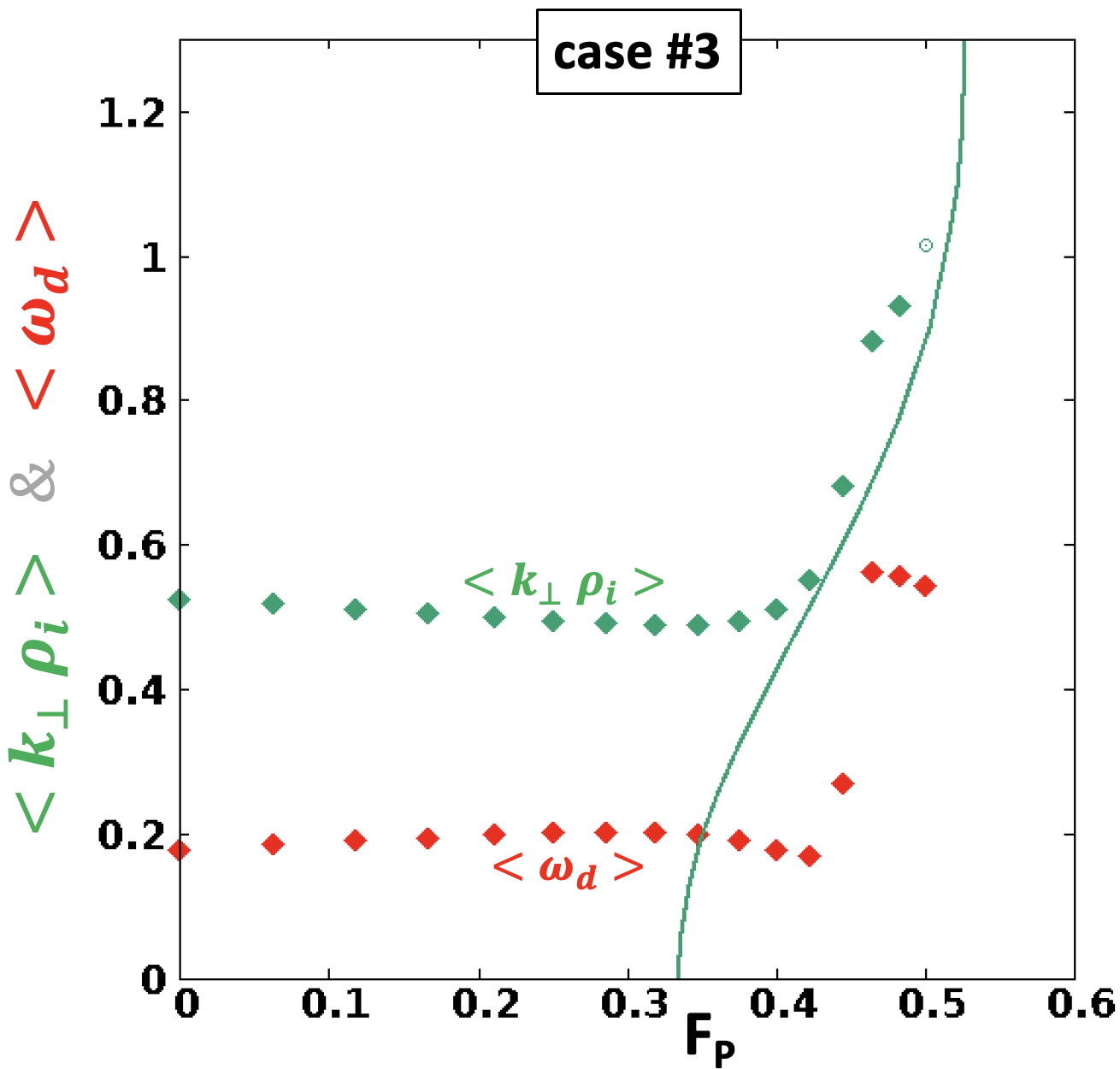}%
}\vfill
\subfloat[]{%
  \includegraphics[width=.33\linewidth]{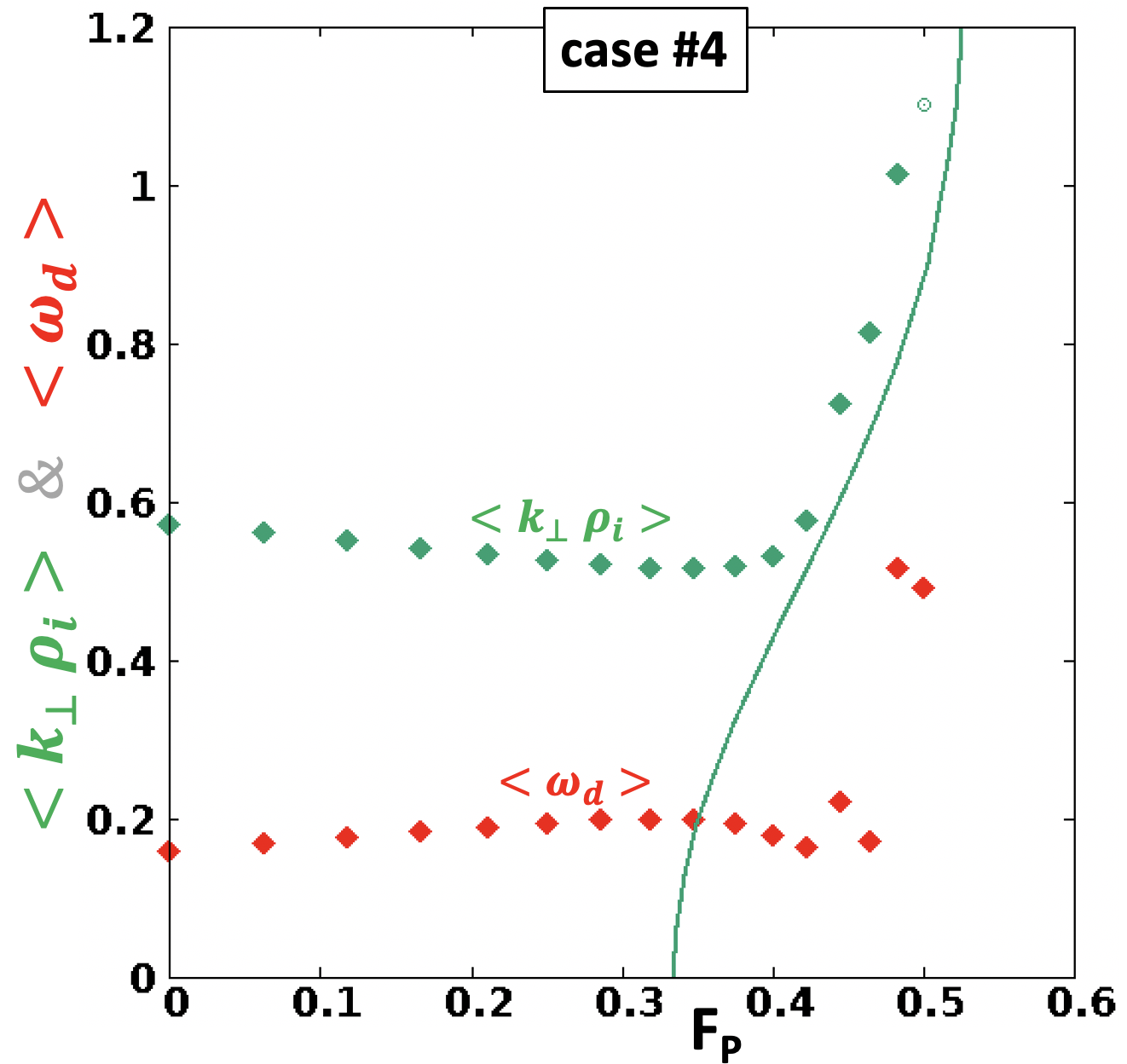}%
}\hfill
\subfloat[]{%
  \includegraphics[width=.33\linewidth]{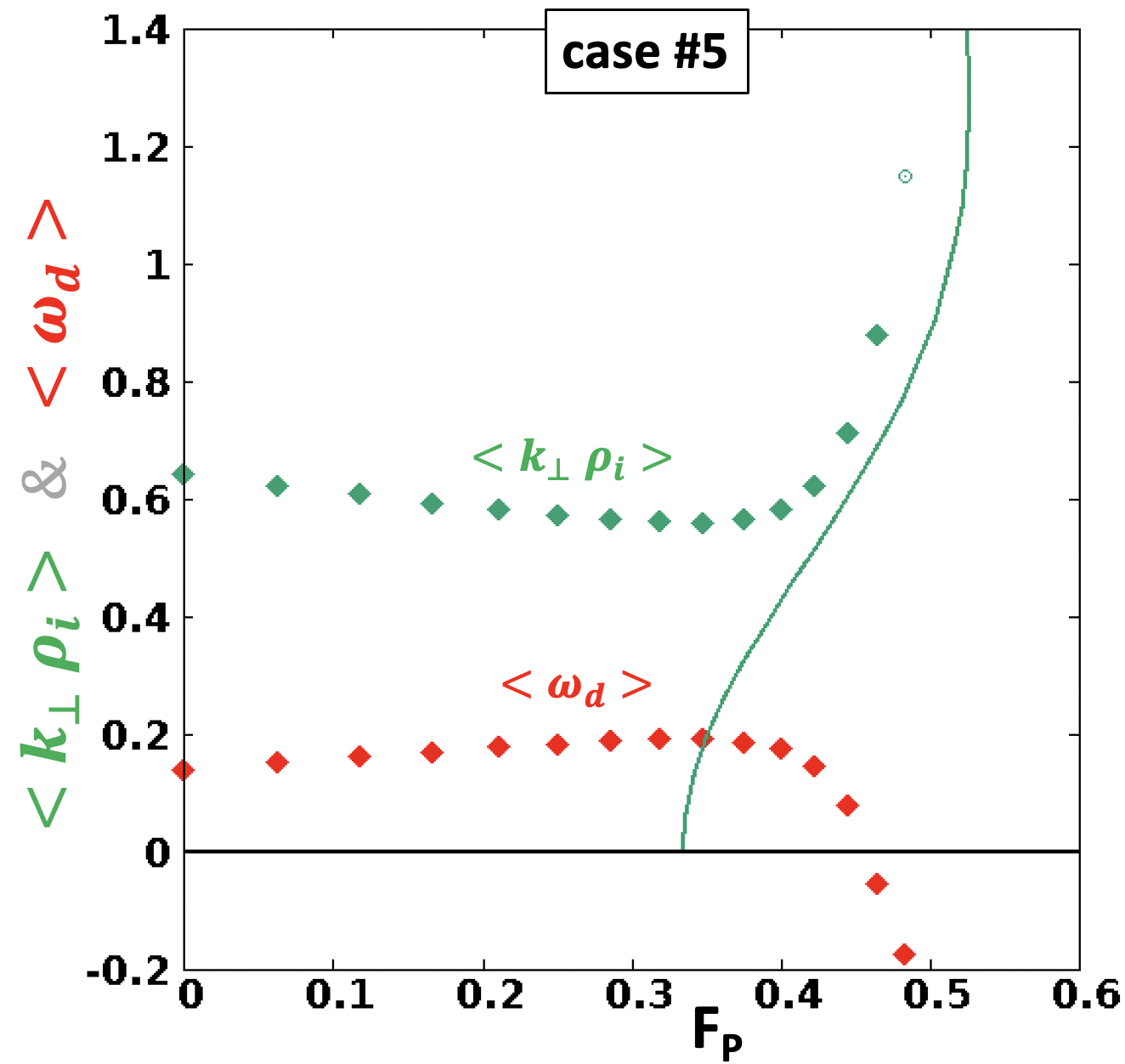}%
}\hfill
\subfloat[]{%
  \includegraphics[width=.33\linewidth]{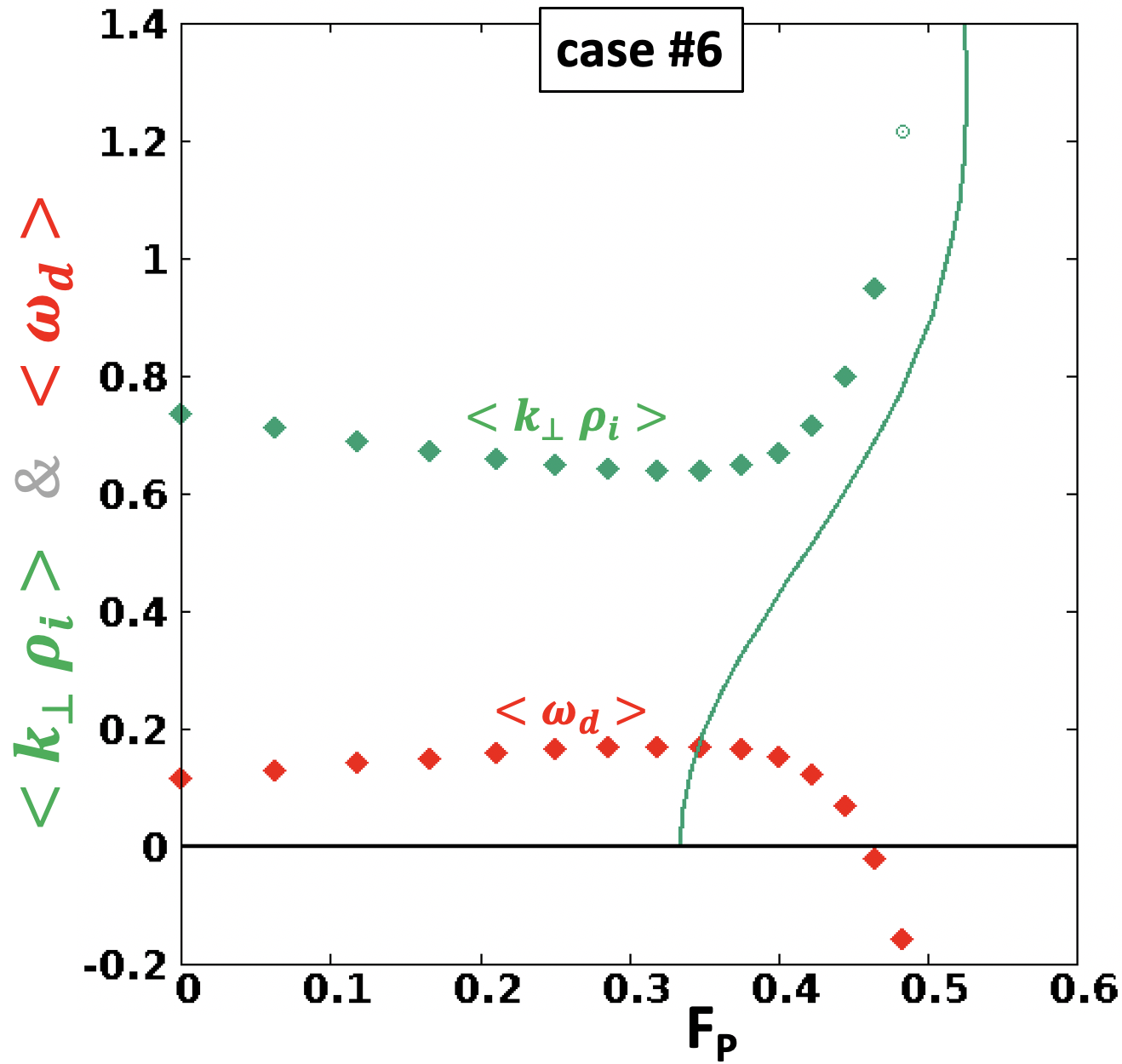}%
}
\caption{\label{fig:curv_ae2} Plots of the eigenfunction averaged $<k_\perp \rho_i>$ and average curvature $<\omega_d>$  for cases 1-6. The $<k_\perp \rho_i>$ follows the analytic bound for all cases. The average curvature remains positive except when the eigenfunction is called upon to simultaneously adapt to the energy and the FC while the curvature is  extremely preponderantly negative.}
\end{figure*}

Despite the enormously stabilizing curvature over most of $\theta$, the eigenfunction, by strongly concentrating in the bad curvature region, still remains unstable; the eigenfunction-averaged curvature Eq(\ref{eq:omdi})remains positive, although it does decrease somewhat (Fig 11c). The contrast in Figs 11a and 11c is quite amazing! Without such adaptability, the hugely negative curvature would have directly stabilized the modes, and density gradients would be unnecessary. We also show in figure 11c what \emph{would} happen if the mode had \emph{not} adapted. We do this via a thought experiment- we compute the average curvature for different equilibria but using the eigenfunction from case $\# 1$.  In this hypothetical situation, the average curvature would drop to sharply negative values. Using the SKiM dispersion relation and typical values for the other parameters, we plot the growth rate for various average curvatures. Sufficiently negative eigenfunction averaged curvatures stabilize the mode, as seen in fig\ref{fig:curv_ae1}d, for \emph{all} $F_P$. Such values \emph{would have} arisen if the mode structure had not adapted to stay predominantly in the bad curvature region.

\begin{figure*}
\subfloat[]{%
  \includegraphics[width=.32\linewidth]{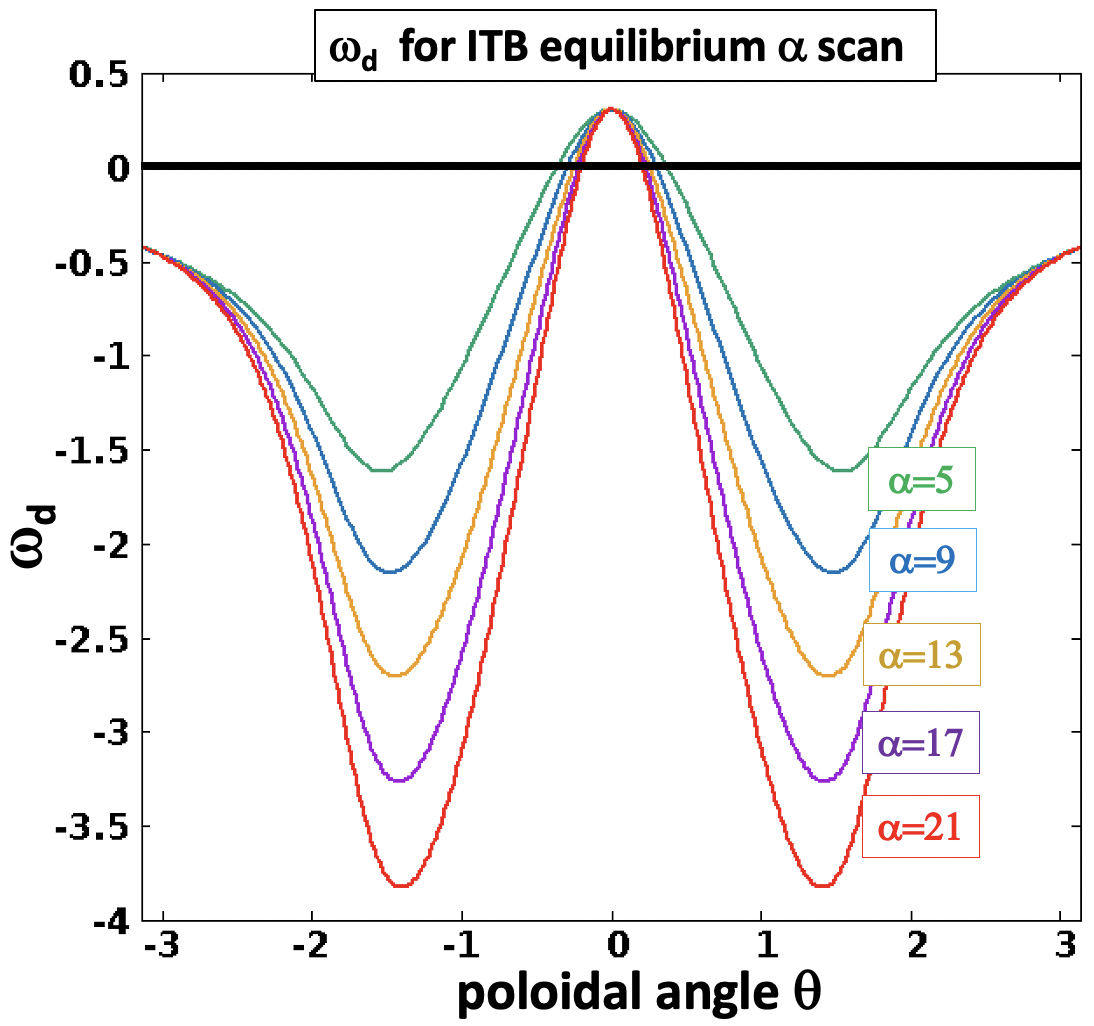}%
}\hfill
\subfloat[]{%
  \includegraphics[width=.32\linewidth]{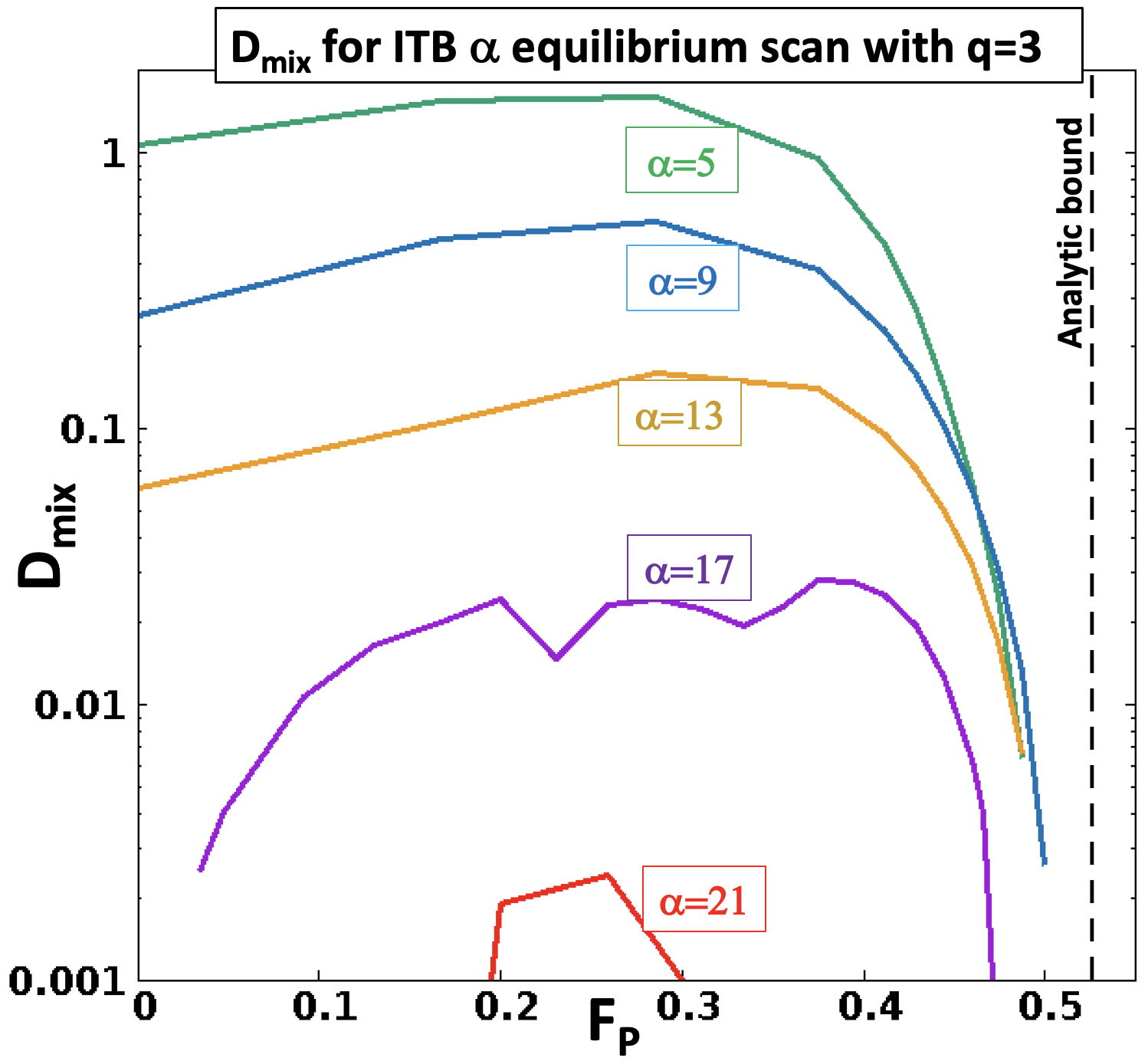}%
}\hfill
\subfloat[]{%
  \includegraphics[width=.32\linewidth]{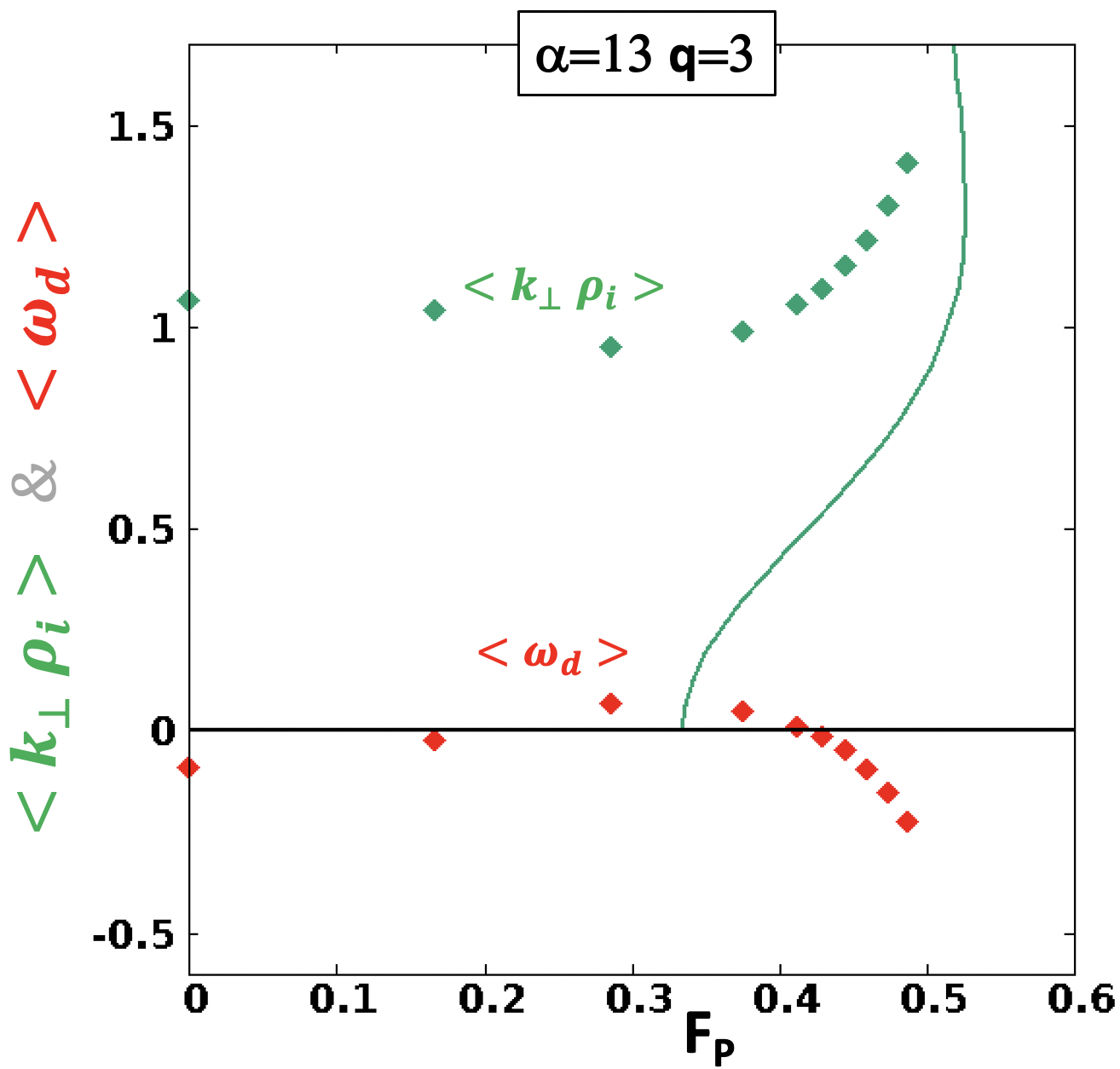}%
}\vfill
\subfloat[]{%
  \includegraphics[width=.32\linewidth]{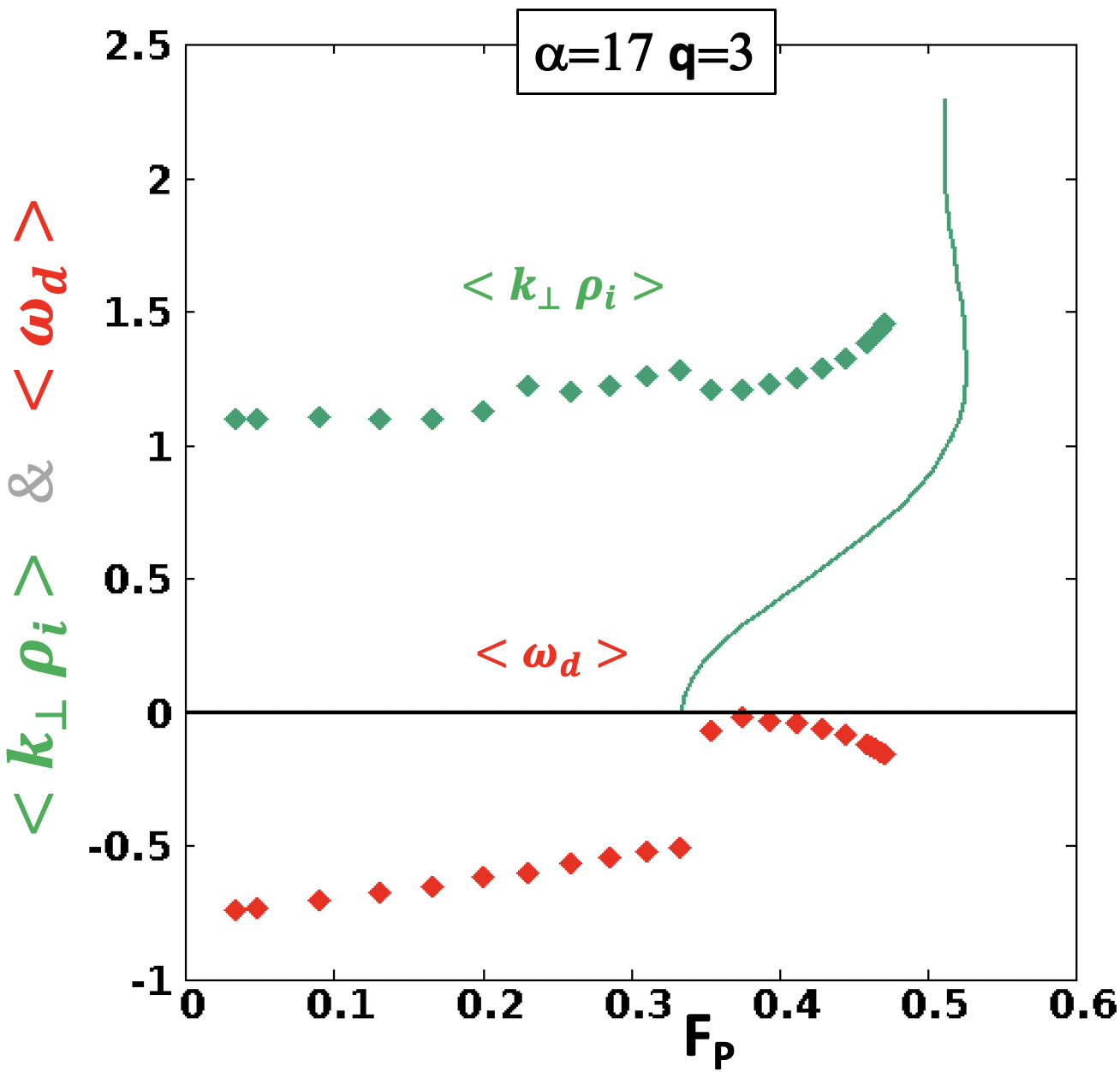}%
}\hfill13
\subfloat[]{%
  \includegraphics[width=.32\linewidth]{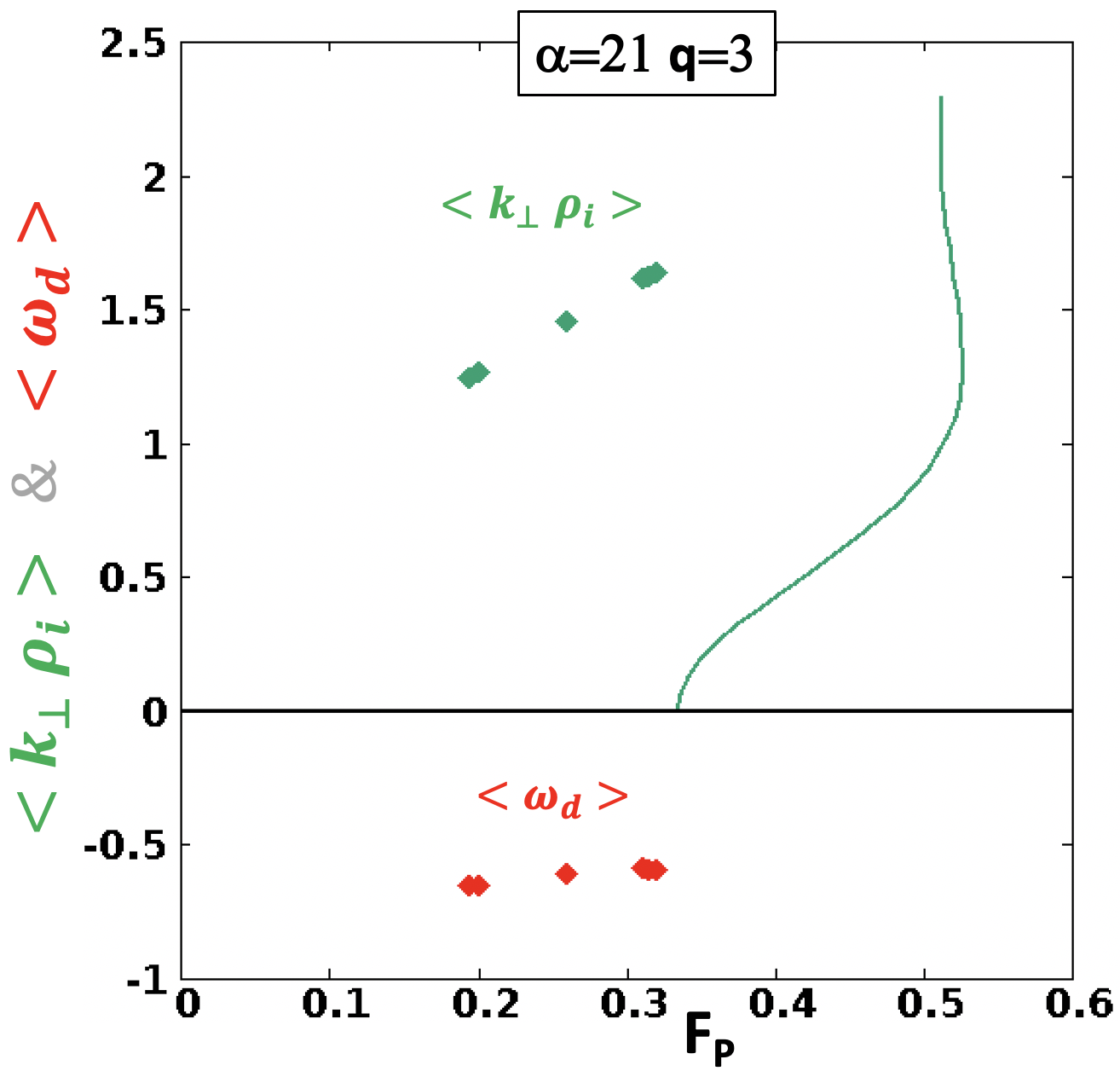}%
}\hfill
\subfloat[]{%
  \includegraphics[width=.32\linewidth]{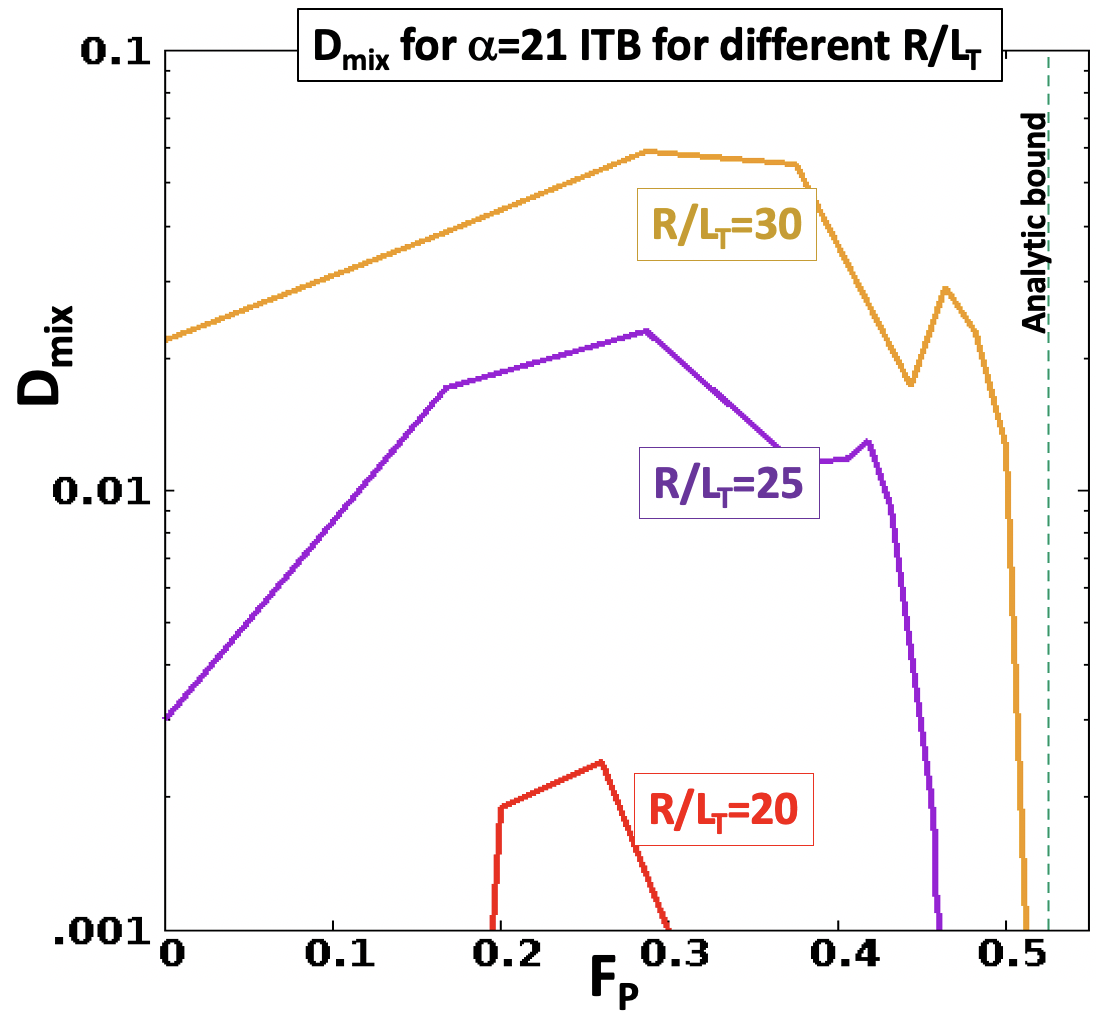}%
}
\caption{\label{fig:curv_ae3} a) The curvature $\omega_di(\theta)$ for the equilibrium sequences shown, for increasing $\alpha$ to extreme values. It becomes even more overwhelmingly stabilizing than in fig(\ref{fig:curv_ae1}) b) $D_{mix}$ from simulations for $k_{\theta} \rho_i=0.35$; the maximum $D_{mix}$ for three values of $\eta_0$ are shown c) the eigenfunction averaged $<k_\perp \rho_i>$ and average curvature $<\omega_d>$  for cases with c) $\alpha=13$,  d) $\alpha=17$ and  f) $\alpha=21$. The adaptability of the eigenfunction is being exceeded by the extreme curvature, so the average curvature is becoming quite negative over a large range of $F_P$ e)   $D_{mix}$ for $\alpha=21$ for increased temperature gradient $R/L_T$ from the value $R/L_T=15$ in b). This results in sufficient free energy so that, once again, the mode can only be stabilized by the FC.}
\end{figure*}

\begin{figure*}
\subfloat[]{%
  \includegraphics[width=.33\linewidth]{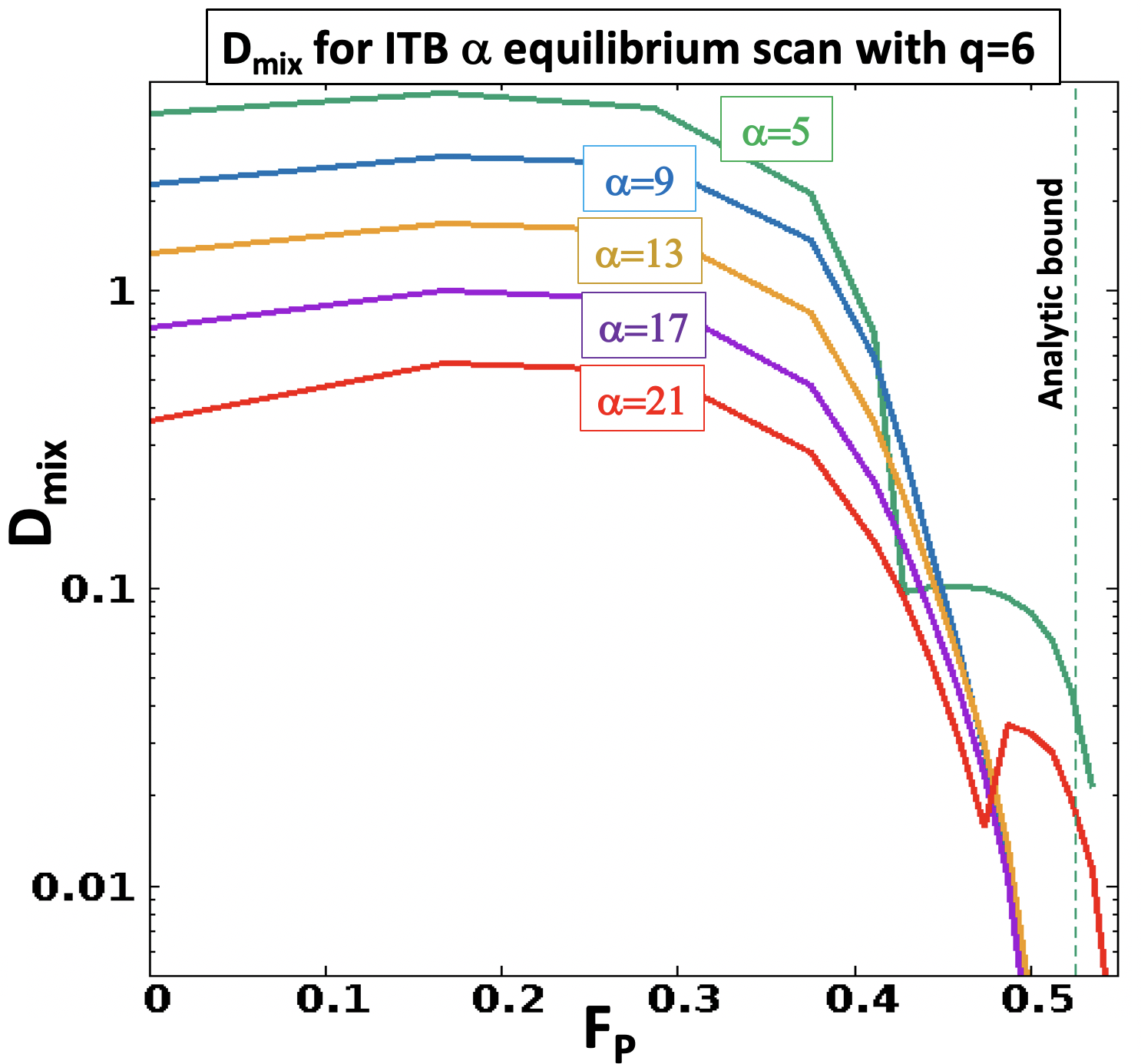}%
}\hfill
\subfloat[]{%
  \includegraphics[width=.33\linewidth]{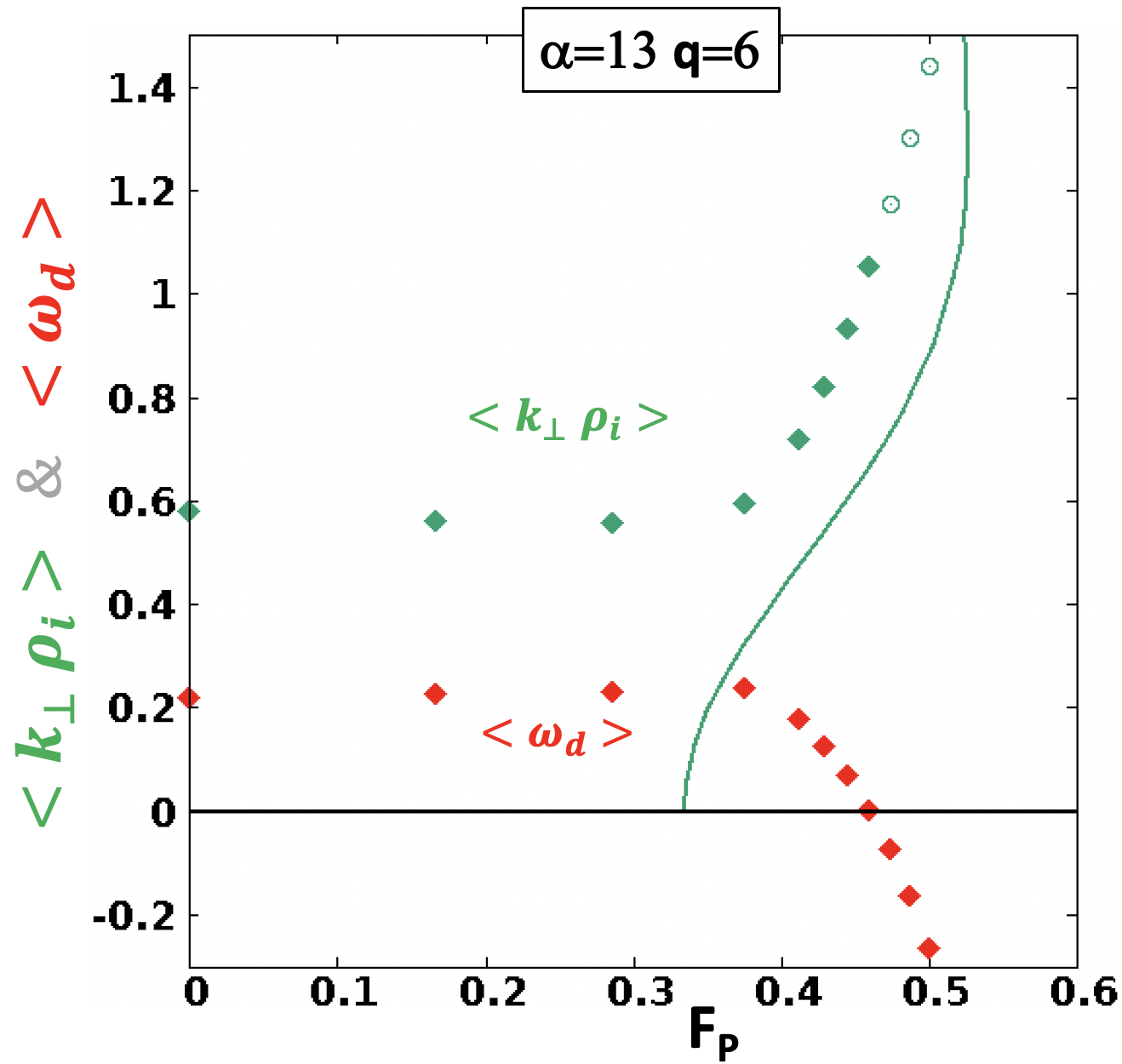}%
}\hfill
\subfloat[]{%
  \includegraphics[width=.33\linewidth]{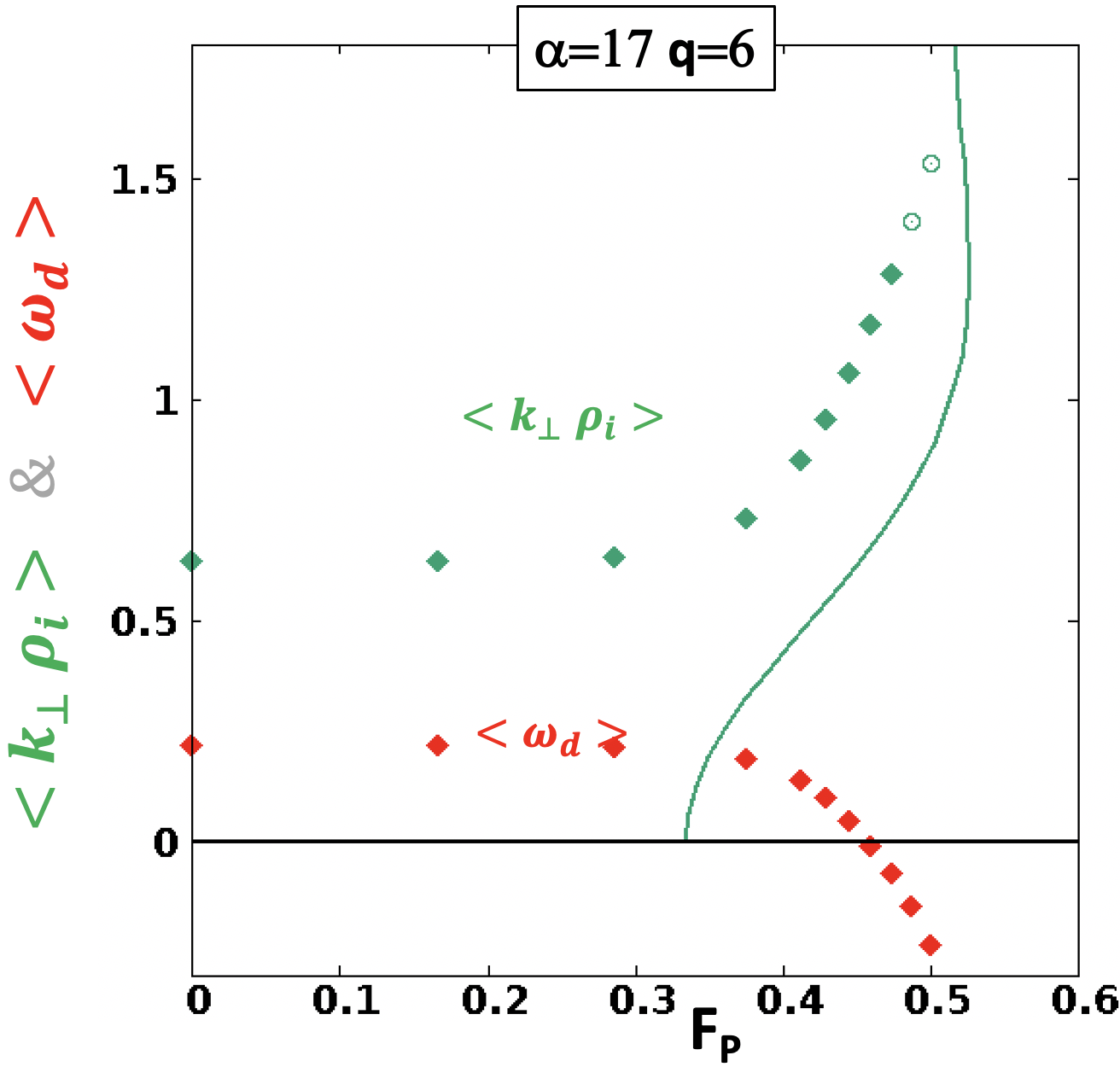}%
}\vfill
\subfloat[]{%
  \includegraphics[width=.33\linewidth]{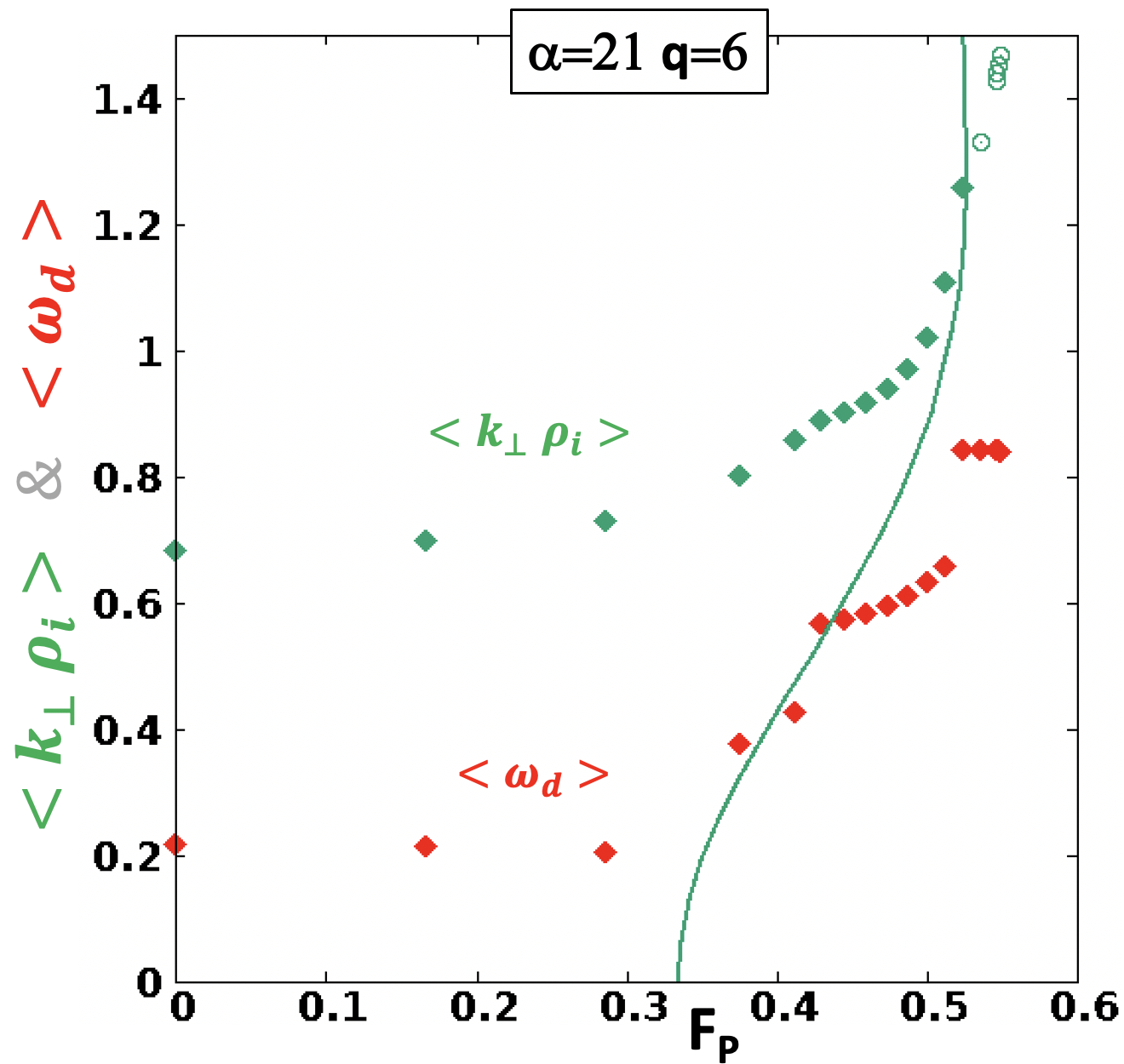}%
}\hfill
\subfloat[]{%
  \includegraphics[width=.33\linewidth]{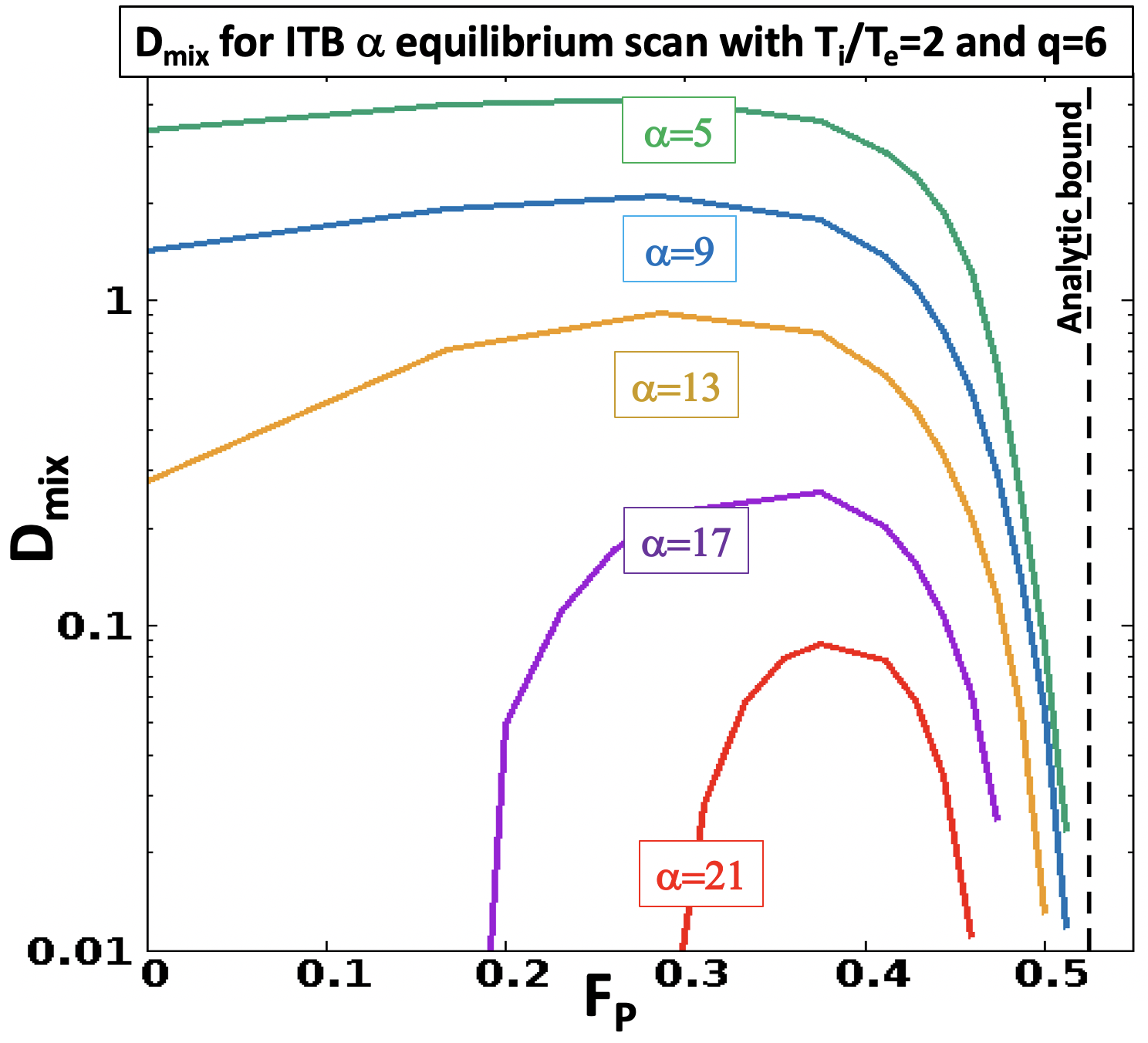}%
}\hfill
\subfloat[]{%
  \includegraphics[width=.33\linewidth]{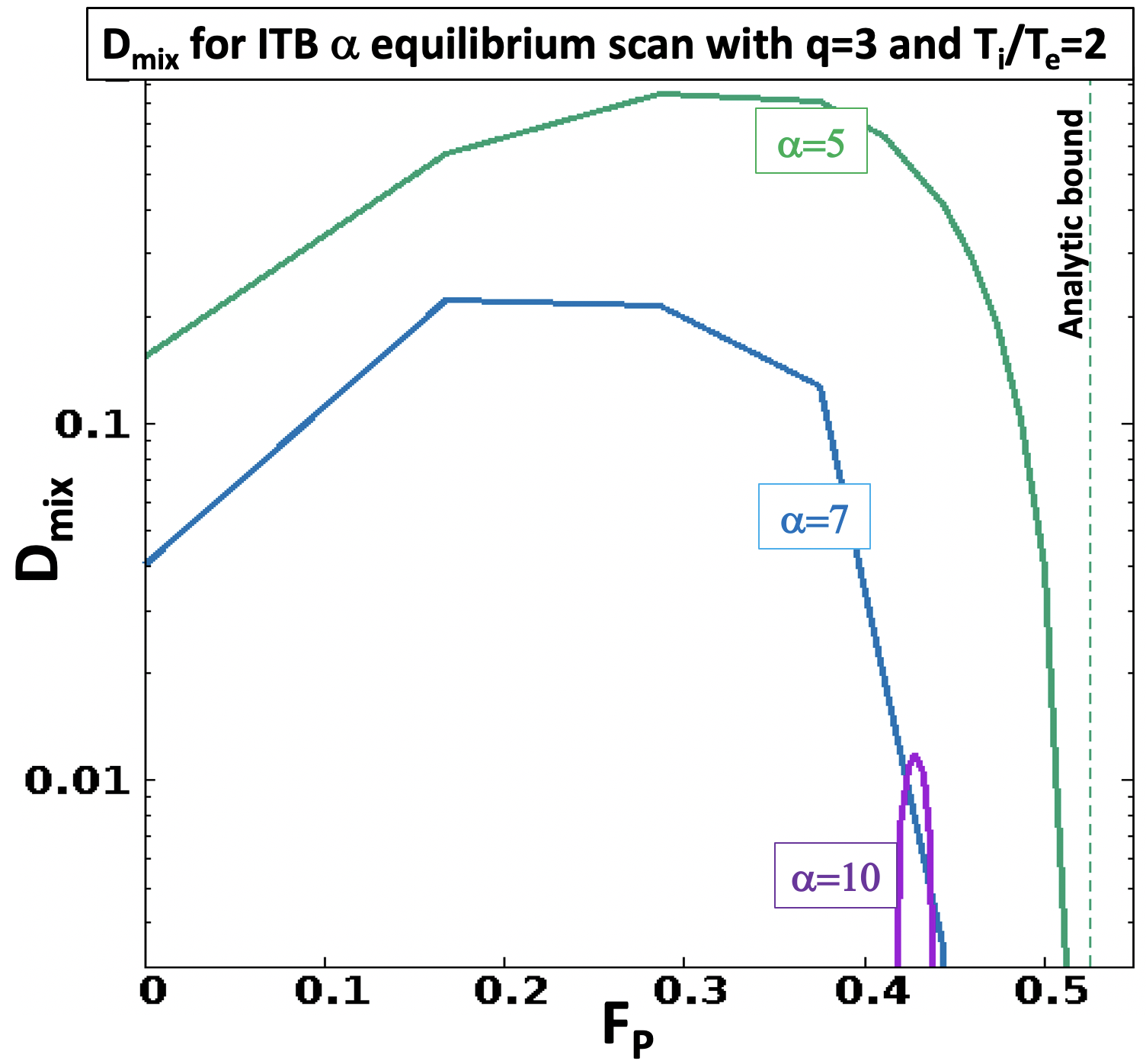}%
}
\caption{\label{fig:curv_ae4} a) $D_{mix}$ from simulations for the geometry in fig(\ref{fig:curv_ae3}) but now with $q=6$. Again, the maximum $D_{mix}$ for three values of $\eta_0$ are shown b) the eigenfunction averaged $<k_\perp \rho_i>$ and average curvature $<\omega_d>$  for cases with b) $\alpha=13$, c) $\alpha=17$ and  d) $\alpha=21$. The adaptability of the eigenfunction is restored by $q=6$, as described in the text. Negative average curvature results only when the eigenfunction is called upon to simultaneously adapt to the energy and the FC while the curvature is extremely preponderantly negative e) Increasing $T_i/T_e=2$ puts a further energetic load in the free energy balance, so the mode becomes energetically stabilized at lower $\alpha$ f) Since these affects of curvature and $T_i/T_e$  arises from basic properties of the free energy balance equation, it is expected to be typical. It is found for the $q=3$ case }
\end{figure*}

Notice that SKiM finds that somewhat before the point of complete stabilization by curvature, the critical $F_P$ is unrelated to the constraint prediction: one stability point occurs at low $F_P \sim 0$, and it also occurs for other $F_P$ that are much less than the solubility boundary of the constraint. Depending upon the particular value of negative average curvature, the critical $F_P$ can be anything from 0 up to the constraint value. In other words, curvature stabilization is through the free energy equation, and so the stability point is unrelated to the FC solubility bound. 

\emph{Simulations find that stabilization independent of $F_P$ rarely happens for TB level gradients ($R/L_T > 15$ or so). The simulation results in fig \ref{fig:curv_ae1} are typical of many cases we have examined. Because of adaptability to stabilizing curvature, only the FC can give stability for most ITB parameters. Hence, (sufficient) density gradients are usually a crucial ingredient for suppressing ITG modes in ITBs, at least those with sharp gradients.} This is, at least qualitatively, in full agreement with the JET experimental results for cases with low velocity shear; density gradients were, indeed, crucial to the ITB in the experiment. In fact, simulations find that the crucial role of density gradients persists even for much more negative shear than is typically found in experiments. 

To emphasize: \emph{ the mode adapts to avoid curvature stabilization; It is only stabilized by the FC.}

 As $F_P$ increases, \emph{eigenmode structure in TBs must adapt to both the curvature and to the FC solubility to remain unstable. }

Interestingly, as the FC solubility boundary is approached, the mode must undergo two metamorphisms simultaneously: It must increase $k_{\perp}$ to remain unstable, and at the same time strongly peak in whatever bad curvature region that is available. The mode tries to do exactly that, as  shown in simulations Fig.~\ref{fig:curv_ae2}. The eigenfunction adapts to have positive eigenfunction average curvature except for cases with truly extreme curvature and large FP near the solubility limit. This simultaneous combination exceeds the adaptive capacity of the system. 
%That the mode does exactly that is amply shown in simulations (fig \ref{fig:curv_ae2}).  {\color{red} some additional explanation may be needed}. 
At larger $F_P$, the region of destabilizing curvature region is not near $\theta=0$, but is located at $\theta> \pi$ (this region is not shown in fig(\ref{fig:curv_ae1}a)).The eigenfunction average curvature stays positive even as the mode structure changes to increase $k_{\perp}$, except very near the solubility limit for the cases with the most extreme curvature (case $\#5$ and case$\#6$).

 \subsection{Exploring Extreme cases for advancing theoretical foundations}

But in these extreme curvature cases, whether the curvature is positive or slightly negative at large $F_P$, the stability point is always close to the analytic solubility bound. This is seen even more dramatically by considering the following case that is even more extreme. The example below further displays and reinforces the differences between stabilization by the FC, and stabilization by energetic effects. 

This case is motivated by the following reasonable supposition: one would expect that there is \emph{some} limit to adaptivity to the curvature. For example, if the region of bad curvature becomes sufficiently narrow, and the mode isolates to that region, the $k_{\parallel}$ becomes large enough that the ions begin to become adiabatic. Since no unstable modes can exist if both electrons and ions are adiabatic, energetics would not permit extreme localization for growing modes. Thus the unstable mode will  leak out to the stabilizing curvature region, and ultimately become stabilized. 

To investigate this theoretically most interesting expectation, we consider an equilibrium sequence where the Shafranov parameter is increased to extreme values that are rarely seen in experiments, $\alpha =21$. We simultaneously keep the safety factor q at a fairly low value, $q=3$, in order to make $k_{\parallel}$ relatively large.  In such conditions, there is no mode structure that can avoid curvature stabilization: either the mode narrows to avoid stabilizing curvature but then becomes stabilized by adiabaticity, or, it keeps $k_{\parallel}$ small enough but then is broad enough to experience highly stabilizing curvature. The geometry is so challenging that no adaptation to stay unstable will succeed. 

The curvature for the $q=3$ scan is shown in fig \ref{fig:curv_ae3}a. Simulation results for $D_{mix}$ are shown in fig  \ref{fig:curv_ae3}b. At all but the highest $\alpha$ value, the behavior is the one we are familiar with from previous examples: the critical $F_P$ is very close to the solubility limit of the FC. This is so even though the eigenfunction averaged curvature varies quite strongly among the cases. 

For the highest $\alpha=21$, the stability resembles the SKiM result for highly negative eigenvalue averaged $\omega_d$: stability occurs for much lower $F_P$ than the bound on the constraint, and   $D_{mix}$ becomes very small for all $F_P$ values. The eigenfunction averaged curvature  becomes strongly negative- more so than the other cases; the mode becomes stable far from the FC solubility curve (see \ref{fig:curv_ae3}e).

For further confirmation that this is an energetic effect, we increase the temperature gradient. As we have seen, this affects the energetics but has no effect upon the FC solubility bound. We see in fig (\ref{fig:curv_ae3}f) that even a small increase in $R/L_T$ greatly increases $D_{mix}$, and the critical $F_P$, once again, becomes consistent with FC bound. 

Such extreme conditions are atypical in tokamak experiments, because \emph{global} ideal MHD instabilities would usually arise. However, parameters approaching this have occurred in, for example, in the JT-60U discharge with record fusion triple product. The sequence below is roughly based upon this shot, where $\alpha$ reached values $\sim 10 -15$. More typically, though, very high $\alpha$ occurs experimentally at higher $q$. (This is qualitatively due to the parametric dependencies upon plasma current of $\alpha$ and the global MHD stability parameter for tokamaks, $\beta_{Normal}$. Roughly, $\alpha \sim q \beta_{Normal}$. Hence, staying within the global stability limits on $\beta_{Normal}$ it is easier to reach high $\alpha$ at high $q$.) Conditions of high $\alpha$ and high $q$ arise in high $\beta_{poloidal}$ discharges on, for example, DIII-D. Therefore, we will also consider a scan to $\alpha =21$ for a case with higher $q=6$ and compare it to the lower $q$ scan. Note that curvature drift $\omega_d$ differs by only a few percent from the $q=3$ cases. 

A continuing theme in this work is the distinguishably different effect of the FC and energetic considerations. We continue with a few more illustrative examples. 

Next, we check whether the energetic stabilization is related to large $k_{\parallel}$ by increasing $q$ from three to six. For a given mode structure, this roughly halves $k_{\parallel}$. We see in fig \ref{fig:curv_ae4} that energetic type stabilization no longer occurs, and the stability point in $F_P$ is once again near the FC solubility bound. From the analytic derivation, we know that variations in $q$ (and thus $k_{\parallel}$) have no effect upon the FC solubility bound. But the safety factor $q$ is well known to affect the critical temperature gradient of the $ITG_{ae}$ in conventional regimes, and, $k_{\parallel}$ should affect the free energy equation (by making the ions closer to adiabatic, thus reducing the ion flux - a stabilizing effect). The sensitivity of the stabilization seen in this scan to $q$ further confirms this as an energetic effect.

Finally, we increase $T_i/T_e$, which does not affect the FC bound for the $ITG_{ae}$, but it does strongly affects the energetics. This increases the energetic burden to be both unfavorable curvature and high $T_i/T_e$. Simulations find that energetic-type stabilization behavior occurs for substantially lower $\alpha$. But before this energetic stabilization happens, stability in $F_P$ occurs near the analytic bound. Since this arises from basic properties of the free energy balance equation, we expect this to be typical. For example, simulations find this for the $q=3$ case: higher $T_i/T_e$ results in energetic stabilization for lower $\alpha$ than for $T_i/T_e=1$.

Note that when trapped electrons are introduced, the ITG/TEM mode also shows strong eigenfunction adaptation in response to curvature (to avoid energetic stabilization) just as in $ITG_{ae}$. Eventual stabilization, however, occurs when $F_P$ is increased; density gradients are, thus, a crucial ingredient of weakening ITG/TEM.

In conclusion, these examples are illustrative of the differences between stabilization due to the FC and stabilization due to free energy considerations. Crucially, distinctly different types of stability behavior are expected from each type of stabilization, with very different controlling parameters.  Knowing what process is relevant when, advances our understanding of  the physics and phenomenology of TB sustainment.

In the next section, we derive the analytic bound for the FC that applies for low $k_y$ modes.

\section{Derivation of the Flux Constraint for low k modes}

Examining Eq(\ref{eq:FC0}), the mode  number appears in two places: though $<\omega_{di}>$  and through the arguments of Bessel function $J_0$. In the limit of vanishing $k$, the FC becomes insoluble for $F_P> 1/3$ (it differs from the limiting value $F_P> .6$ due to the contribution of the second term in the square brackets). However, simulations show:1) low $k$ instabilities regularly exceed  $F_P=1/3$, and 2) instabilities occur at  $<k_{\perp}>$ that cannot be considered small. But almost always,  $<\omega_{di}>$ remains small compared to $<k_{\parallel}>$. We therefore analytically derive a new FC bound by taking $<\omega_{di}>$ small compared to $<k_{\parallel}> v$, but allow $<k_{\perp}> \rho_i$ to have any value. By neglecting $<\omega_{di}>$ in the denominator, one can integrate the resonant denominator in $v_{\parallel}$ for low $\gamma$. We are particularly interested in the region near marginal stability, where a delta function approximation to the resonant denominator suffices. The integral in $v_{\parallel}$ becomes trivial, and Eq(\ref{eq:FC0}) becomes 

\begin{eqnarray}
&&\int d{v_{\perp}} v_{\perp} e^{(v_{\perp}^2+\omega_r^2/k_{\parallel}^2)/v_{th}^2} J_0(k_{\perp}\rho_i)^2   \nonumber\\
&&\big[ 1/L_n+(v_{\perp}^2+\frac{\omega_r^2}{k_{\parallel}^2}-3/2)/L_T\big] = 0 \nonumber \\ 
\label{eq:FC1}
\end{eqnarray}

For $\omega_r=0$, the flux solubility boundary, 

\begin{eqnarray}
&&  \frac{1}{L_T} \frac{\int d{s_{\perp}} s_{\perp} e^{-s_{\perp}^2/2} J_0(k_{\perp norm}s_{\perp})^2 (3/2-s_{\perp}^2/2)}{\int d{s_{\perp}} s_{\perp} e^{-s_{\perp}^2/2} J_0(k_{\perp norm}s_{\perp})^2}  \nonumber\\
&&= \frac{1}{L_n} \nonumber \\, 
\label{eq:FC1B}
\end{eqnarray}
an expression relating $1/L_n$ and $1/L_T$ in terms  of $k_{\perp norm}=k_{\perp norm} v_{thi}/\Omega_i$, was used to obtain the analytic curve in the previous section (see Eq(\ref{eq:FC1B})). 

Note that Eq(\ref{eq:FC1B}), an approximation of Eq(\ref{eq:FC0}),is not as accurate as the original expression. And indeed, simulations do find points that lie slightly over the boundary in the nominally insoluble regime. One should note various ways the approximation could break down:

\begin{figure*}
\subfloat[\label{sfig:2a}]{%
  \includegraphics[width=.99\linewidth]{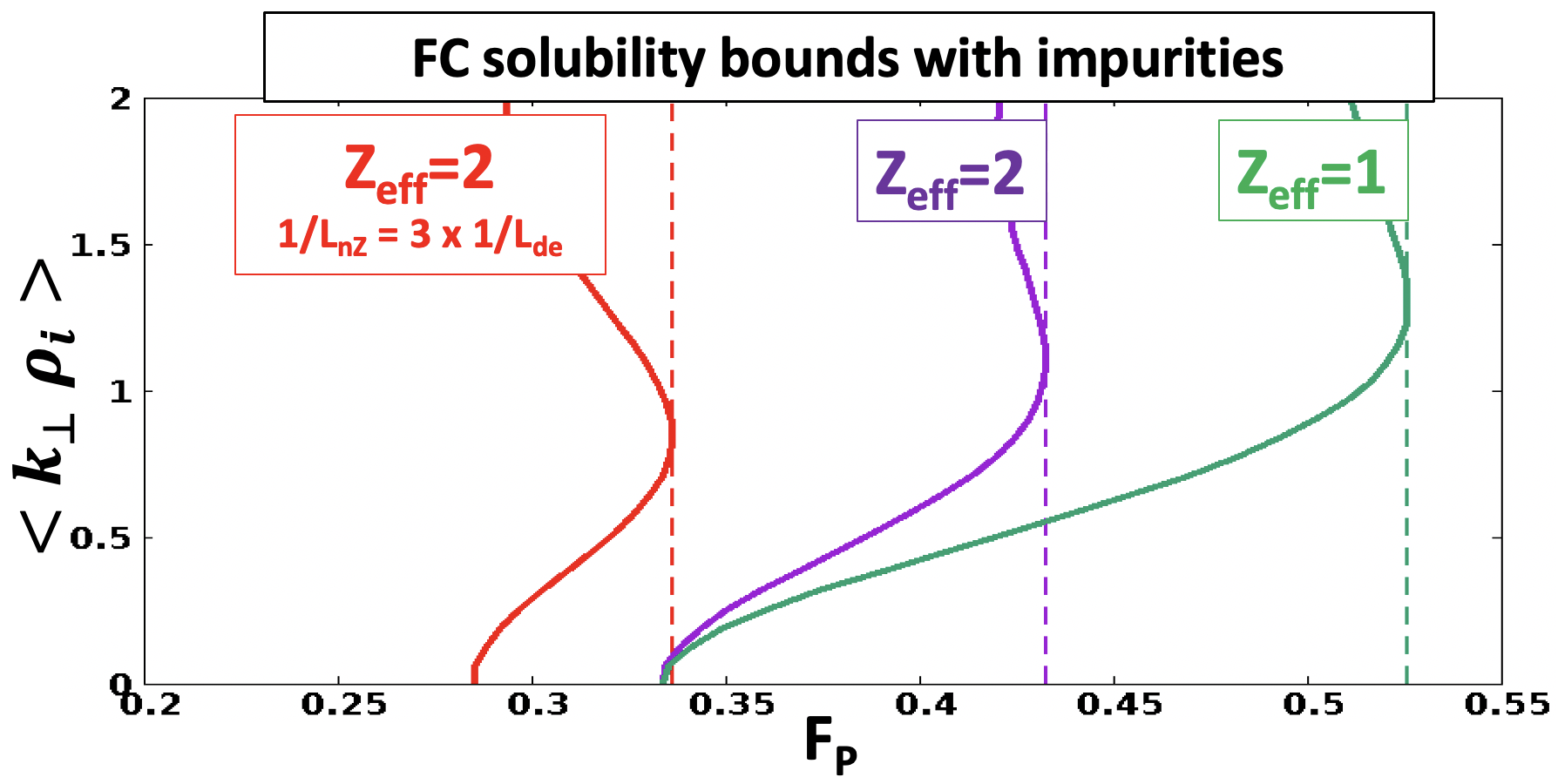}%
}
\hfill
\caption{\label{fig:FCbounds}  FC solubility boundaries including carbon impurity in a deuterium plasma. The case with $Z_{eff}=2$ (green) has the same density gradient scale length for all species, whereas the case $Z_{eff}=2$ $1/L_{nZ}=3/L_{nd}$ has three times steeper impurity gradient than dueterium. The $F_P$ is defined in terms of the electron density gradient, $F_P= 1/L_{ne}/(1/L_{ne}+1/L_{Ti})$ which is the most readily measured experimental quantity. }

\end{figure*}

\begin{itemize}

\item 	Even if the expansion is fairly accurate for the FC bound, there will be next order corrections, so that the boundary could be somewhat shifted from the computed value
\item 	The expression for $<k_{\perp}>$ is intuitive and can be derived in the limit of small $<k_{\perp}>$, but it is only an approximation; there could be corrections.
\item 	Trapped ions are not included in the kinetic treatment, and as $\omega$ and $\gamma$ become small, these could affect the results.
\item 	Because of any combination of these effects, even if the over-all theory is reasonably accurate, residual, weak instabilities are potentially possible beyond the computed solubility bound. 

\end{itemize}

\section{The effect of impurities on solubility of the flux constraint for low k modes}

Since our flux constraint demands the charge weighted trurbulent flux to be zero, impurities, because of their higher charge, could have a disproportionately large effect on the solubility condition. We will, therefore, generalize, Eq(\ref{eq:FC1B}) to include impurity species treating the impurity terms exactly in the way as the main ions; all ionic species have the same temperature. Equation(\ref{eq:FC1B}) modifies to  

\begin{eqnarray}
&&\sum_s \frac{n_sZ_s^2M_s^{1/2} \int d{s_{\perp}} s_{\perp} e^{-s_{\perp}^2/2} J_0(k_{\perp }\rho_s)^2 (3/2-s_{\perp}^2/2)}{L_{Ts}}  \nonumber\\
&&= \sum_s \frac{n_sZ_s^2M_s^{1/2}\int d{s_{\perp}} s_{\perp} e^{-s_{\perp}^2/2} J_0(k_{\perp}\rho_s)^2}{L_{ns}}        \nonumber \\ 
\label{eq:FC1BZ}
\end{eqnarray}

where $\rho_s= \rho_{0}M_s^{1/2}/Z_s$, $\rho_{0}$ is the Larmor radius of the main ion species $\rho_{0}= s_{\perp} v_{th0}/\Omega_{0} $, $Z_s$ is the impurity charge state and $M_s$ is the mass ratio of the impurity to the main ion species. Graphs of the computed solubility bound for sample parameters are shown in fig \ref{fig:FCbounds}. We see that impurities shift the solubility boundary to lower $F_P$. Many experimental cases for pedestals and ITBs find that the impurity gradients $1/L_n$ are steeper than the electron values. This shifts the curve to even lower $F_P$, i.e., it is a stabilizing effect. (In the opposite case where the impurity gradient is more shallow than the electron value, this shifts the curve to higher $F_P$ values.) 

For common impurities and relevant impurity levels, e.g. $Z_{eff} = 2 \sim3$, a significantly lower density gradient is needed to stabilize ITG modes. This brings significantly more experimental TB into the range where the ITG is stable without velocity shear. 

Before we show, through simulations, the heightened importance of impurity stabilization, let us contrast the true operative physics here with the usual invocation regarding impurities: the so called dilution. 

In the context of TBs, conventional approaches attribute stabilization by impurities to dilution.  In fact, for the free energy equation and curvature driven modes, stabilization via dilution is a reasonable conclusion because it can be shown that higher Z impurities contribute much less free energy than the bulk ions they displace. The effect is then proportional to the impurity to ion density fraction.

Impurities affect the FC strongly, way more than what could be explained by the effects of dilution. Quite unlike dilution, the shift in $F_P$ seen in fig(\ref{fig:FCbounds}) arises because impurities contribute a \emph{disproportionately large} charge flux. If the impurities had been inert, then the FC boundary would be unchanged, since in Eq(\ref{eq:FC1B}) the numerator and denominator would be reduced by the same impurity dilution factor. Impurity dilution, then, would result in the same solubility boundary as without impurities. Instead, quite the opposite happens: the impurities contribute \emph{more} strongly because their denominator is smaller, since their velocities are smaller (by $\sim \sqrt(m_{bulk}/m_{impurity}$). Hence, they give a larger contribution to the FC relative to their charge fraction in the equilibrium- exactly opposite to how they effect the free energy equation.

Hence, for ITG stabilization by density gradients, impurities are not a dilution effect at all; they stabilize by substantial contributions to charge flux. 

We now show how strongly the simulations agree with the analytically computed bounds for the FC with impurities.

\section{Impurity Effects on FC - Theory and Simulations }

In this section, we find that simulations corroborate the analytical predictions:  when impurities are included, the density gradient needed to stabilize the $ITG_{ae}$ is reduced. For experimental parameters, this reduction can be up to a factor of $\sim 2$. This brings a far larger number of experimental TBs into the regime where the $ITG_{ae}$ is stable without velocity shear. And in TB, it is frequently the case that impurity gradients scale lengths are shorter than the electron density scale length further reducing the needed density gradient. 

These results on impurities being facilitators of TB formation could be of great practically importance. These results also constitute important theoretical/scientific evidence of the fundamental processes that control such formation. The analytically predicted solubility bounds on the FC are found to agree with simulations  fortifying the validity of the FC based approach.

\begin{figure*}
\subfloat[\label{sfig:7a}]{%
  \includegraphics[width=.33\linewidth]{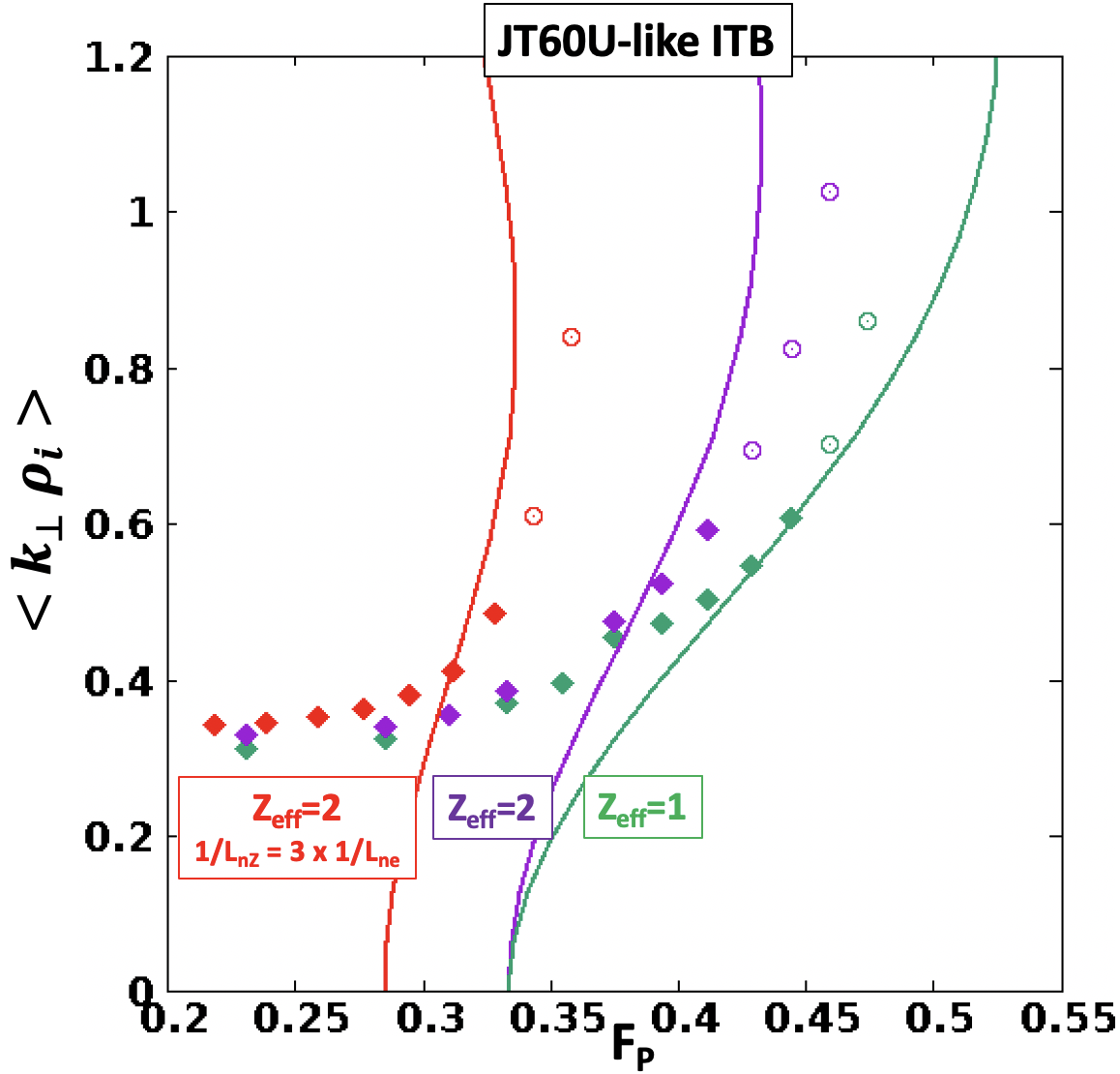}%
}\hfill
\subfloat[\label{sfig:7b}]{%
  \includegraphics[width=.33\linewidth]{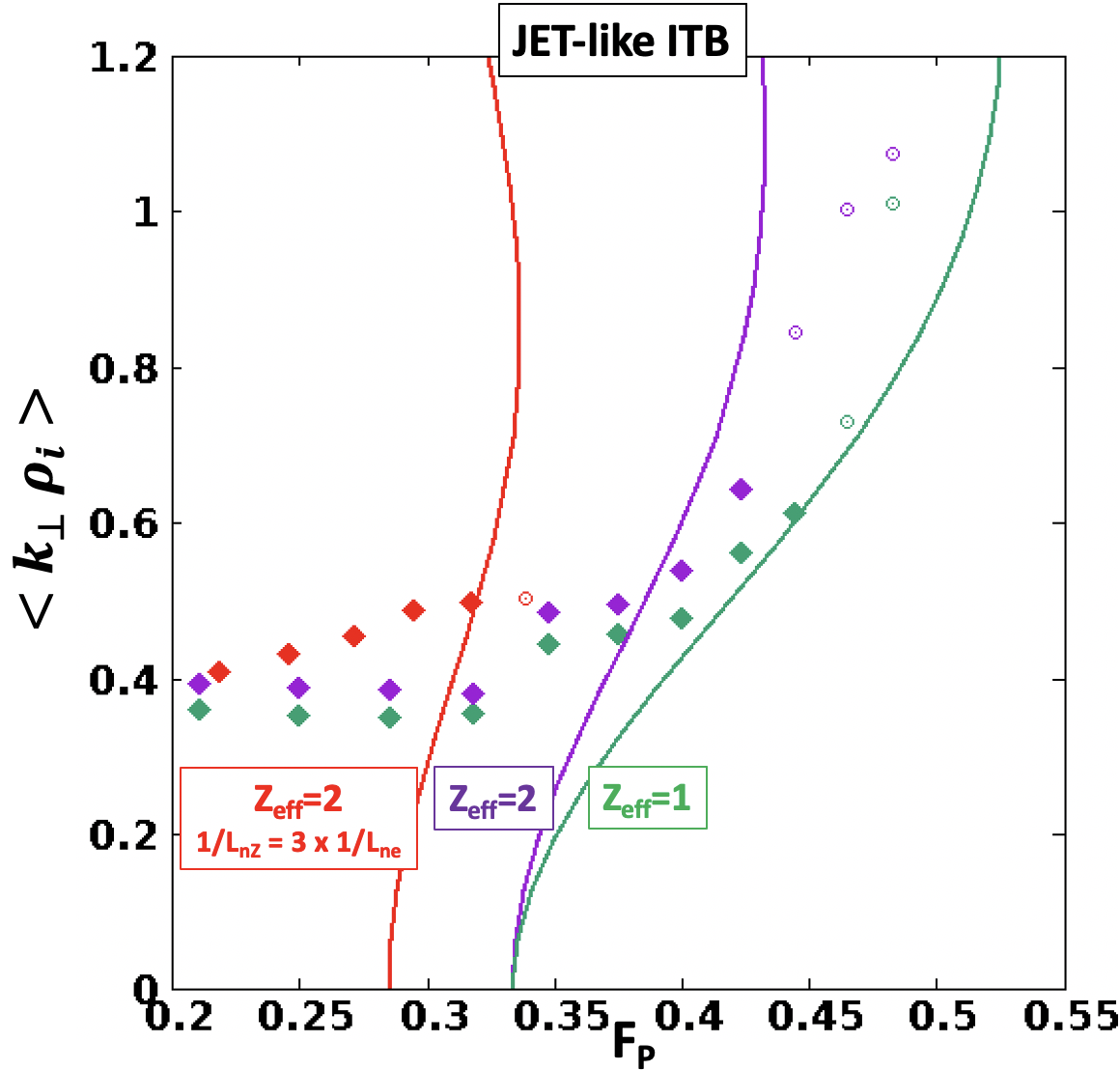}%
}
\hfill
\subfloat[\label{sfig:7c}]{%
  \includegraphics[width=.33\linewidth]{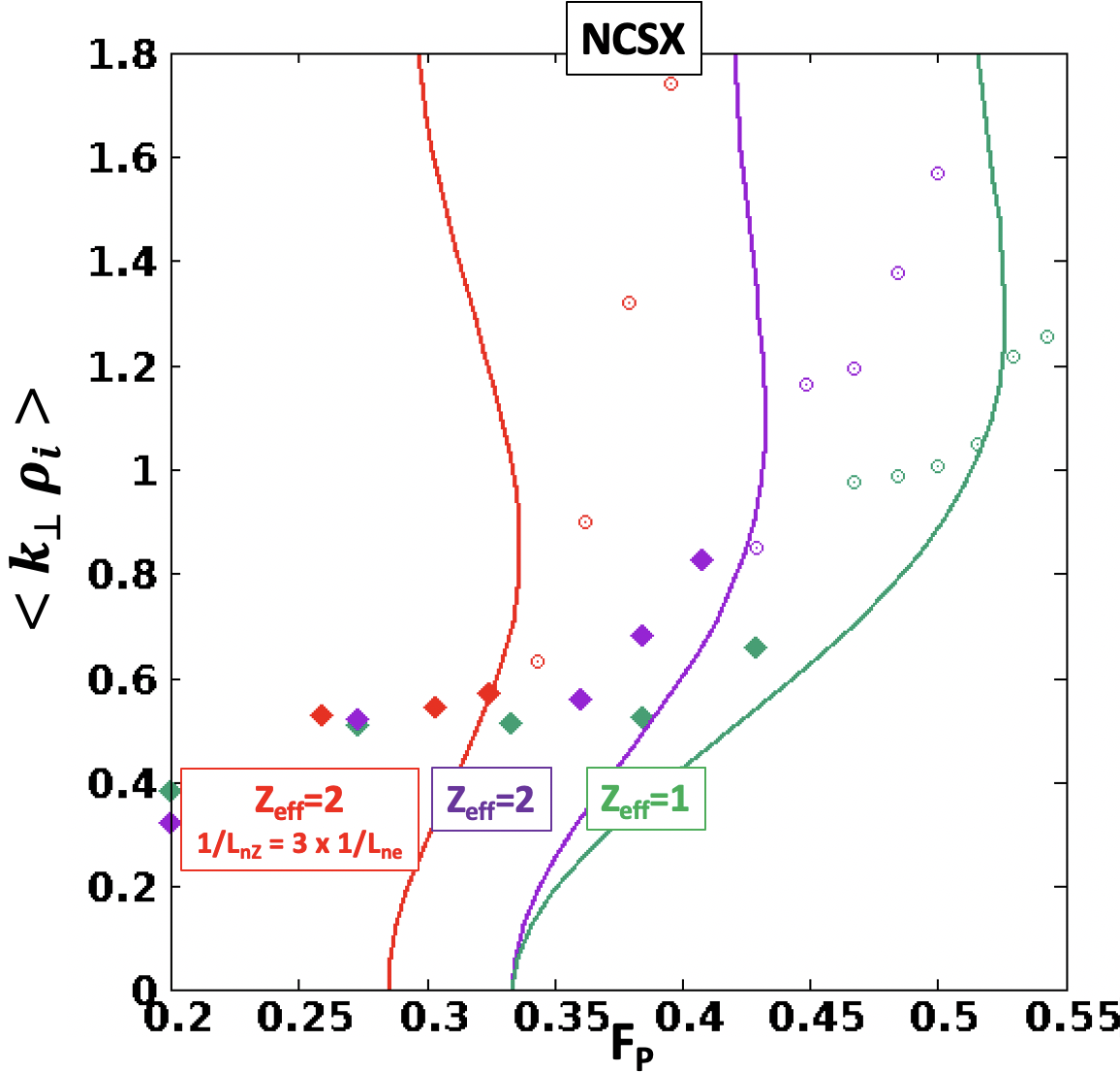}%
}\vfill
\subfloat[\label{sfig:7d}]{%
  \includegraphics[width=.33\linewidth]{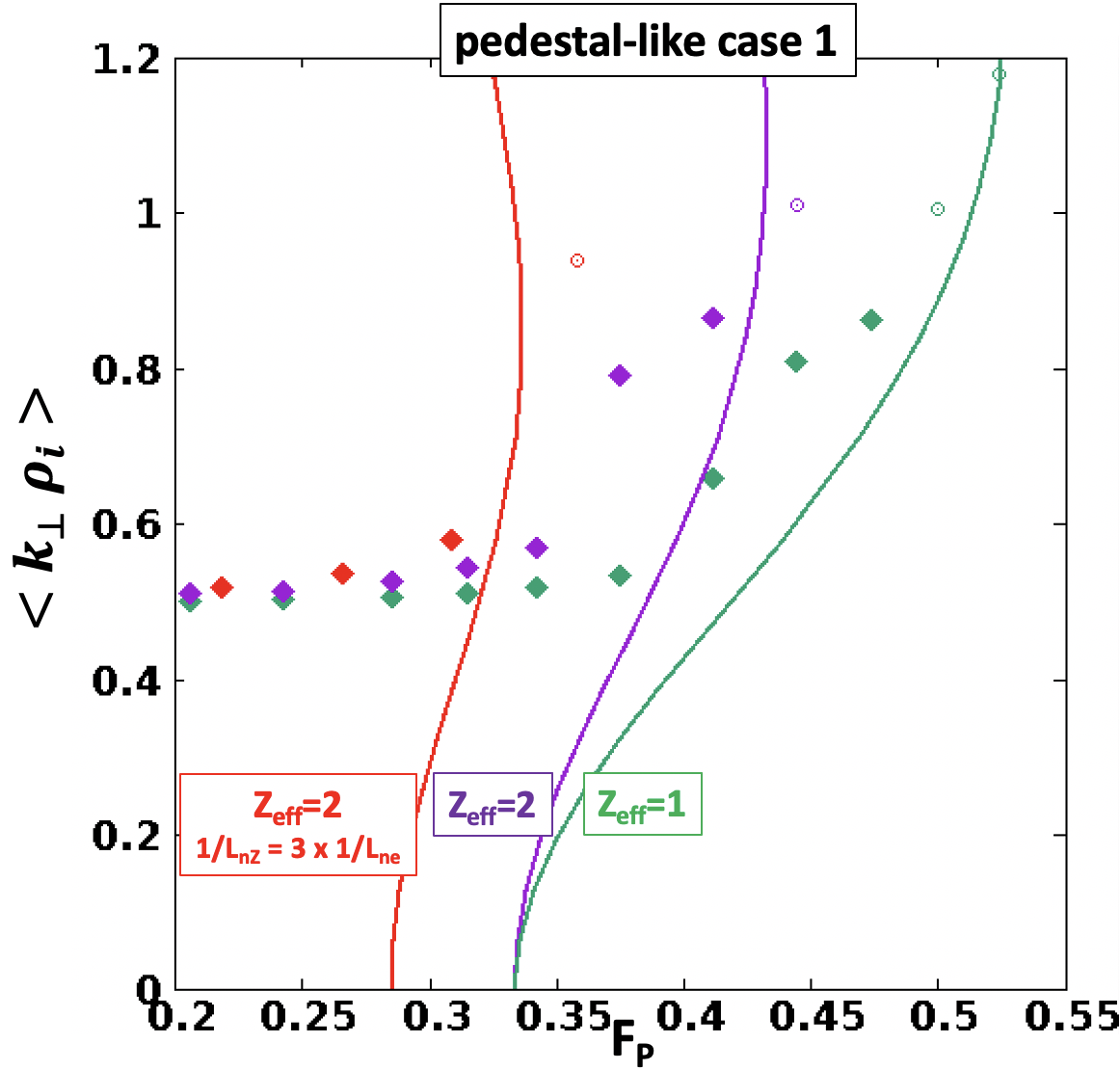}%
}
\hfill
\subfloat[\label{sfig:7e}]{%
  \includegraphics[width=.33\linewidth]{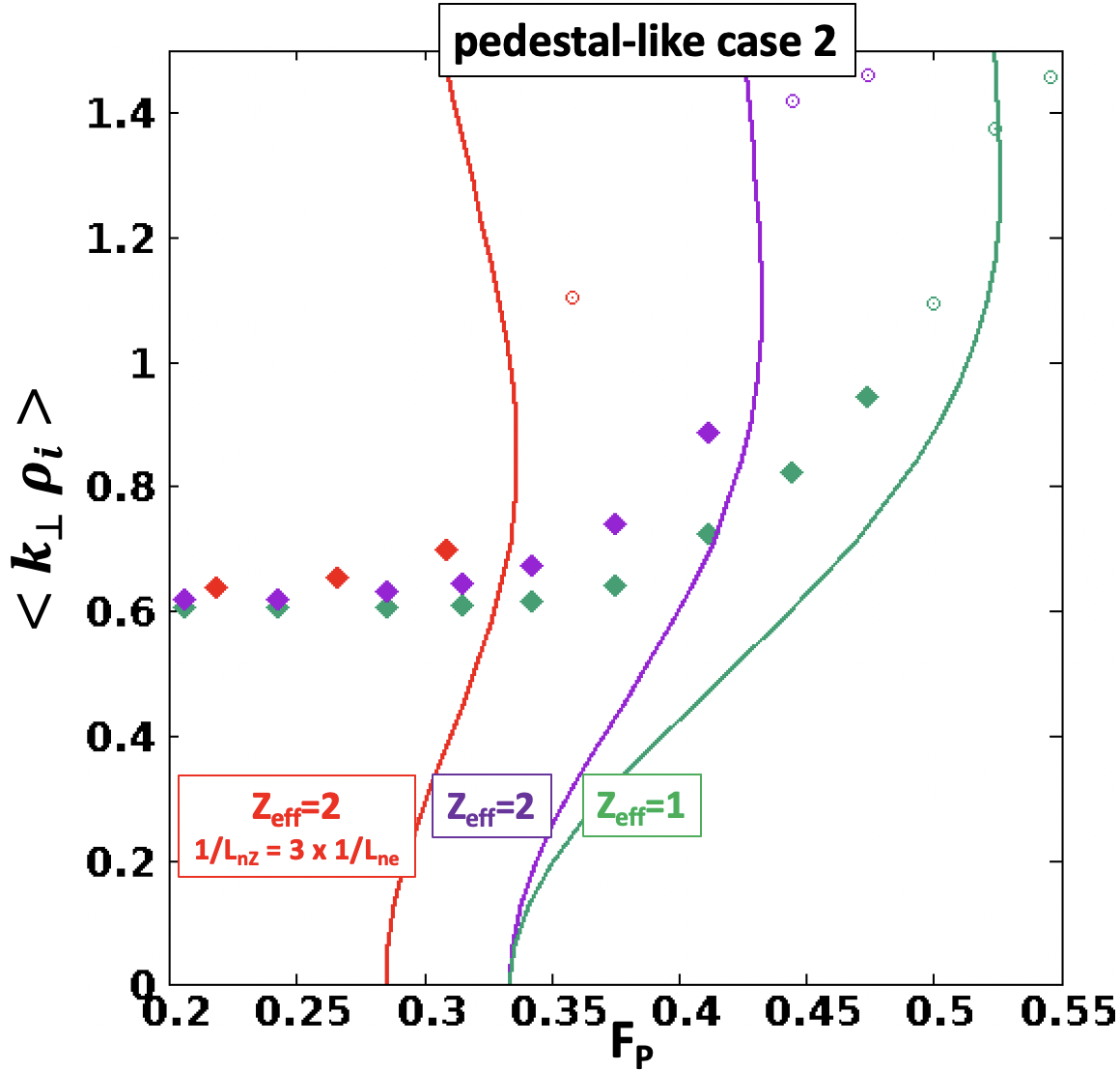}%
}\hfill
\subfloat[\label{sfig:7f}]{%
  \includegraphics[width=.33\linewidth]{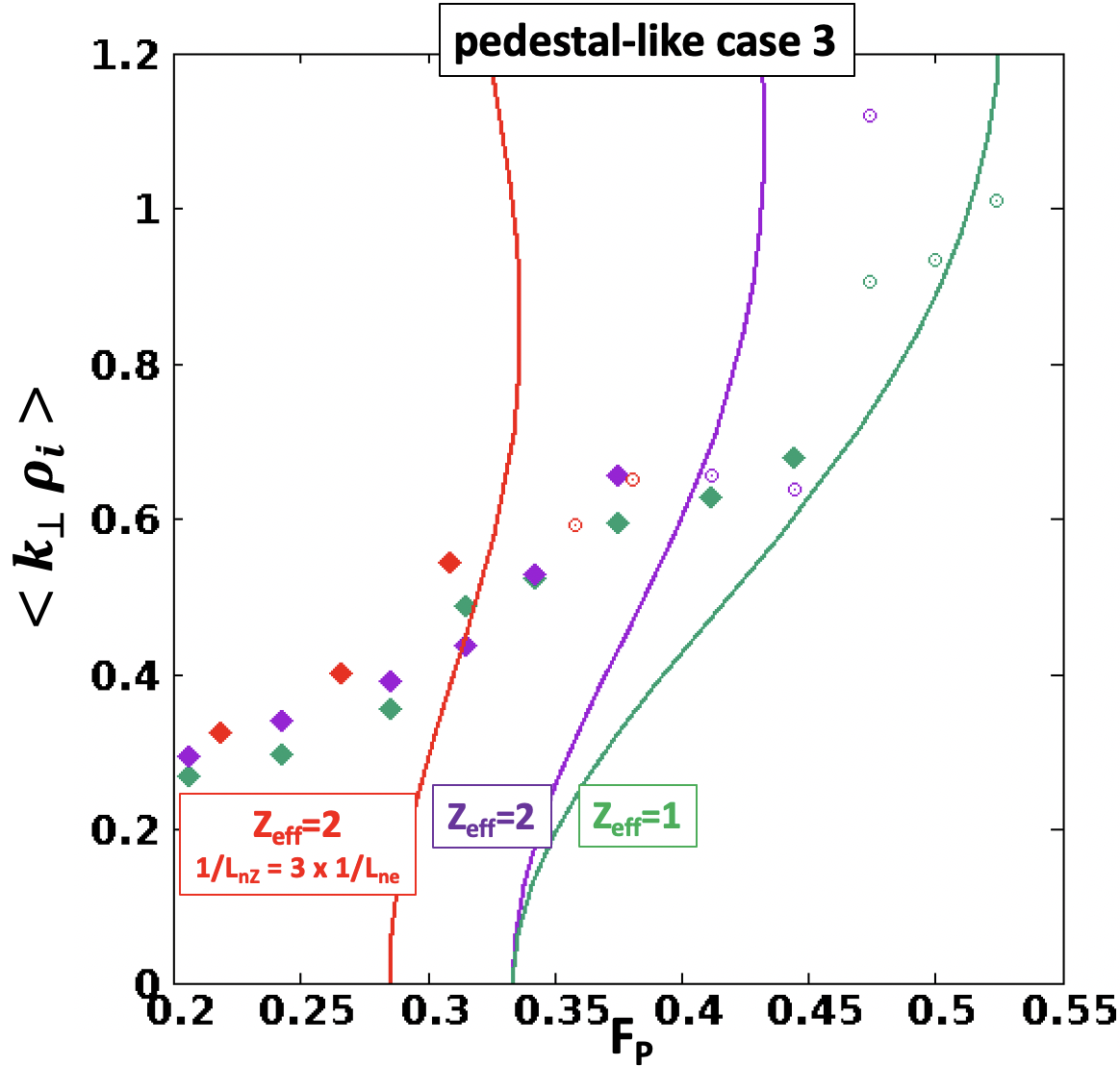}%
}
\caption{\label{fig:seven} The simulated $<k_{\perp}> \rho_i$ vs $F_P$ for several equilibrium, including impurites, plotted with the analytic FC solubiity bound. The impurity parameters and analytic predictions are identical to those in fig(\ref{fig:FCbounds}) (recall this is for carbon impurity). Green is $Z_{eff}=1$, purple is  $Z_{eff}=2$ and red is $Z_{eff}=2$ but the impurity density gradient scale length is 1/3 of the deuterium gradient.}
\end{figure*}

In fig (\ref{fig:seven}), Simulations, with diverse geometries and temperature gradients, track the analytic bounds, except that quite weak instabilities (with $D_{mix} < 2\%$ of the maximum value for the scan, denoted by an open circle) persist into the nominally insoluble region. As for the plots in previous sections, three values of the $k_x$  were simulated ($k_x=k_y \hat{s} \eta_0$),for $\eta_0 =0$,$\pi/2$ and $\pi$, and the lowest value of $<k_{\perp}> \rho_i$ (which is the one that most challenges the analytic boundary). Simulations also track the analytic bounds predicted for different impurity levels and impurity gradients, with the same parameters as in fig(\ref{fig:FCbounds}).

It is striking that, despite tremendous changes in geometry and temperature gradient, the behavior of all but these very weak modes respect the FC bound of low $k$ modes with the impurity modifications. This validates the analytical theory. Consistent with the theory, the impurities cause stability for a lower $F_P$, and this tracks the analytic curve: in the insoluble domain, only feeble modes are present where $D_{mix} < 2\%$ of the maximum value for that scan. 

\begin{figure*}
\subfloat[\label{sfig:7a}]{%
  \includegraphics[width=.33\linewidth]{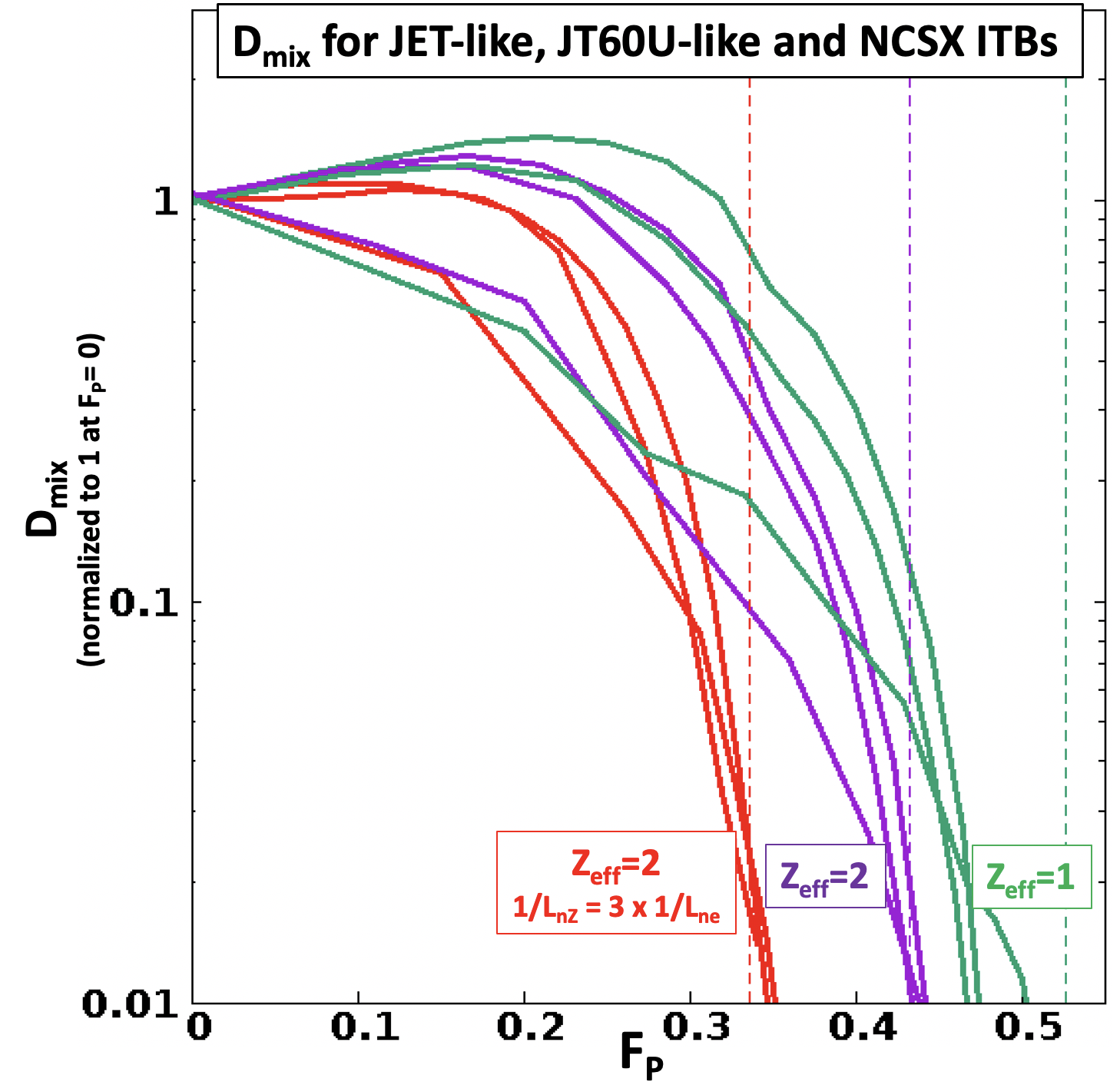}%
}\hfill
\subfloat[\label{sfig:7b}]{%
  \includegraphics[width=.33\linewidth]{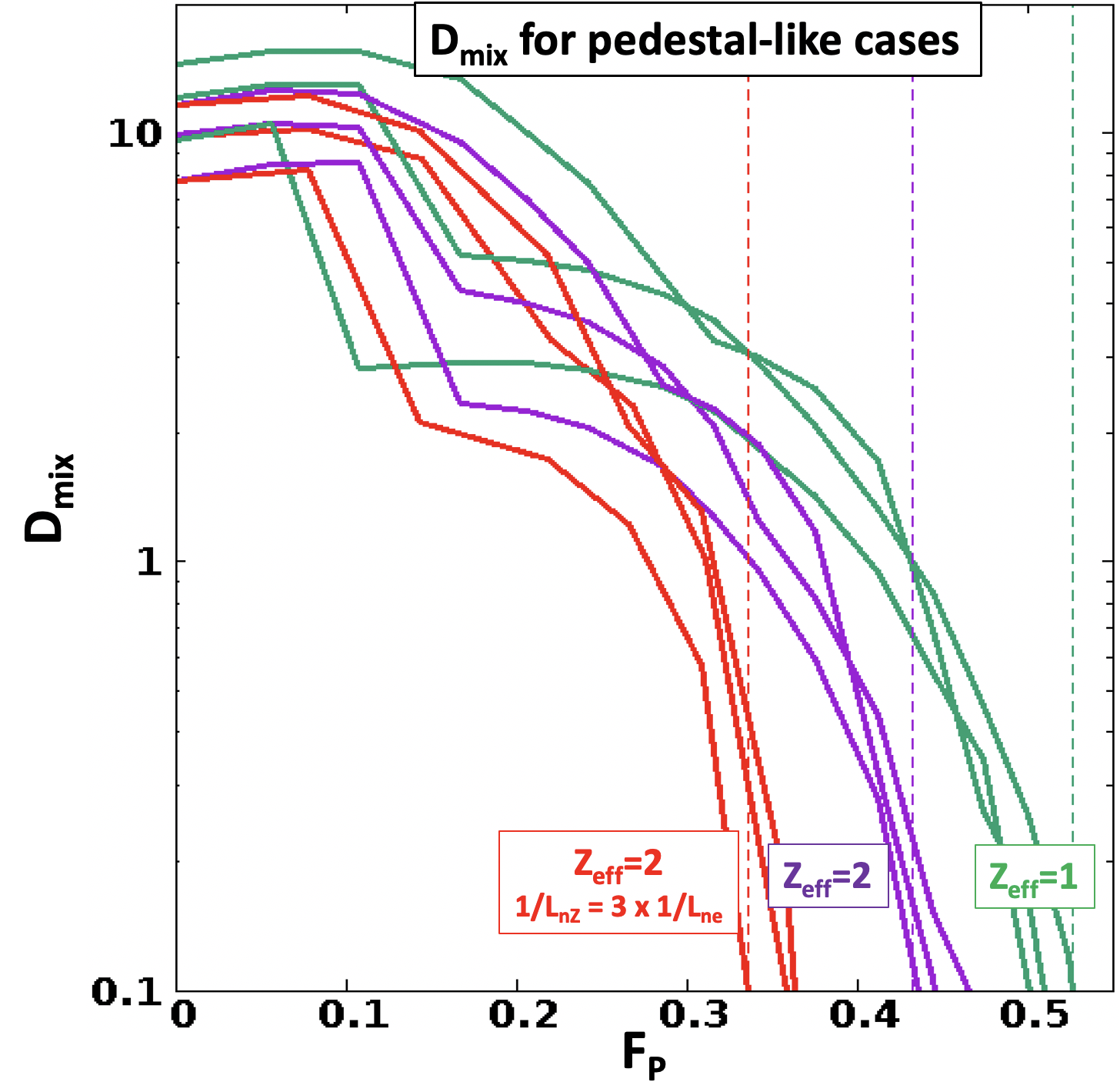}%
}
\hfill
\subfloat[\label{sfig:7c}]{%
  \includegraphics[width=.33\linewidth]{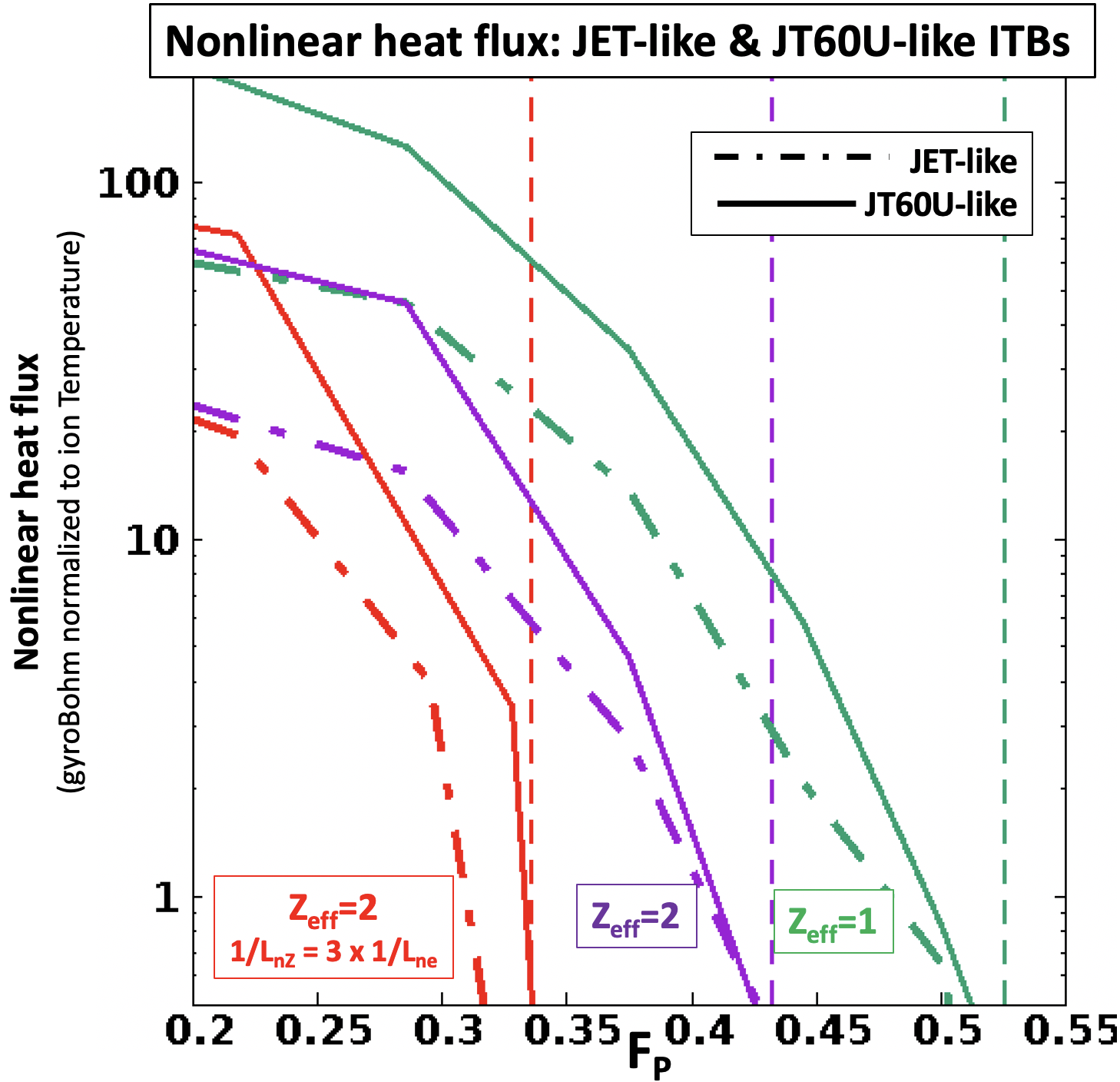}%
}\vfill
\subfloat[\label{sfig:7d}]{%
  \includegraphics[width=.33\linewidth]{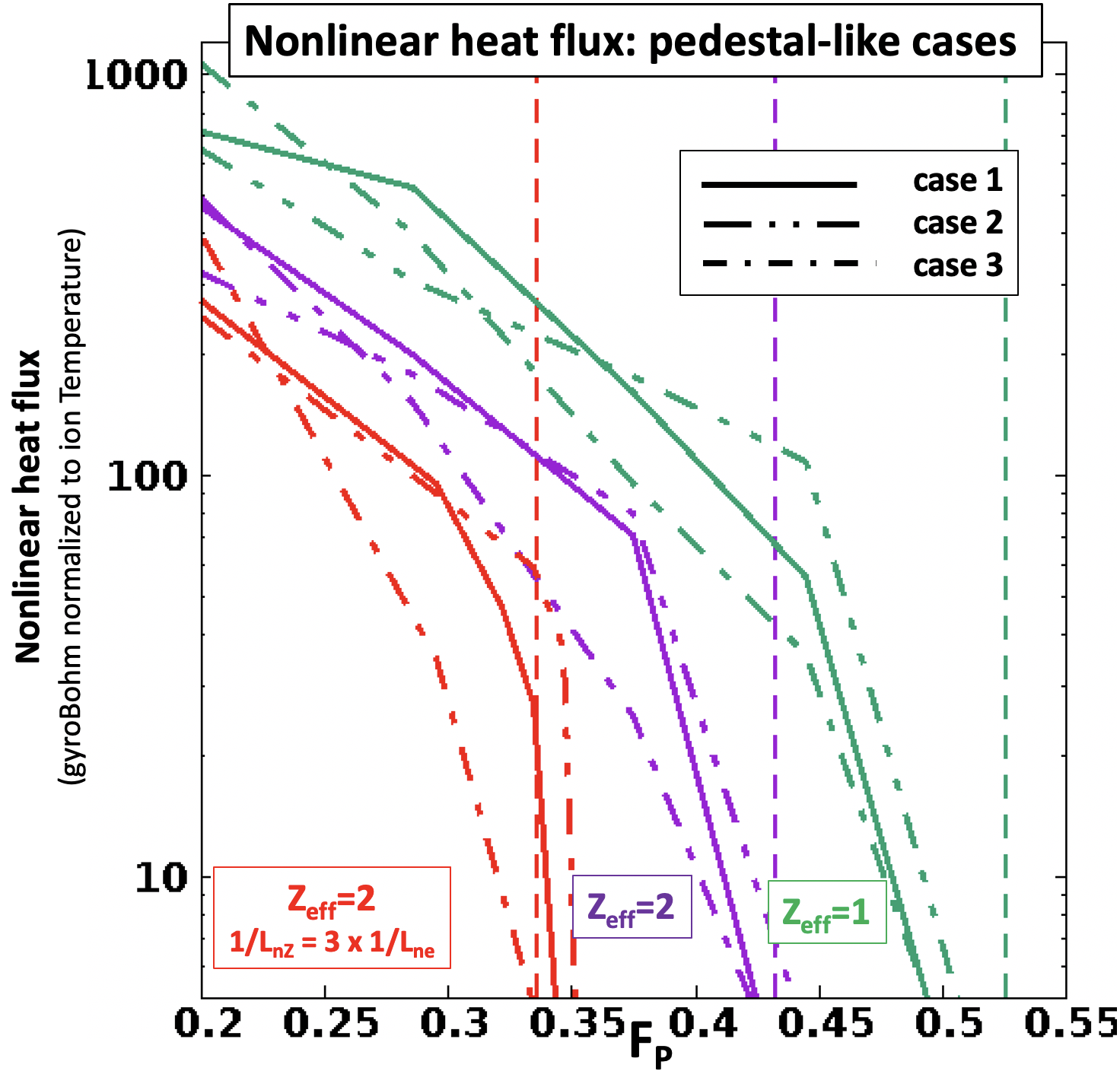}%
}
\hfill
\subfloat[\label{sfig:7e}]{%
  \includegraphics[width=.33\linewidth]{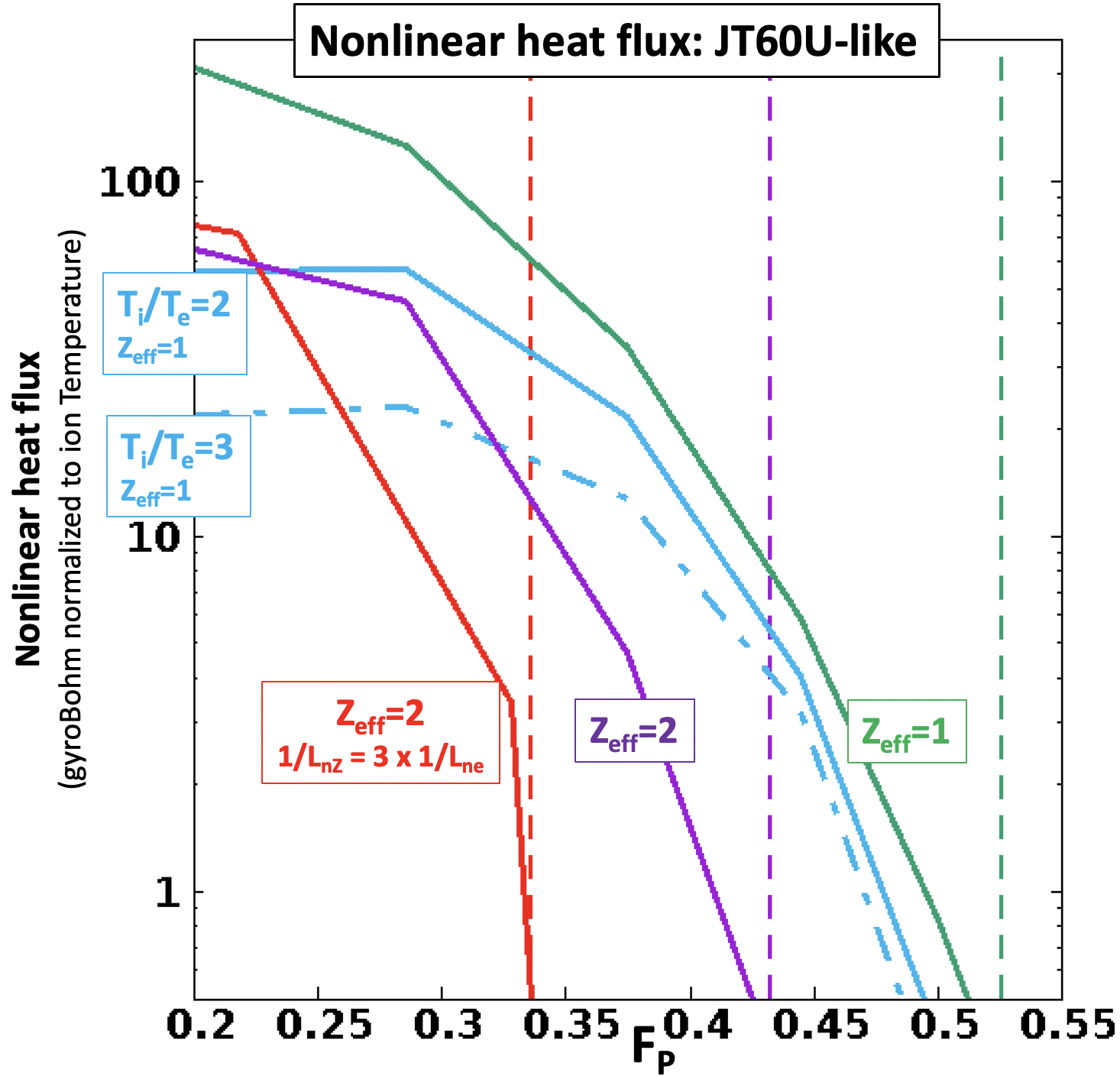}%
}\hfill
\subfloat[\label{sfig:7f}]{%
  \includegraphics[width=.33\linewidth]{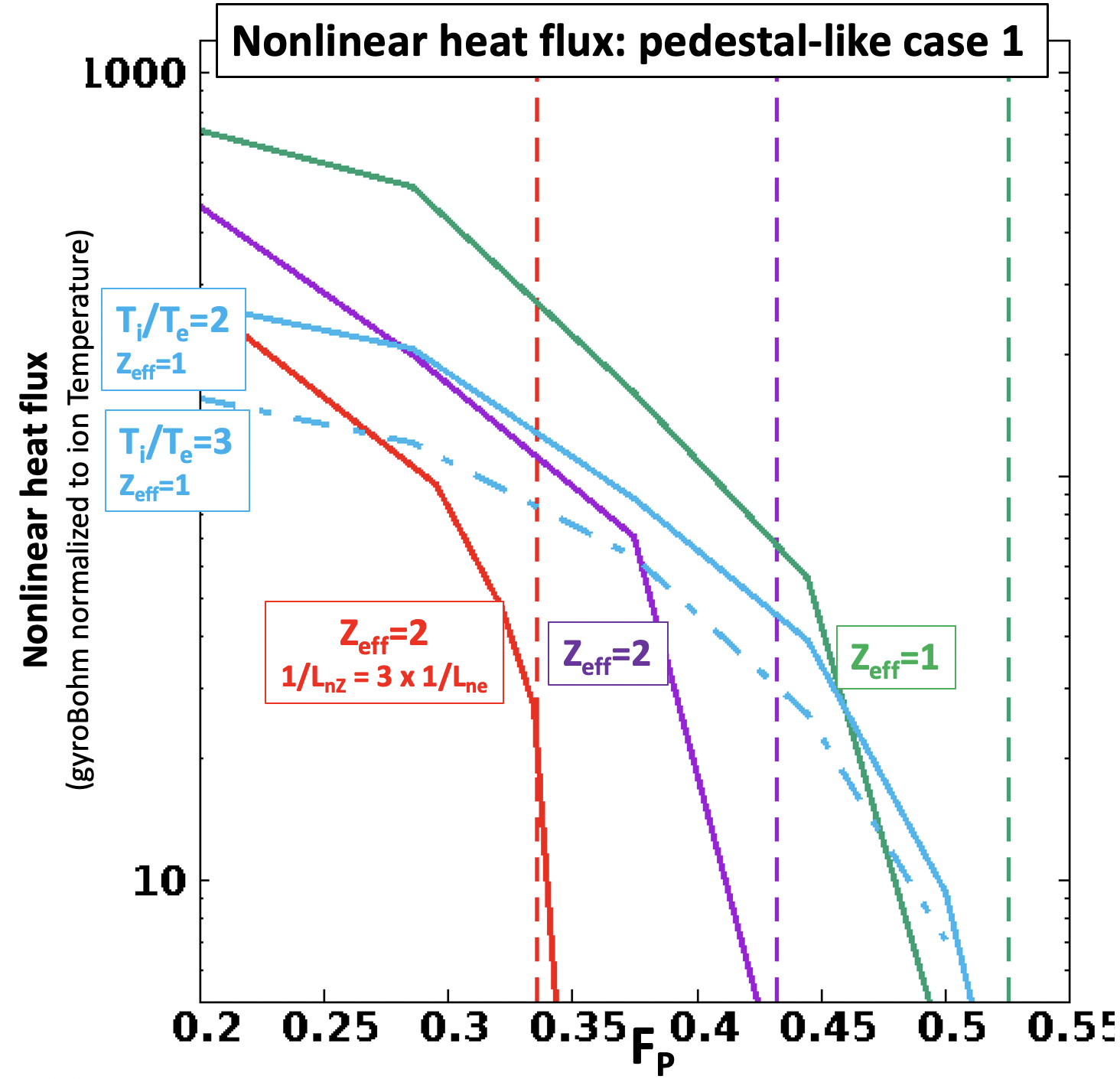}%
}
\caption{\label{fig:eight} The simulated $D_{mix}$ and nonlinear heat flux for diverse cases with impurities.}
\end{figure*}

These residual instabilities have little impact on the behavior of $D_{mix}$ for the low $k$ modes, as seen in Fig.~\ref{fig:eight}a-b. Despite the tremendous differences in geometry and gradients $R/L_T$, there is dramatic and rapid reduction in $D_{mix}$ near the analytic solubility bound for the FC. Specifically, the cases with impurities drop at the analytically predicted $F_P$ value for their respective impurity parameters, validating the analytic bounds. The differences in parameters are manifest in somewhat different trajectories, but the endpoint of stability (or near stability) is consistent with the predictions.

Nonlinear simulations of the heat flux follow the same behavior as $D_{mix}$ (see fig 8c-d), and the flux drops by about two orders of magnitude at the analytically predicted $F_P$ bound. This occurs for both ITB-like and pedestal configurations with much steeper gradients. 

We examine the effects of $T_i/T_e$ in Fig.~\ref{fig:eight}e-f . We compare the nonlinear fluxes for cases with $T_i/T_e =2$ and $3$ and $Z_{eff}=1$ and the ones with impurities. (Fluxes are in gyroBohm normalized units in the ion temperature). Recall that $T_i/T_e >1$ does not affect the FC bound, but it does have a stabilizing effect in the free energy equation. The cases with $T_i/T_e =2$ and $3$ have far lower flux than the case with $Z_{eff}=1$ and $T_i/T_e =1$ at low $F_P$. This is the expected energetic effect. But the FC solubility limit for all these cases is the same, so they all drop to low values at the same $F_P$.

Now, let us turn on the impurities. For low $F_P=0.2<1/3$, the FC is not very restrictive. The impurity cases have a larger flux than the cases $T_i/T_e =2$ and $3$. But as $F_P$ increases, this reverses. The presence of impurities imply a lower FC bound, and so the flux drops below the cases with $T_i/T_e >1$. Once again, the FC dominates the behavior at large $F_P$ near the solubility limit, and is more important than energetic effects. Once again, the behavior of the system cleanly distinguishes the roles of the free energy equation and the FC. 

Thus, even in nonlinear simulations, the behavior follows the theoretical understanding based upon separating the dynamics of the FC and the free energy equation. The nonlinear behavior manifests the underlying structure of the gyrokinetic system, as it should.

These results reinforce the power of the FC concept/analysis, and simulations validate analytic derivations of the solubility criterion.  Since experimental levels of impurities can significantly reduce the electron density gradient necessary for insolubility of the FC, they can play an outsize, even crucial role in TB formation.

 \section{ Interpretation of insolubility of the FC as arising from basic concepts of statistical mechanics, and generalization to non-adiabatic electrons}
 % Interpretation of insolubility of the FC as arising from basic concepts of statistical mechanics}

 The theoretical structure of the gyrokinetic system, as described in this work, is clearly supported by extensive gyrokinetic simulations for diverse cases using the GENE code~\cite{jenko_gene}. Let us step back and examine the conceptual foundations.

The FC is derived for \emph{essentially} all perturbations(and not just eigenmodes). Conceptually then, one must face the fact the FC must have its origin as well as its operational manifestation in rather general physical principles, more general than are invoked in the conventional eigenmode-stabilty analysis. Since the FC is a (charged) particle -flux constraint, it naturally valorizes the density gradients. Let us consider this in more detail. 

 In non-equilibrium systems the  thermodynamic forces drive the corresponding thermodynamic flux; the density gradients, thus, drive particles fluxes. Consequently, a sufficiently large ion density gradient can be expected to result in a net ion particle flux no matter what the overall dynamics may be. For $ITG_{ae}$ (with no electron flux), this ion flux cannot be compensated. Hence there will be a non-zero charge flux, so the FC is insoluble. Hence, density gradients eventually precludes growing fluctuations.

In the so called Onsager region, the thermodynamic forces are linearly related to the thermodynamic fluxes. It is well known that this classic region does not, generally, pertain to transport from gyrokinetic instabilites. But for the simpler problem of calculating the charge flux for a \emph{given} fluctuation, methods can be used that are completely analogous to the Onsager region. Hence, the solubility of the FC can be understood through such concepts.

In the classic Onsager theory, the gradient $dQ_i/dx$ of a physical quantity $Q_i$ represents a thermodynamic force. The flux  $\Gamma_i$ of $Q_i$ is given by 
\begin{equation}
\Gamma_{i} = \sum_i  M_{ij} dQ_j/dx  
\label{eq:Onsager}
\end{equation}
where the matrix $M_{ij}$ often has off-diagonal components. (i.e., temperature gradients could drive particle flux, for instance). The second law of thermodynamics demands that this matrix must have positive definite eigenvalues. For our purposes, the important corollary is that the diagonal components must be positive - thermodynamic forces drive their fluxes down the gradients.

In the gyrokinetic case of interest here, the $dQ_i/dx$ are gradients of temperature or density for each species. In the linear gyrokinetic equation, the non-adiabatic part of the distribution function is driven by these gradients; there is also an additional term proportional to frequency. And crucially, in the context of showing insolubility of the FC to general fluctuations (as in the sections above), that frequency can be taken to be arbitrary. In other words, it is not considered to be an eigenfrequency, and the fluctuation structure also need not be an eigenmode. 

In the presence of a general fluctuating potential $\phi$, the induced quasilinear flux, then, has an additional term (due to the finite frequency) modifying Eq(\ref{eq:Onsager}) to,

\begin{equation}
\Gamma_{i} = \sum_i  G(\phi,\omega)_{ij} dQ_j/dx + \Gamma0(\phi,\omega)_{i}
\label{eq:gyroresponsematrix}
\end{equation}

For example, in the simple $ITG_{ae}$ with a single species, the ion particle flux will be

\begin{equation}
\Gamma_{i} =  A dT/dx +B dn/dx + C  
\label{eq:particleflux1}
\end{equation}

In general, A and B have similar magnitudes but typically $A< 0$, and necessarily $B>0$. So, for some density gradient that is large enough, the ion particle flux will always be positive, and the FC is insoluble. This insolubility is unrelated to the amount of free energy in the equilibrium, which is to say, how strong the gradient is. The requisite (stabilizing) density gradient is determined by the temperature gradient; the solubility condition, thus, can be expressed in terms of a ratio like $\eta$ or $F_P=  1/L_n/(1/L_T+1/L_n)$.

The fact that diagonal terms (like $B$) in the response matrix are positive definite is a consequence of entropy considerations, \emph{qualitatively} like the classic case, but different in some important details. For example, the classic considerations do not apply to highly non-equilibrium situations like perturbations that are increasing exponentially, or, where collisions are zero so there is no entropy production. Nonetheless it can be shown to be true that the diagonal components of the gyrokinetic response matrix $G$ are positive, even for exponentiating fluctuations and without collisions. For this, we use the free energy equation Eq(\ref{eq:two}), which describes entropy considerations for the gyrokinetic system.  See Appendix A.

%{\color{red} the following para must be a footnote}

%(For the interested reader, the argument goes thus. Even without collisions, and for an exponentially growing fluctuation, Eq(\ref{eq:two}) still implies that free energy is transferred from the equilibrium scales to the fluctuation scales. Hence, the free energy of the equilibrium scales decreases (which is to say, the entropy of those scales increases.) So the behavior of $G_{ij}$ is like the classic case from statistical mechanics. Furthermore, although the external $\phi$ does work on system, this can be included in the analysis, and it does not change the result that the response matrix $Q_{ij}$ must be positive definite, so the diagonal elements are positive. So the gyrokinetic response matrix, defined for an arbitrary fluctuation, has the same properties as the classic Onsager matrix, and for essentially similar reasons.) 

%So in general, one expects there to be a density gradient beyond which the FC is insoluble for the $ITG_{ae}$. The fundamental reason for this is the general statistical mechanical principle that thermodynamic forces drive thermodynamic fluxes. 

\textit{It should be emphasized that the generality of the preceding arguments transcend the simplified model $ITG_{ae}$. For example, even with non-adiabatic electrons, and electromagnetic fluctuations, the matrix $Q_{ij}$ will have qualitatively the same properties. }

The FC might be readily satisfied by having the ion and electron charge fluxes cancel each other. That is, even though the equilibrium density gradient will drive a particle flux of each species, the net charge flux will vanish, and the FC is soluble. So in the general case, there need not be a regime for TBs.

However, if one of the species (say electrons) is sufficiently close to adiabatic, it will cause relatively little contribution to the charge flux. And in that case, a large enough density gradient will lead to a net charge flux, i.e., the FC will be insoluble.

To the best of our knowledge, no one has recognized/investigated that electrons tend to be near adiabatic for the instabilities that must be quenched for TB formation. We will do this in the following sections. 

The next section will begin by modifying the FC when electrons are non adiabatic.

It is time to add a fourth item to the evolving statistical mechanical ansatz:

\begin{itemize}

\item 	In situations with large free energy (large gradients), the insolubility of the FC is often the operative dynamic that leads to stability. 
\item 	As is observed in very many systems with a large number of degrees of freedom, the gyrokinetic system manifests apparent adaptive behavior, and "finds a way" to dissipate free energy until a "hard" dynamical constrain prevents it- here, the FC.
\item 	This apparent adaptivity also applies to curvature drive-the eigenmode structure avoids regions of stabilizing curvature, even when these are strongly predominant
\item 	The basic trends of stability due to the FC can be understood from the very general statistical mechanical propensity for thermodynamic forces to drive the corresponding thermodynamic fluxes. In cases where one species is sufficiently close to adiabatic, this propensity eventually makes the FC insoluble as density gradients are increased.

\end{itemize}

\section{Inclusion of non-adiabatic electrons}

\subsection{ The flux constraint including non-adiabatic electrons}

The section above showed how the basic concepts of the FC generalize to the case of non-adiabatic electrons. Here we consider the full range of pertinent dynamics in detail.  

Non-adiabatic electrons often play a crucial role in gyrokinetic instabilities. For the ITG/TEM, this arises primarily from trapped electrons, e.g., these can lead to the TEM. But passing electrons can also play a role. Since non adiabatic electrons contribute charge flux that will diminish the ion (plus impurities) charge flux, violation of FC becomes more difficult. We will see, however, that even with non-adiabatic electrons, the flux constraint may still be potent enough to suppress instabilities and allow TBs formation.  We find this applies to diverse parameters similar to experimental TBs. Electron contribution, however, modifies the solubility conditions. 

%Now electron charge fluxes must be taken into account in the FC. Thermodynamic forces drive thermodynamic fluxes for electrons just as for ions. So electron density gradients will drive a corresponding electron flux, which will be part of the FC. Crucially, since the electron charge is opposite to the ions, this tends to cancel ion charge fluxes driven by density gradients. \emph{Hence, non-adiabatic electrons make the FC more easily soluble as density gradients are increased}. 

\begin{figure*}
\subfloat[\label{sfig:7a}]{%
  \includegraphics[width=.5\linewidth]{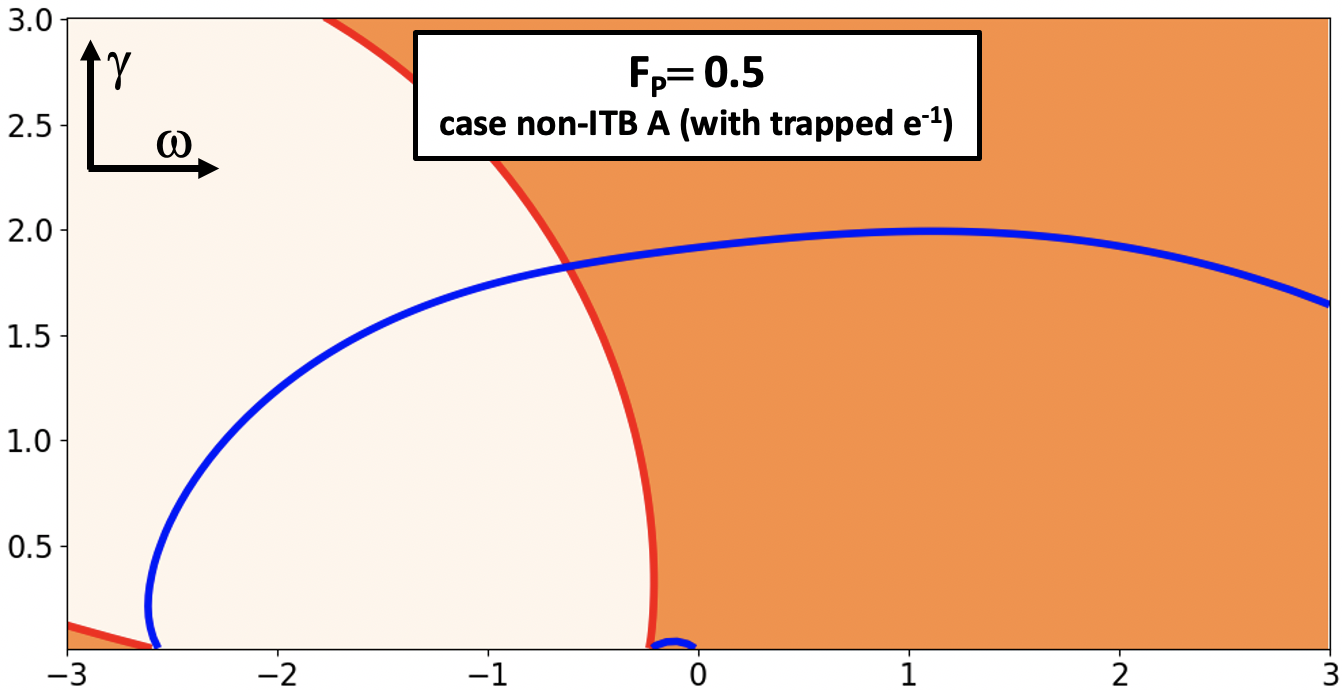}%
}\hfill
\subfloat[\label{sfig:7b}]{%
  \includegraphics[width=.5\linewidth]{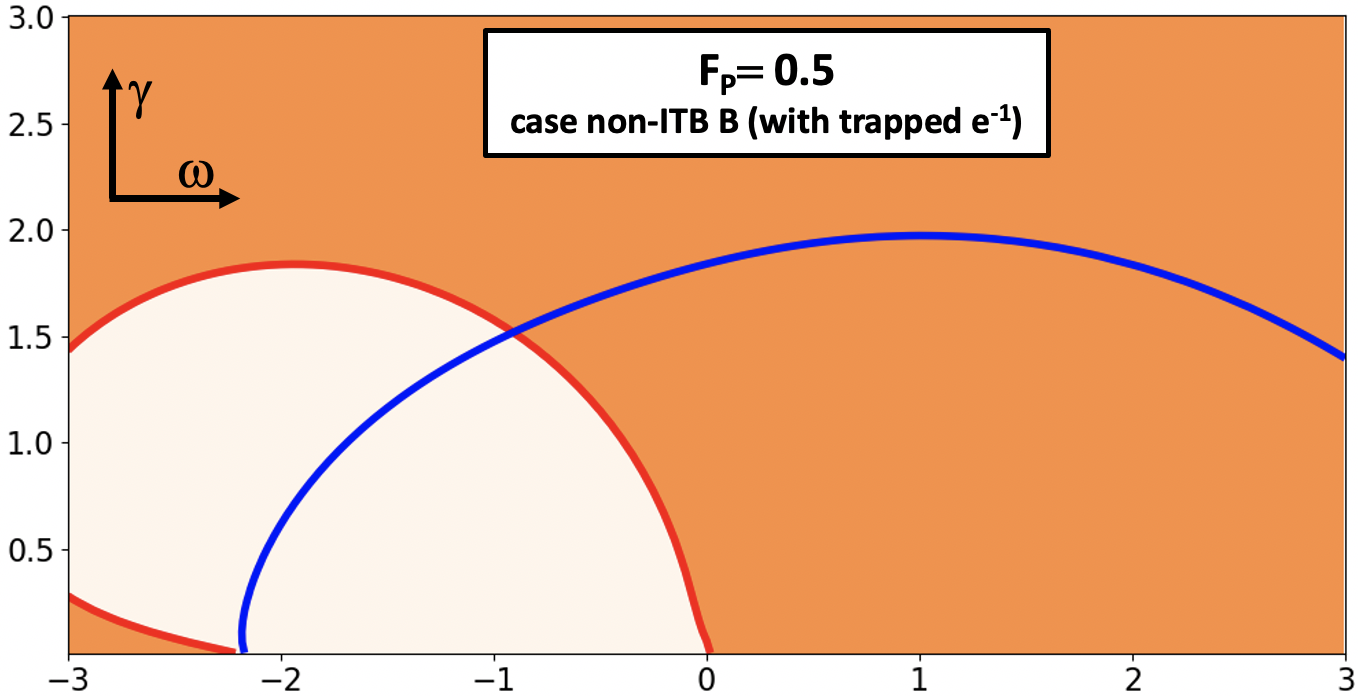}%
}
\vfill
\subfloat[\label{sfig:7c}]{%
  \includegraphics[width=.5\linewidth]{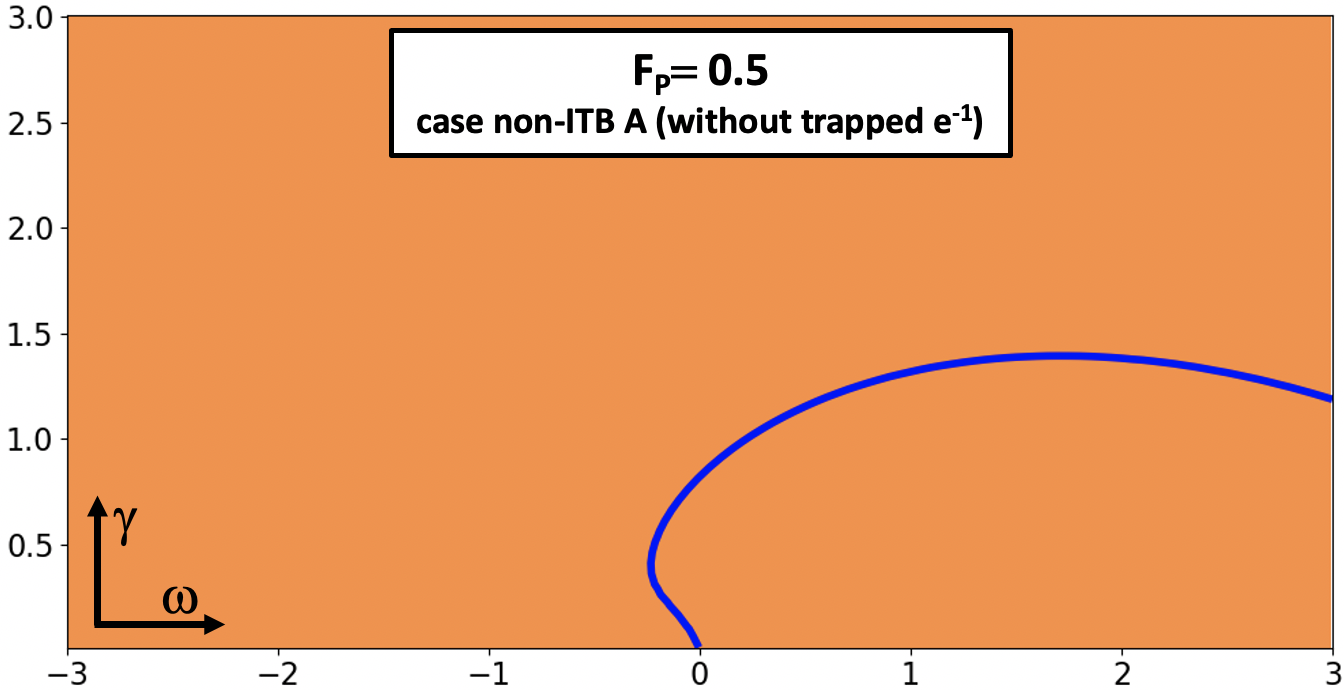}%
}\hfill
\subfloat[\label{sfig:7d}]{% 
  \includegraphics[width=.5\linewidth]{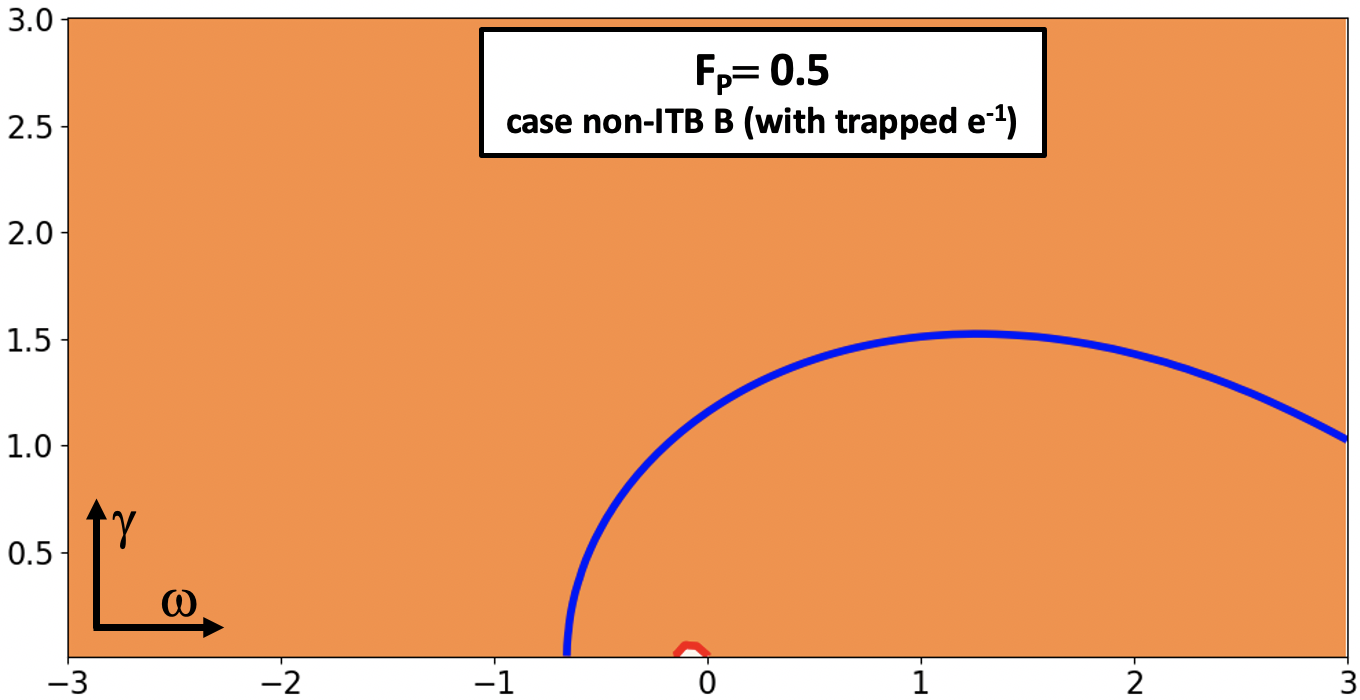}%
}
\caption{\label{fig:BP0} E FC plot for cases trapped electrons compared to adiabatic electrons. The red curve is again the solution of the FC, and the blue curve is the free energy balance. The eigenvalue lies at the intersection. The electron flux makes the FC soluble for high growth rates, when it is insoluble for adiabatic electrons. This allows a strong instability for the full electron cases. It also makes the free energy curve go to higher $\gamma$, but the qualitative change is to the FC. The free energy curves indicate that the ITG drive contributes a substantial fraction of the ITG/TEM free energy. In fact, even without the boost in free energy from trapped particles, there would still be a substantial instability just due to the FC curve modified by trapped electrons. }
\end{figure*}

It is, however, possible that when the non-adiabatic electron charge flux is large enough, density gradients do not, necessarily, lead to insolubility. This situation occurs for many typical tokamak geometries, in fact. In such geometries, the stabilization seen in the previous sections is either greatly reduced or eliminated because the FC remains soluble for much larger density gradients. 

Within the conventional framework of dispersion relations, this is interpreted as the onset of electron instabilities, e.g. the TEM, and this is attributed to accessing new free energy from density gradients. 

However, the dominant factor is not so much an increase in the free energy as that the constraint becomes soluble. We show this for the two non-TB equilibria in Table 1 (case non-ITB A and B). We can include trapped electrons in the SKiM model. Details are given below; here let us discuss the conceptually important results. We use simulation eigenfunctions to evaluate the parameters in SKiM. We plot the E-FC plots for each case at $F_P=0.5$ in Fig.~\ref{fig:BP0}. We compare to the plots with the trapped electron terms neglected- the latter being $ITG_{ae}$. Although the trapped electrons do indeed quantitatively increase the free energy significantly, the \emph{qualitative} difference is in the FC. Without trapped electrons the FC is insoluble except for $\gamma \sim 0$. With trapped electrons, the FC is soluble in a large region, up to high $\gamma$. This allows a strong instability. 

In fact, even if the free energy curve were the same as the $ITG_{ae}$ case, the fact that the constraint was rendered soluble by non-adiabatic electrons would have produced an instability with a high growth rate.

 When the FC does not constrain the system to a low growth, there is nothing to prevent the relaxation of free energy. Then the gyrokinetic system behaves like the vast majority of physical systems: instabilities arise to relax the free energy. Strong gradients will drive strong hybrid instabilities of ITG and TEM, whatever name one might wish to call it.  

Irrespective of the details of the instability, we expect that a highly adaptive system like gyrokinetics will produce a strong instability once the FC becomes soluble. For example, the $ITG_{ae}$ results showed an \emph{extreme} adaptability of the eigenmode structure in order to produce an instability that satisfies the FC and avoids stabilizing curvature. With such adaptability, the inclusion of non-adiabatic electrons that make the FC soluble at high $\gamma$ should be expected to lead to strong instability, whatever the specific details.  Just as for the ITGae, we expect that the gyrokinetic system will display extreme adaptability to avoid curvature (energetic) stabilization, by changing its mode structure to avoid stabilizing curvature. We find that this is, in fact, the case, when trapped electrons are added.  

\emph {In order to restore the conditions for TB formation, then, we must probe into mechanisms/conditions that would still violate FC  even when the non-adiabatic electron contribution of negative charge flux is pushing towards solubility. How do we do this?}

Unlike previous analysis of TB regimes, we will not start by analyzing dispersion relations of any particular mode type (ITG, TEM, or ITG/TEM).  In the larger framework that we have been developing in this paper, all these modes must satisfy two fundamental equations, the free energy equation and the FC.  The already discussed mechanism of adaptivity equips the gyrokinetic system to optimally satisfy these equations to produce an instability. In fact, the fluctuations can couple together aspects of the ITG and TEM in myriad possible ways, to produce whatever hybridization is needed to give instability. The only way to avoid instability is if the character of the equilibrium and parameters preclude it, due to some strong and unavoidable dynamic. 

That strong dynamic that must come into play for stability is the FC constraint. Since density gradients will drive ion and impurity fluxes in one direction, while electrons go in the opposite direction, one can think of at least two ways an imbalance ( FC violation) can occur: 1) Electrons are adiabatic - the ion-impurity flux is nonzero at sufficiently large density gradients, 2) Electrons are non-adiabatic but weakly so; their flux is still not enough to compensate the ion-impurity combined flux.

Following the FC based analysis, we can understand the behavior of simulations from a more fundamental perspective than previous interpretations. In fact, it is by resorting to the constrained dynamics that we may make sense of the surprising behavior found in the simulations. Conventional interpretations are, at best, inadequate.

In this section, we extend the  general concepts identified for $ITG_{ae}$ to include non-adiabatic electrons. The relative simplicity of the $ITG_{ae}$ allowed us to trace its behavior back to basic and general principles of statistical mechanics. \textit{The inclusion of more complete electron dynamics is not expected to vitiate such basic statistical mechanical principles. And indeed, we find it does not. } And in fact, we believe the framework developed here provides the first consistent understanding of the TB regime for many relevant parameters seen in experiments.

 Perhaps, the overwhelming practical drive for this paper is to search for regimes ( geometries) where velocity shear is not the deciding instrument for TB formation. Such regimes have been discovered well before this work, in experiments and in simulations of tokamak geometries with low or negative magnetic shear and/or high Shafranov shift.  Therefore, the basic intellectual drive for this study comes from a desire to create a deep and detailed framework to understand and interpret this amazing phenomenon. \emph{Equally important is the corollary that this understanding could guide us to access even more promising TB formations.}
 
 Based upon results in this section, this effort could be summarized as: 

\begin{itemize}

\item 	Understand the fundamental origin of TB formation from processes unrelated to velocity shear
\item 	Show that the simulation behavior in such cases is better explained by a collection of insights that we call the statistical mechanical ansatz:
\item 	The simulation behavior has the characteristics of stabilization due to the Flux constraint (FC)
\item 	It usually is not due to energetic stabilization- free energy considerations
\item         The realization that stabilization is due to the FC  implies different mode behavior and different requirements for stabilization as compared to energetic stabilization
\item. 	In particular, density gradients are a crucial aspect of the stabilization. Impurities and impurity gradients are just as relevant; can lead to TBs in unexpected regimes.

\end{itemize}

What are the practical consequences of such a new understanding? 

The FC dynamics valorizes the density gradients as a stabilizing force. In the framework of conventional stability theory, this is a bit odd since density gradients are just as much a source of free energy as, for instance, temperature gradients. It could happen that some particular class of instabilities may not access their free energy, but it is quite inconceivable they emerge as stabilizers of all fluctuations when they are sufficiently large-that is, they prevent access to all other sources of free energy. This is probably the most profound scientific phenomenon that this study uncovers and dwells on. 

Stabilization by energetics, on the other hand, requires extreme geometries (at least for the tokamak). These tend to be very challenging from the viewpoint of \emph{global} ideal MHD stability and other practical considerations. 

This is, therefore, the first practical advantage that the new understanding brings- with sufficiently strong density gradients TBs could be formed in less extreme and challenging magnetic geometries (global deal MHD stability); the range of accessibility could be vastly increased.

In the stellarator context, a practical consequence of this understanding is to develop better informed optimization strategies for the magnetic geometry, and, to give a different interpretation than has sometimes been adopted.  

An additional practical implication is that stabilization due to the FC means that density gradients and/or impurity gradients can make it easier to initiate TBs from a geometry that did not initially have one. 

We turn now to simulation results for tokamaks. This displays the characteristics of the statistical mechanical ansatz with non-adiabatic electrons quite well. (The stellarator has similar general considerations, but will not be considered here due to space.)

\subsection{Simulation results with non-adiabatic electrons}

For quite some time, it has been realized that tokamak geometries with low or negative magnetic shear and high Shafranov shift lead to weak ITG/TEM, and hence, TBs can result.  By understanding the mechanism for this in terms of the general dynamics delineated above, we can recognize and explain various qualitative trends for the first time. These trends are often directly relevant to experiments. 

To review, four pivotal, general behaviors uncovered in previous sections were:

\begin{itemize}

\item 	In situations with large free energy (large gradients), the insolubility of the FC is often the operative dynamic that leads to stability. 
\item 	The basic trends of stability due to the FC can be understood from the very general statistical mechanical propensity for thermodynamic forces to drive the corresponding thermodynamic fluxes. In cases where one species is close to adiabatic, this propensity eventually makes the FC insoluble as density gradients are increased.
\item 	As is observed in very many systems with a large number of degrees of freedom, the gyrokinetic system manifests apparent adaptive behavior, and "finds a way" to dissipate free energy until a "hard" dynamical constrain prevents it- here, the FC.
\item 	The apparent adaptivity was explored in the context of the curvature drive that could be stabilizing or destabilizing.  We found that in order to stay unstable, the eigenmode structure avoids regions of stabilizing curvature, even when these are strongly predominant.

\end{itemize}

%We will refer to this as the "statistical mechanical ansatz".

\begin{figure*}
\subfloat[]{%
  \includegraphics[width=.49\linewidth]{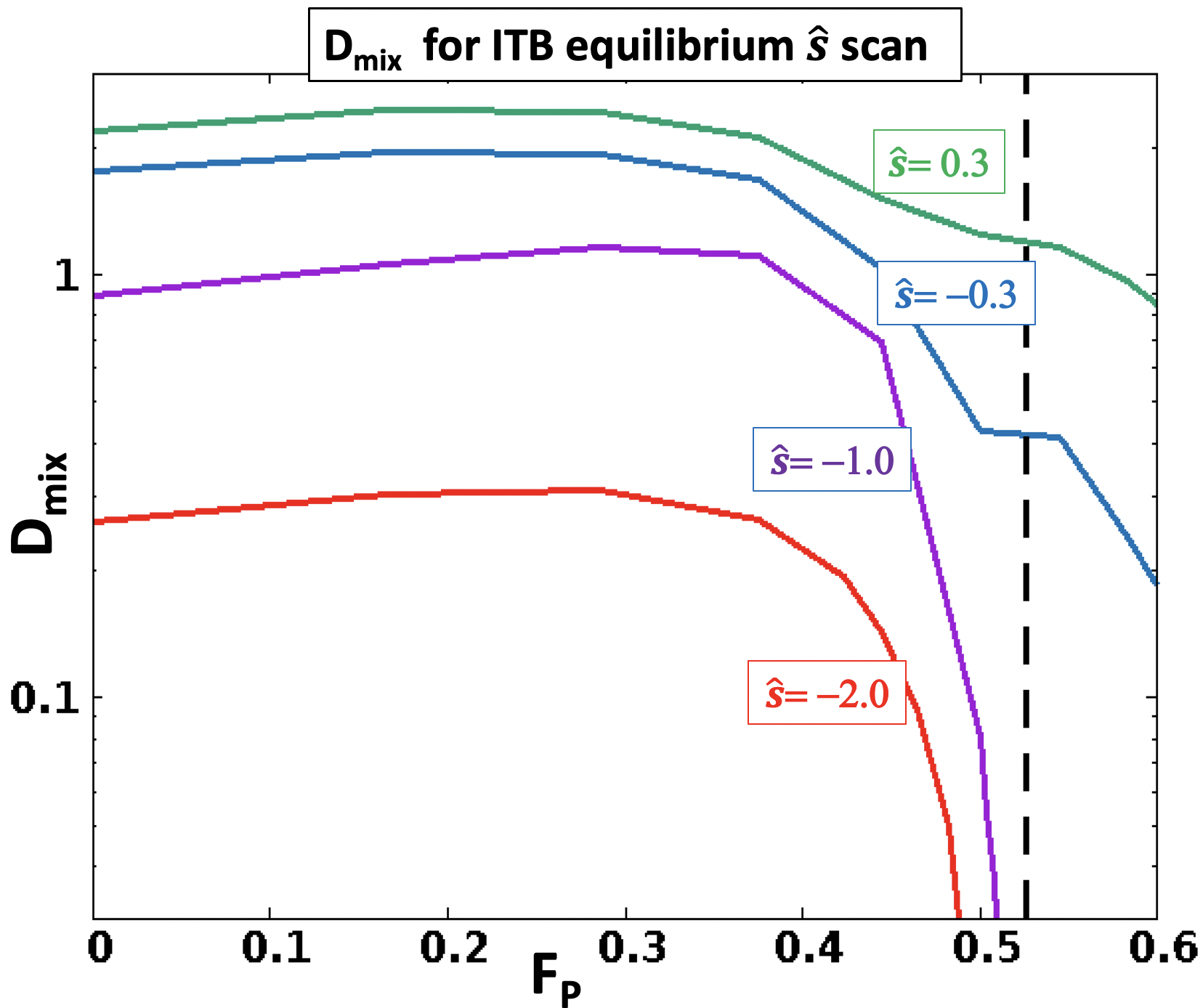}%
}\hfill
\subfloat[]{%
  \includegraphics[width=.51\linewidth]{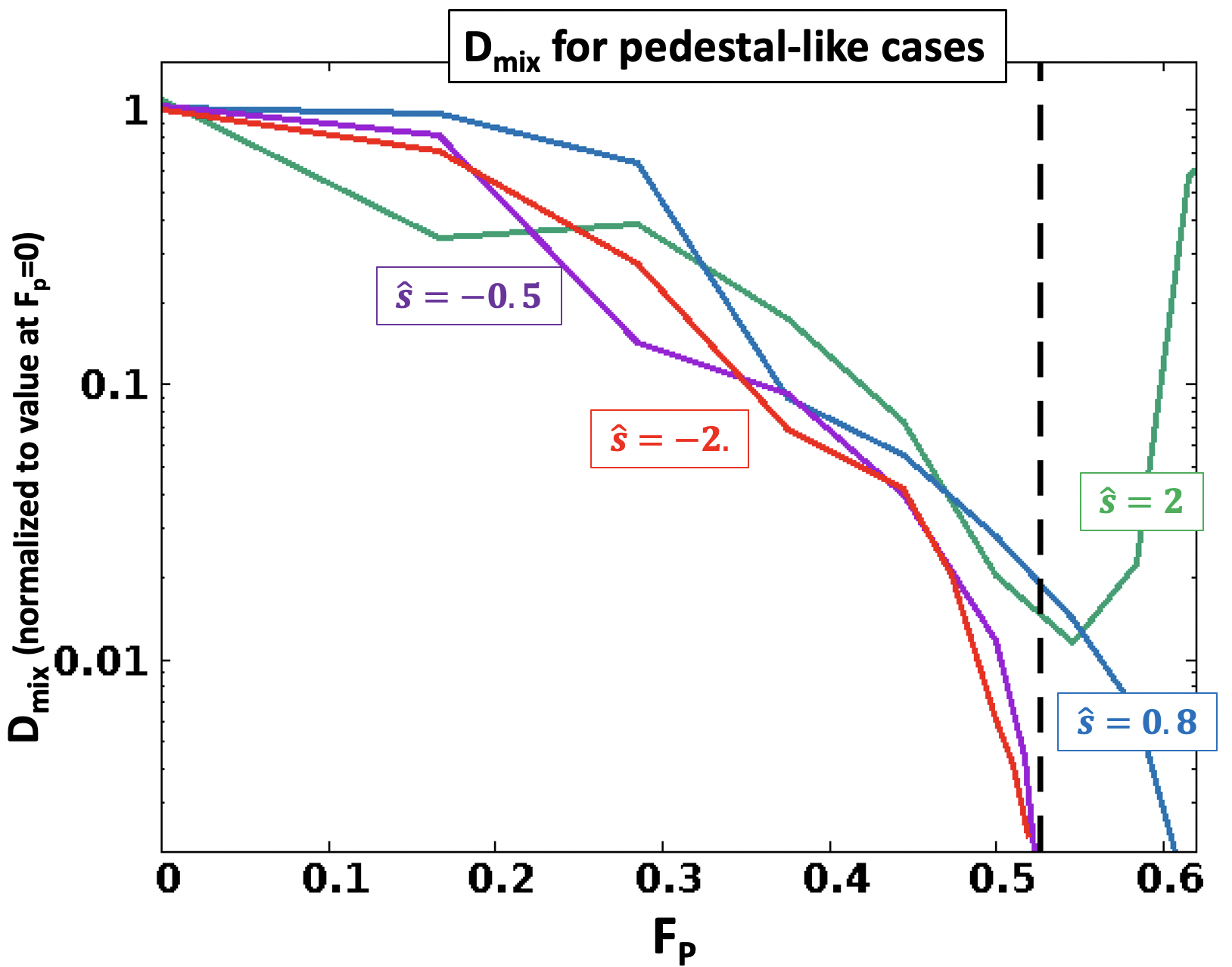}%
}
\vfill
\subfloat[]{%
  \includegraphics[width=.5\linewidth]{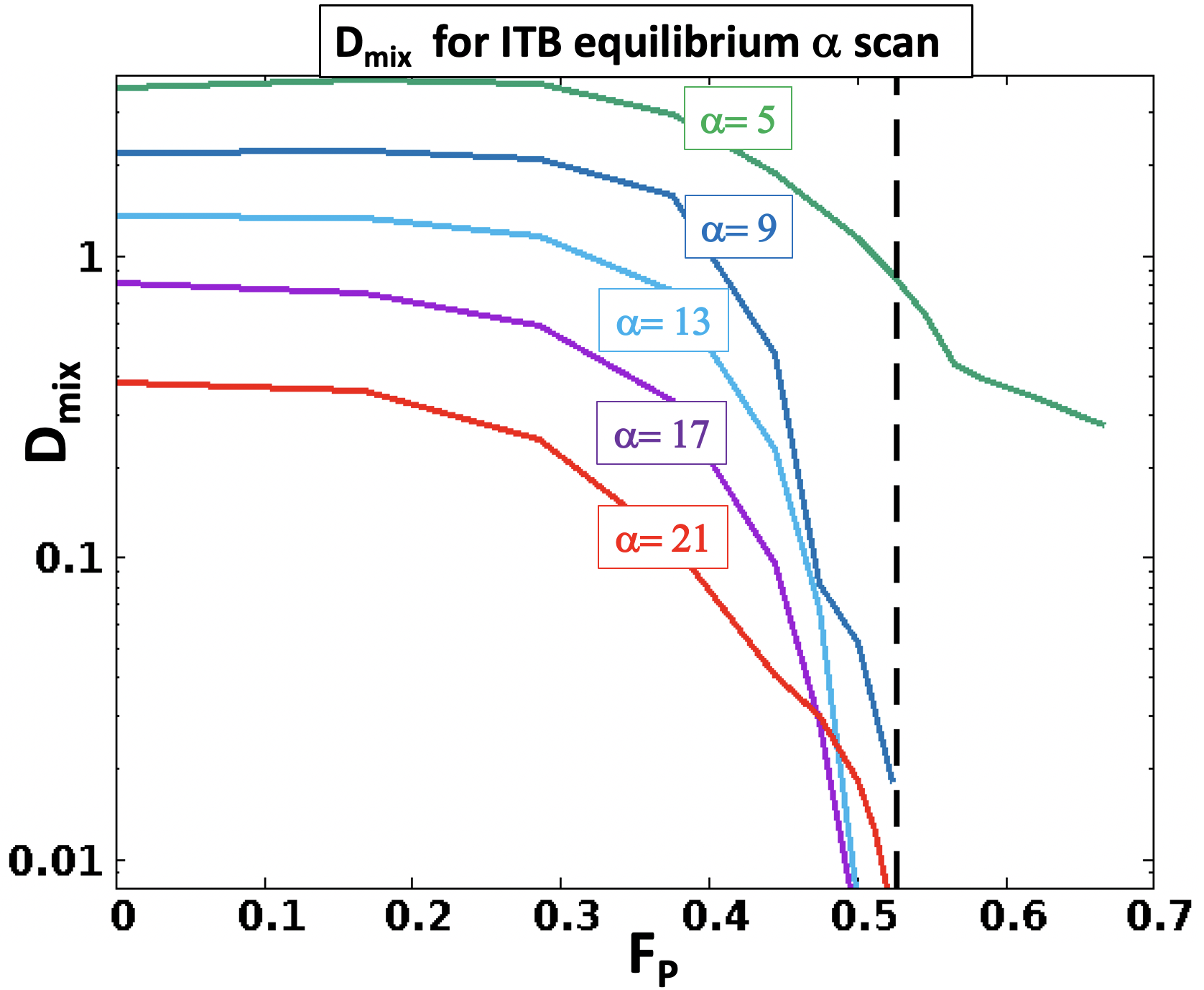}%
}\hfill
\subfloat[]{%
  \includegraphics[width=.5\linewidth]{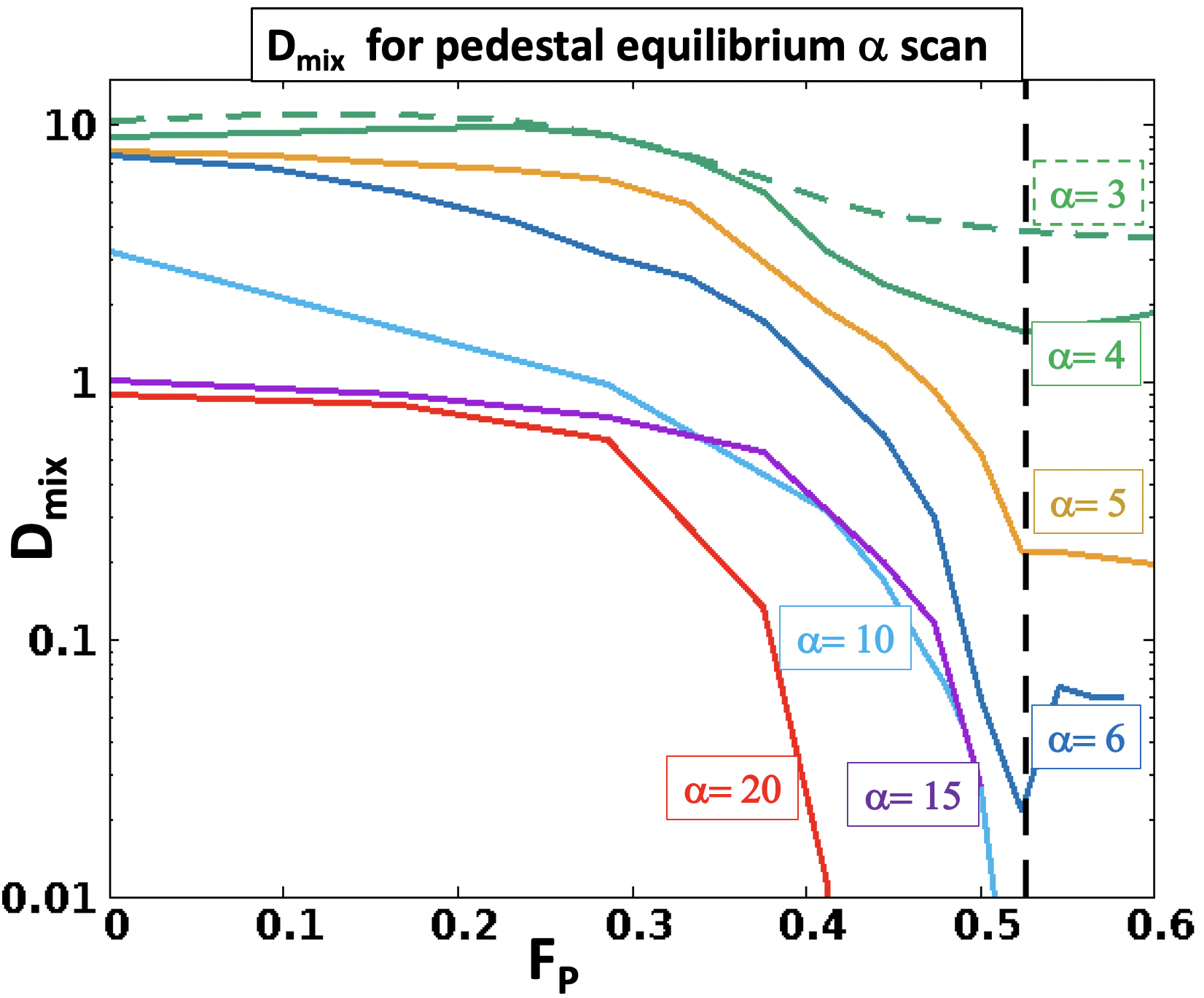}%
}
\caption{\label{fig:GEOSCAN} A) $D_{mix}$ from simulations with full electron dynamics and electromagnet effects, in scans of $\hat{s}$ and $\alpha$. The corresponding curvatures are shown in fig(\ref{fig:GEOSCANcurvs}). In all cases, increases in $\hat{s}$ or $\alpha$ eventually reduce the threshold $F_P$ needed for very low  $D_{mix}$ down to the limit of the $ITG_{ae}$, but even large increases beyond that do not materially reduce that $F_P$ }
\end{figure*}

Let us now view the results of electromagnetic gyrokinetic simulations with full electron dynamics in diverse geometries and equilibria. In Fig(\ref{fig:GEOSCAN}), we plot $D_{mix}$ as a function of $F_P$ for two different equilibria (for ITB and pedestal gradients) doing a parameter scan in $\hat{s}$ and $\alpha$. (Note all these cases have electron collisionality $nu^* \sim 0.05-0.1$. Higher collisionality produces a similar result but stability is reached for a less extreme $\hat{s}$ or $\alpha$.)
 
\begin{figure*}
\subfloat[]{%
  \includegraphics[width=.50\linewidth]{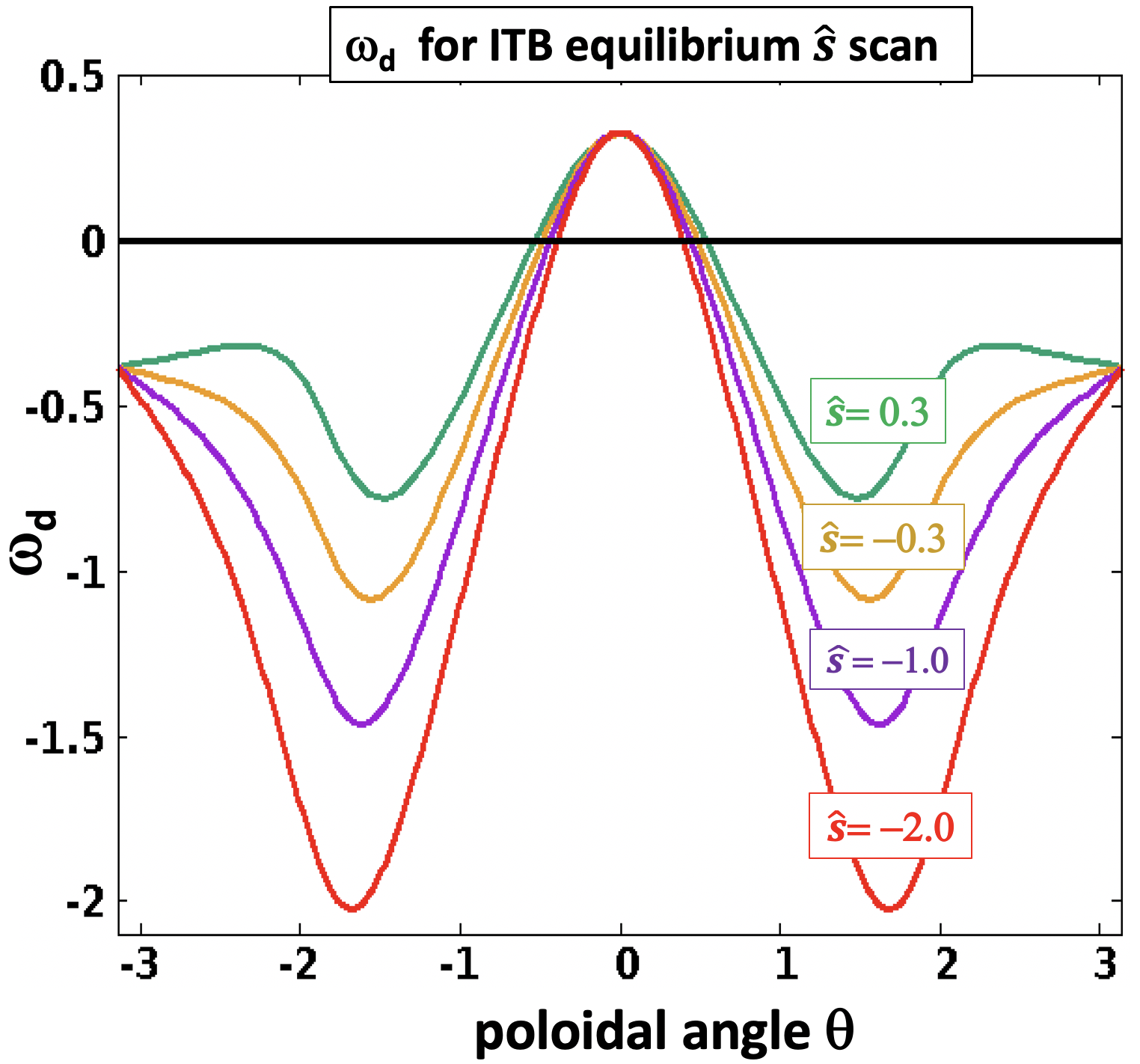}%
}\hfill
\subfloat[]{%
  \includegraphics[width=.50\linewidth]{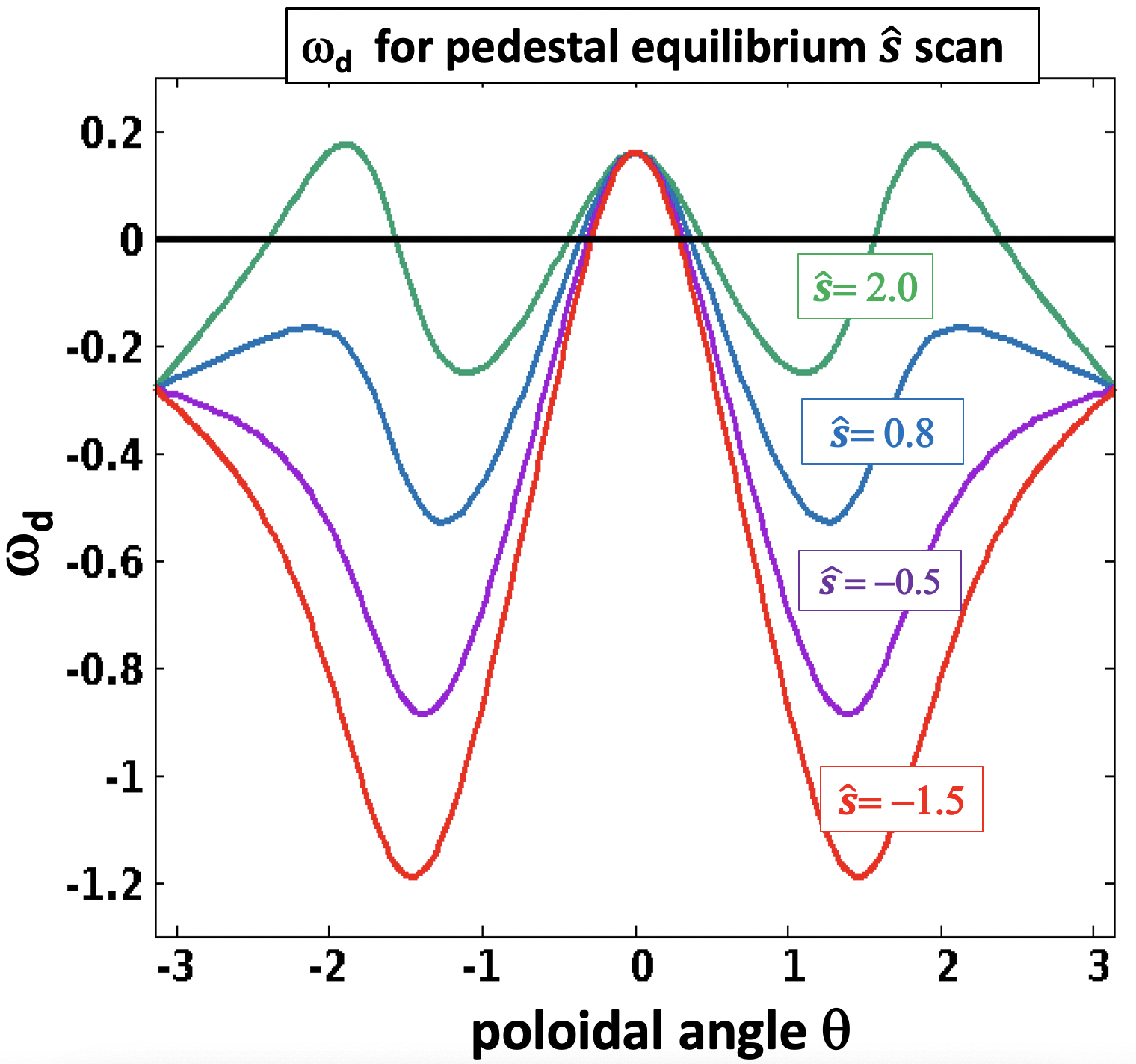}%
}
\vfill
\subfloat[]{%
  \includegraphics[width=.5\linewidth]{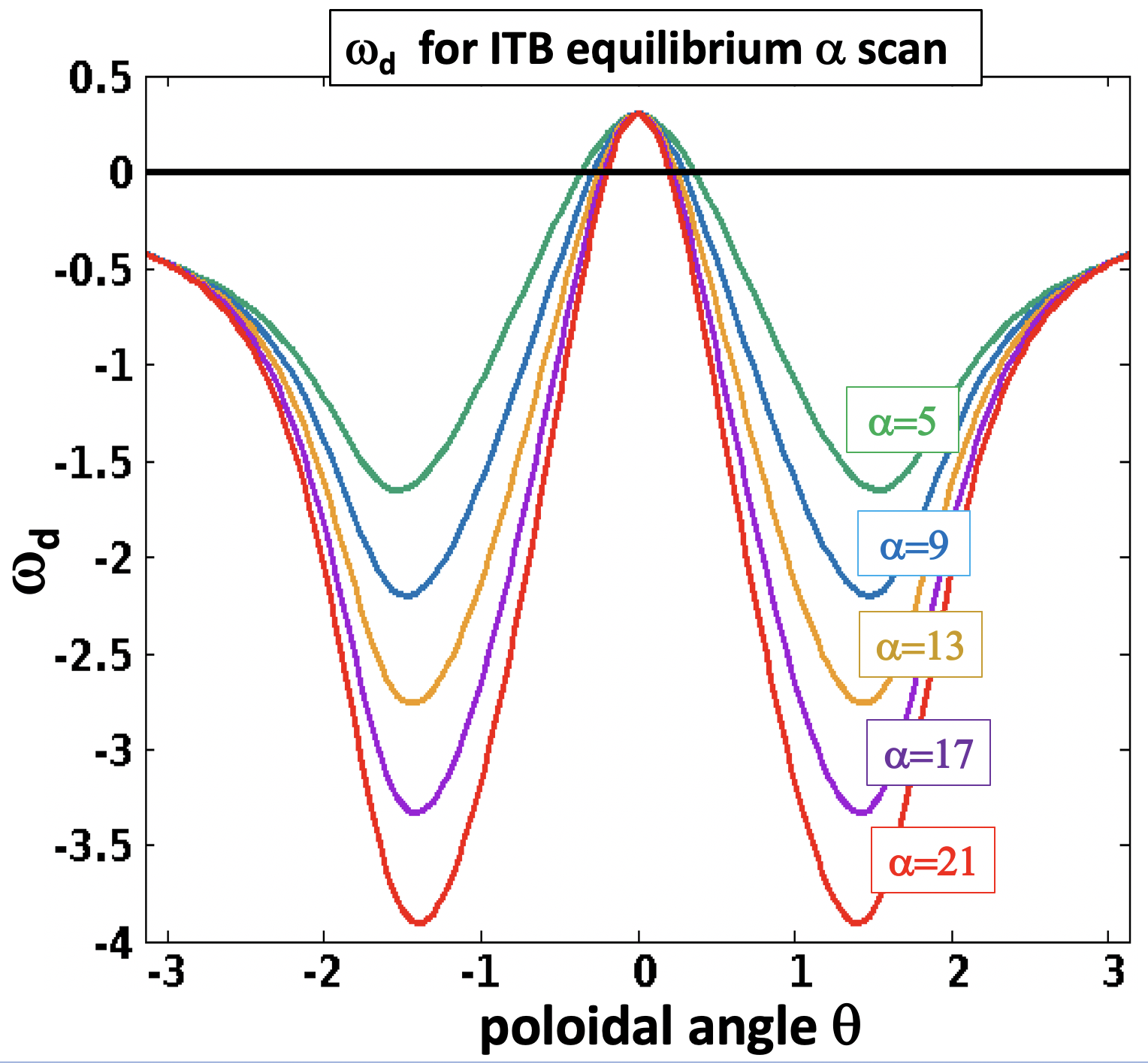}%
}\hfill
\subfloat[]{%
  \includegraphics[width=.5\linewidth]{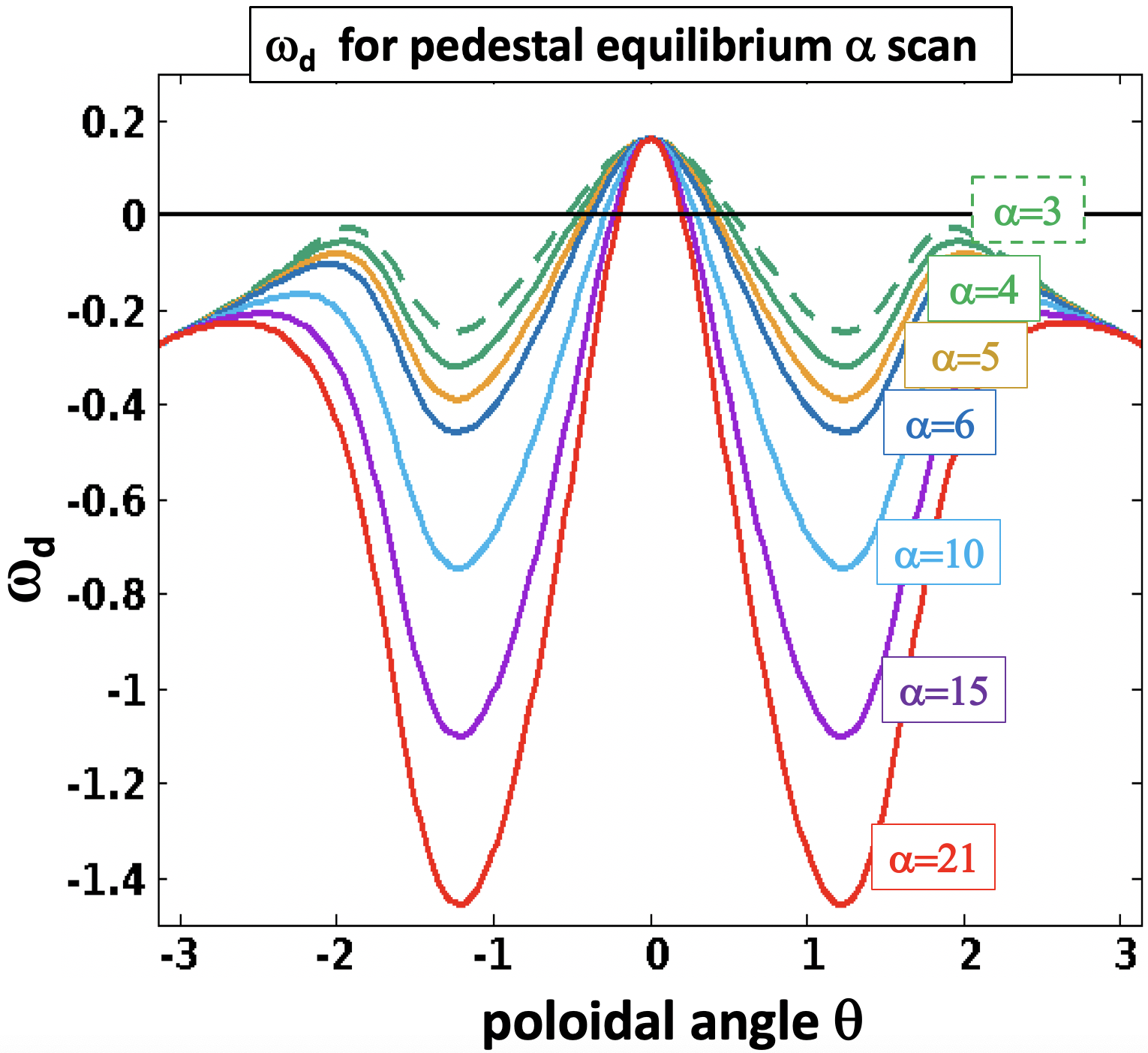}%
}
\caption{\label{fig:GEOSCANcurvs} A) $\omega_d(\theta)$ for the cases in fig(\ref{fig:GEOSCAN}. The curvatures become very strongly preponderantly negative, but this does not affect the $F_P$ bound found in simulations in fig({fig:GEOSCAN}). }
\end{figure*}

The figures show that an initial increment in $\hat{s}$ or $\alpha$ brings the mode close to $F_P$ stability point of the $ITG_{ae}$ (the vertical dashed line). Further, \emph{much larger} increases do not significantly improve the stability point in $F_P$.  As can be seen in Fig(\ref{fig:GEOSCANcurvs}), the curvature structure becomes extraordinarily heavily weighted toward stabilizing curvature (except for a small region of destabilizing curvature), and continues to become more stabilizing. 

But eventually, the huge preponderance of good curvature hardly improves the stability in $F_P$

In all cases, the requisite $F_P$ to cause stability approaches the solubility bound for the $ITG_{ae}$. This behavior is roughly generic. How does this result square with the prevailing picture before now?

The prevailing point of view would, roughly, go like this: the large region of stabilizing curvature of TB geometries (negative $\hat s$ or large Shafranov shift for tokamaks) leads to an orbit-averaged stabilizing curvature for the trapped electrons. This stabilizes the TEM mode. Then, some combination of density gradients and stabilizing curvature stabilizes the ITG. We'll call this the "conventional" picture of ITG/TEM stability in a TB. 

This picture is inconsistent with the displayed simulation results. To see why, we must first recognize the fact that the ITG and TEM are inevitably tightly coupled together. The ITG mode produces an electrostatic potential fluctuation, and the trapped electrons react to this with a bounce average response in exactly the same way as for a "pure" TEM, including the bounce averaged curvature effects. And similarly, a TEM mode produces ExB fluctuations that convect the ion temperature, and this couples that free energy into the fluctuation, just as with an $ITG_{ae}$. So the ITG and TEM are not like two "elementary particles", each with an independent existence, rather, each is more like a quark in the sense that it is only found in nature bound to another quark. The ITG/TEM is nearly always a composite mode (though for various parameters one aspect might be predominant). 

Consequently, an extraordinarily stabilizing curvature for a TEM should affect the energetics of the ITG/TEM similarly. If curvature stabilizesTEM, ITG/TEM should be stabilized too- consequently it should be  more stable than the ITG alone; and this relative stabilization would increase as the curvatures become more extreme. Simulation results are completely to the contrary:  The incremental stabilizing effect on the critical $F_P$ gets weaker as the curvature becomes more extreme. In fact the $F_P$ stability point approaches the adiabatic $ITG_{ae}$ most closely when the stabilizing trapped electron curvature is overwhelmingly large, even when it is far larger than the destabilizing curvature in the tiny region near $\theta=0$. 

Fortunately, the new dynamic explored in this paper does explain the simulations: just like what we found in $ITG_{ae}$, the mode structure adapts, to avoid interacting with the stabilizing regions of phase space. Specifically, its spatial structure "decouples" from the trapped electron response for particles with highly stabilizing orbit-averaged curvature, so it does not interact much with such particles. The mode might only couple strongly with the minority fraction of trapped electrons that have destabilizing orbit average curvature. Because there is little coupling to most trapped electrons, the "effective" trapped electron fraction is considerably smaller than the actual fraction of trapped electrons. So the non-adiabatic electron response is small. Hence, the electrons produce little charge flux,  pushing the $F_P$ limit for ITG/TEM close to that of $ITG_{ae}$.

 The FC based stabilization is qualitatively different from the conventional stability analysis controlled by energetic factors; consequently, different parametric behaviors will be predicted. FC predicts that increasing density gradients lead to stabilization; whereas we will see that the conventional picture of curvature stabilization requires extreme geometries, and in this case, density gradients play an insignificant role.
 
In short, the essential physics of stabilization (as revealed in the FC context) is the same as that of $ITG_{ae}$; note that  the same behavior vis-a-vis curvature was found in section VIII.

\subsection{Analysis using SKiM including trapped electrons, and the fallacy of considering the ITG and TEM independent}

These qualitative differences can be explicitly demonstrated with the mean field theory SKiM  with trapped electrons included.  The resulting dispersion relation, 

%calculation that led to the eigenfunction averaged curvature expressions can be generalized to also include trapped electrons. Will will describe some details of this in the next section. The result is the addition of a new term to the previous SKiM dispersion relation to include trapped electron effects, on the right in Eq(\ref{eq:SKtot}):

\begin{eqnarray}
&&\sum_{ion \ species} \left( \int \mathrm{d} \vec{v} \frac{ F_{Ms} J_0( k_{\perp} \rho_i)^2(\omega-\omega_s^{\star})}{\omega-k_{\parallel}v_{\parallel}-\omega_{ds}}+1 \right) \frac{q_s^2}{T_s} =  
\label{eq:SKtot} \nonumber \\
&& -f_{trap}  \frac{e^2}{T_e}   \int 4\pi v^2 \mathrm{d}v  \frac{ F_{Me}(\omega-\omega_e^{\star})}{\omega-\omega_{d~orbit~e}(v)+i\nu_{eff}(v)}   +\frac{e^2}{T_e}  \nonumber \\
\label{eq:SKiMDRtrap}
\end{eqnarray}
has an additional term on the on the right hand side of Eq(\ref{eq:SKtot}).

Qualitatively, this term is of a familiar from early investigations of the ITG and TEM modes. It contains an integral over energy, curvature resonances and an effective collision term. In SKiM, two new quantities are introduced for trapped electrons, which are defined and discussed in Appendix B: 1) an eigenmode average of the orbit averaged curvature $<\omega_{d~orbit~e}>$, appearing in the resonant denominator. It is similarly to the ions, but the weighting involves orbit averages of $\phi$ rather than $\phi$ itself. And 2) an overall strength of the coupling to the trapped particles, the "effective" trapped electron fraction $<f_{trap}>$. The expression, independent of curvature, depends only upon the magnitude of orbit averages of $\phi$ (relative to $\phi$ itself). It describes the eigenfunction averaged strength of coupling to trapped particles, irrespective of whether the trapped response is stabilizing or destabilizing as regards curvature. For some eigenmode structures, it can be considerably less than the actual fraction of trapped particles in the geometry. 

We now use Eq(\ref{eq:SKtot}) to show the qualitative difference in the prediction of the conventional picture and that of the statistical mechanical ansatz.

\begin{figure*}
\subfloat[\label{sfig:1a}]{%
  \includegraphics[width=.50\linewidth]{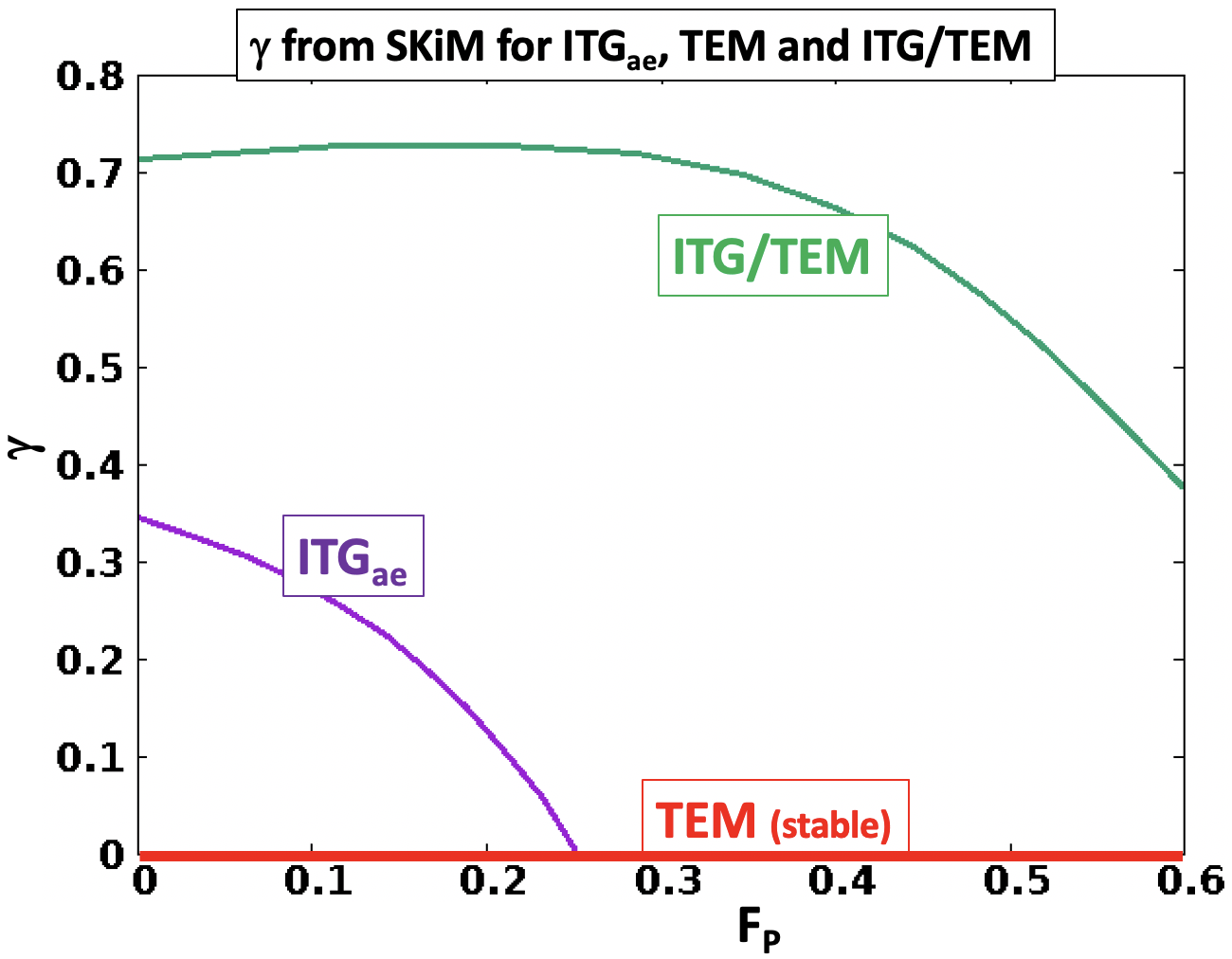}%
}\hfill
\subfloat[\label{sfig:1b}]{%
  \includegraphics[width=.50\linewidth]{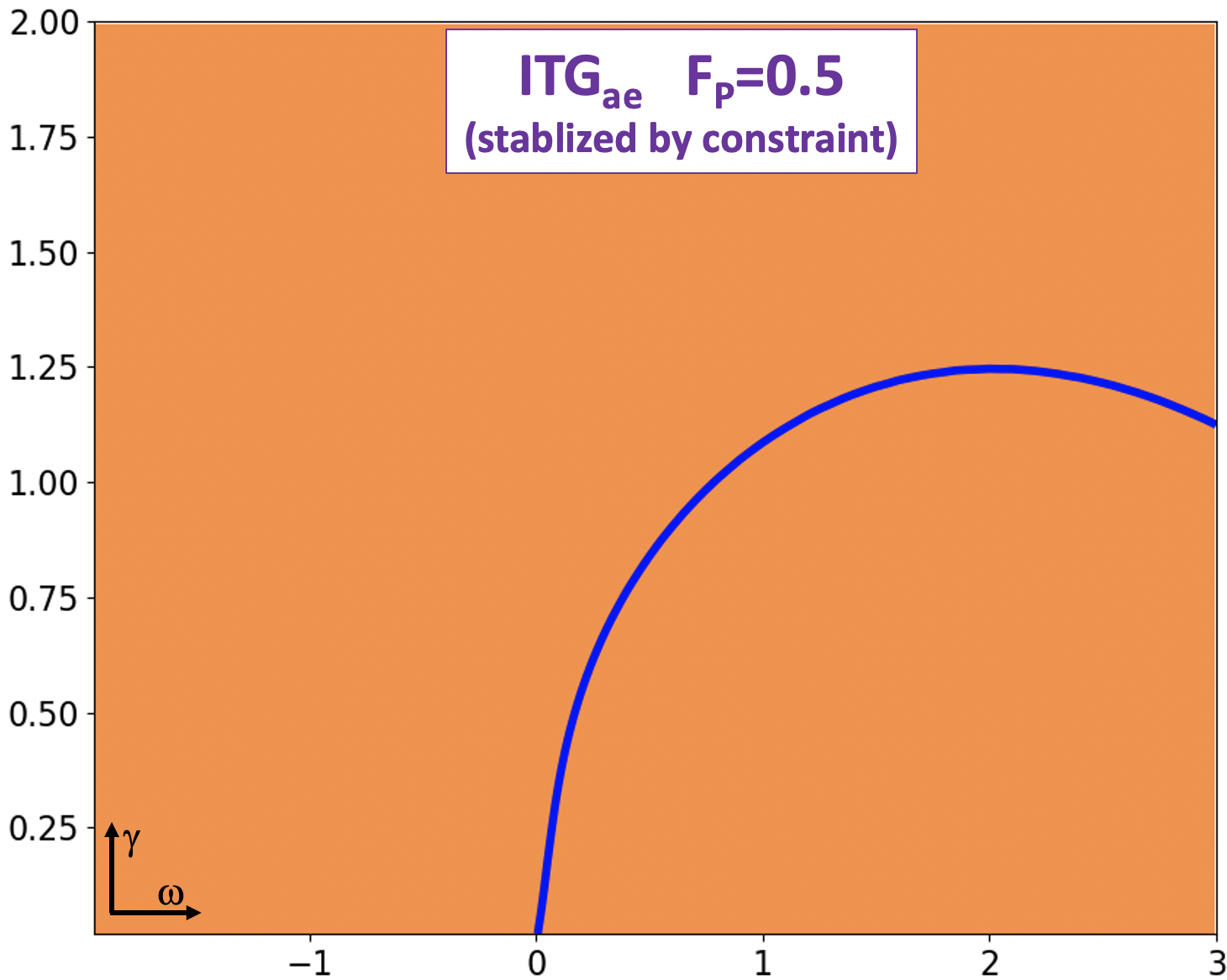}%
}
\vfill
\subfloat[\label{sfig:1c}]{%
  \includegraphics[width=.5\linewidth]{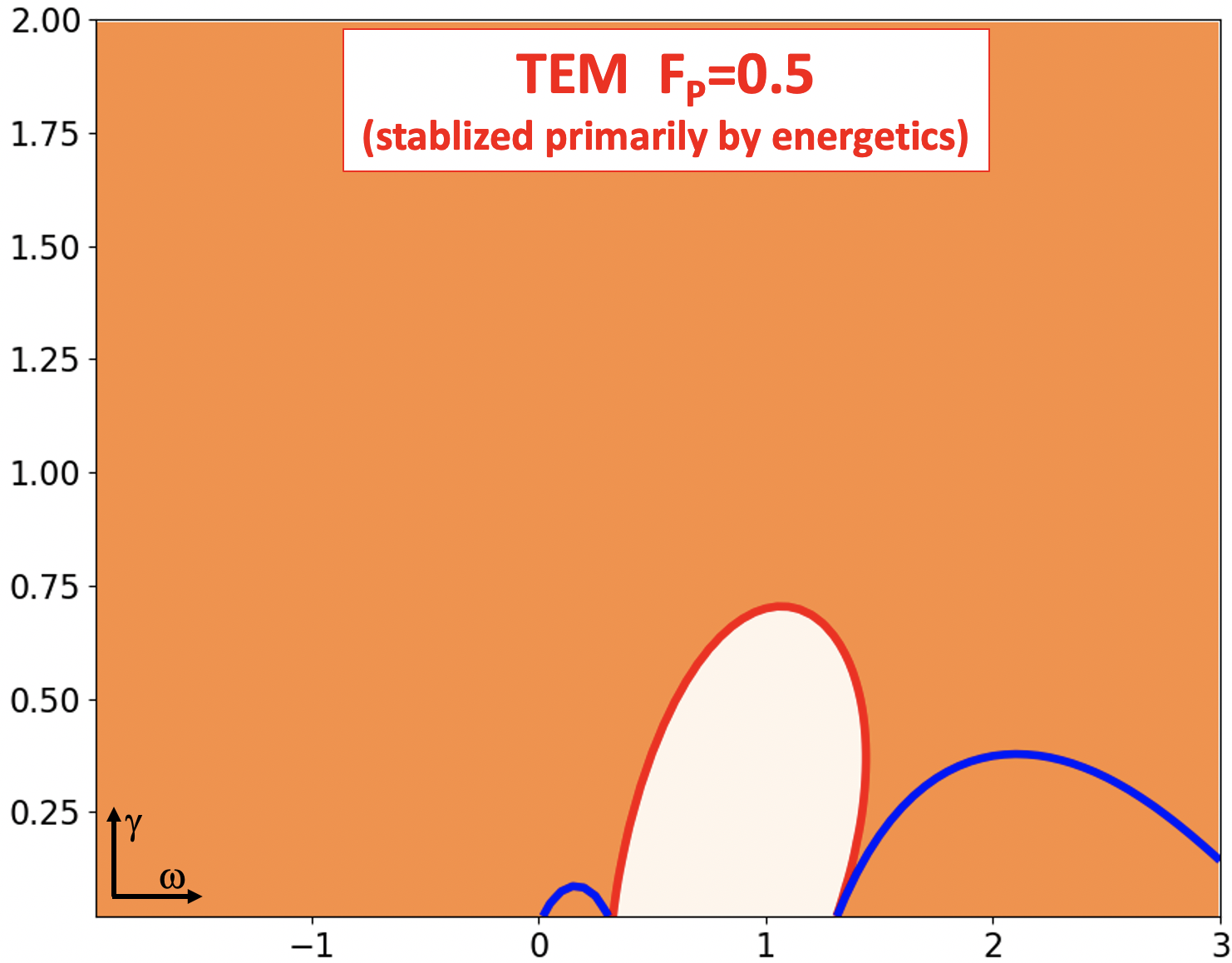}%
}\hfill
\subfloat[\label{sfig:1d}]{%
  \includegraphics[width=.5\linewidth]{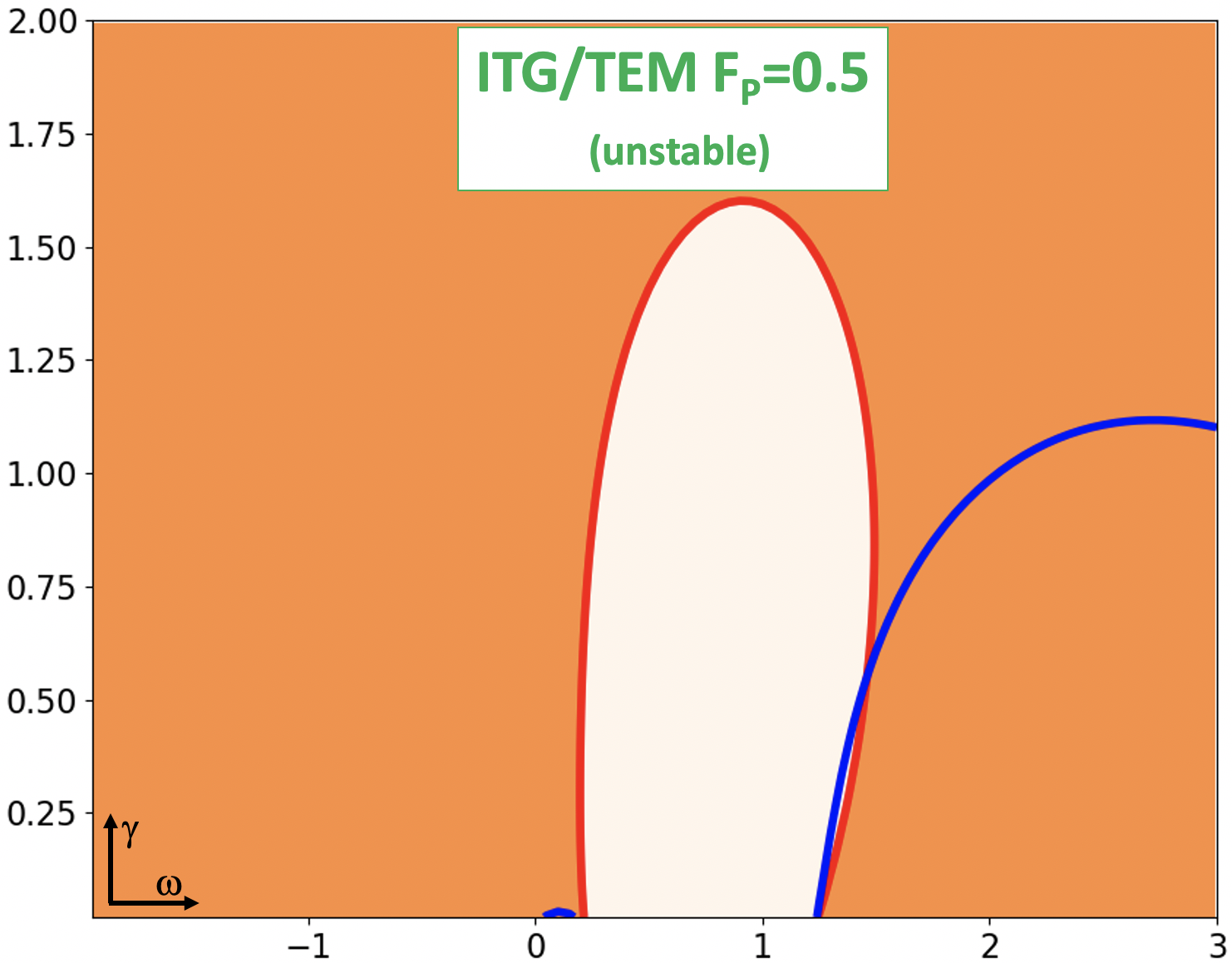}%
}
\caption{\label{fig:SKiMfe} The fallacy of understanding the ITG/TEM as decoupled ITG and TEM a) $\gamma$ in SkiM for representative parameters with large trapped particle fraction $<f_{trap}>$ and sufficient negative average electron curvature $<\omega_{d~orbit~e}>$ to stabilize the pure $TEM$ mode (with ion temperature gradient equal zero.) However, the coupled ITG/TEM (with $R/L_{Ti} = R/L_{Te}$) is \emph{not}  stabilized by $F_P$ even though the ITG and TEM is stable. (The $ITG_{ae}$ is stabilized by $F_P$ as usual.)  The reason for this is revealed by the E-FC plots for $F_P=0.5$  b) the  $ITG_{ae}$ is stabilized by the constraint as always c) quite the contrary, the TEM is stabilized mainly energetically: the free energy curve is so close to $\gamma=0$ that it does not intersect the FC curve, so, there is no unstable curve.  However the FC alone would allow a substantial growth rate. For the ITG/TEM in c) the free energy curve, now supplemented by the free energy of ion temperature gradients, allows high growth rates, unlike the TEM. The FC curve also allows high growth rates, since electron particle fluxes can balance the ions. Hence strong instability results for the ITG/TEM even though the corresponding ITG and TEM are stable. }
\end{figure*}

Let us begin by examining the notion that the stabilization of the TEM can explain the behavior of the ITG/TEM; we will find that it is definitely not so! 

Reasonable parameters can allow the ITG/TEM to be quite unstable even though the TEM is stabilized by curvature. This is because the non-adiabatic electron response allows the FC to be satisfied for large $F_P$, by having an electron charge flux to balance the ions. This electron flux requires an expenditure of free energy, however, due to the stabilizing orbit averaged curvature. We'll see that the ITG drive can supply more than enough energy for this, and have enough left over to attain a large growth rate. 

Consider a representative example. In Fig(\ref{fig:SKiMfe}), we consider typical  parameters corresponding to the convention mode of thinking, where the average trapped electron curvature is quite stabilizing. 

We compare the stability of the TEM, the $ITG_{ae}$, and the ITG/TEM. We display two kind of graphs (all with $R/L_{Te}=20$, as in a TB): Fig(\ref{fig:SKiMfe})a is a $\gamma-F_p$ plot while the other three are E-FC plots in the $\omega-\gamma$ plane.  Recall these plots from section XII: they show the curve where the FC is satisfied (red), and the curve where the free energy balance is satisfied (blue). (The orange region is where the FC is not satisfied because of too much ion flux, the white region is where the FC is not satisfied because of the opposite situation, and the boundary between them where the charge flux vanishes is the red curve.)

There are three separate regimes to study:

1) The TEM specific regime- set  ion temperature gradient to zero and keep only density and electron temperature gradients. Now as in the conventional scenario:

\begin{itemize}

\item 	We adjust the electron curvature to be sufficiently negative; TEM is always stable for all density gradients. (We use $R/L_{Te}=20$)
\item 	We assume a strong coupling to trapped electrons:  $<f_{trap}>$ is a typical physical fraction of trapped particles.

\end{itemize}

2) $ITG_{ae}$ specific regime: set $<f_{trap}>=0$ and use the same  $R/L_{Ti}=20$. 

3) ITG/TEM hybrid regime- combine the ITG and TEM- include the ion temperature gradient in the TEM case with trapped electrons ($R/L_{Te}=R/L_{Ti}=20$). 

All these cases are run at the same $k_{\perp}$ and $k_{\parallel}$, at representative values typical of the eigenfunctions above ($k_{\perp} \rho_i \sim 0.5$ and $k_{\parallel}=0.71$ and $<\omega_{di} = 0.24$ ).

Growth rates from Eq(\ref{eq:SKtot}) are shown in Fig(\ref{fig:SKiMfe}). As expected, the $ITG_{ae}$ is stabilized at low values of $F_P$. The TEM is always stable. But the ITG/TEM is far more unstable than the $ITG_{ae}$, up to $F_P$ values much greater than the $ITG_{ae}$.

\emph{Stabilizing the (pure) TEM does not cause the ITG/TEM to be stable; it remains much more unstable than $ITG_{ae}$ for all $F_P$ .}

This surprising behavior becomes comprehensible when we contrast the effects of trapped particles on the FC, and on free energy dynamics. We plot the E-FC curves for these three cases at $F_P=0.5$. The $ITG_{ae}$ (Fig.20b) is stable (all orange) because no point in the upper half plane satisfies the FC with adiabatic electrons. But there is considerable free energy, i.e, growth rates would be quite large due to free energy balance alone. The TEM (Fig.20c) is stable primarily because the free energy is so low: the stabilizing curvature has so reduced it, that balance is only achieved for low growth rates in rather tiny regions. These regions are so shrunken that they do no intersect the FC for $\gamma > 0$, so there is no instability. 

And distinct from $ITG_{ae}$, the FC could, in principle, be satisfied up to much higher growth; the electron flux due to the non-adaibatic electron response can easily balance the ion charge flux; stability is due to the lack of free energy

Now consider the ITG/TEM in the last panel. The trapped electrons (the TEM part) provide the particle flux to balance the ion flux ( to satisfy the FC) and ITG part supplies the free-energy drive. Thus large $\gamma$ is allowed even at $F_P=0.5$.  The intersection of the red and blue curves yields the unstable eigenfrequency of the dispersion relation.

Note that the trapped electron response does, indeed, reduce the free energy available to the perturbation: the free energy ( blue) curve in the ITG/TEM is limited to lower $\gamma$ compared to that of 
 ITG alone, and by a large margin for $\omega_r \lesssim 1$. \emph{But the greatly increased region where the FC is soluble allows a substantial instability.} 

In other words, the energetic cost of exciting the stabilizing trapped electrons to balance the FC is one that the ITG/TEM instability can and does pay.  

This example shows the fallacy of considering the TEM as somehow independent from the ITG. One must understand the behavior of the coupled system; it is really a hybrid mode!

\subsection{Using SKiM to understand the stabilization of ITG/TEM seen in simulations: the statistical mechanical ansatz vs the conventional picture}

\begin{figure*}
\subfloat[\label{sfig:1a}]{%
  \includegraphics[width=.50\linewidth]{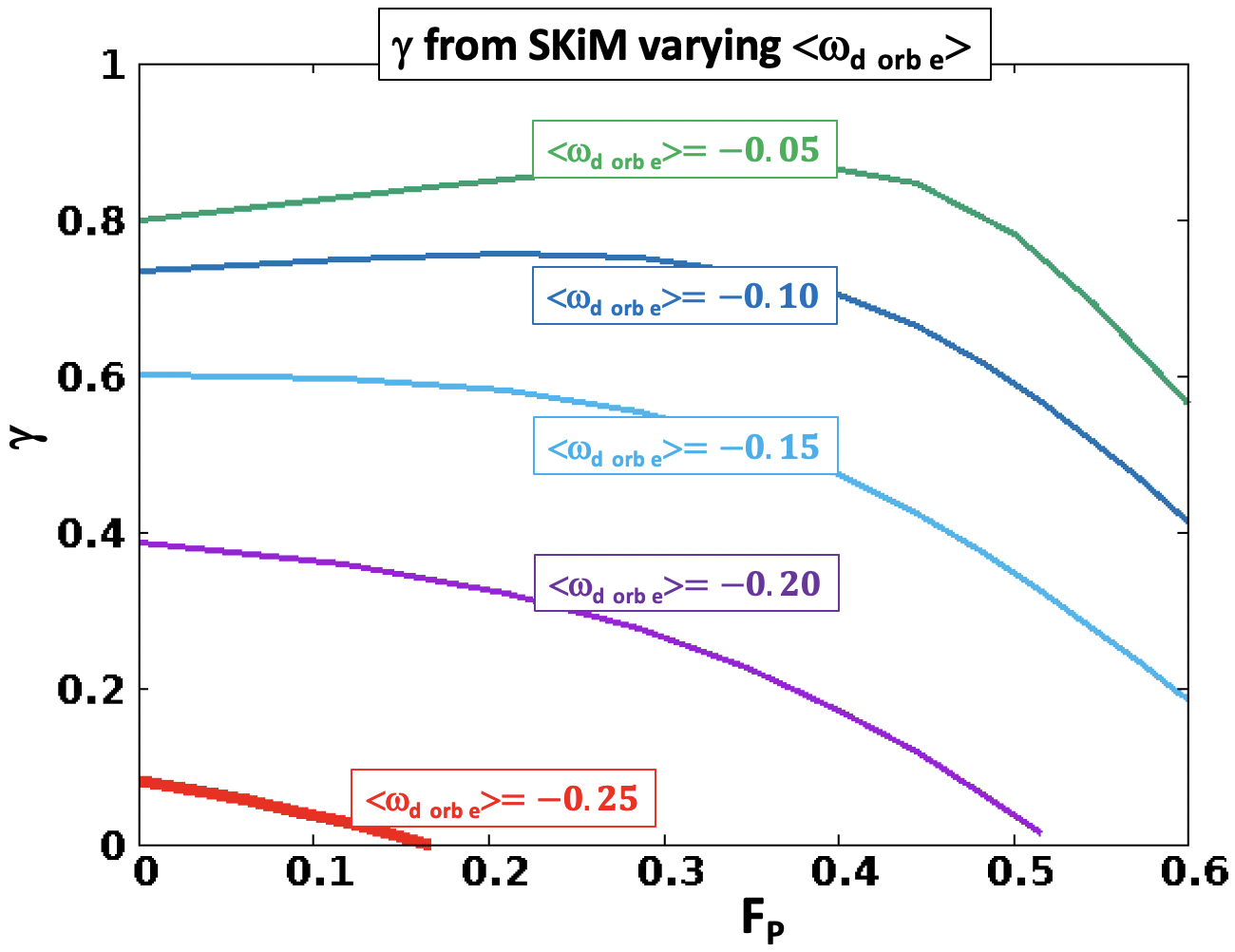}%
}\hfill
\subfloat[\label{sfig:1b}]{%
  \includegraphics[width=.50\linewidth]{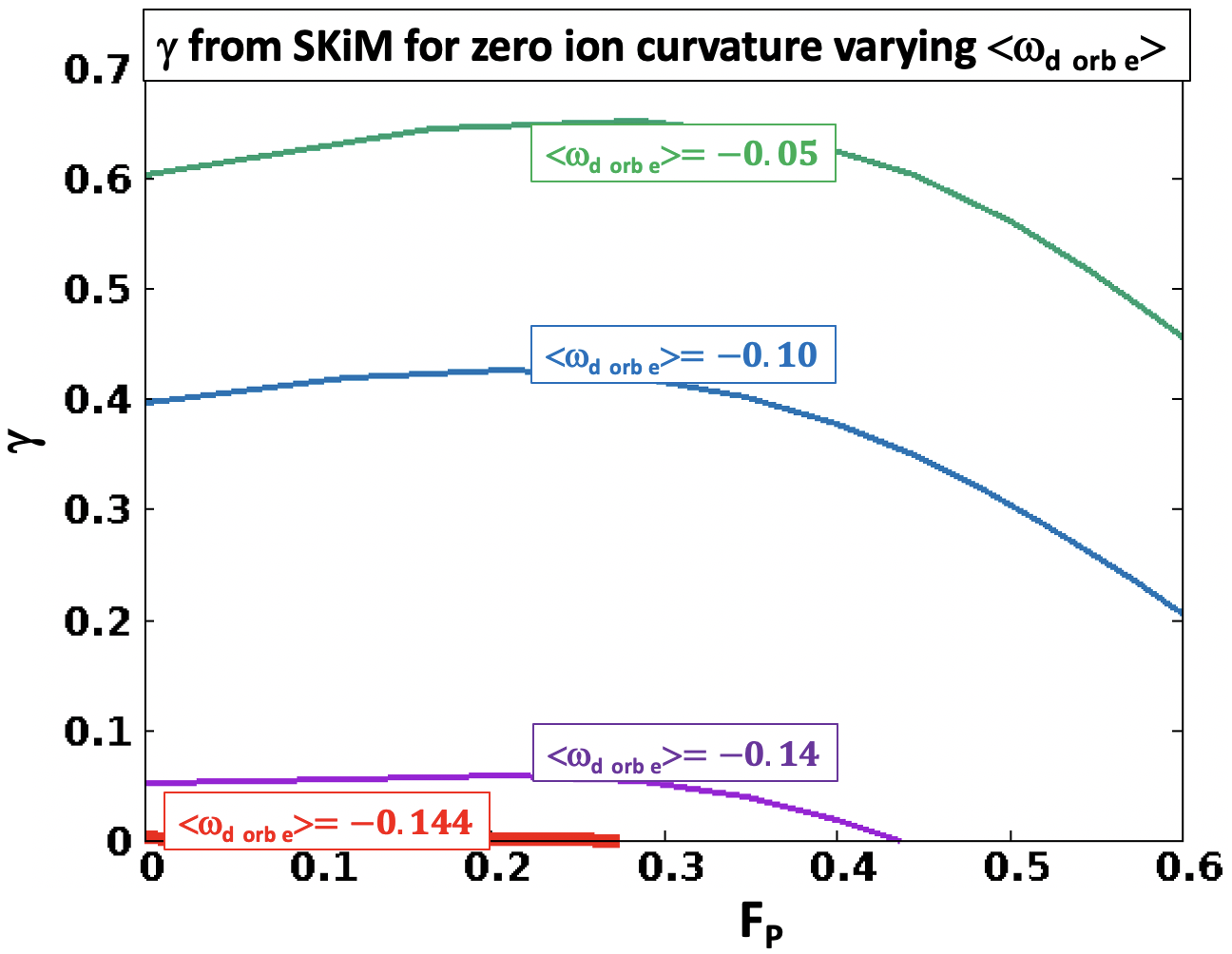}%
}
\vfill
\subfloat[\label{sfig:1c}]{%
  \includegraphics[width=.5\linewidth]{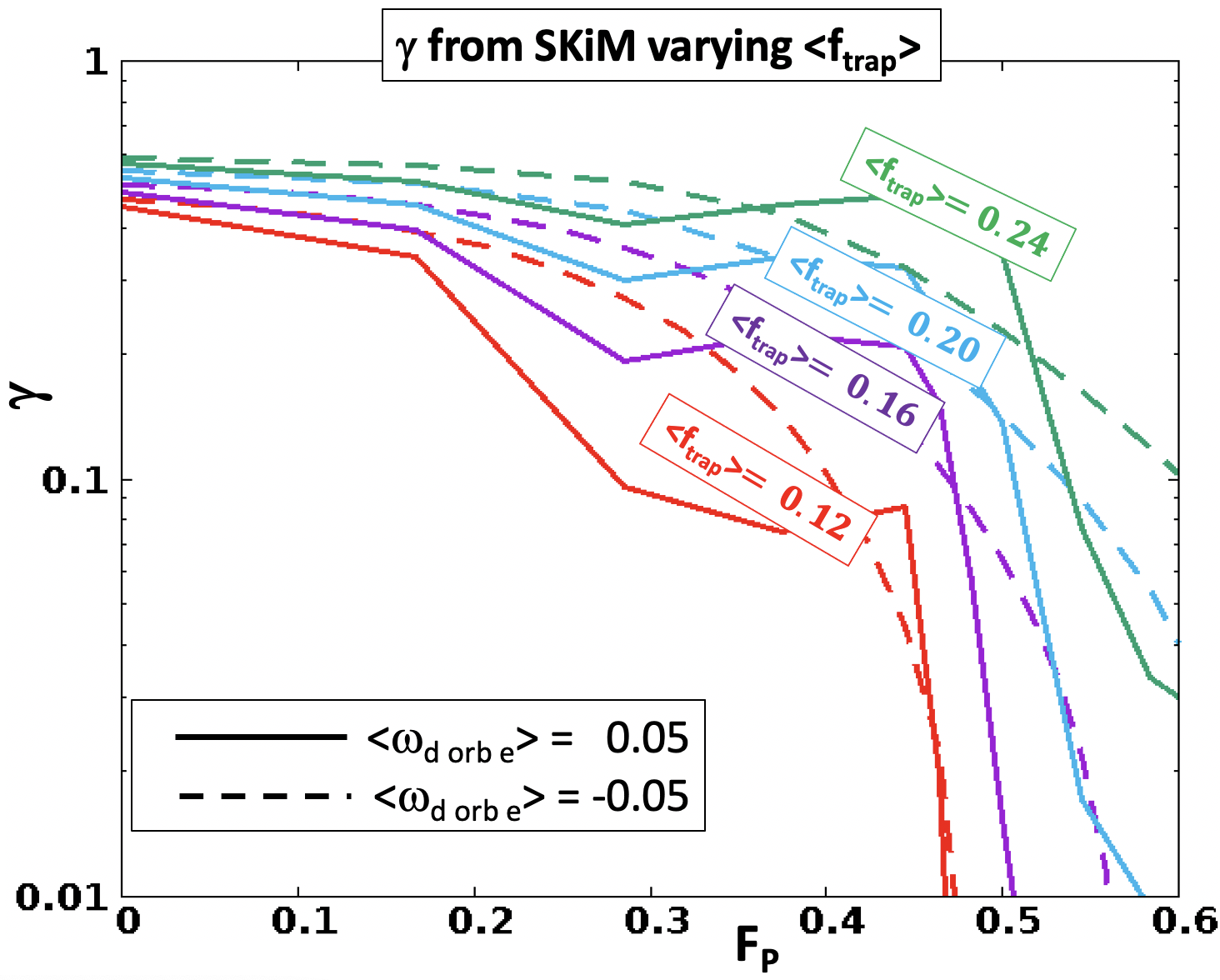}%
}\hfill
\subfloat[\label{sfig:1d}]{%
  \includegraphics[width=.5\linewidth]{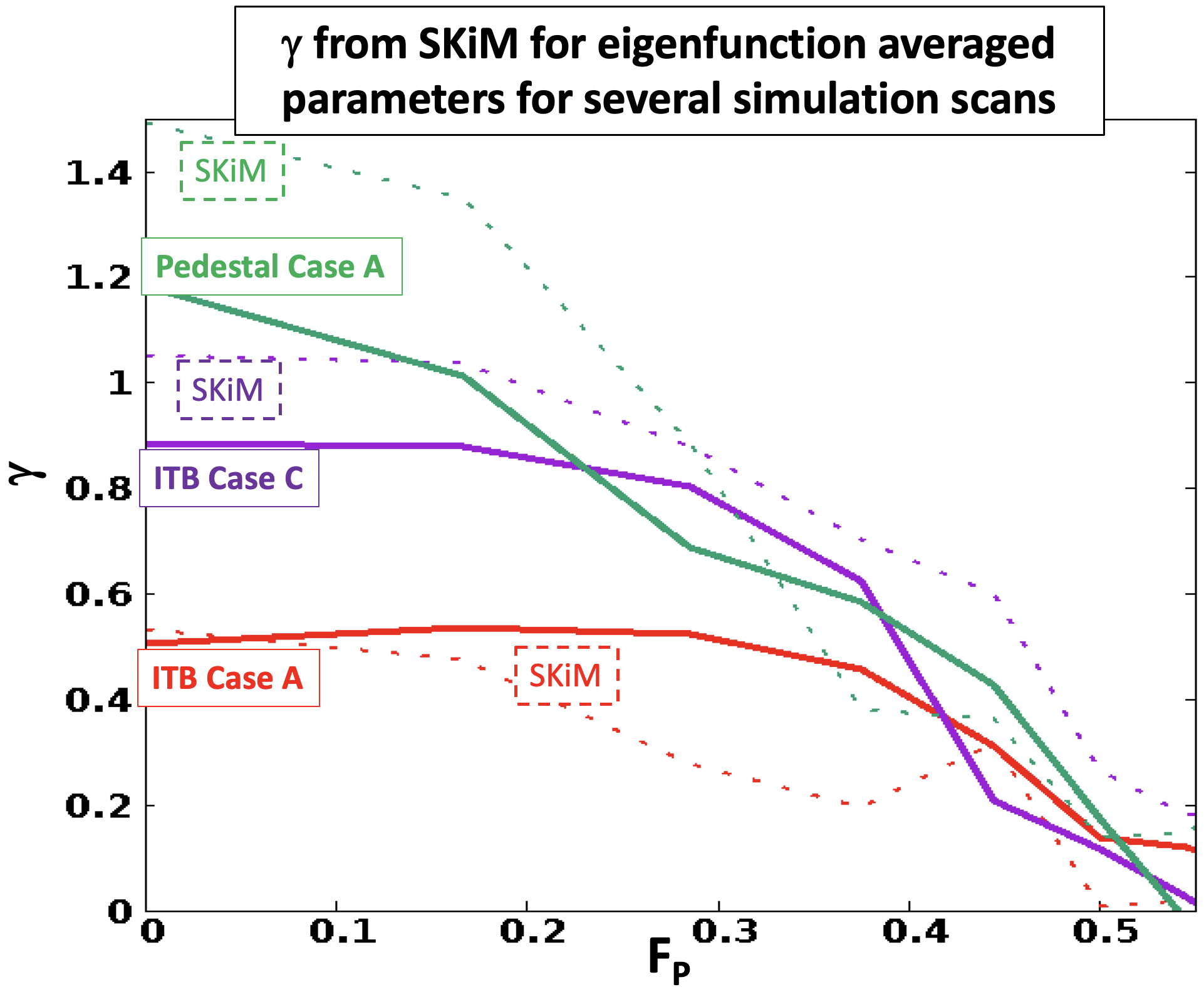}%
}
\caption{\label{fig:SKiMfe2} Contrasting mode behavior in the conventional picture with behavior from the statistical mechanical ansatz, using SKiM. a) In the conventional picture $<f_{trap}>=0$ remains large and the electron curvature is made more negative. Stabilization results for any value of $F_P$ depending upon how negative the curvature is. This is quite unlike the simulation results for progressively more negative curvature b) This qualitative behavior is generic energetic stabilization as in the convenitonal picture: a similar scan with the ion average curvature set to zero behaves similarly c) The growth rate in SKiM for a parameter scan like that in the statistical mechanical ansatz, where $<f_{trap}>=0$ is reduced by eigenfunction adaptation to the progressively more negative curvature. For either positive or negative electron curvature, stability is reached a particular $F_P$ near the FC solubility limit for $ITG_{ae}$. d) Growth rates from SKiM compared to simulation, where all eigenfunction averages are taken from the simulation eigenfunction and the SKiM $\gamma$ is from its dispersion relation eq(\ref{eq:SKiMDRtrap}). The agreement is fairly close. Hence SKiM is a valid model to understand the simulations. }
\end{figure*}

In the conventional picture, we expect that sufficiently large negative curvature( $<\omega_{d~orbit~e}>$) will stabilize the ITG/TEM. This is, indeed, what is found in SKiM (See Fig(\ref{fig:SKiMfe2})a). For a certain small range of negative orbit average curvature, growth rate dependence upon $F_P$ is similarly to the simulations: stabilization occurs near $F_P \sim 1/2$.  But \emph{unlike} the simulations, the stability point in $F_P$ varies strongly as the curvature is varied. Even a $20\%$ change in the negative curvature causes either complete stability for \emph{all $F_P$}, or, an instability up to large $F_P \sim 0.6-0.7$. 

\emph{This strong variation of stabilizing $F_P$ with negative curvature is the typical behavior found by SKiM when there is a large trapped particle fraction.}     For example, as a further test, we reduce the ion curvature to zero (i.e. a slab-like ion mode) in fig(~\ref{fig:SKiMfe2})b. Again, stabilization at $F_P \sim 1/2$ is only possible for a small range of negative $<\omega_{d orbit}>$, and the mode becomes unstable for small changes beyond this. 

\emph{This is quite unlike the typical behavior in simulations (see fig\ref{fig:GEOSCAN}) which find, that over an enormous range of curvature (fig(\ref{fig:GEOSCANcurvs})), $F_P \sim 1/2$ always produces stability.  } 

To understand simulations, let us appeal to adaptivity, a major element of what what we have called the statistical mechanical ansatz: \emph{the eigenfunction adapts to avoid curvature stabilization by changing its structure. Simulations find that adaptation largely decouples from the trapped electron orbits with strongly negative curvature. By decoupling from most trapped electrons, the $<f_{trap}>$ becomes small.}

Going back to SKiM, let us track $\gamma$ as $<f_{trap}>$ is lowered (fig(\ref{fig:SKiMfe2})c) for two values of . $<\omega_{d~orbit~e}>=0.05, -0.05$, As we would expect, for low $<f_{trap}>$ the stability becomes similar to the $ITG_{ae}$ . The average $<\omega_{d~orbit~e}>$ is taken from simulation eigenfunctions; it is often slightly positive, because the eigenfunction decouples from the trapped electrons with the most stabilizing curvature.  Notice that the stability point stays near $F_P \sim 1/2$ for low $<f_{trap}>$.  This qualitative behavior is the same if the sign of the electron curvature is changed, i.e., it is slightly negative. This happens because the non-adiabatic response is vanishing for either sign of curvature; the FC stabilization is not strongly affected by curvature drive, unlike energetics. Note the similarity of these results to the ITGae: curvature affects the magnitude of the growth rate (here by a factor of $\sim 2$) but the stability point in $F_P$ is not strongly affected.

%But the stabilization does not qualitatively depend upon the sign of $<\omega_{d~orbit~e}>$, so we display results for negative $<\omega_{d~orbit~e}>$ as well (fig(\ref{fig:SKiMfe2})c); the stability point stays near $F_P \sim 1/2$ for low $<f_{trap}>$. This is because the non-adiabatic response is vanishing, and not because the curvature is either slightly positive or slightly negative. 

%Plotted on a log scale, these SKiM results are qualitatively similar to the simulation results plotted on a log scale in fig(\ref{fig:GEOSCAN}). The type of dynamics in the statistical mechanical ansatz qualitatively reproduces the simulation behavior. 

Finally, we can evaluate the coefficients in SKiM for the actual eigenfunctions of the simulations. When we do, SKiM gives good qualitative agreement with the simulation growth rates. \emph{These simulation eigenfunctions all give small $<f_{trap}>$ as we will see in the next section, and frequently $<\omega_{d~orbit~e}>$ is small and positive (destabilizing) }. This, qualitatively, is just like the ITGae in section VIII, where the mode structure adapted to keep the average curvature drive positive; stabilization was due to the FC.  

\begin{figure*}
\subfloat[\label{sfig:7a}]{%
  \includegraphics[width=.5\linewidth]{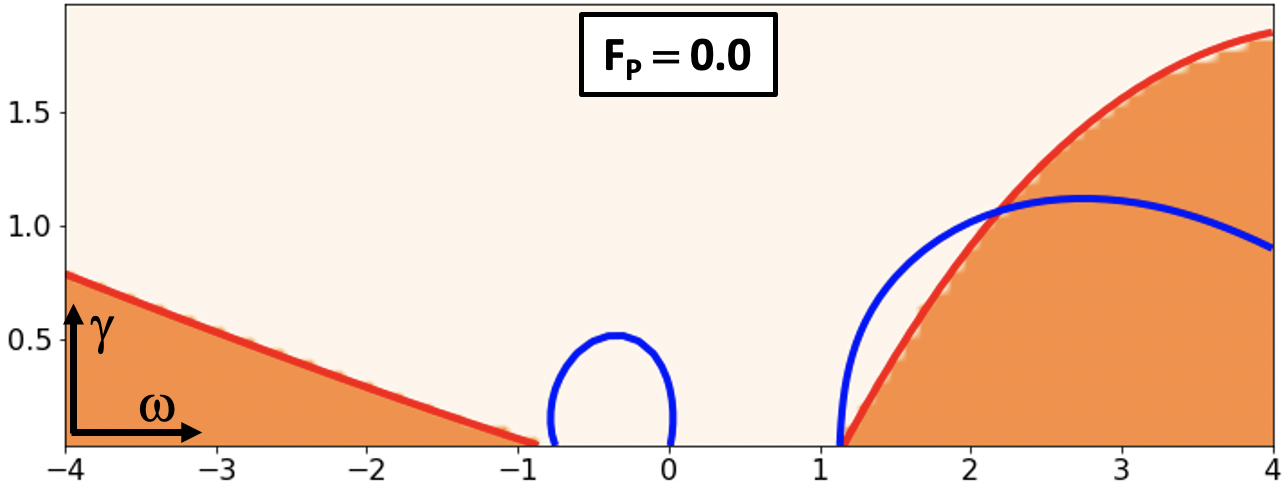}%
}\hfill
\subfloat[\label{sfig:7b}]{%
  \includegraphics[width=.5\linewidth]{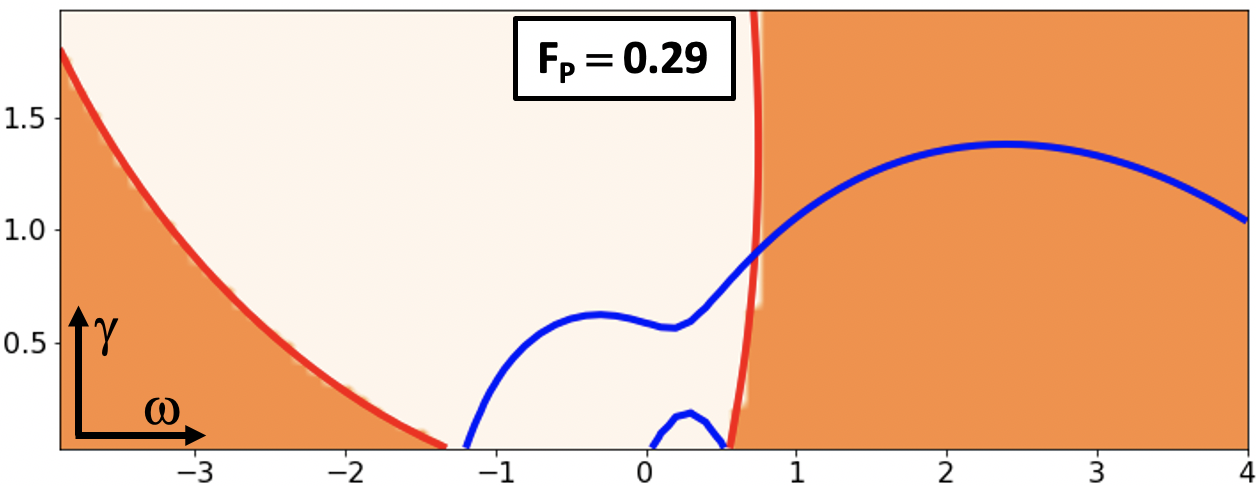}%
}
\vfill
\subfloat[\label{sfig:7c}]{%
  \includegraphics[width=.5\linewidth]{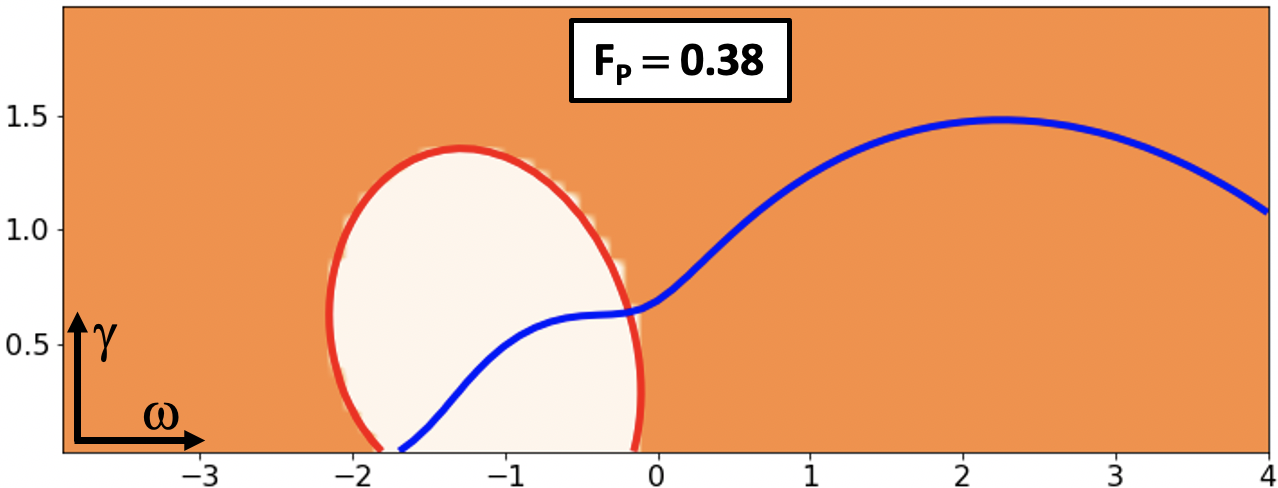}%
}\hfill
\subfloat[\label{sfig:7d}]{%
  \includegraphics[width=.5\linewidth]{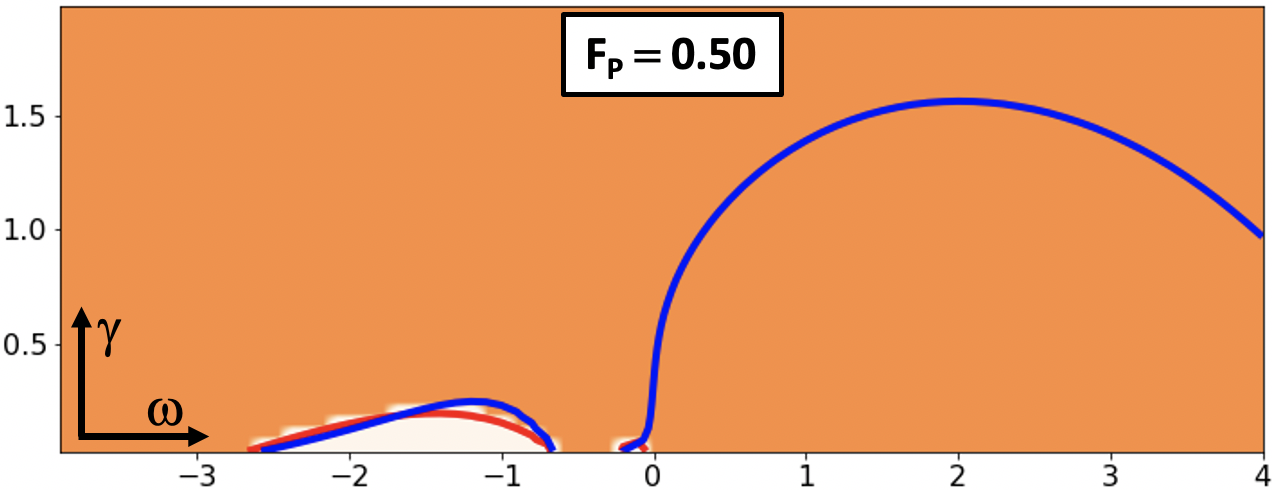}%
}
\caption{\label{fig:BP3} E FC plot for the representative TB scan (ITB case C in fig(\ref{fig:SKiMfe2}d)), with eigenfunction averaged parameters computed from the simulation eigenfunction. As $F_P$ is increased in a)-d), the FC curve shrinks while the free energy curve does not. The stabilization seen in the simulations is clearly from the flux constraint. }
\end{figure*}

Let us now turn to E-FC plots. For typical SKiM cases with parameters averaged over from simulation eigenfunction, the stabilization with $F_P$ follows the $ITG_{ae}$path. The ITB case C is shown in fig(\ref{fig:BP3}). Like the $ITG_{ae}$, stabilization results because the FC curve shrinks to have low $\gamma$ as $F_P$ is increased. And also like the $ITG_{ae}$, the free energy curve alone, however, goes to large $\gamma$. On the other hand, plots of the conventional picture with curvature stabilization and stabilizing electron curvature look exactly the opposite-  fig(\ref{fig:SKiMfe2})c. As we artificially increase the negative electron curvature, the \emph{free energy curve} shrinks to be very close to $\gamma \sim 0$, but the FC can go to high $\gamma$.

These E-FC plots are qualitatively like those for the ITGae, in the respective cases of FC stabilization and energetic (curvature) stabilization. 

This difference in underlying physics would manifest experimentally: if the FC embodied physics controls the underlying physics, density gradients play a dominant role, whereas if energetics control stability, one will need extremely strong curvature and density gradients are not required.

In summary, the results with electron dynamics are qualitatively the same as for the $ITG_{ae}$ in section VIII. \emph{From the statistical mechanical point of view, the addition of non-adiabatic electrons does not introduce any fundamental new element}. 

It is just that the solubility of the FC now depends upon the electron dynamics, so stabilization requires the non-adiabatic electron response to be small. This, in turn, requires that the eigenfunction must adapt to give such a small response, despite the fact that the equlibrium has a large fraction of trapped particles. This adaptation results from the \emph{eigenfunction avoiding the regions of stabilizing curvature, as found for the $ITG_{ae}$.}

\emph{ The same fundamental dynamic operates for both the $ITG_{ae}$ and the ITG/TEM: the eigenfunction adapts to avoid energetic stabilization in the case of stabilizing curvature. But having done so, it  is only the insolubility of the FC that can impose stability. }

\begin{figure*}
\subfloat[\label{sfig:7a}]{%
  \includegraphics[width=.5\linewidth]{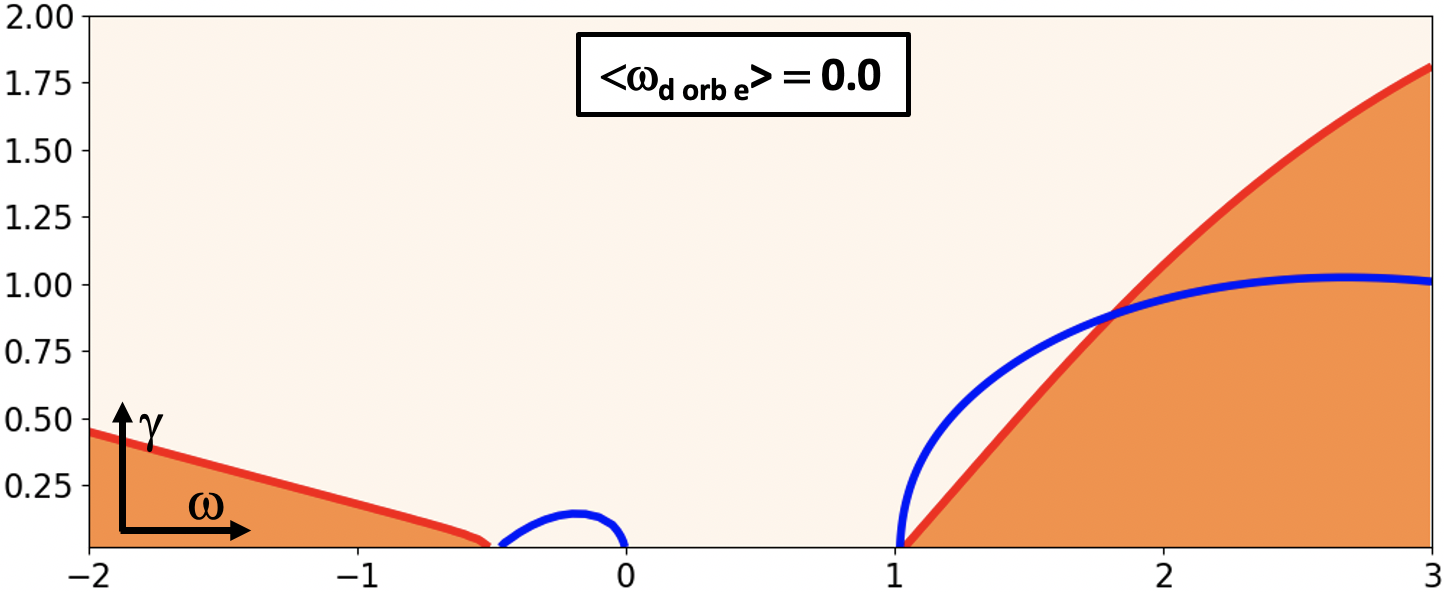}%
}\hfill
\subfloat[\label{sfig:7b}]{%
  \includegraphics[width=.5\linewidth]{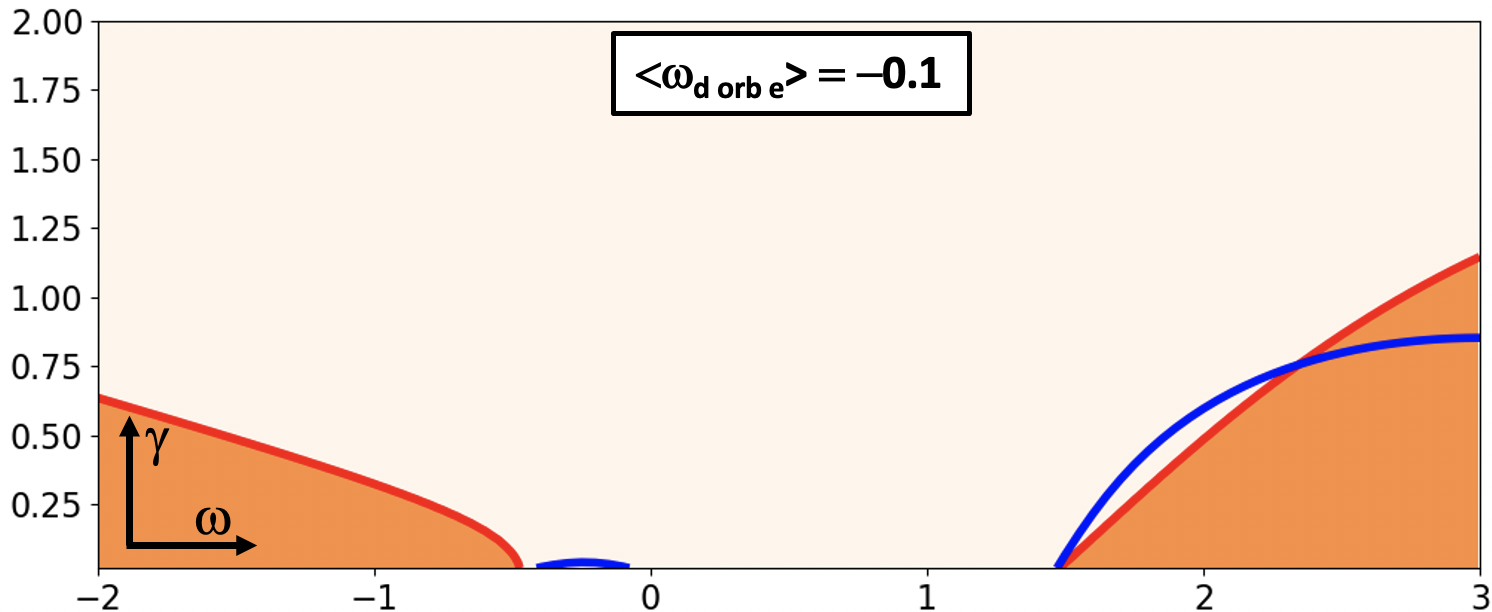}%
}
\vfill
\subfloat[\label{sfig:7c}]{%
  \includegraphics[width=.5\linewidth]{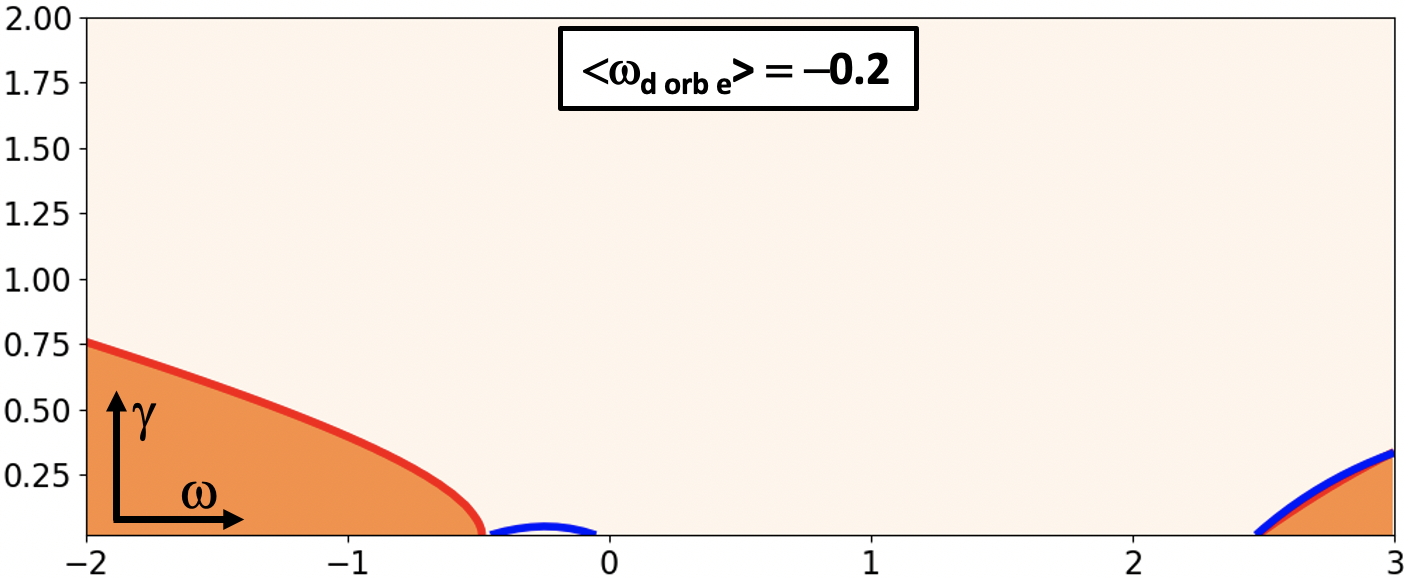}%
}\hfill
\subfloat[\label{sfig:7d}]{%
  \includegraphics[width=.5\linewidth]{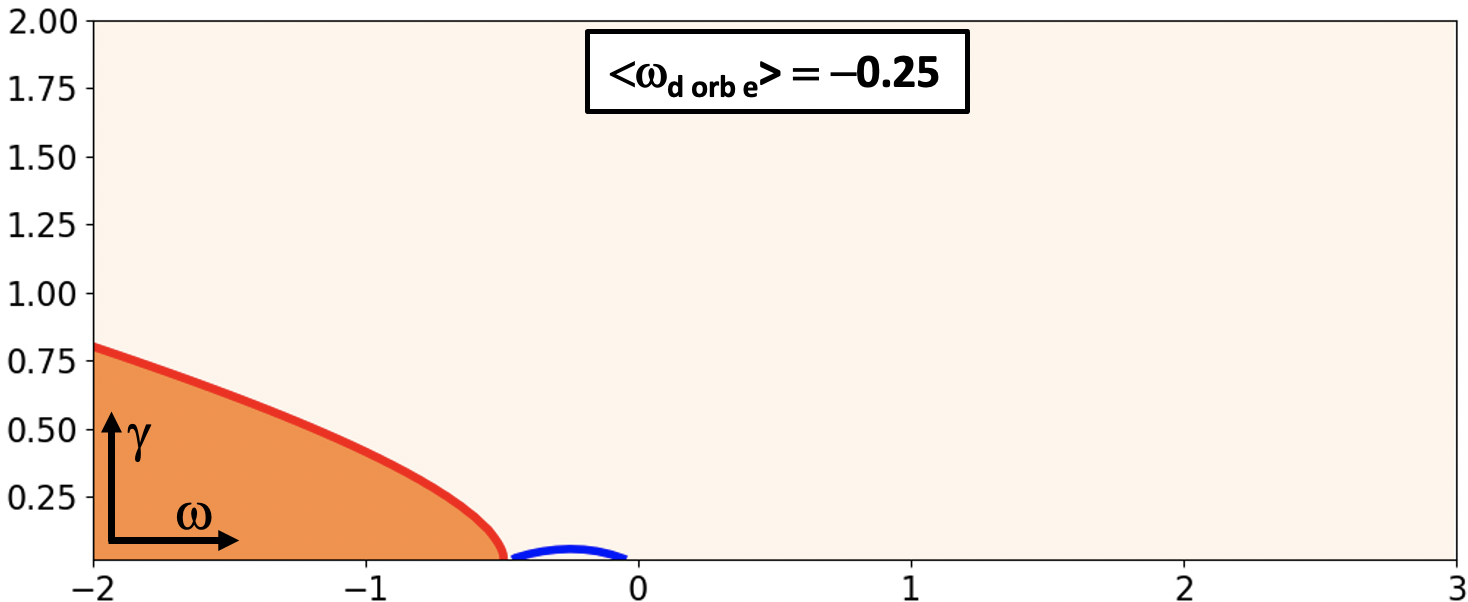}%
}
\caption{\label{fig:BP4} E FC plot for a case like the conventional picture of curvature stabilization (fig(\ref{fig:SKiMfe2}, for constant $F_P=0.5$ . The curvature becomes more negative in a)-d). The free energy curve shrinks till the mode becomes stable. Stability is primarily from the free energy balance forcing $\gamma$ to be near zero. }
\end{figure*}

\subsection{Pure TEM in the framework of the statistical mechanical ansatz }

Studying the behavior of pure TEM (no ion temperature gradients) is theoretically interesting because it is an even more striking (than $ITG_{ae}$ and the ITG/TEM) manifestation of what we have called the the statistical mechanical ansatz; it predicts that there are regimes where the TEM is stabilized by density gradients, and simulations find the predicted regime.

Ideally, any new theoretical understanding should predict some new and unexpected regime. Within the conventional picture, TEM are driven unstable by the free energy in density gradients. So to find a regime where the opposite happens is indeed striking. 

Since the TEM has considerably less free energy drive than the ITG/TEM, it can be stabilized by curvature more easily. But just before that happens, there is a regime where:

\begin{itemize}

\item 	The eigenfunction adapts to decouple from the trapped electrons that sample highly stabilizing curvature 
\item 	Hence the non-adiabatic response is small
\item 	Increasing density gradients, then, will tend to result in a much higher ion charge flux than the balancing   electron charge flux
\item		Consequently, The FC will become insoluble: the TEM will be stabilized by increasing density gradients, \emph{irrespective of energetic considerations}

\end{itemize}

\begin{figure*}
\subfloat[]{%
  \includegraphics[width=.5\linewidth]{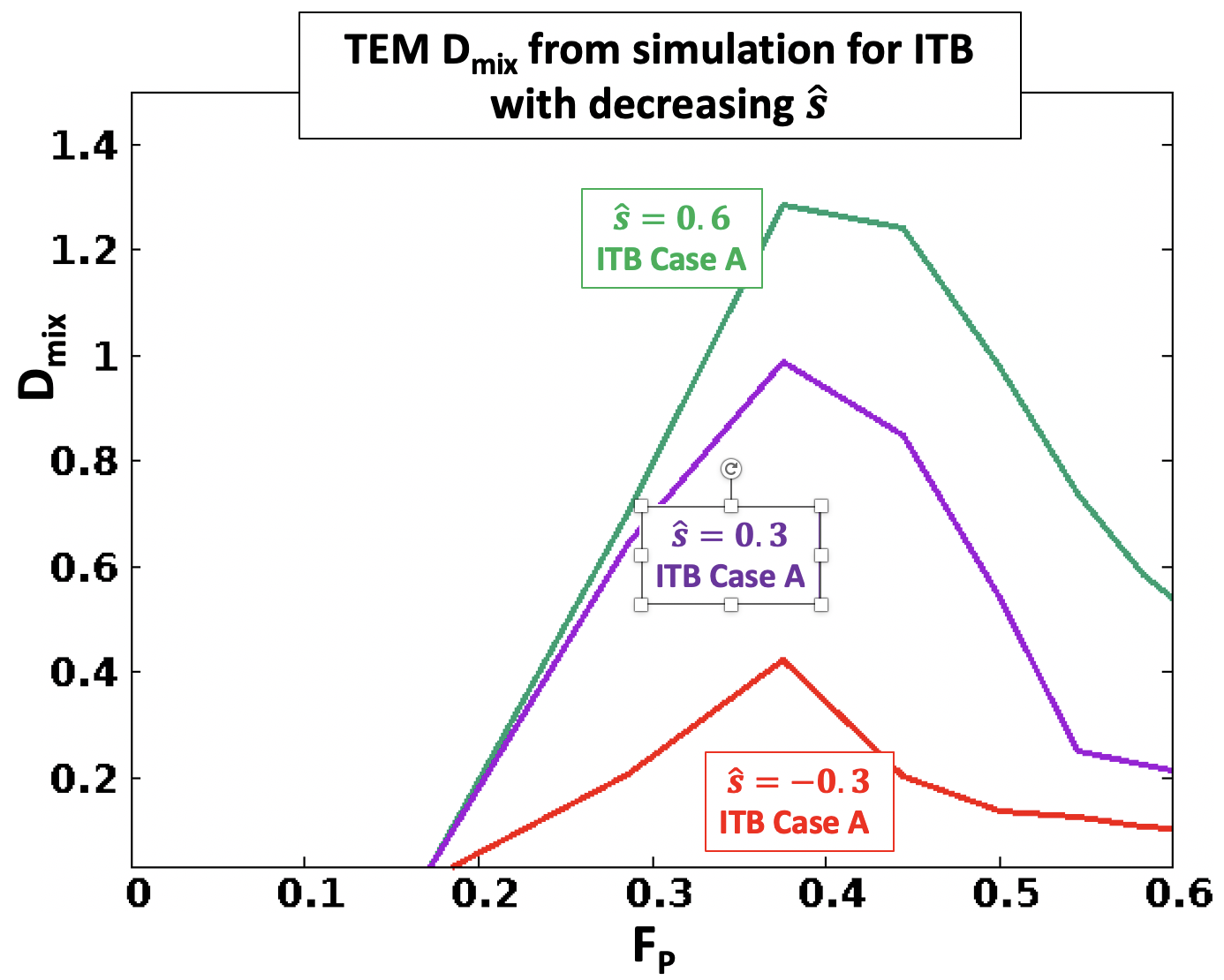}%
}\hfill
\subfloat[]{%
  \includegraphics[width=.5\linewidth]{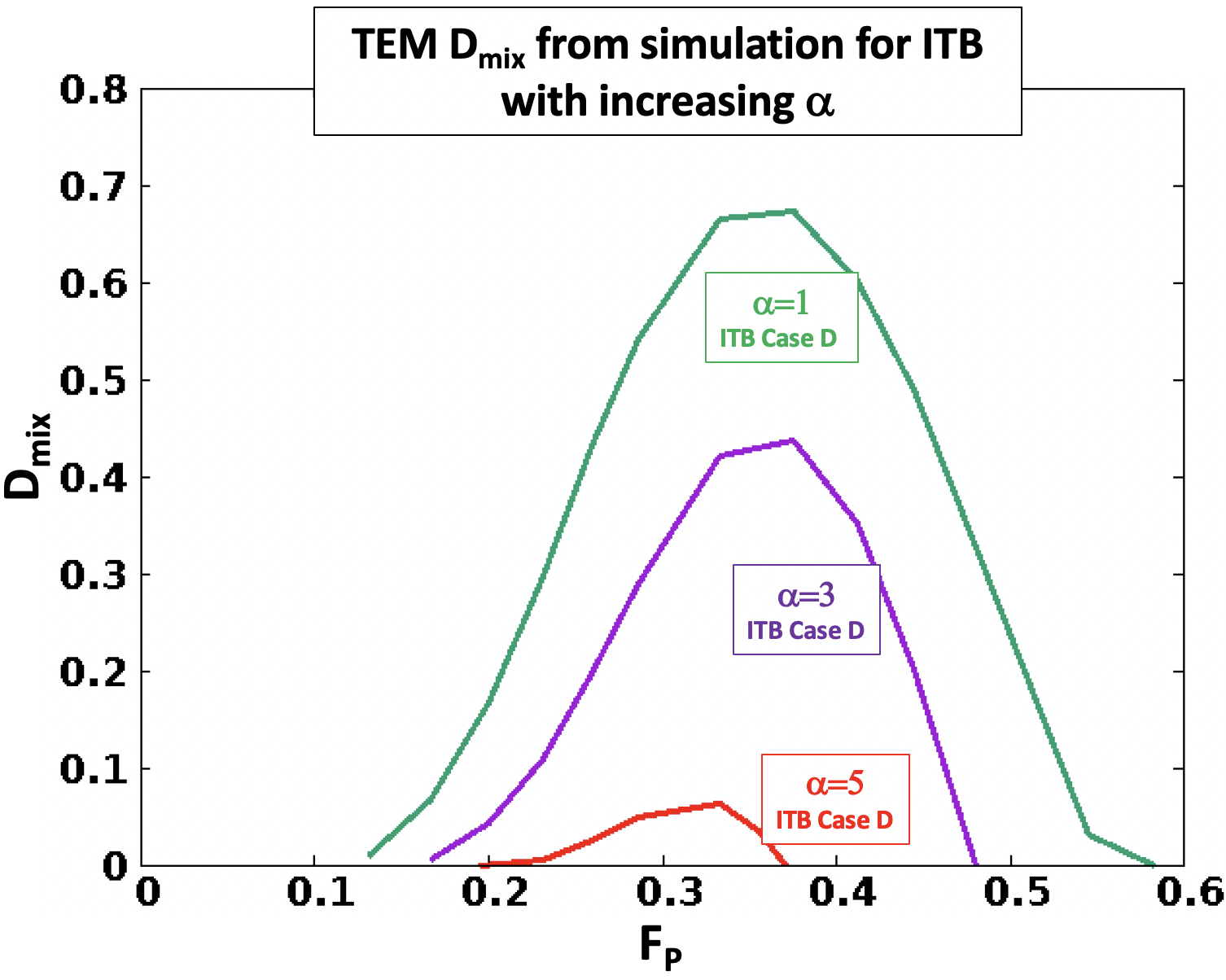}%
}
\vfill
\subfloat[]{%
  \includegraphics[width=.33\linewidth]{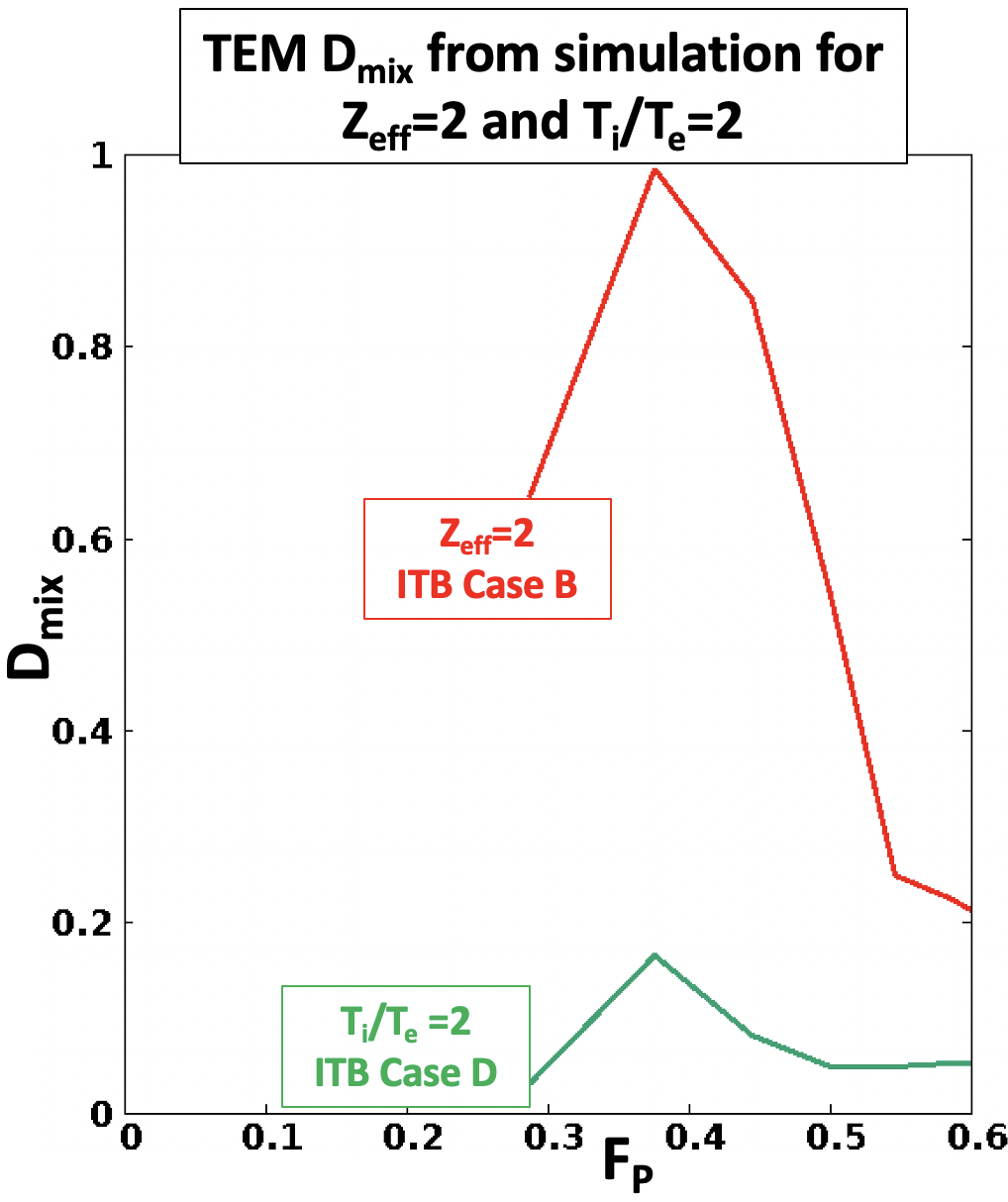}%
}\hfill
\subfloat[]{%
  \includegraphics[width=.33\linewidth]{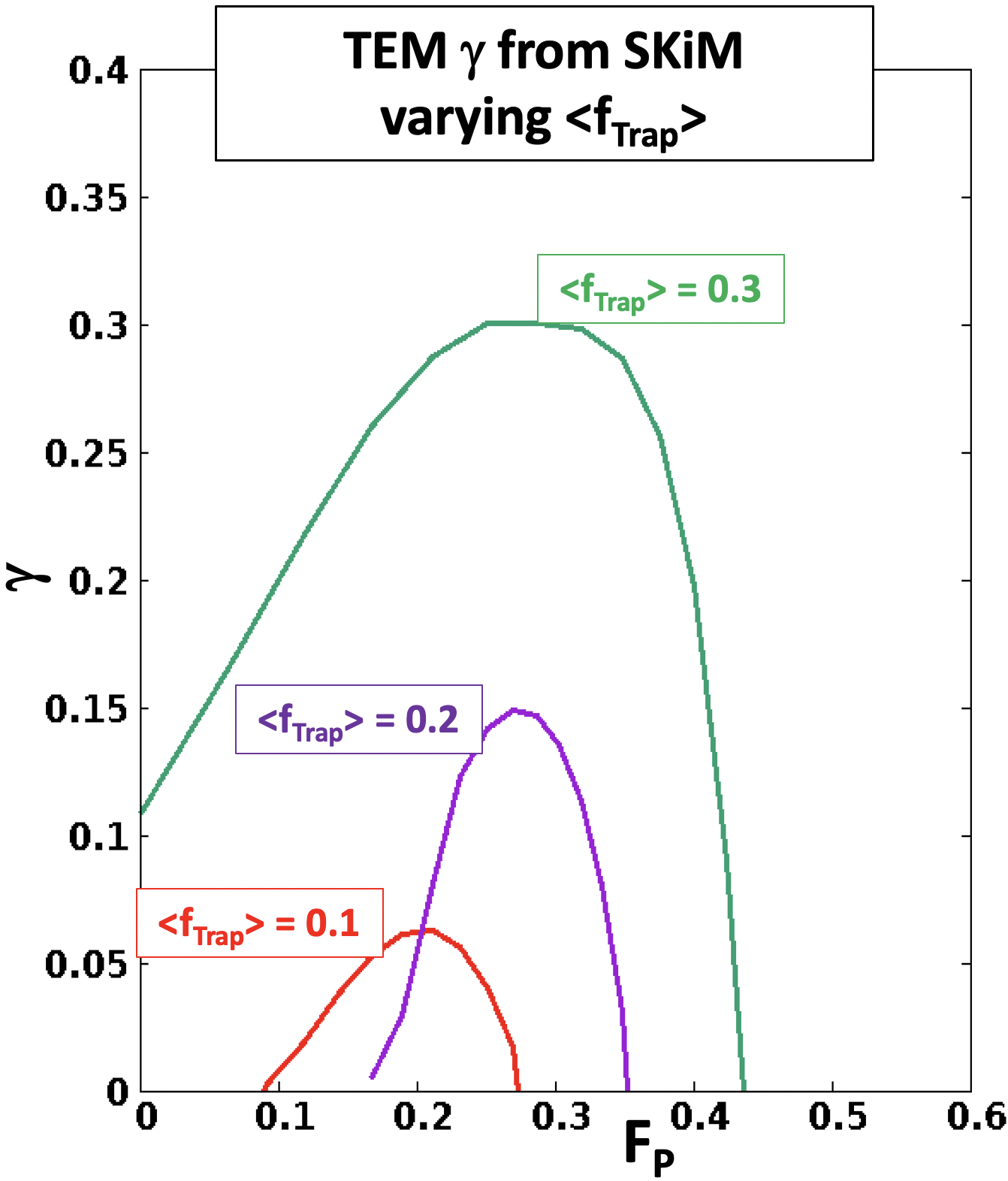}%
}
\hfill
\subfloat[]{%
  \includegraphics[width=.33\linewidth]{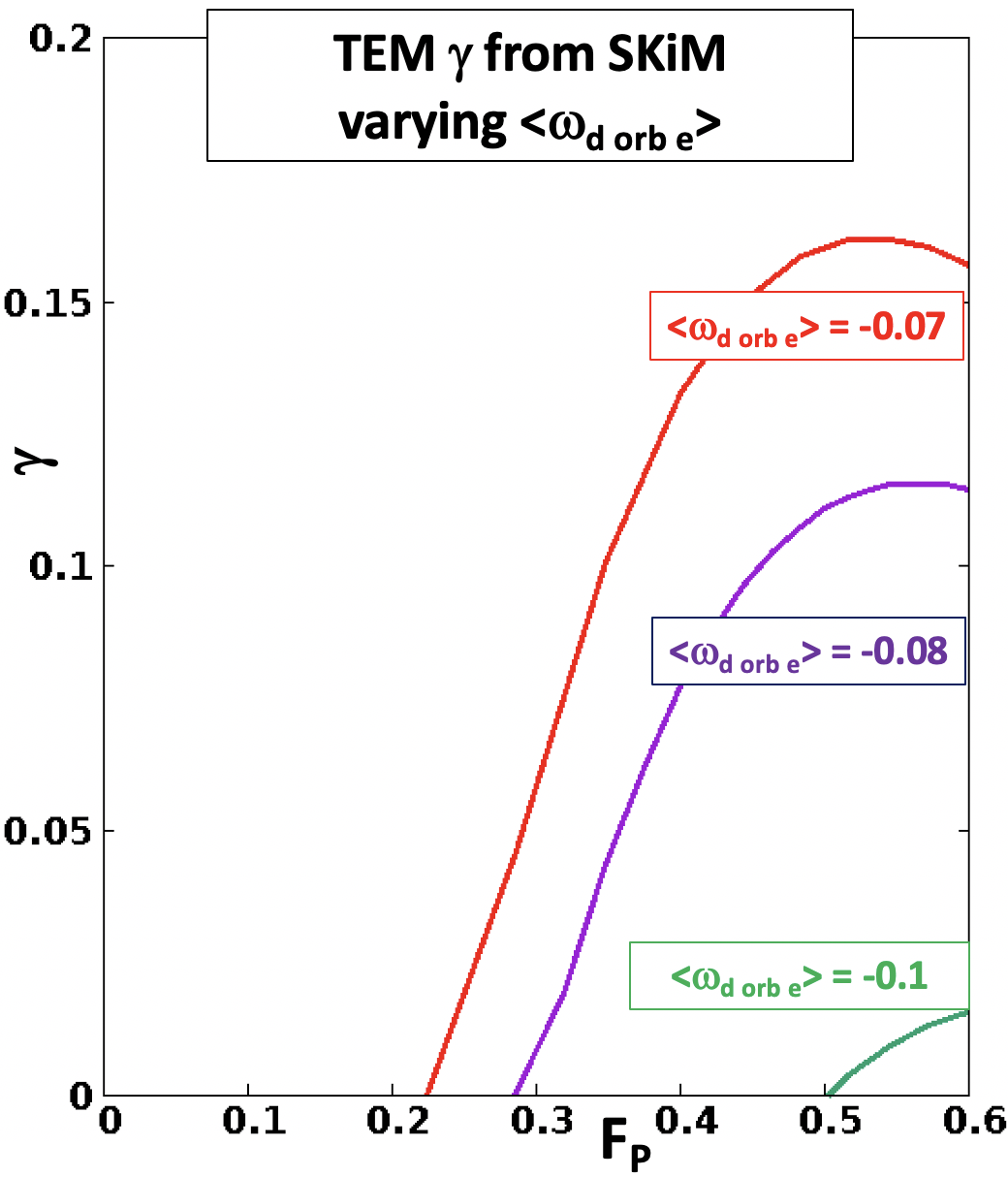}%
}
\vfill
\subfloat[]{%
  \includegraphics[width=.5\linewidth]{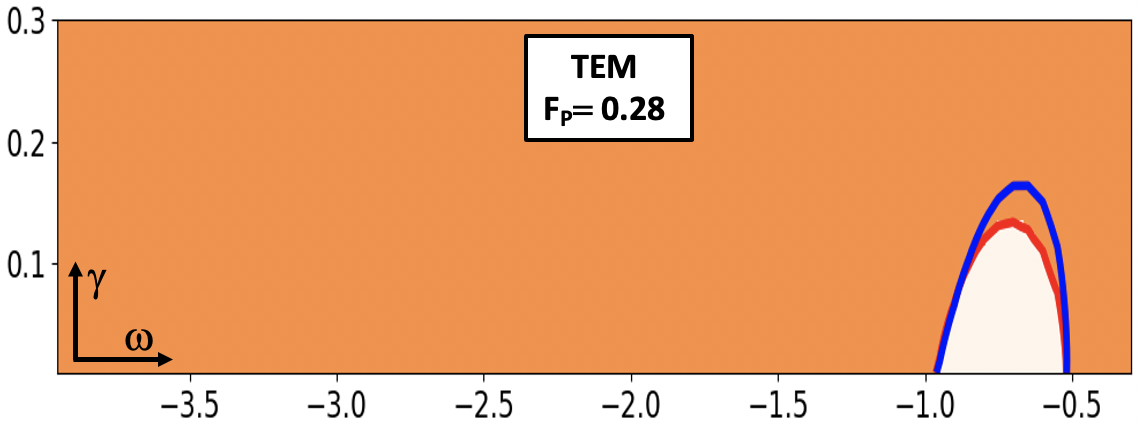}%
}\hfill
\subfloat[]{%
  \includegraphics[width=.5\linewidth]{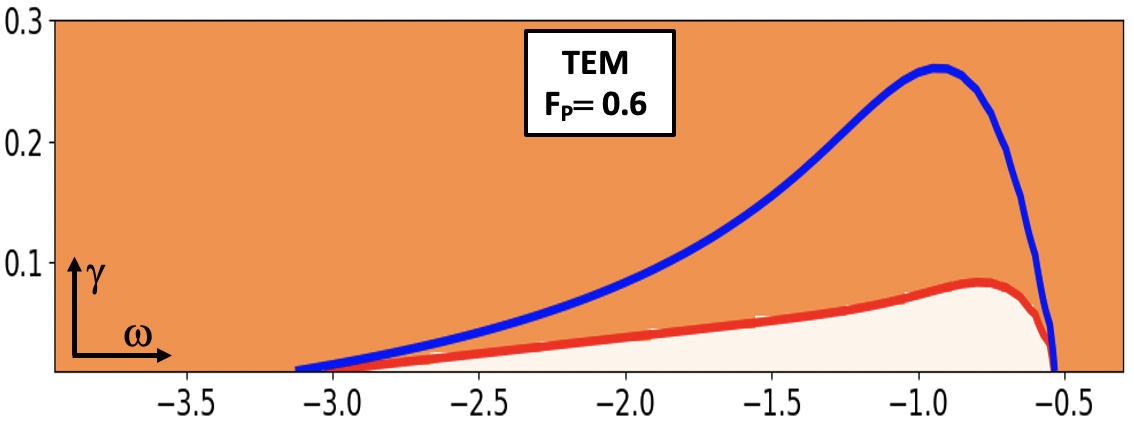}%
}
\caption{\label{fig:TEMSKiM} The regime where density gradients ultimately stabilize the TEM, an unexpected regime predicted by the statistical mechanical ansatz. a) Simulations of ITB case A with varying $\hat{s}$ Increasing density gradients first destabilize, and then stabilize, the TEM, as predicted b) Similar behavior from ITB case D with varying $\alpha$ c) Similar behavior arises with impurities and $T_i/T_e$ not equal to one d) Results from SKiM for the TEM and representative eigenfunction averaged parameters, showing that at small $<f_{trap}>$ this behavior manifests e) Results from SKiM showing that such qualitative behavior does not result from stabilizing electron curvature and large $<f_{trap}>$. Hence, only the statistical mechanical ansatz produces this behavior. Next we elucidate this regime using E-FC plots f) for a case in d at modest density gradient $F_P=0.28$, showing instability g) At large density gradient $F_P=0.6$, the free energy is indeed larger than at $F_P=0.28$: free energy balance extends to higher $\gamma$, and over a larger extent in $\omega_r$ However, the FC shrinks, as predicted by the statistical mechanical ansatz. The shrinkage is sufficient that no intersection between the FC and free energy balance occurs, so there is no instability.}
\end{figure*}

The results of simulations, performed with geometries in fig(\ref{fig:GEOSCANcurvs}) (where the TEM is not yet stable but is becoming weak), are shown in fig(\ref{fig:TEMSKiM}). %{\color{red} Eq. number is not printing}

When density gradients are increased, the TEM at first becomes unstable. But then, with increasing density gradients, it becomes weaker or even stable. This behavior has been found for all the ITB geometries considered above. \emph{This situation is, surely, incomprehensible in previous pictures of the TEM, where only free energy drive is included in the physical interpretation.}

Guided by SKIM, two possible scenarios of TEM stabilization could emerge

 \begin{itemize}

\item 	reduce $<f_{trap}>$ maintaining slightly positive $<\omega_{d~orbit~e}>$
\item 	the conventional picture without adaptation: $<f_{trap}>=0.45$ and $<\omega_{d~orbit~e}>$ is scanned to progressively more stabilizing values

\end{itemize}

Results are shown in fig(\ref{fig:TEMSKiM}). For the first case, SKIM and simulations are alike: the TEM is initially destabilized by increasing density gradients and is, then, stabilized, reflecting the characteristic behavior of FC stabilization. This behavior (frames fig(\ref{fig:TEMSKiM})a-d) occurs in all the TB geometries we have tested (when close enough to stabilization) and persists when plasma parameters are changed: impurities are included or/and when $T_i/T_e > 1$. 

In the second case, the growth rates increase with increasing density gradient. But for sufficiently negative curvature, the mode is stable for all density gradients. This is the characteristic behavior for stabilization by  energetic considerations that we also see in the ITGae, which is supposedly a very different type of mode. But when examined under the lens of the dual equations of the FC and free energy balance, the behavior of the ITGae and TEM is quite similar.

To nail down the origin of stabilization, we examine the E-FC contour plots. First consider the case with low $<f_{trap}>=0.2$. Near the peak growth rate at $F_P =0.28$ the energy curves and FC curves are similar, and intersect near their peak growth rate. However, at $F_P =0.6$ the situation is much different. The energy curve has gotten considerably larger both in maximum height and its horizontal extent. As expected for a TEM, increasing the density gradient has increased the free energy. The FC curve has gotten much lower however, and is limited to low growth rate. The energy and FC curves have pulled apart so much, in fact, that there is no intersection at positive growth rate. \emph{In short, the FC is stabilizing this mode}.

This detailed  study of pure TEM is very edifying- it shows how the mode behavior that is in striking violation of the conventional stability notions becomes perfectly understandable in terms of the  FC based larger statistical mechanical perspective. We must remember that stability considerations must respect the simultaneous dictates of free energy and the FC; the FC could give stability despite very strong free energy drive (from whatever source) just as for the ITGae.

%The problem with such previous pictures is that free energy drive is the only statistical mechanical consideration that is accounted for. But from the statistical mechanical perspective, there is nothing surprising at all in the behavior of this casethat the FC could give stability despite very strong free energy drive. This statement is in no way limited to free energy from ion temperature gradients. \emph{it applies to free energy from any equilibrium gradient, including density and electron temperature}. 

When one species is nearly adiabatic (here the electrons), the fundamental tendency for thermodynamic forces to drive thermodynamic fluxes causes the FC to become insoluble as the density gradient is increased. This occurs in the simulations because the adaptive eigenfunction decoupled from the highly stabilizing regions of phases space, \emph{just like the behavior of the $ITG_{ae}$}

The stability framework developed in this paper is general, and is by no means limited to a given set of modes. The statistical mechanical ansatz applies to all, and with the same consequences for TB equilibrium like the ones in fig(\ref{fig:GEOSCANcurvs}).

% $ITG_{ae}$. \emph{From a statistical mechanical perspective, the TEM is not fundamentally different from the $ITG_{ae}$ or the ITG/TEM.}

\begin{figure*}
\subfloat[\label{sfig:7a}]{%
  \includegraphics[width=.5\linewidth]{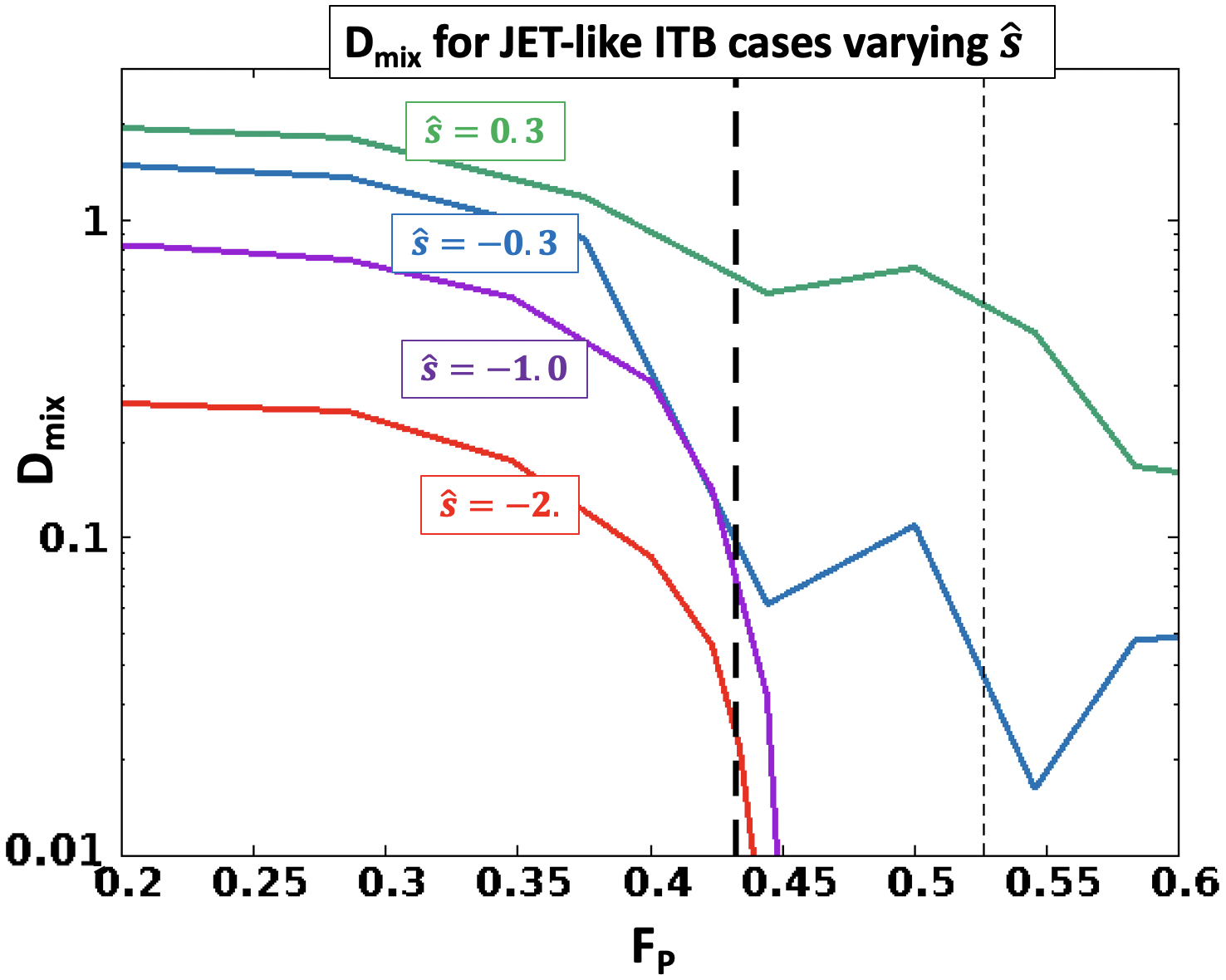}%
}\hfill
\subfloat[\label{sfig:7b}]{%
  \includegraphics[width=.5\linewidth]{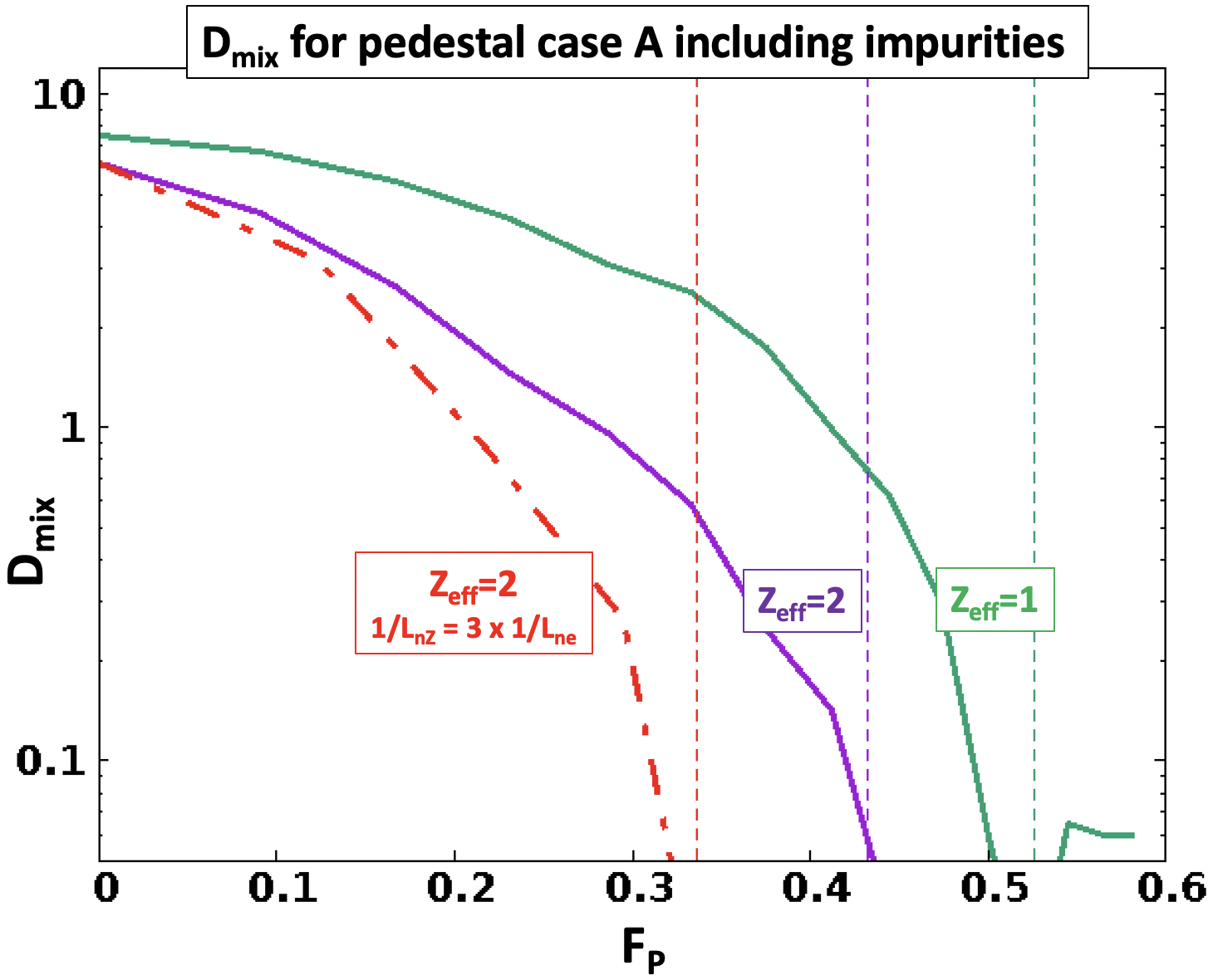}%
}
\vfill
\subfloat[\label{sfig:7c}]{%
  \includegraphics[width=.5\linewidth]{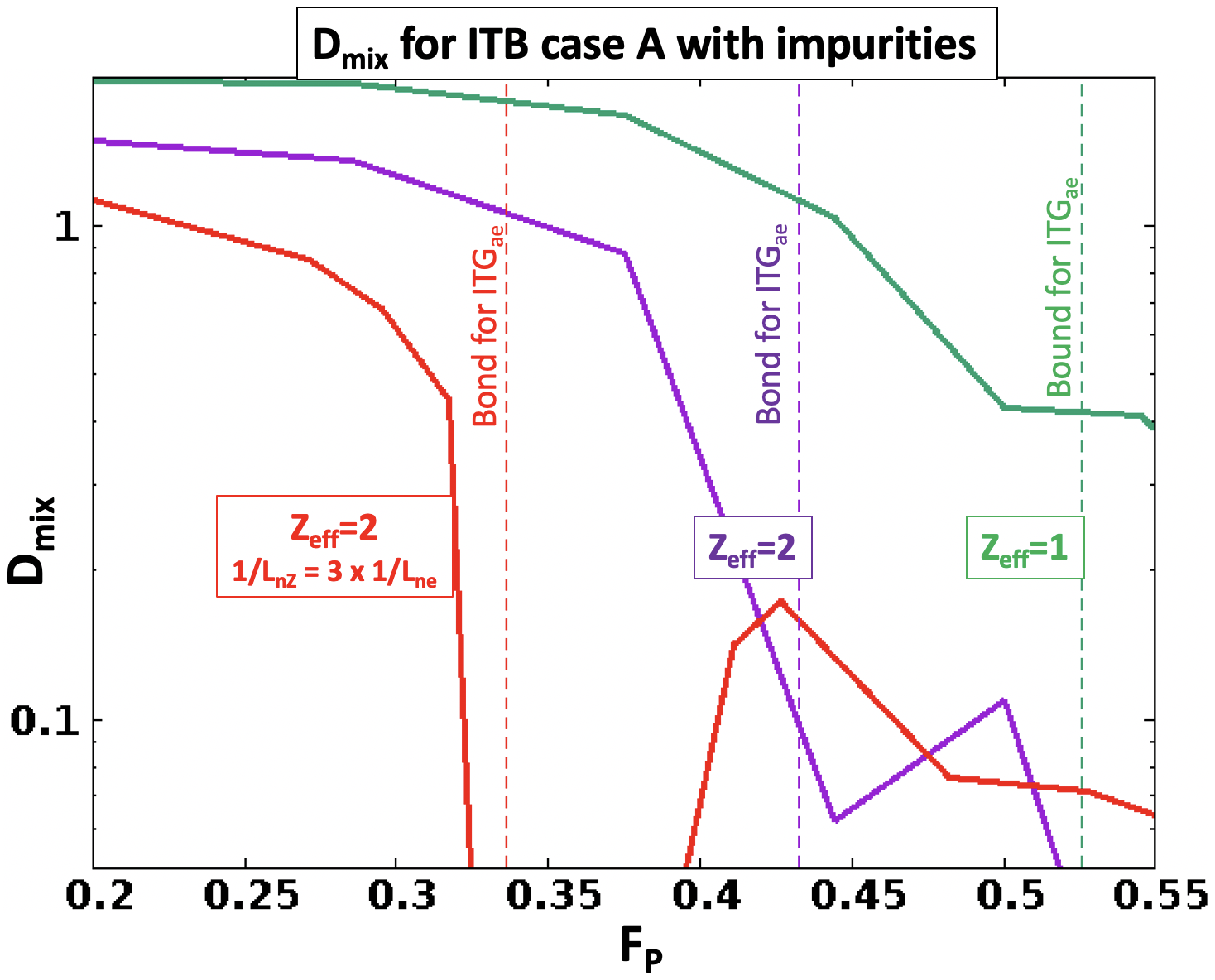}%
}\hfill
\subfloat[\label{sfig:7d}]{%
  \includegraphics[width=.5\linewidth]{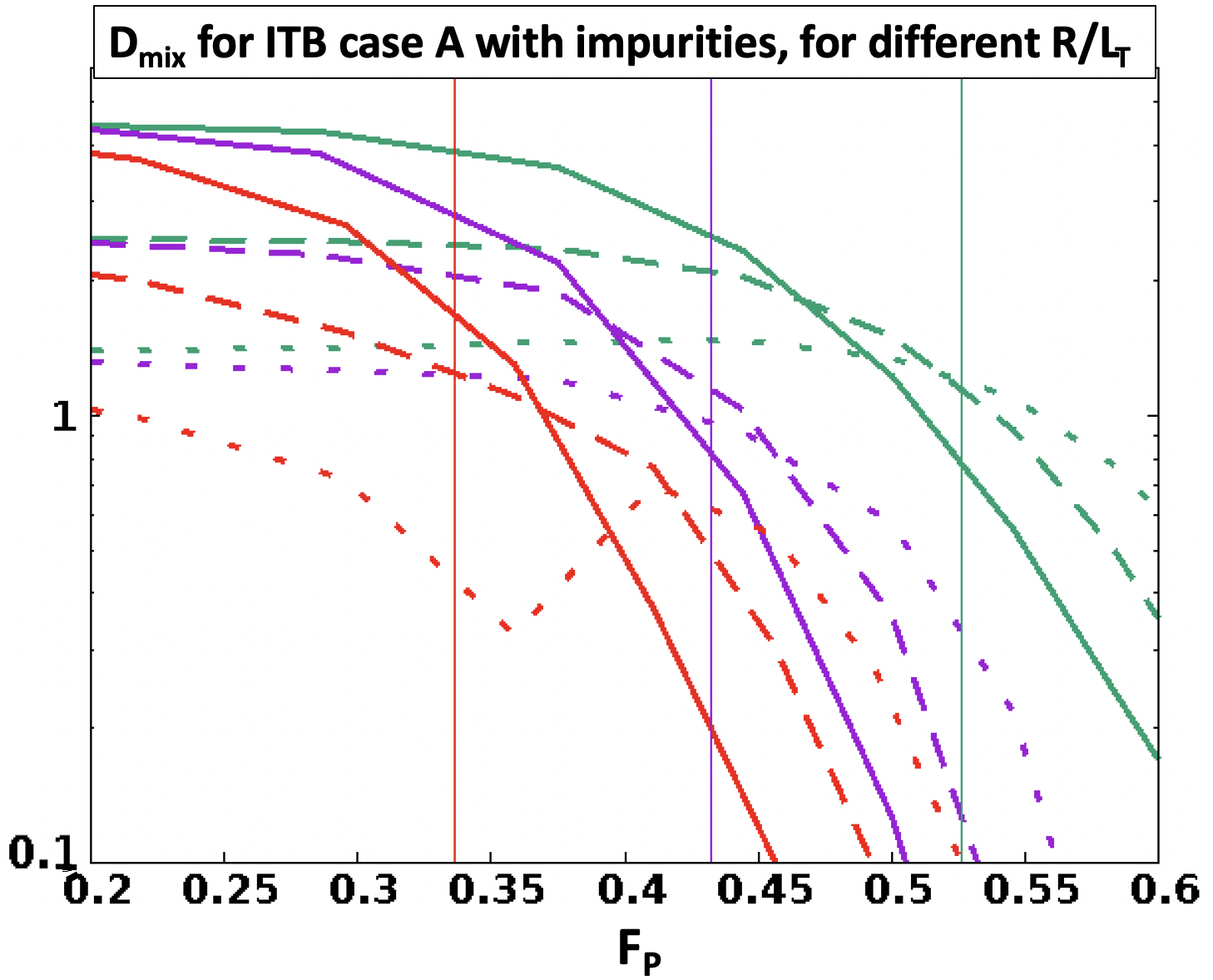}%
}
\caption{\label{fig:feimp} Simulation results including impurities in cases with full electron dynamics and electromagnetic effects a) For $Z_{eff}=2$, as $\hat{s}$ is made more negative, the stabilization occurs at the predicted FC  bound for the $ITG_{ae}$ including impurities. b) A pedestal case, showing stabilization at the analytically predicted FC bound for the respective impurity parameters c) ITB case A, where non-adiabatic electrons are small but large enough to extend the FC solubility limit. Nonetheless, qualitative trends with impurities have similarities to the $ITG_{ae}$ d) This same case varying the temperature gradient. The trends with $F_P$ are similar, verifying that the stabilization in the simulations is from FC insolubility and not from energetic effects. }
\end{figure*}

\begin{figure*}
\subfloat[\label{sfig:7a}]{%
  \includegraphics[width=.5\linewidth]{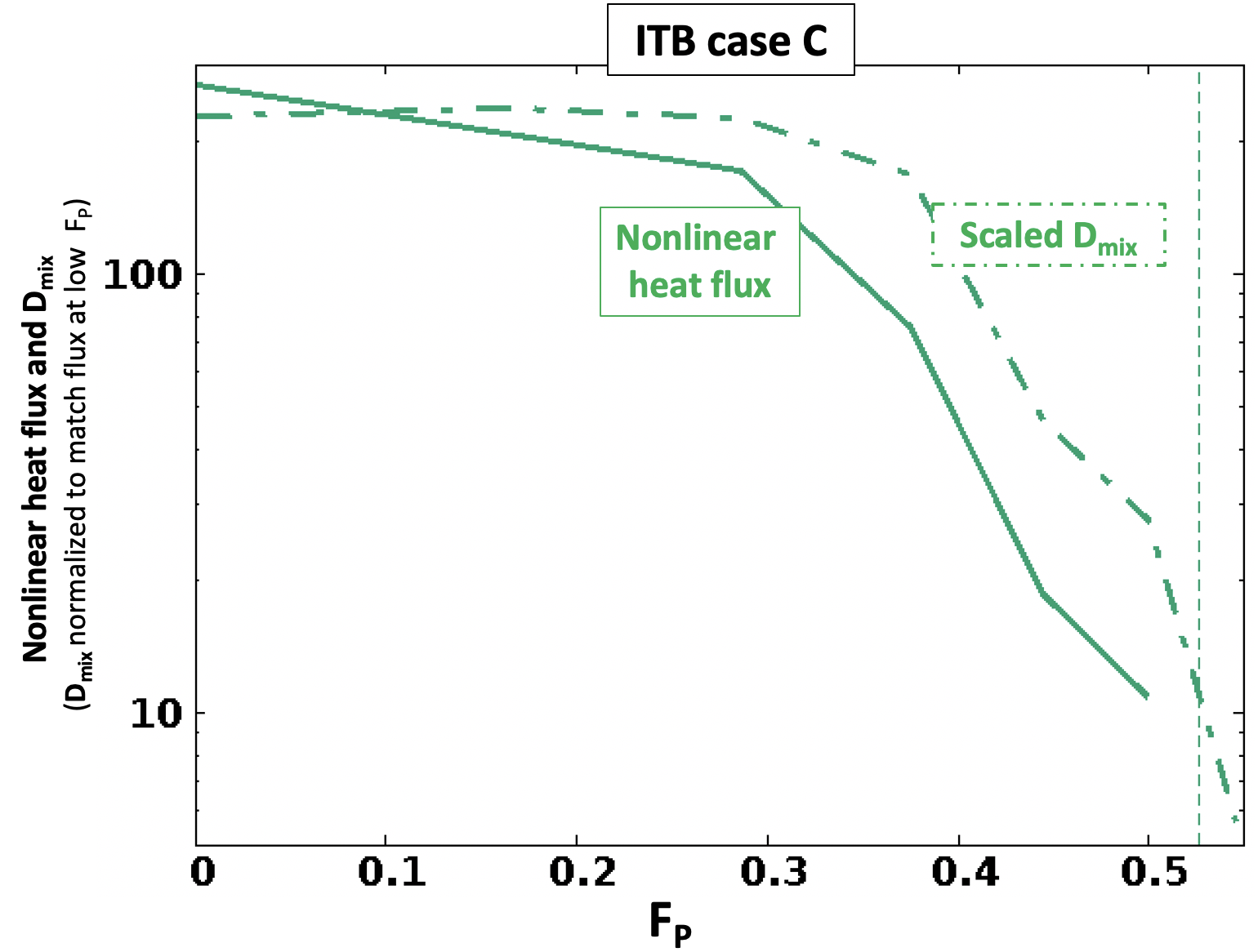}%
}\hfill
\subfloat[\label{sfig:7b}]{%
  \includegraphics[width=.5\linewidth]{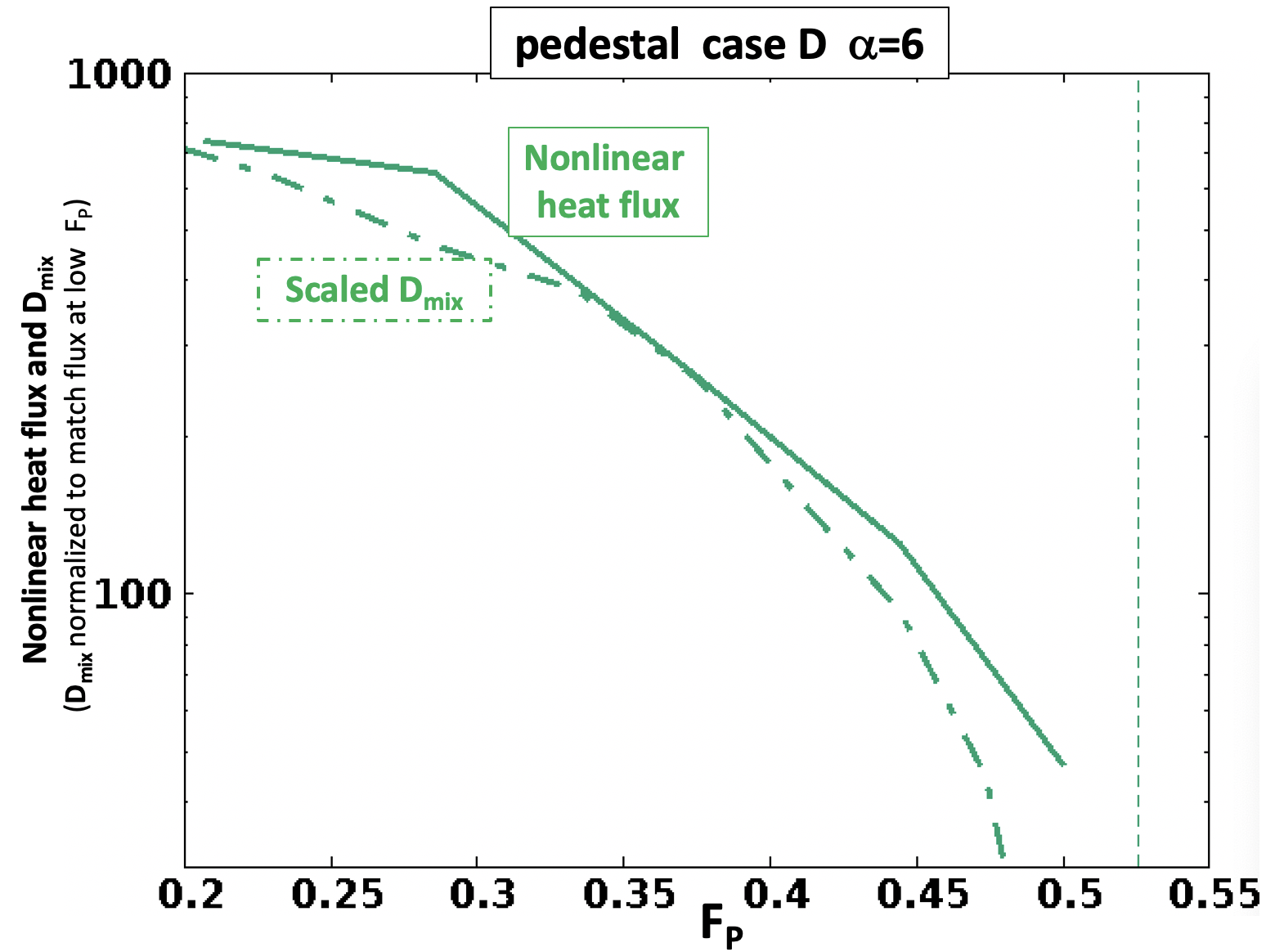}%
}
\vfill
\subfloat[\label{sfig:7c}]{%
  \includegraphics[width=.5\linewidth]{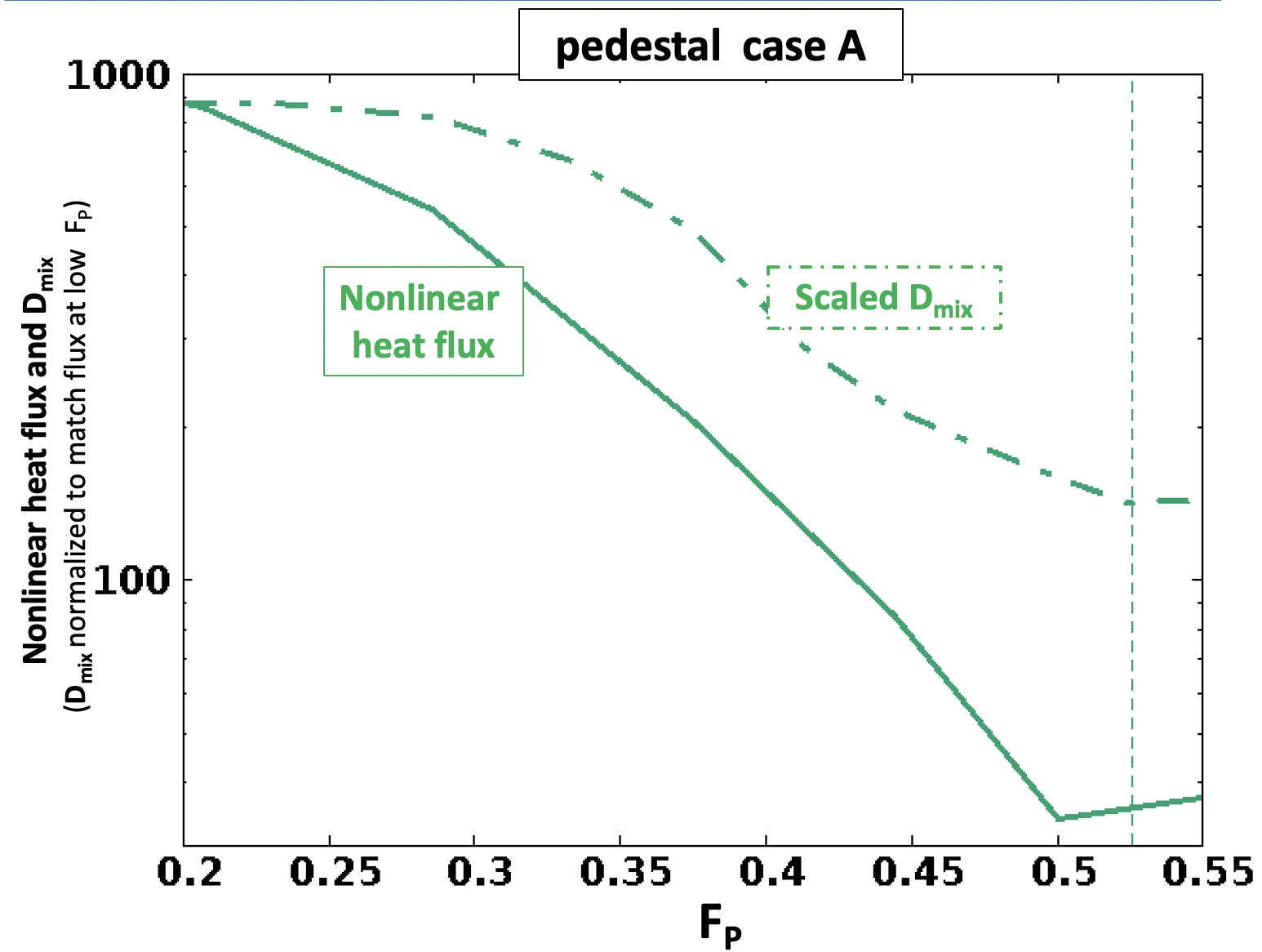}%
}\hfill
\subfloat[\label{sfig:7d}]{%
  \includegraphics[width=.5\linewidth]{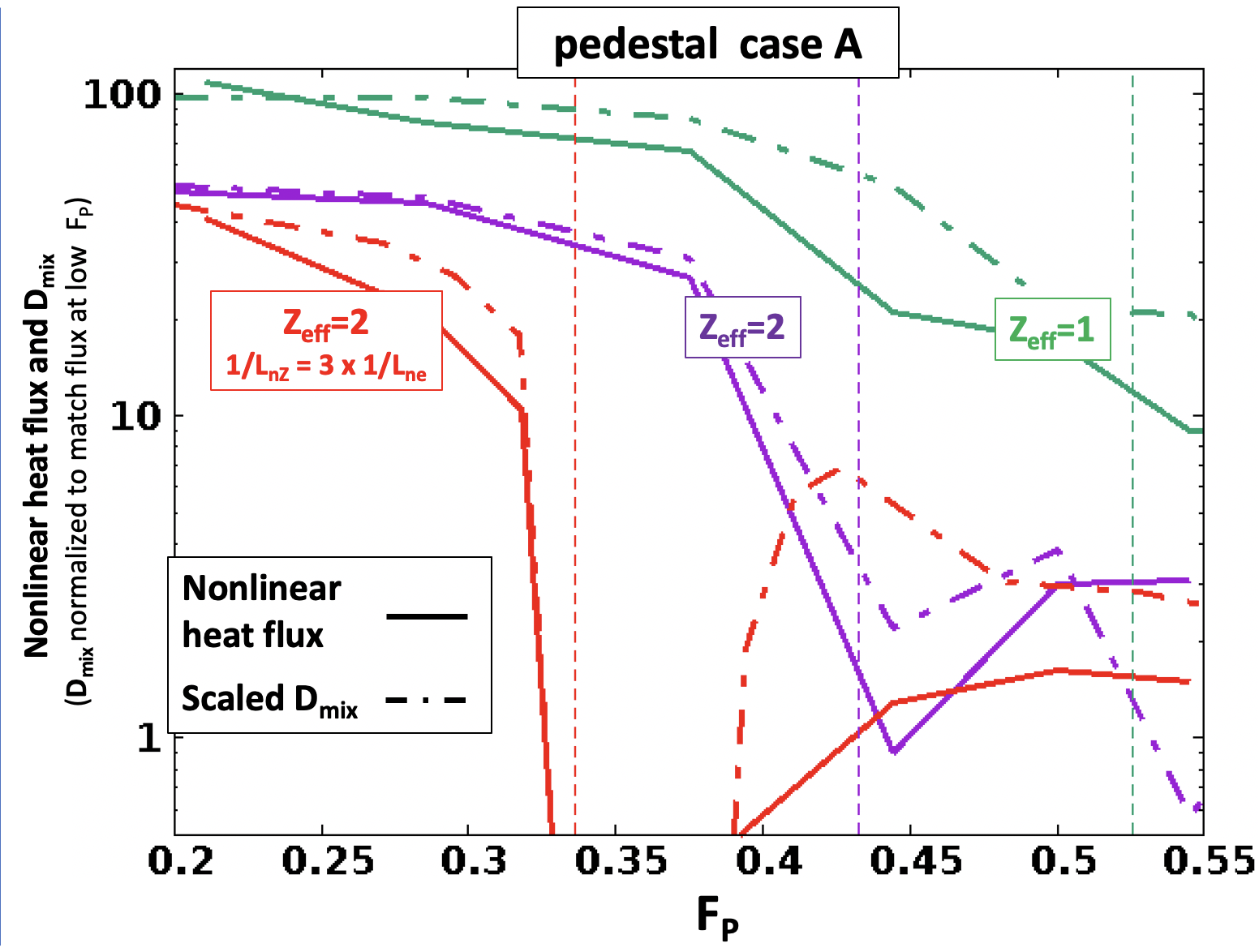}%
}
\caption{\label{fig:fenlDm} Nonlinear simulation results including full electron dynamics and electromagnetic effects compared to $D_{mix}$ . The $D_{mix}$ is scaled to the nonlinear result at low $F_P$ to see if the trends are similar as $F_P$ is increased. First, cases without impurities. When non-adiabatic electrons are weak in a) and b) the trends of of the nonlinear heat flux match the trends in $D_{mix}$ rather closely. c) A pedestal case where non-adiabatic electron effects are non-negligible, the trends are qualitiatively quite similar but not as close quantitiatively. d) An ITB case with non-negligible non-adiabatic electron effects, and where impurities are included. The trends in nonlinear heat flux match the trends in $D_{mix}$  fairly well. }
\end{figure*}

\subsection{Summing up the lessons  of pure TEM analysis}

The results in the TEM section, very strongly,  drive home the point: the flux constraint is true for all fluctuations independent of what they are and what drives them and how strongly. And if the plasma dynamics insures that one species is nearly adiabatic (contributes little charge flux), there will be regimes where the FC insolubility will dictate stability. The resulting stability persists no matter what equilibrium gradients contribute to the free energy.

And, from a practical point of view, the FC stabilized regimes become accessible, for instance, for curvature structures that are not nearly as extreme as those that are required for energetic stabilization.

% from a statistical mechanical perspective, there is no fundamental difference between the $ITG_{ae}$, ITG/TEM and the TEM. The statistical mechanical ansatz applies to all, with qualitatively similar consequences for equilibrium like the ones in fig(\ref{fig:GEOSCANcurvs}). 
% 
% There are regimes where the FC can be driven into insolubility, because one species has a small non-adiabatic response. And this leads to stability, no matter what equilibrium gradients contribute to the free energy. And, this regime occurs for curvature structures that are not nearly as extreme as those that are required for energetic stabilization. 

\section{ Impurities and nonlinear heat fluxes for ITG/TEM}

In earlier sections, we showed that the addition of impurities significantly affected the FC for the $ITG_{ae}$, and  altered the $F_P$ value for stability. In the equilibrium scans above, we found that the eigenfunction adapts to reduce the non-adiabatic electron response. Hence, we expect that the addition of impurities to cases with electron dynamics would alter the $F_P$ value where $D_{mix}$ is strongly reduced, like the $ITG_{ae}$. This is indeed the case.

In the scan of $\hat{s}$ for ITB-like parameters (specifically: fig(\ref{fig:GEOSCAN})a), let us include carbon impurities with $Z_{eff}=2$. We find that as $\hat{s}$ becomes more negative, $D_{mix}$ is, now, strongly reduced at the commensurate $F_P$ of the $ITG_{ae}$ FC solubility limit (fig(\ref{fig:feimp})a). This $F_P$ is distinctly different from the expected bound when $Z_{eff}=1$. In fig(\ref{fig:feimp})b, we consider a pedestal equilibrium with large enough $\alpha$ that the $F_P$ behavior was close to the $ITG_{ae}$ (specifically: pedestal case D, with $\alpha=6$ in  fig(\ref{fig:GEOSCAN})d). With impurities, the $D_{mix}$ is strongly reduced at $F_P$ similar to the solubility limit of the $ITG_{ae}$. 

Now let us consider equilibrium where the mode behavior is not quite as stable as the $ITG_{ae}$, i.e., when the non-adiabatic electron response is not small and plays a significant role. Specifically, consider the ITB case A with $\hat{s}=-0.3$ in fig(\ref{fig:GEOSCAN}a), where the drop in $D_{mix}$ is not as sharp as for adiabatic electrons. What do impurities do in this equilibrium? impurities shift the curves to the right (fig(\ref{fig:feimp}) c) qualitatively like the adiabatic electron case. And it makes the $D_{mix}$ substantially smaller than when there are no impurities, also qualitatively like the $ITG_{ae}$. But the non-adiabatic electrons still allow instabilities up to significantly larger $F_P$ than for the corresponding $ITG_{ae}$. 

Nonetheless, the observed stabilization is very likely still due to the FC. As a test, we increase $R/L_T$ considerably, from value of $15$ in fig(\ref{fig:feimp})c to $R/L_T = 20, 30$ and $50$ in fig(\ref{fig:feimp})d. This increases the $D_{mix}$ by factors of 2-4, but the $F_P$ value for a precipitous drop in $D_{mix}$ changes only by $\sim 10-15\%$. This is characteristic behavior for stabilization due to insolubility of the FC. So despite the fact that the drop in $D_{mix}$ is beyond the solubility limits of the $ITG_{ae}$, the drop is still due to the FC, it is just that the non-adiabatic electrons are playing a significant role in allowing the FC to be more easily soluble ( towards destabilization).

Next, we show that the nonlinear simulations for the general ITG/TEM yield results similar to linear $D_{mix}$ just as for the $ITG_{ae}$. First, in fig(\ref{fig:fenlDm}), consider two equilibria where the $D_{mix}$ showed a strong drop at $F_P$ close to the bound for the $ITG_{ae}$. Figure(\ref{fig:fenlDm})a is an ITB case and fig(\ref{fig:fenlDm})b is a pedestal case. As we see, the total electrostatic nonlinear heat flux (the sum of electrons plus ions) behaves similarly to the $D_{mix}$. 

In these figures, \emph{the simulations were electromagnetic, but we have plotted only the electrostatic component of the heat flux.} In some of these cases, electromagnetic modes (probably MTM) were present which produced a small heat flux. Such modes have not been part of our theoretical consideration, and obviously any $F_P$ bound for such modes is not expected to be related to the $ITG_{ae}$ bound. Such modes, and their role in TB physics, will be considered in future. 

Now, let us consider nonlinear results for configurations where the $F_P$ stabilization was not quite as strong as for  $ITG_{ae}$. For the pedestal case A ( fig(\ref{fig:GEOSCAN})d with $\alpha=6$ ), the nonlinear behavior is again qualitatively similar to $D_{mix}$, but the drop in the nonlinear flux is even faster ( fig(\ref{fig:fenlDm})c). Finally, the ITB case A, also did not show stabilization as strongly as the $ITG_{ae}$. With impurities included, the drop in the total electrostatic nonlinear heat flux follows $D_{mix}$, but the drop is more rapid (especially for the case with strong impurity gradients).

A common  feature of all these cases is that the nonlinear flux drops faster than the estimate $D_{mix}$. Hence, the strong reductions in the $D_{mix}$ imply even stronger reductions in the nonlinear heat flux (most relevant to experimental confinement). 

\section {Summary-Conclusions}

We will now conclude our somewhat long enquiry into the physics of highly thermodynamic states like the transport barriers that are, by definition, endowed with strong gradients. States of high confinement are possible only when the turbulent transport originating in plasma instabilities can be minimized. Energetically, a plasma with high gradients
should be violently unstable due to enormous amount of free energy that can feed instabilities. And yet, transport barriers (both internal and edge) are routinely created in almost all tokamaks. 

For a long time, the existence of (at least) the edge barriers was attributed to velocity shear that controlled the mode 
( ITG/TEM) that was expected to cause the most turbulent transport. Since many current experiments do manage to create them in configurations that have too little velocity shear, there must be something else in the plasma dynamics that lets these highly non-thermodynamic states to emerge and stay.

Unearthing that ``something'', different from velocity shear, was the primary driver of this enquiry. In a detailed and comprehensive study, using analytical as well as computational tools, we set out to find how states with strong gradients (therefore large free energy) manage to avoid feeding fluctuations that, in turn, could destroy them (by wiping out the gradients). 

Before now, this situation has been understood as the properties of specific instabilities, the ITG and TEM, each with their unique stability properties. And when these unique and complex instabilities are coupled together, fortuitously, they happen to have a stability domain arising from their unique properties, that experiments have accessed. The accomplishment of this work is that this situation can be viewed very differently, and in the process, understood more clearly. Both these instabilities (and likely many others) are subject to basic dynamical relations in the gyrokinetic system:  the restraining influence of the FC, and the destabilizing influence of the free energy balance equation. And both the ITG and TEM are fluctuations of a dynamical system with very many degrees of freedom, which is expected to behave adaptively to tap free energy wherever possible-based upon experience with many other physical systems. And so, when viewed through the lens of the dual dynamical equations, the $ITG_{ae}$, the ITG/TEM and the TEM behave essentially the same: their stabilization can be understood via what we have called the statistical mechanical ansatz.  And because their stabilization usually derives ultimately from the FC, the conditions needed for stability can be understood by its basic character: since thermodynamic forces drive their corresponding thermodynamic fluxes, when one species is close to adiabatic, the FC becomes insoluble at high enough density gradient. And so stabilization results. Let us review our deliberations in more detail.

The most important new concept that we have developed and exploited in this paper originates in a  characteristic property of fluctuations in a gyrokinetic system: the system will sustain only those classes of fluctuations for which the charge weighted (radial) particle flux vanishes. The flux constraint (FC), though known in the literature, was never fully harnessed; we believe that it is the first time that its enormous stabilizing potential has been marshaled to develop what could qualify to be a ``theory'' of fluctuations in a transport barrier.  It is based upon fundamental physical principles, i.e., those that are general to many systems in physics.

We were, thus, forced to investigate the constrained (rather than conventional) free-energy dynamics. Fluctuation behavior ( stability) was to be determined as a consequence of these two very different dynamics operating simultaneously: the FC and the conventional free-energy. Naturally we have concentrated on investigating the stability of the canonical ITG/TEM- we have studied it extensively: in its various limiting cases, using different geometries, shaping, and plasma parameters (all with strong gradients ). We were able to distinguish the working of these two different dynamics, their separate roles, and their very different characteristics. In particular, we were able to identify what aspects of our extensive simulation results could be attributed to what dynamics. 
 
In the process, we encountered an extremely interesting aspect of linear gyrokinetic instabilities - their adaptability;
the corresponding eigenmodes change their structure so as to remain maximally unstable when subjected to stabilizing conditions (like negative curvature) that will reduce free energy. In principle, that would prevent the realization of such states unless some other mechanism were to drive the system back to stability. Adaptability, 
as emphasized in the text, is observed in very many systems with a large number of degrees of freedom.

the gyrokinetic system manifests apparent adaptive behavior, and "finds a way" to dissipate free energy until a "hard" dynamical constrain prevents it- here, the FC.

Understanding adaptability is, theoretically, rather interesting.  Understanding adaptability in a gyrokinetic system, however, is an absolute must for buildings a comprehensive theory of transport barriers. We have, therefore, attached an entire section (appendix C) to show its manifestation in a variety of magnetic geometries and plasma conditions for the principal mode invoked in this paper, the ITG/TEM. It is adaptability that renders the standard free- energy based stability analysis even more incapable of explaining the why of experimentally observed transport barriers (particularly those without shear).

Because the constraint pertains only to particle flux, the density gradients get, automatically, valorized in the theory. 
The simplest manifestation of the FC violation is seen in the suppression/elimination of fluctuations when density gradients exceed a certain threshold. Of course, the FC violation takes place when the electron flux cannot balance the combined flux of ion and impurities. By investigating the ITG/TEM dynamics in a large variety of situations ( different geometries, different plasma parameters )we were able to demonstrate the universal working of the constraint; there is a large enough $F_P$ (the normalized density gradient fraction) beyond which the mode is either fully stable or stable enough that the turbulent transport is drastically lowered.

We note that the same physics applies to stellarators, although that has not been the focus of this paper aside from the brief demonstration shown in Fig.~\ref{fig:one}.  Notably, recent results on W7X have demonstrated the key role of density gradients in enabling high ion thermal confinement~\cite{stellarator3,stellarator4}. %(cite these two papers:  High-performance plasmas after pellet injections in Wendelstein 7-X   (DOI 10.1088/1741-4326/ab7867), Turbulence Mechanisms of Enhanced Performance Stellarator Plasmas (https://doi.org/10.1103/PhysRevLett.125.075001)).  
In stellarators there are additional tactics (beyond those employed in tokamaks) available to realize the transport barrier geometry.  We expect that the framework outlined in this paper may find fruitful applications to optimizing stellarators for turbulent transport.
 
We will not repeat here all the ``results'' discussed in detail in the text. We will, however, list the essence of our findings:
\begin{itemize}

\item It is the constrained free energy dynamics- the conventional free energy considerations subjected to the zero flux constraint - that determines the gross stability of gyrokinetic fluctuations.
\item In most cases relevant to transport barriers ( high gradients), the flux constraint is the crucial and more efficient stabilizer-  it succeeds in reducing turbulent transport, for instance, in much less demanding geometries than would be required by the free-energy route. 
\item The violation of the flux constraint at some level of $F_P$ seems to be possible in substantial regions of plasma -parameter space suggesting an equally large  parameter space in which transport barriers are possible 
\item The importance of the flux constraint in suppressing fluctuations is even more impressive in the context of their adaptability that lets them counter stabilizing influences, and remain maximally unstable.  Thus reducing free energy, say by extreme geometries, may not work as efficiently as intended.
\item The flux constraint is violated whenever the negative Flux (= charge weighted flux) of electrons cannot balance the positive Flux of ions plus impurities. So if the electron response to fluctuations is more adiabatic than of electrons,
then it is relatively easier to violate FC.
\item Since the Flux is charge weighted, impurities, by virtue of their higher charge can play an outsize role in creating conditions hospitable for transport barriers. They can be the deciding factor even when the electron and ion Fluxes were to balance; this provides us a powerful knob to induce transport barriers.
\item The ultimate knob for inducing stabilization by the FC, however, is the density gradient.  % - Although in some cases smaller density gradients could be destabilizing ( after all they are also a source of free energy), they eventually (after crossing some threshold)emerge as the stabilizing force due to the imposition of FC. 

\end{itemize}
The reader can find in the text a very large survey of the parameter space through detailed simulations. The aim was to chart out the existence and access route to territories where transport barriers ( without velocity shear) could reside.  In order to be able to scan parameters easily to elucidate the essential physics, we conducted these simulations in simplified Miller-like geometries and controlled scans were highly salutory for this purpose. The large number of variations which confirm the theoretical framework can be taken as showing that it has a substantial domain of validity.  The next obvious step is to carry out this combination of analysis- simulations for reconstructed equilibria of experimentally observed transport barriers. It is likely, then, that we would be well equipped to find/perfect recipes to engineer these high confinement configurations in current and future machines.

\section{Appendix A}

%For the interested reader, the argument goes thus. Even without collisions, and for an exponentially growing fluctuation, Eq(\ref{eq:two}) still implies that free energy is transferred from the equilibrium scales to the fluctuation scales. Hence, the free energy of the equilibrium scales decreases (which is to say, the entropy of those scales increases.) So the behavior of $G_{ij}$ is like the classic case from statistical mechanics. Furthermore, although the external $\phi$ does work on system, this can be included in the analysis, and it does not change the result that the response matrix $Q_{ij}$ must be positive definite, so the diagonal elements are positive. So the gyrokinetic response matrix, defined for an arbitrary fluctuation, has the same properties as the classic Onsager matrix, and for essentially similar reasons. 

For the interested reader, the argument goes thus. Even without collisions, and for an exponentially growing fluctuation, Eq(\ref{eq:two}) still implies that free energy is transferred from the equilibrium scales to the fluctuation scales. Since the free energy of the fluctuation scales is increasing (exponentially), the free energy of the equilibrium scales decreases (which is to say, the entropy of those scales increases.) So the behavior of $G_{ij}$ is like the classic case from statistical mechanics: the matrix must be positive definite, so the diagonal components are positive. Furthermore, although the external $\phi$ does work on system, this can be included in the analysis, and it does not change the result that the response matrix $Q_{ij}$ must be positive definite, so the diagonal elements are positive. So the gyrokinetic response matrix, defined for an arbitrary fluctuation, has the same properties as the classic Onsager matrix, and for essentially similar reasons.

 \section{Appendix B :A more advanced version of SKiM}
 
Here we explore why so simple a model as SKIM is so remarkably accurate, and why it can predict what simulations find in a far more complicated geometry. We will also derive a somewhat more complete FC relation that has the same solubility bounds. 

The FC bound in SKiM follows from the fact that the perturbed distribution function can be written as a resonant denominator multiplying the driving term in the gyrokinetic system that arises from equilibrium gradients. The latter is the part due to $[\omega_{ns}^{\star} + \omega_{Ts}^{\star}(mv^2/2T -3/2)]\sim [1/L_n -3/(2 L_T)+ (v/v_{thermal})^2/L_T)]$. The resonant denominator gives a weighting, in velocity space, of the plasma response to the gradient driving term. 

For the flux to be zero, the resonance must weight both the positive and negative regions of the driving term equally. This becomes very challenging as $F_P \rightarrow 0.6$, since then, the constant in the driving term $1/L_n -3/(2L_T)$ become close to zero and slightly negative. Hence, over most of velocity space, the driving term $ [1/L_n -3/(2 L_T)+ (v/v_{thermal})^2/L_T)]$ is positive.  To solve the FC, the resonant plasma response must have a high concentration it the small region $v/v_{thermal} \sim 0$. 

And to attain such high levels of concentration as $F_P \rightarrow 0.6$, the resonance must be progressively more sharp, so the growth rate $\gamma \rightarrow 0$.

The requisite concentration can happen by some combination of parallel resonances $\sim v_{\parallel}k_{\parallel}$ and perpendicular resonances $\sim \omega_d$.

In the case of a more realistic geometry than SKiM, the plasma response would be different in detail, \emph{but as long as the plasma response could still concentrate near $(v/v_{thermal} \sim 0$ the maximum value of $F_P$ would be the same}. Hence, SKiM would correctly predict the solubility bound, despite having a different resonance structure. 

Let us further consider the case of the FC bound for low $k_y$ perturbations where $\omega_d$ is small compared to $k_{\parallel} v_{\parallel}$. In the resonant denominator, the plasma response is concentrated in a region where $(v_{\parallel} \sim 0$.  At low $k_{\perp} \rho_i$, this is \emph{no} tendency to concentrate in the perpendicular direction. Hence the FC bond is considerably lower than in the case above: $F_P= 1/3$. However, if $k_{\perp} \rho_i$ becomes larger, the Bessel function $J_0$ reduces the response at high $k_{\perp} \rho_i$. This is another way of saying that the response becomes \emph{relatively} more concentrated it at low $v_{\perp}$. This allows $F_P$ to increase beyond $F_P= 1/3$. But the Bessel function does not concentrate as strongly as a resonance might. So the the maximum $F_P$ is less the $0.6$, and the SKiM calculations above finds it is about $0.53$.

Therefore, for a more accurate model than SKiM, as long as  the plasma response is due to parallel ion motion could concentrate near $(v_{\parallel} \sim 0$, and as long as Bessel function effects reduced the response in the perpendicular direction, the maximum value of $F_P$ would be similar to $SKiM$. Naturally the plasma response could be quite different in detail.    

To derive an improved version of SKiM in the collisionless limit, one can formally solve the 1D gyrokinetic equation in ballooning coordinates for the non-adiabatic distribution function $h$ by integrating along particle orbits. 

\begin{equation}
h(l,v)= \int_{-\infty}^{0} \mathrm{d}\tau \phi_J[l'(l,v,\tau)] e^{-i\int_{\tau}^{0} \mathrm{d}\tau\{\omega-\omega_d[l'(l,v,\tau)]\}}
\label{eq:hexact}
\end{equation}

where $\phi_J=\phi J_0(k_{\perp} \rho_i)$, and $l'(l,v,\tau)$ is the particle trajectory that begins at $l$ at $\tau=0$ and satisfies $ \mathrm{d}l'/ \mathrm{d}\tau = v_{\parallel}$. From this, the perpendicular flux is

\begin{equation}
Im (\int (\mathrm{d}l /B)  \mathrm{d}v \phi_J^{\star} \int_{-\infty}^{0} \mathrm{d}\tau \phi_J[l'(l,v,\tau)] e^{-i\int_{\tau}^{0} \mathrm{d}\tau\{\omega-\omega_d[l'(l,v,\tau)]\}} )
\label{intFC}
\end{equation}

This expression is too complicated to give useful results. It can be greatly simplified if we take $k_{\perp}$, $\omega_d$ and $v_{\parallel}$ to be constant. Introducing the Fourier transform in l, $\phi(l)=\int \phi_{k_{\parallel}} e^{ik_{\parallel}l}$, we obtain for the FC:

\begin{eqnarray}
&&\int \mathrm{d}k_{\parallel} |\phi_{k_{\parallel}}|^2 \int d\bm{v} f_M J_0(<k_{\perp}>\rho_i)^2  \nonumber\\
&& \big[ \frac {\gamma \big[ 1/L_n+(v^2-3/2)/L_T\big] } {\gamma^2+(\omega_r-<k_{\parallel}>v_{\parallel}-<\omega_d>)^2}\big]  = 0 \nonumber \\ 
\label{eq:FCkpar}
\end{eqnarray}

This is just the SKiM FC Eq({\ref{eq:FC0}) for a given $k_{\parallel}$  integrated over the values in the spectrum, weighted by $|\phi_{k_{\parallel}}|^2$. In SKiM, there is only a single average $k_{\parallel}$, which is a spectrum average (the SKiM expression is equivalent to $<k_{\parallel}>^2 \sim \int k_{\parallel}^2 |\phi_{k_{\parallel}}|^2$).  Equation(\ref{eq:FCkpar})} is more realistic, in that it has a full spectrum of $k_{\parallel}$ contributions, weighted by the Fourier amplitudes $|\phi_{k_{\parallel}}|^2$.  \emph{Crucially, the FC solubility bounds for Eq(\ref{eq:FCkpar}) are identically the same as SKiM.} (When the SKiM FC is insoluble, it means the flux is positive definite for all $k_{\parallel}$. Then the integral of such positive definite contributions weighted by the positive quantity $|\phi_{k_{\parallel}}|^2$ is also positive definite, and the converse.) This also applies to the FC bounds for the low $k_y$ case as well.

Once again, the same philosophy of obtaining a solubility bound applies to Eq({\ref{eq:FC0}): if $F_P$ is large enough, then any choice of $k_{\perp}$, $\omega_d$ and $|\phi_{k_{\parallel}}|^2$ will make the FC insoluble because the flux is positive definite. 

For any choice of $\phi(l)$, a reasonable estimate for the average $k_{\perp}$ and $\omega_d$ are the SKiM values, that is average in $l$ weighted by $|\phi(l)|^2$. 

Even without the approximations leading to Eq(\ref{eq:FCkpar})}, the basic properties of the plasma response will give a similar result. For example, if the response is concentrated to low $v_{\parallel}$, and  $k_{\perp} \rho_i$ is small, the FC bound will be $\sim 1/3$, and it will increase due to the $J_0$ as $k_{\perp} \rho_i$ increases. In other words, even the exact response will have similar properties to SKiM insofar as the solubility of the FC. The actual response will include numerous complex effects such as trapped ion orbits, etc. This could have a major effect on a dispersion relation based upon Eq(\ref{eq:hexact}). But if is not so surprising that the maximum bound on solubility will be similar to that from SKiM. 

 \section{Appendix C - definitions of the electron terms in SKiM, and their interpretation}

Now we define the parameters $<\omega_{d~orbit~e}>$ and $<f_{trap}>$ used in the figures above. These can be derived in various simplifying limits, but they are sufficiently intuitive that we simply present them here. In SKiM, the parameters for trapped particles are:

\begin{equation}
<\omega_{d~orbit~e}> = \frac {\int \mathrm{d} l^\prime \,  \int \mathrm{d} \Omega_v \, <\phi>_{orbit}^2 <\omega_{d,e}>_{orbit}} 
{\int \mathrm{d}l \,  \int \mathrm{d} \Omega_v \, <\phi>_{orbit}^2}
\label{omde}
\end{equation}

\begin{equation}
\langle f_{trap} \rangle_{eigen} = \frac{\int \mathrm{d}l^\prime d\Omega_v \langle \phi 
\rangle^2_{orbit}}{\int dl d\Omega_v \phi^2}
\label{ftrap}
\end{equation}

Where $<>_{orbit}$ is the orbit average or "bounce average", i.e., the time average over a bounce orbit. (For mathematical simplicity we take $<>_{orbit}$ to vanish outside the trapped region, to make eq\eqref{ftrap} and eq\eqref{omde} more transparent to write.) The $ \mathrm{d}l^\prime \mathrm{d}\Omega_v $ is a weighting over phase space, where $ \mathrm{d}l^\prime$ is the configuration space weighting, and the relevant velocity coordinate for trapped particles is the velocity pitch angle, with a weight proportional to the solid angle,  $\mathrm{d}\Omega_v $.

The average trapped electron curvature $<\omega_{d orbit e}>$ is evidently the orbit averaged curvature $<\omega_{d,e}>_{orbit}$ weighted by the magnitude of the orbit averaged response of that orbit $<\phi>_{orbit}$. It is the obvious generalization of the ion expression $<\omega_{di}>=\int  \mathrm{d}l^\prime |\phi|^2 \omega_d / \int  \mathrm{d}l^\prime |\phi|^2$ to a trapped species, where all quantities are orbit averaged. 

The effective trapped particle fraction $f_{trap}$ averages the magnitude of the response $\sim <\phi>_{orbit}^2$ over phase space and compares it to $\phi^2$. If the bounce average response were as large as possible $<\phi>_{orbit}^2 = \phi^2$, this expression would be the actual physical fraction of trapped particles. But because the orbit averaged response can be considerably less than this, the "effective" trapped fraction $f_{trap}$ for a particular eigenfunction can be considerably less than the actual fraction.

Heuristically, the magnitude of the non-adiabatic electron response from trapped particles is $\sim f_{trap}$, whereas $<\omega_{d orbit e}>$ gives whether the curvature acting on that response is  energetically stabilizing or destabilizing (destabilizing is positive). 

First, the eigenfunction averaged electron curvature will be considerably more destabilizing than the curvature plot would indicate. And second, since many of the orbits have destabilizing curvature, and the eigenfunction averaged trapped fraction will be considerably smaller than the actual fraction of trapped particles. 

How can eigenfunction structure affect the electron quantities $<\omega_{d~orbit~e}>$ and $f_{trap}$? Consider typical eigenfunctions and trapped electron orbits In fig(\ref{fig:eigtrap}). In fig(\ref{fig:eigtrap})a, we have a typical example from the simulations that concentrates in the bad curvature region. Because the curvature is preponderantly stabilizing, the width of the eigenfunction is narrow. Hence, it couples only weakly to typical trapped electron orbits. The orbit average of $\phi$ is, then, small because orbits sample mostly the small $\phi$ region. An additional factor that reduces $<\phi>$ is that $\phi$ is positive near $\theta \sim 0$, but is slightly negative for the rest of the range. This negative region subtracts from the positive and further reduces the magnitude of $<\phi>$ for typical trapped orbits. On the other hand, the small orbits near $\theta \sim 0$ couple quite strongly to the eigenfunction: for them the $<\phi>$ is relatively large. Such orbits have bad orbit averaged curvature. Typical orbits have stabilizing curvature. But due to the different coupling of typical orbits and small orbits, the small orbits are disproportionately weighted. Hence in eq(\ref{omde}) the average over all obits, weighted by the coupling $<\phi>^2$, is destabilizing. 

In summary, the eigenfunction shape preferentially decouples from the stabilizing orbits, but maintains strong coupling to the destabilizing orbits. Also, because many orbits have a weak coupling, the effective trapped fraction $<f_{trap}$ is significantly reduced from the actual physical fraction of trapped particles. This corresponds to many of the cases in fig(\ref{fig:evol1}-\ref{fig:evol2}). 

Very different eigenfunctions arise after there is a mode jump in fig(\ref{fig:evol1}-\ref{fig:evol2}). A representative eigenfunction is in fig(\ref{fig:eigtrap})b. In such cases, the effective trapped fraction is considerably smaller than the case above. And, the orbit average curvature is quite stabilizing, but the ion average curvature is quite destabilizing to make up for this. These cases also have a larger $k_{\perp} rho_i$, which helps to help satisfy the FC when $F_P$ is large. These eigenfunctions are usually localized in a field period away from the usual one $-\pi<\theta<pi$, as indicated. Furthermore, the eigenfunction is localized around a point other than  the midplane where $|B|$ is minumum.  Hence, many trapped particle orbits don't even make it to the peak of the eigenfunction. In addition, the eigenfunction is quite narrow, as above. For both these reasons, the average trapped particle fraction is very low (as in such cases in fig(\ref{fig:evol1}-\ref{fig:evol2})). The ion average curvature is highly destabilizing, as is apparent in fig(\ref{fig:eigtrap})b. However, any trapped particle orbit that interacts with $\phi$ averages over regions where the curvature is very stabilizing. Thus, the weak trapped particle interaction has a strongly stabilizing curvature. 

These are the examples of how the eigenfunction shapes achieve the values shown in fig(\ref{fig:evol1}-\ref{fig:evol2}). 
 
\section{Appendix D-Eigenfunction adaptations for TB cases with non-adiabatic electrons}

Since the linear behavior is found to be representative of the nonlinear behavior, we consider the latter in detail here. 

Recall that the $ITG_{ae}$ eigenfunction adapts in response to both FC solubility and energetic considerations ("striving" for positive $<\omega_{di}> $.) If the curvature is mostly but moderately stabilizing, it can satisfy both. As the curvature became very preponderantly stabilizing, the energetic considerations became "too hard" to meet as the FC bound was approached, and so $<\omega_{di}> $ dropped below 0.

The simulations with non-adiabatic electrons are qualitatively quite similar. Simulations in geometries such as the scans in fig(\ref{fig:GEOSCANcurvs}) reveal that there are broadly two specific strategies followed by the modes. In both of them, the eigenfunction substantially decouples from the trapped orbits that are highly stabilizing. This is the aspect of striving for strong energetic drive, like the $ITG_{ae}$. Then, in addition:

\begin{itemize}

\item        If there is still interaction with orbits with destabilizing curvature,   $<\omega_{d~orbit~e}>$ will be positive or close to zero. The $<f_{trap}>$ will be small relative to the physical fraction of trapped particles, but since the eigenfunction still couples to some orbits, non adiabatic response is still finite
\item     As $F_P$ is increased, the eigenfunction must adapt to maintain FC solubility - $k_{\perp} rho_i$ is increased.
\item        There is still little interaction with electron orbits with bad curvature. But now, there is also little interaction with orbits with good curvature as well. So $<f_{trap}>$ becomes much smaller, i.e, interaction with the trapped electrons is quite weak. Also, the $<\omega_{d~orbit~e}>$ can be quite stabilizing. Simulations find that, to compensate for the stabilizing electron curvature, the ion curvature becomes \emph{more} destabilizing. 

\end{itemize}

The  $<\omega_{d~orbit~e}>$ and $<f_{trap}>$ are defined and further discussed in Appendix B.

The ITB $\hat{s}$ scans in fig(\ref{fig:evol1})a-c, show a strong reduction in $D_{mix}$ as $F_P$ is increased. Notice that the average trapped electron curvature is \emph{destabilizing}(positive). Also notice that the effective trapped fraction is considerably smaller than its actual value. The simulations seem to follow the script: the eigenfunction is preferentially decoupling from the trapped electrons with stabilizing curvature.  And as for the ion curvature $<\omega_{di}> > 0$: as with the $ITG_{ae}$, it decreases as $F_P$ is increased for the cases with highly negative $\hat{s}$.

Now let us examine the pedestal-like $\hat{s}$ scans, fig(\ref{fig:evol1})d-f. At low $F_P$ the trapped electron curvature is destabilizing, but sometimes there is a mode jump to quite negative values at larger $F_P$. This is reminiscent of the $ITG_{ae}$, where the ion curvature became negative as the solubility bound was approached: the mode was challenged to \emph{simultaneously} adapt to the FC and to the energetics. But in these cases, whenever $<\omega_{d~orbit~e}>$ becomes significantly negative, two things happen: 1) $f_{trap}$ drops to very low values, i.e., the coupling to trapped electrons becomes small, which minimizes the energetic impact upon the mode of stabilizing electron curvature, and  2) the ion curvature compensates by increasing to much higher, more destabilizing values. Both 1) and 2) are easy to interpret as eigenfunction adaptation to stay as unstable as possible, even though it cannot maintain a positive electron curvature. 

Notice also that the effect of $k_{\perp} \rho_i$ variation follows a path similar to the $ITG_{ae}$ cases, except for ITB -A. Apparently, for  ITB -A, there was a sufficient non-adiabatic electron response so that the FC could be satisfied without increasing $k_{\perp} \rho_i$. But for all other curvatures, the non-adiabatic electron contribution to the FC was small enough that the mode was "compelled" to roughly followed the $ITG_{ae}$ evolution of $k_{\perp} \rho_i$. 

In short, as the curvature becomes more stabilizing (more negative $\hat{s}$ or larger $\alpha$), the mode (with more complete dynamics) evolves very similarly to an $ITG_{ae}$.

\emph{As with the $ITG_{ae}$, the eigenfunction evolution does not follow the path of energetic stabilization.} Energetic stabilization would mean that \emph{both} the trapped electron curvature is highly negative \emph{and} the coupling to the trapped electrons is strong \emph{and} the ion curvature is not extremely highly destabilizing. Such an evolution would result in stabilization roughly independent of  $F_P$, and that was not found in the simulations.  

The scans with increasing $\alpha$ are qualitatively similar to the $\hat{s}$ scans. The equilibrium curvatures in fig(\ref{fig:GEOSCANcurvs}) reach considerably more negative values than in the $\hat{s}$ scans, but nonetheless, the same basic behavior holds. 

\emph{The overall behavior is qualitatively quite similar to the $ITG_{ae}$: the mode structure adapts to avoid stabilization from curvature, but then it is still subject to stabilization by the FC.}

\begin{figure*}
\subfloat[\label{sfig:4a}]{%
  \includegraphics[width=.16\linewidth]{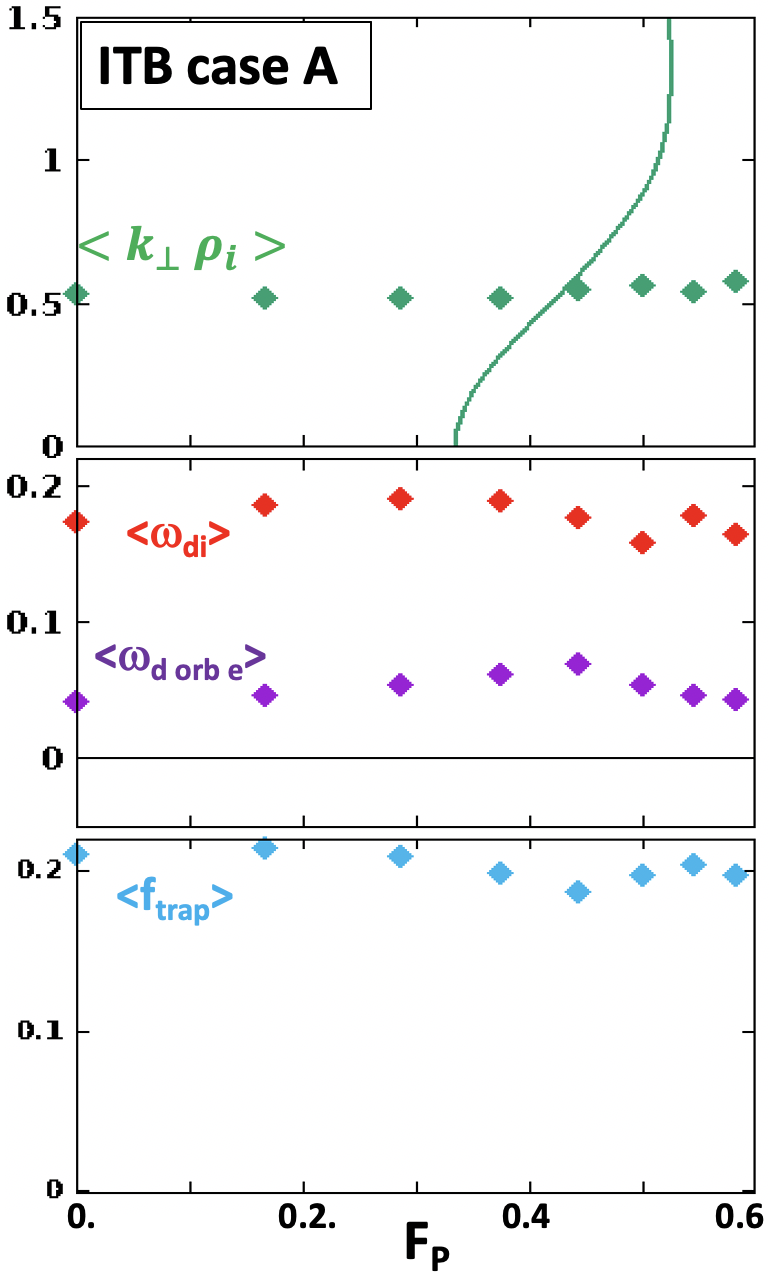}%
}\hfill
\subfloat[\label{sfig:4b}]{%
  \includegraphics[width=.16\linewidth]{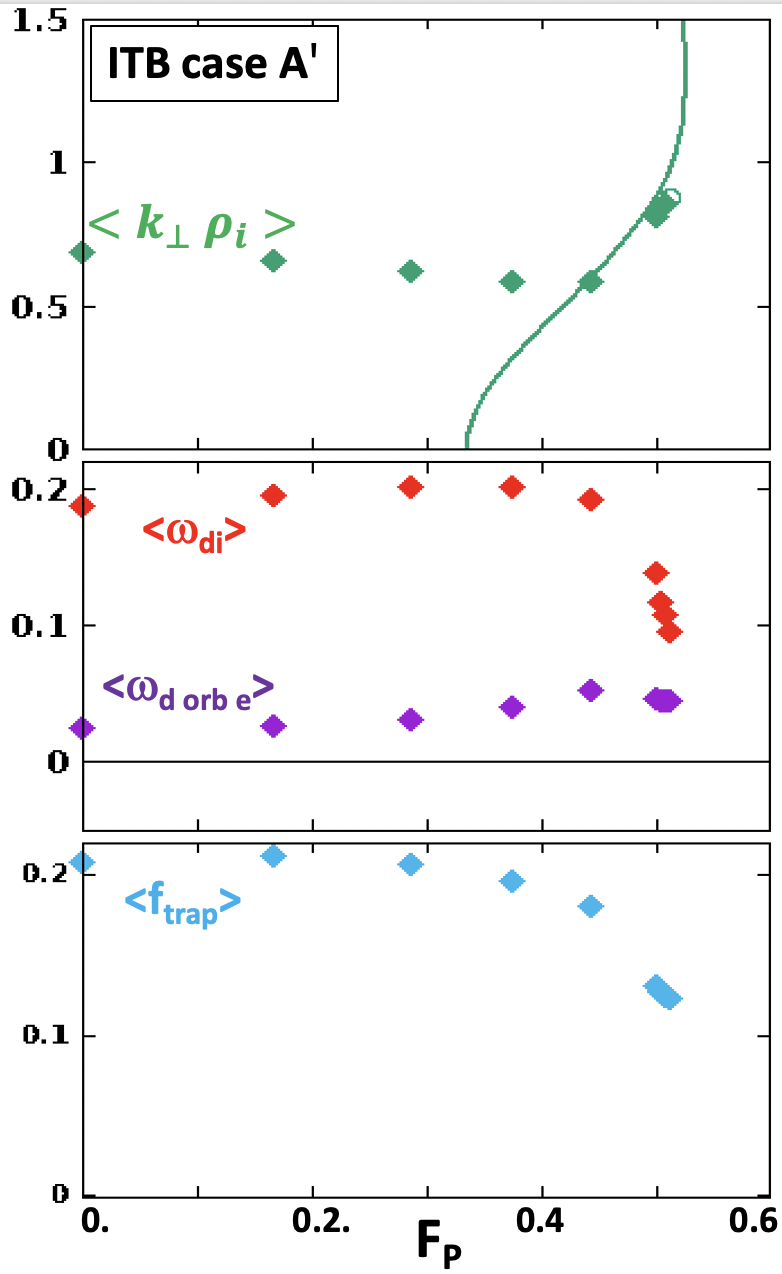}%
}
\hfill
\subfloat[\label{sfig:4c}]{%
  \includegraphics[width=.16\linewidth]{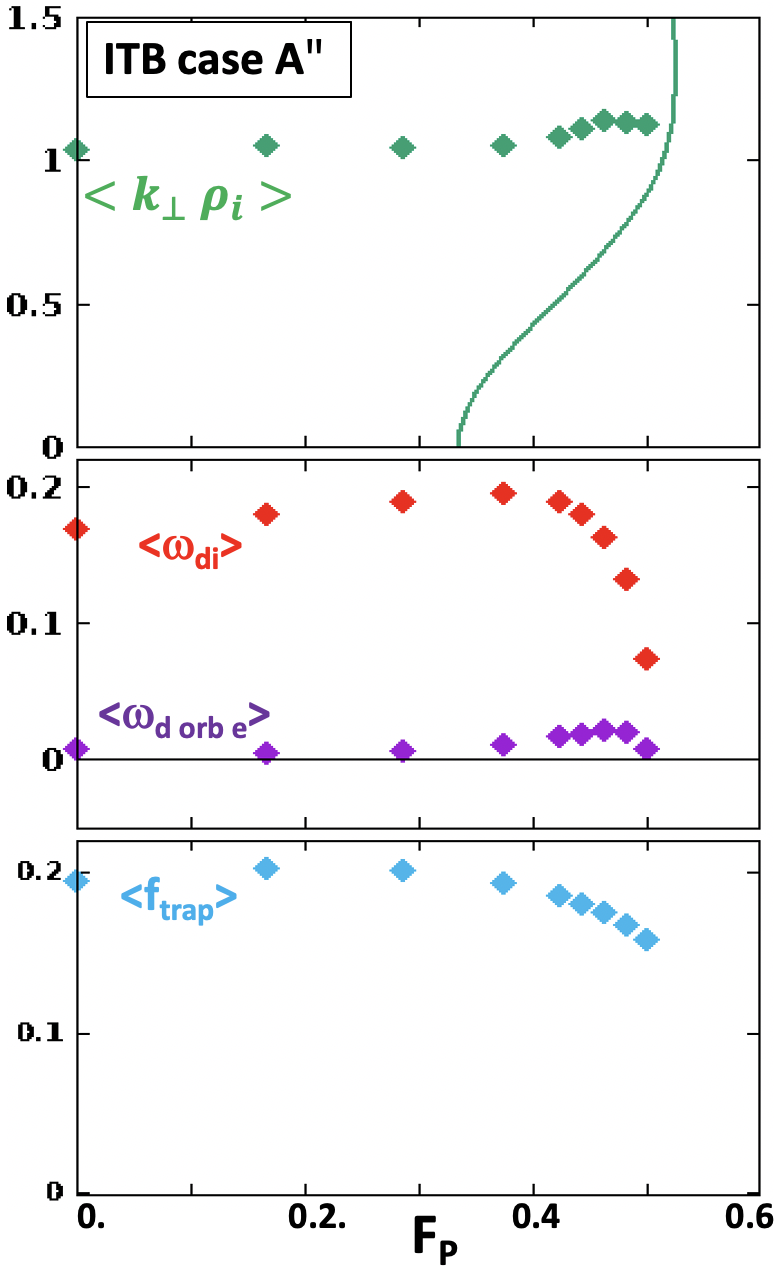}%
}\hfill
\subfloat[\label{sfig:4c}]{%
  \includegraphics[width=.16\linewidth]{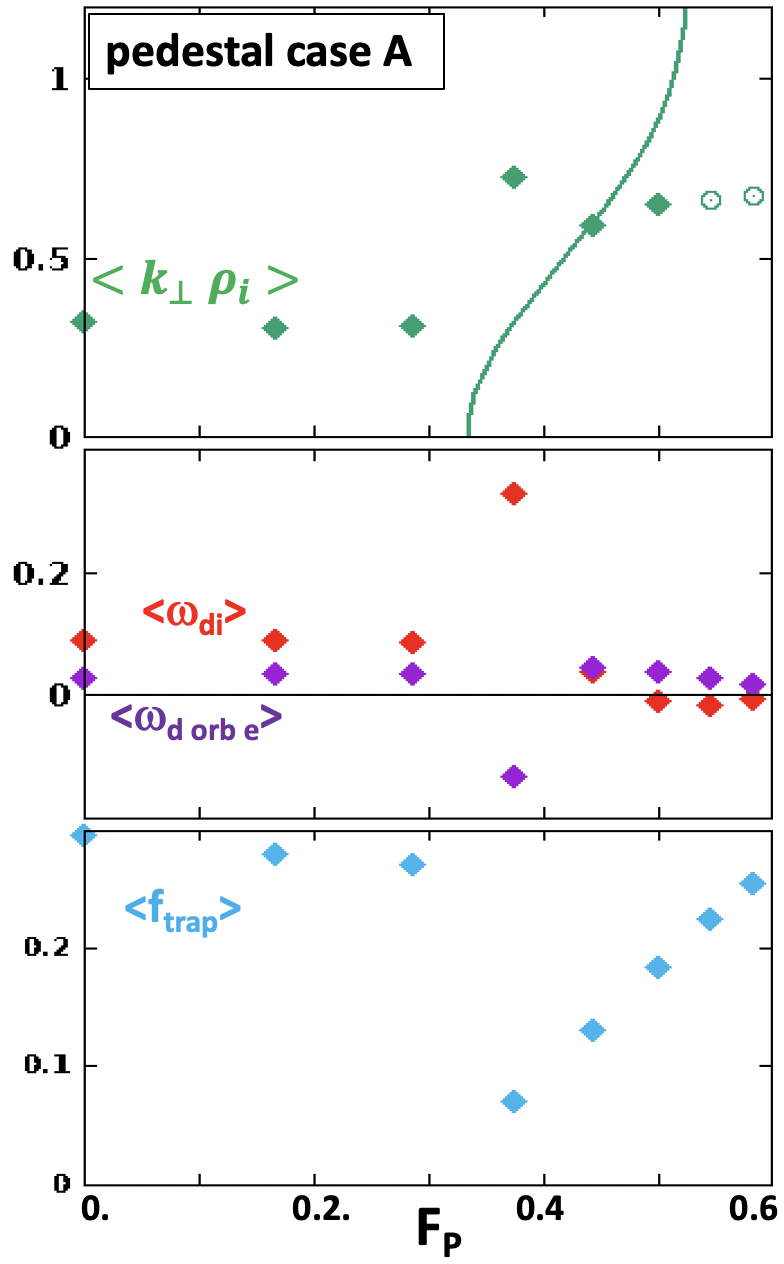}%
}\hfill
\subfloat[\label{sfig:4c}]{%
  \includegraphics[width=.16\linewidth]{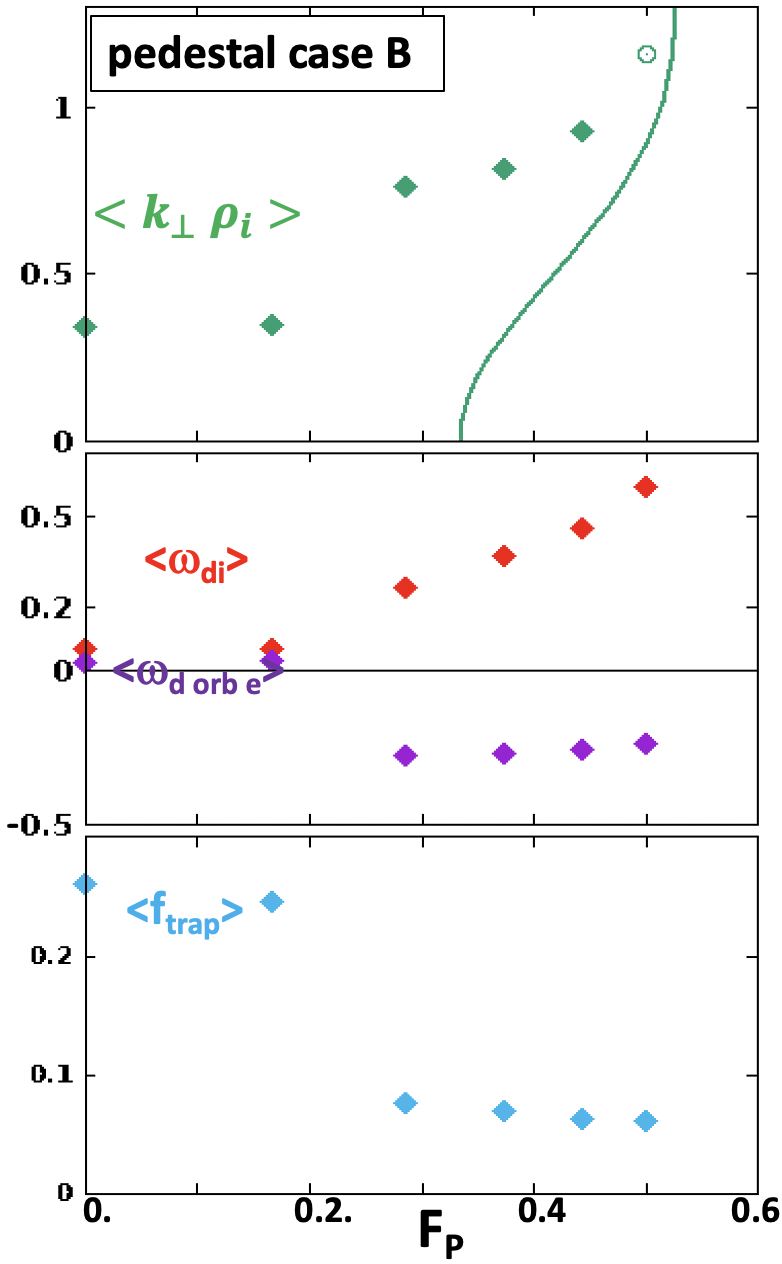}%
}\hfill
\subfloat[\label{sfig:4c}]{%
  \includegraphics[width=.16\linewidth]{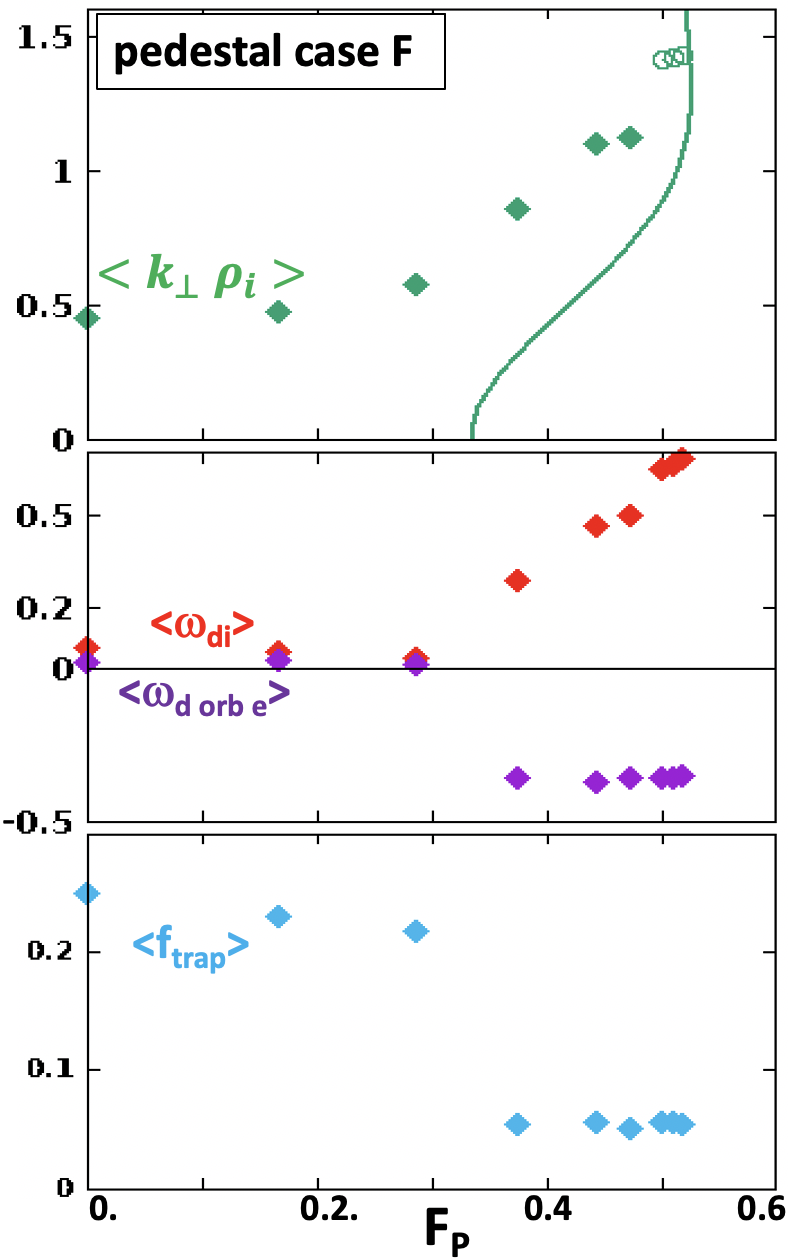}%
}
\caption{\label{fig:evol1} Evolution of eigenfunction parameters as $F_P$ increases for magnetic shear scans in fig(\ref{fig:GEOSCAN}) and fig(\ref{fig:GEOSCANcurvs}) }
\end{figure*}

\begin{figure*}
\subfloat[\label{sfig:4a}]{%
  \includegraphics[width=.16\linewidth]{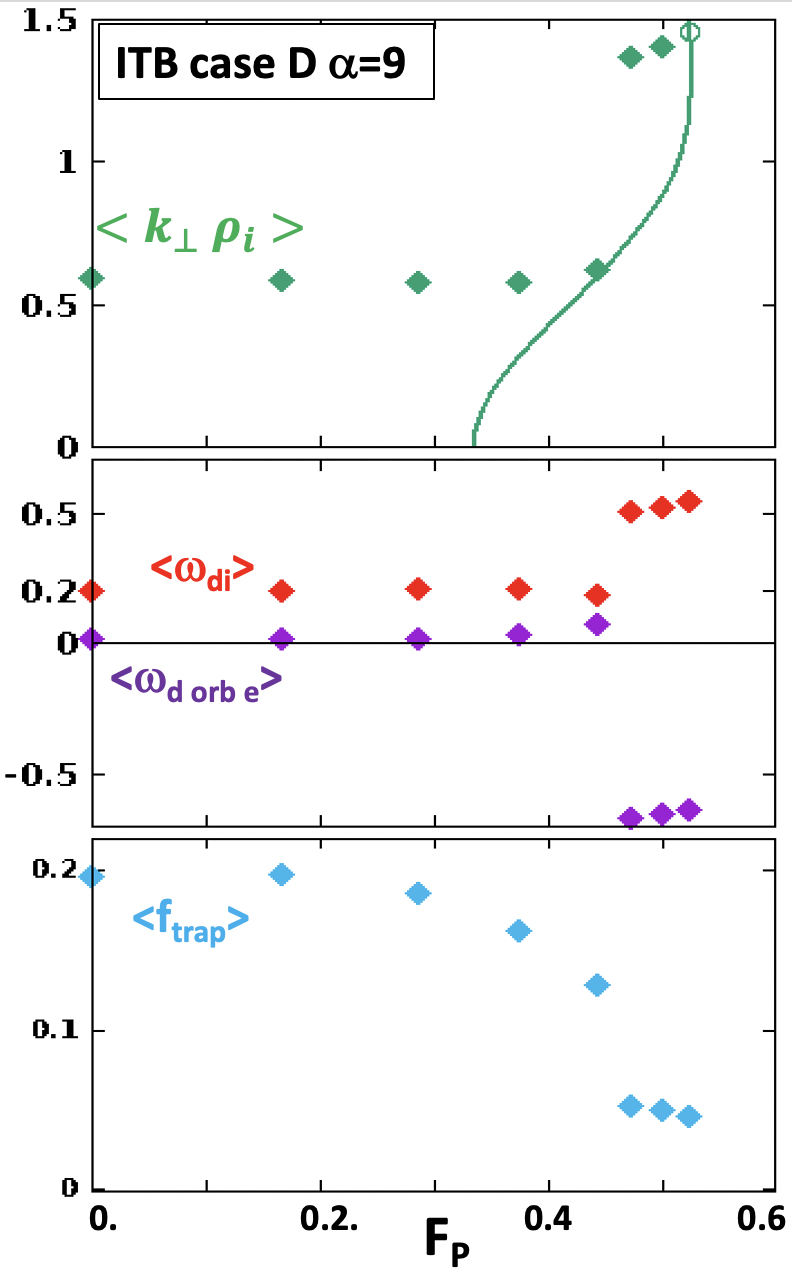}%
}\hfill
\subfloat[\label{sfig:4b}]{%
  \includegraphics[width=.16\linewidth]{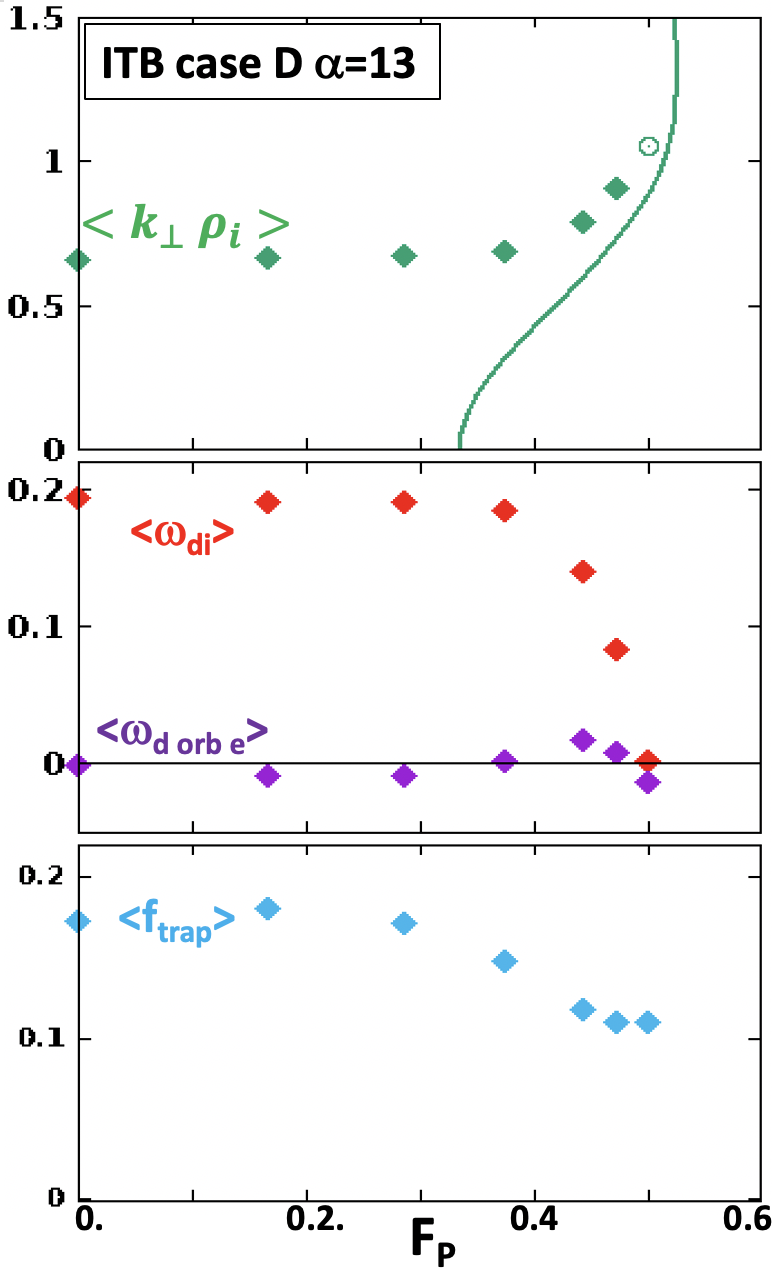}%
}
\hfill
\subfloat[\label{sfig:4c}]{%
  \includegraphics[width=.16\linewidth]{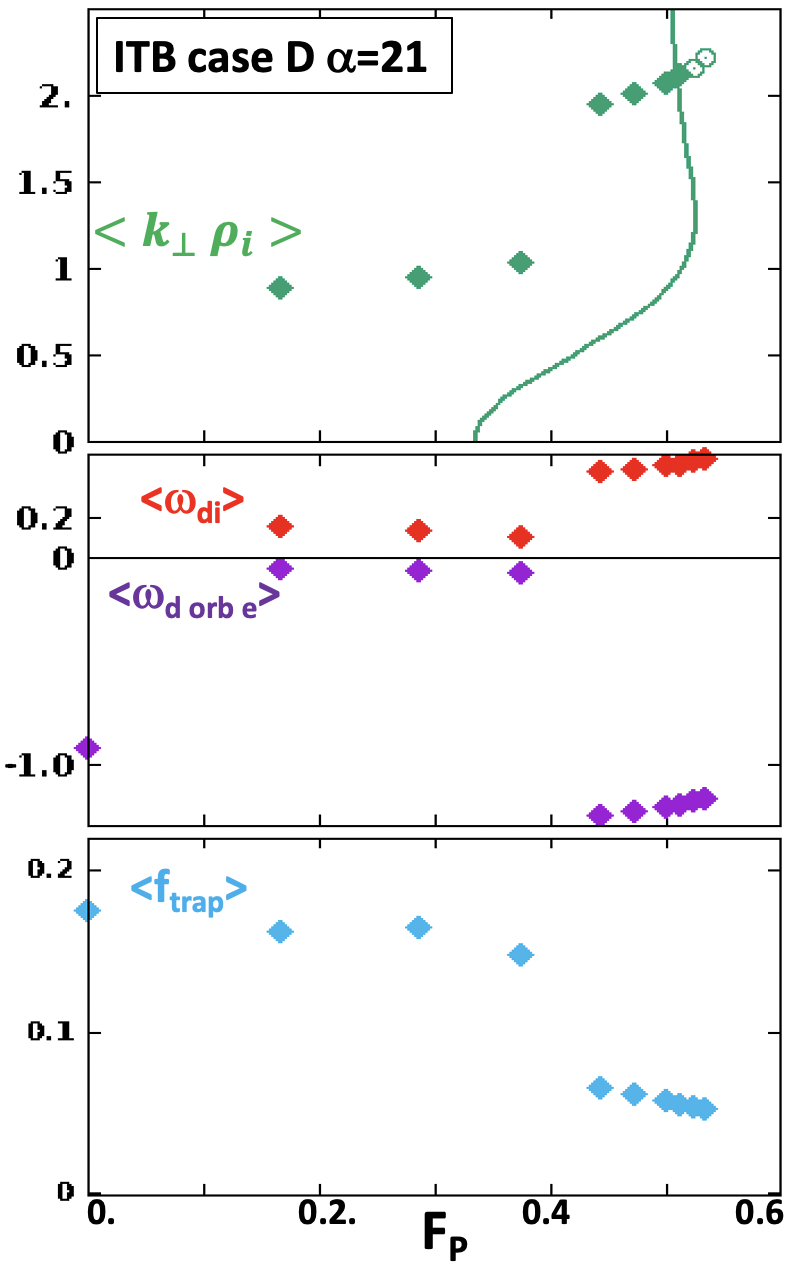}%
}\hfill
\subfloat[\label{sfig:4c}]{%
  \includegraphics[width=.16\linewidth]{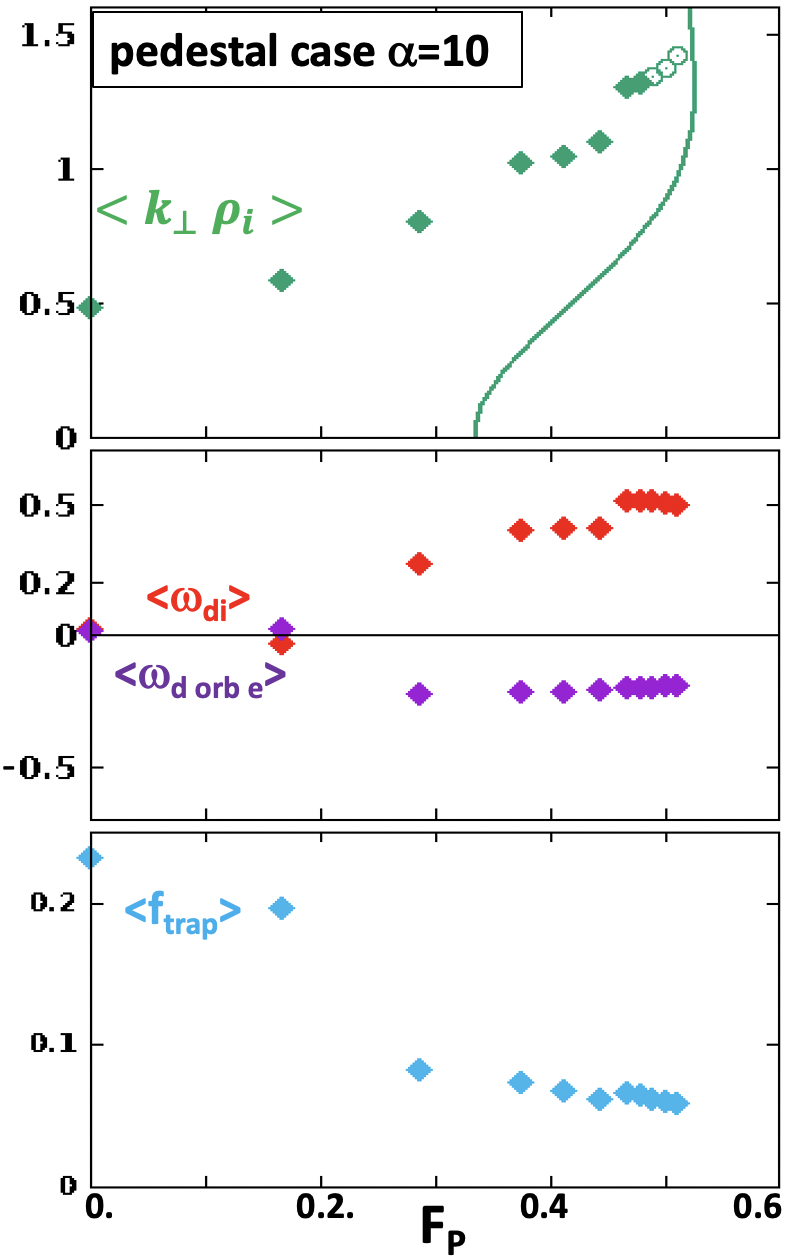}%
}\hfill
\subfloat[\label{sfig:4c}]{%
  \includegraphics[width=.16\linewidth]{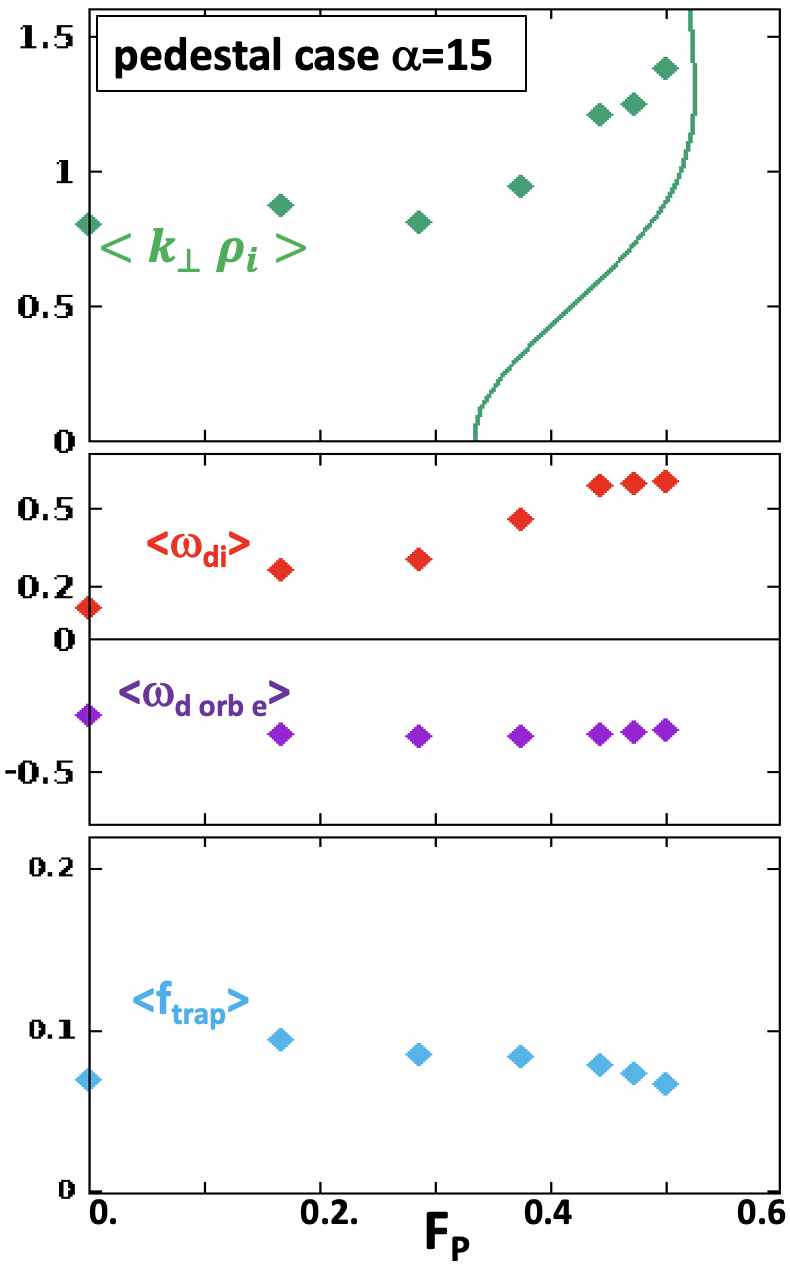}%
}\hfill
\subfloat[\label{sfig:4c}]{%
  \includegraphics[width=.16\linewidth]{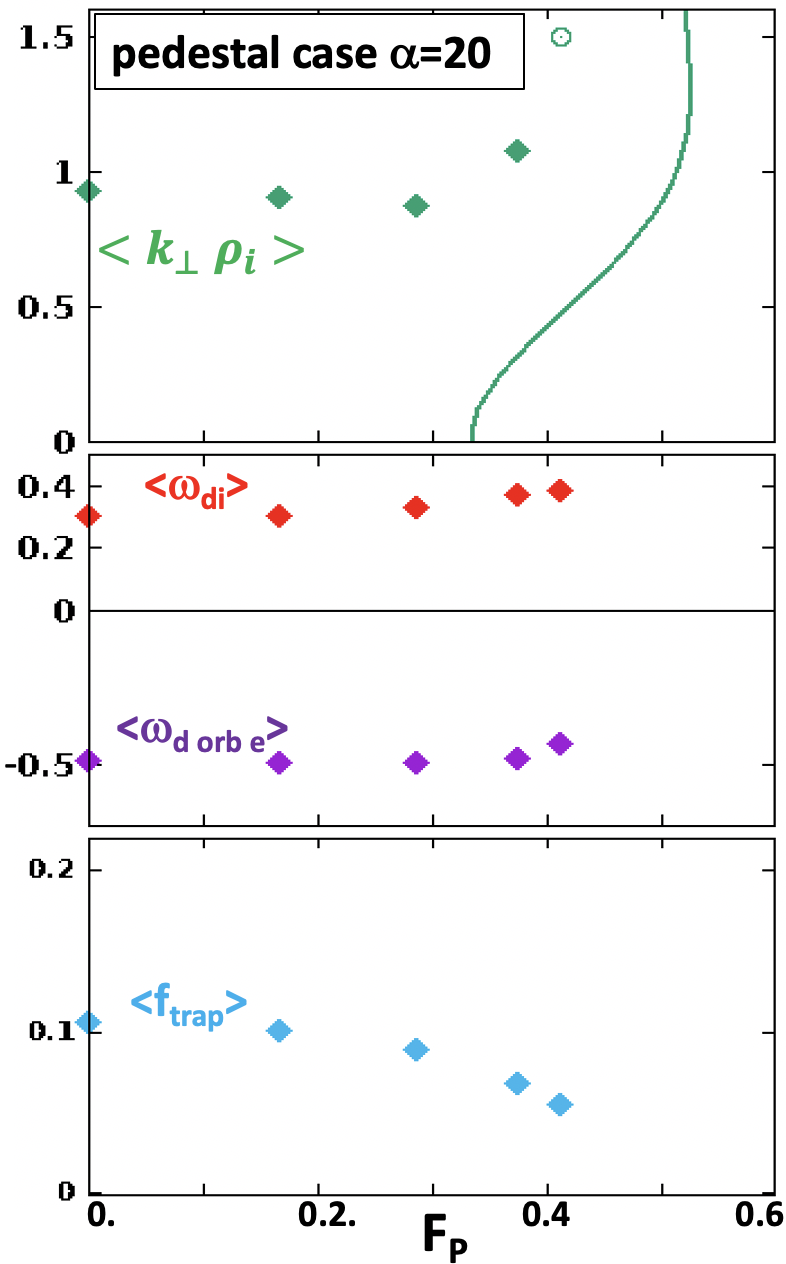}%
}
\caption{\label{fig:evol2} evolution of eigenfunction parameters as $F_P$ increases for $\alpha$ scans in fig(\ref{fig:GEOSCAN}) and fig(\ref{fig:GEOSCANcurvs})}
\end{figure*}

\begin{figure*}
\subfloat[\label{sfig:7a}]{%
  \includegraphics[width=.5\linewidth]{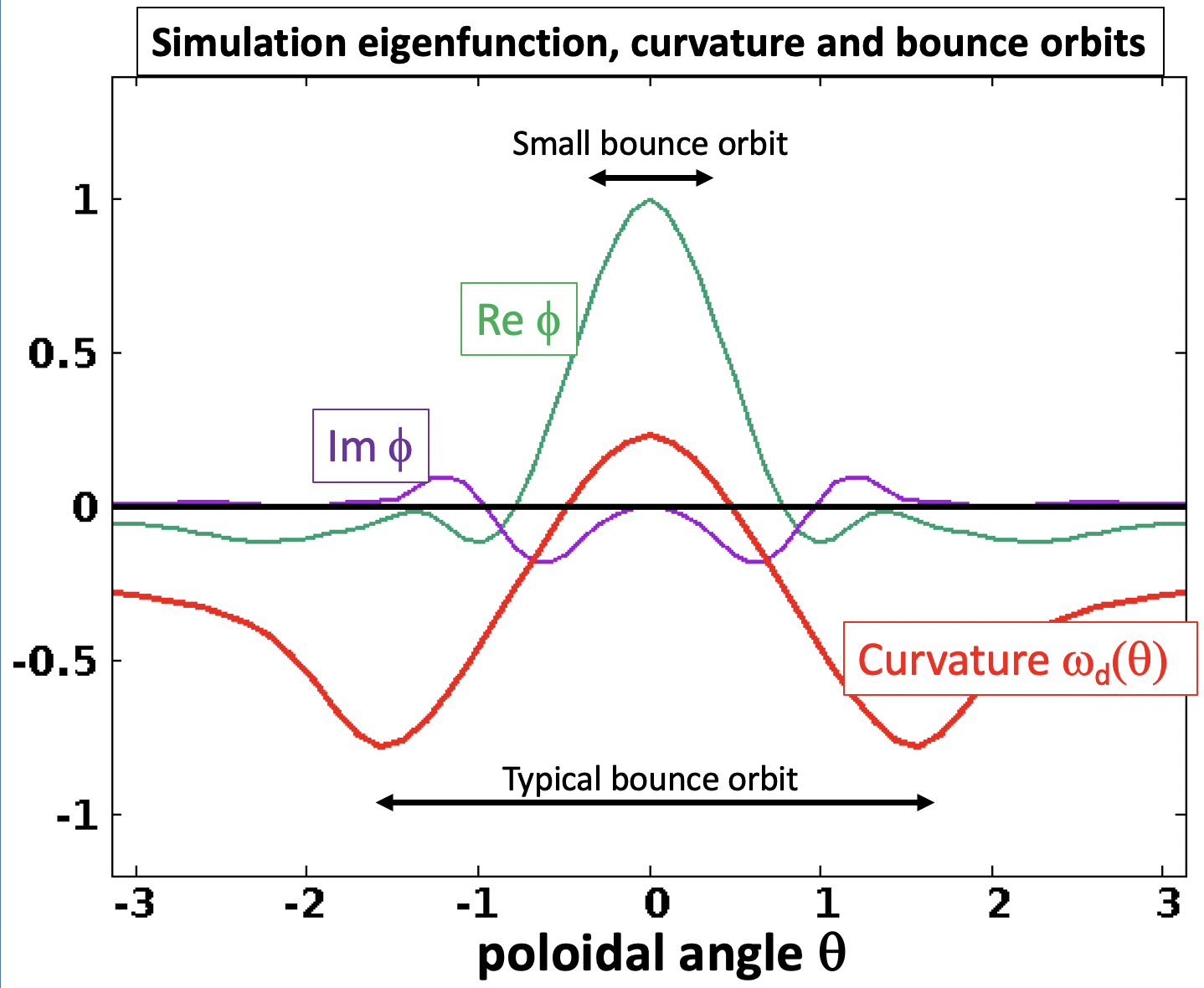}%
}\hfill
\subfloat[\label{sfig:7b}]{%
  \includegraphics[width=.5\linewidth]{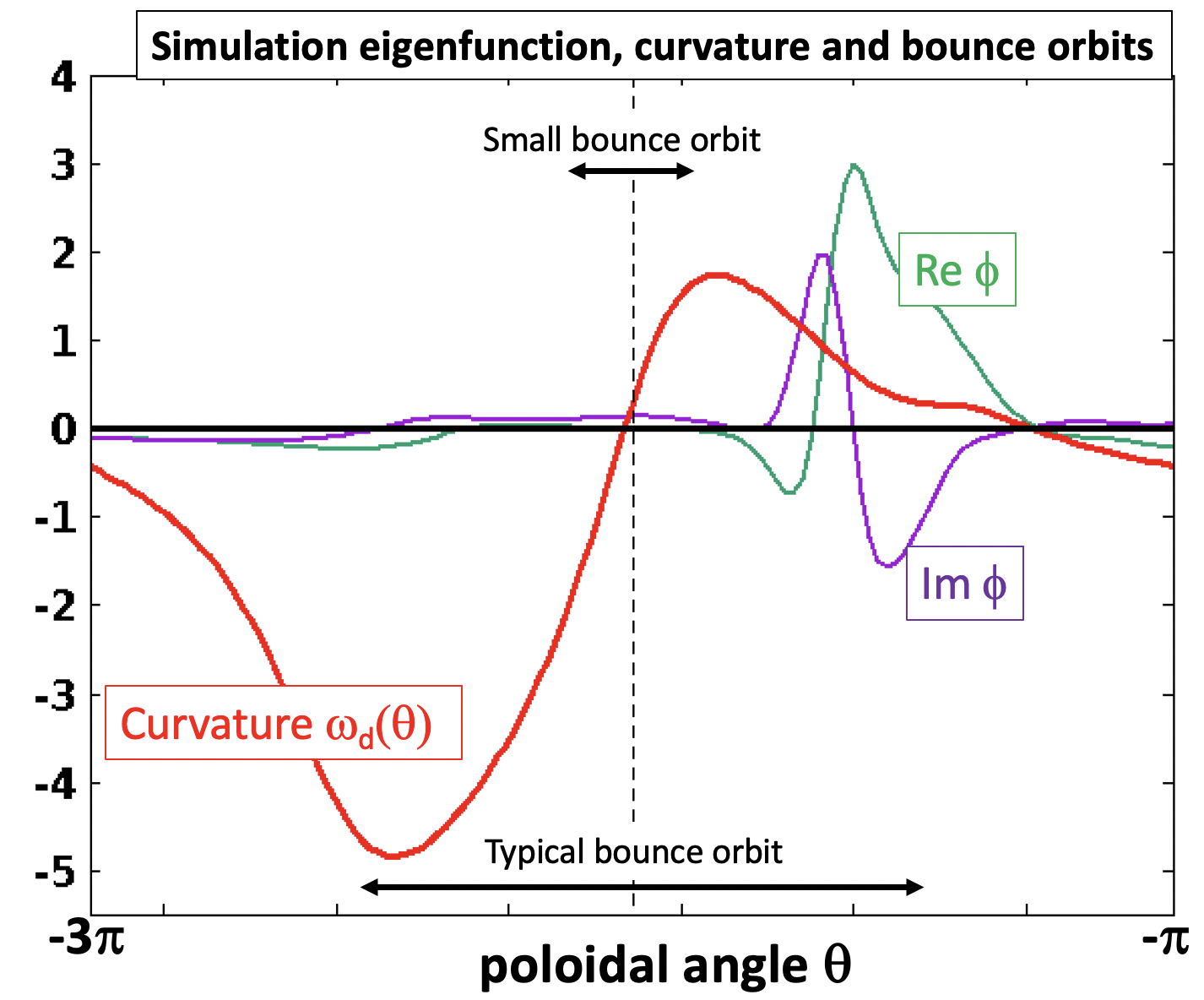}%
}

\caption{\label{fig:eigtrap} eigenfunctions and representative trapped particle orbits. One can see how the bounce average of $\phi$ tends to vanish for particles with large bounce orbit widths, via averaging over large regions where the eigenfunction is nearly zero in case a), or by averaging over regions where $\phi$ alternates signs in b).}
\end{figure*}

\begin{acknowledgments}
We gratefully acknowledge important discussions with R. D. Hazeltine. 
This research is supported by US DOE grants. DE-FG02-04ER54742 and 
DE-AC02-09CH11466, and using the DIII-D National Fusion Facility, a 
DOE Office of Science user facility, under Award DE-FC02-04ER54698.
\end{acknowledgments}

\nocite{*}

%\bibliography{apssamp_save}% Produces the bibliography via BibTeX.
\bibliography{apssamp_save-KSMDRH.bib}

\end{document}